\RequirePackage{snapshot}

\documentclass[11pt]{article}
\usepackage{amsmath}
\usepackage{amssymb}
\usepackage[utf8]{inputenc}
\usepackage{authblk}
\usepackage[square,sort,comma]{natbib}
\usepackage{tikz}
\usepackage{tikz-3dplot}
\usepackage{subcaption}
\usepackage[margin=1in,includefoot]{geometry}
\usepackage[font=small,skip=0pt]{caption}
\usepackage{algorithm}
\usepackage{algpseudocode}
\usepackage{refstyle}
\usepackage{hyperref}
\usepackage{color}
\usepackage[title]{appendix}
\usepackage{etoolbox}
\usepackage{lineno}
\usepackage{setspace}
\patchcmd{\appendices}{\quad}{: }{}{}

\begin{document}

\title{Probabilistic model-error assessment of deep learning proxies: an application to real-time inversion of borehole electromagnetic measurements}

\author[1]{Muzammil Hussain Rammay}
\author[2]{Sergey Alyaev}
\author[3]{Ahmed H. Elsheikh}

\affil[1]{University of Stavanger, Stavanger, Norway}
\affil[2]{NORCE, Bergen, Norway}
\affil[3]{Heriot--Watt University, Edinburgh, UK}

\maketitle

\makeatletter
\def\blfootnote{\gdef\@thefnmark{}\@footnotetext}
\makeatother

\blfootnote{*This is a pre-copyedited, author-produced PDF of an article accepted for publication in Geophysical Journal International following peer review. 
The version of record
"Muzammil Hussain Rammay, Sergey Alyaev, Ahmed H Elsheikh,
Probabilistic model-error assessment of deep learning proxies: an application to real-time inversion of borehole electromagnetic measurements, 
Geophysical Journal International, Volume 230, Issue 3, September 2022, Pages 1800–1817"
is available online at:  \url{https://doi.org/10.1093/gji/ggac147}.\\
Corresponding author: Muzammil Hussain Rammay, muzammil.h.rammay@uis.no
}
\begin{abstract}

The advent of fast sensing technologies allow for real-time model updates in many applications where the model parameters are uncertain. Once the observations are collected, Bayesian algorithms offer a pathway for real-time inversion (a.k.a.~model parameters/inputs update) because of the flexibility of the Bayesian framework against non-uniqueness and uncertainties. However, Bayesian algorithms rely on the repeated evaluation of the computational models and deep learning based proxies can be useful to address this computational bottleneck. In this paper, we study the effects of the approximate nature of the deep learned models and associated model errors {\color{black} during the inversion of borehole electromagnetic measurements, which are usually obtained from logging while drilling (LWD)}. We rely on the iterative ensemble smoothers as an effective algorithm for real-time inversion due to its parallel nature and relatively low computational cost. The real-time inversion of electromagnetic measurements is used to determine the subsurface geology and properties, which are critical for real-time adjustments of the well trajectory (geosteering). The use of deep neural network as a forward model allows us to perform thousands of model evaluations within seconds, which is very useful to quantify uncertainties and non-uniqueness in real-time. While significant efforts are usually made to ensure the accuracy of the deep learning models, it is widely known that the deep neural networks can contain some type of model-error in the regions not covered by the training data, which are unknown and training specific. When the deep learning models are utilized during inversion of electromagnetic measurements, the effects of the model-errors could manifest themselves as a bias in the estimated input parameters and as a consequence might result in a low-quality geosteering decision. We present numerical results highlighting the challenges associated with the inversion of electromagnetic measurements while neglecting model-error. We further demonstrate the utility of a recently proposed flexible iterative ensemble smoother in reducing the effect of model-bias by capturing the unknown model-errors, thus improving the quality of the estimated subsurface properties for geosteering operation. Moreover, we describe a procedure for identifying inversion multi-modality and propose possible solutions to alleviate it in real-time.


\end{abstract}

\newcommand{\todo}[1]{}

\section{Introduction}

Deep Neural Networks (DNNs) are becoming the go-to methods for fast approximation of complex physical systems which have been traditionally modelled by the {\color{black} Partial Differential Equation} (PDE) solvers \citep{HAGHIGHAT2021113552,kani2019reduced,kani2017drrnn}. 
For example, {\color{black} Deep Learning (DL)} has been used in an increasing number of physics-based computational problems like heat transfer \citep{TAMADDONJAHROMI2020113217}, stress concentration \citep{SAHA2021113452} and nano-fluids \citep{SHEIKHOLESLAMI20191}.  
Furthermore, DNNs showed relatively good approximation of the Maxwell's equations required for modeling extra deep electromagnetic (DeepEM) logging-while-drilling (LWD) measurements \citep{alyaev2020modeling}.
 
Fast approximations are specifically important to the fields where real-time inversion is required. 
In drilling operations, the real-time interpretation of subsurface measurements bundled with estimation of relevant subsurface uncertainties could add a significant value by adjusting the well path in real-time (known as geosteering). 
While one can also approximate the inverse operator directly by deep learning  \citep{shahriari2020deep}, recovering relevant uncertainties is non-trivial. 
In many cases these approximations fall short because the inverse problem could be ill-posed or multi-modal.
\citet{shahriari2020error} tried to address the non-uniqueness issue by adding a regularizing term to the training loss function. However, such a regularization term often hides the fact that other solutions of the inverse problem may coexist \citep{shahriari2020error}. {\color{black} These solutions could be representative of the real system and important for real-time geosteering decisions. \citet{shahriari2021design} avoided the regularization and employed a two-step loss function to minimize the effect of non-uniqueness and multi-modality in the deterministic setting. } 

In that context, Bayesian algorithms can be useful for real-time inversion because of the flexibility to account for non-uniqueness and uncertainties. Among those Bayesian algorithms, iterative ensemble filters/smoothers can be the best choice for real-time inversion due to the relatively low computational cost and the parallelizable nature of the algorithm \citep{Chen2015spe,nazainin2021SPWLA}.

While significant efforts are usually made to ensure the accuracy of the DNN forward models, 
it is widely known that the DNNs could contain model-errors in the regions not covered by the training data.
These model-errors are unknown and in general vary with the forward problem, the selected DNN architecture, training setting, and amount of training data. 
Despite that, previous attempts at inverting deep-learning models did not address the approximating nature of the DNN models \citep{wang2020deep,nazainin2021SPWLA}.
During real-time inversion of electromagnetic (EM) measurements, the neglect of model-errors could produce bias in the estimated input parameters of the DNN model and result in sub-optimal decisions. {\color{black} Recently different types of methodologies and algorithms have emerged in the published literature for accounting for model-error during Bayesian inversion. These include; (1) utilization of known prior model-error statistics computed from pairs of high-fidelity and low-fidelity models \citep{rammay2019quantification,rammay2020robust}, (2) addressing complex error statistics using an orthogonal basis generated from pairs of high-fidelity and low-fidelity models \citep{Corinna2019,KOPKE2017}, (3) estimation of unknown model-error from the residual (data mismatch) during the inversion \citep{Oliver2018,rammay2020flexible}.}
To that end, adaptation of flexible iterative ensemble smoother (FlexIES) that accounts for unknown model-errors \citep{rammay2020flexible} can be a promising approach to account for the approximate nature of the DNN models.

In the numerical sections, we present results for real-time probabilistic geophysical inversion of Extra-Deep EM measurements transmitted during drilling using iterative ensemble smoothers. 
The measurements are simulated using a high-fidelity model provided by the vendor.
However, for the inversion we use an imperfect pre-trained DNN forward model where we focus on model-error as one of the main challenges associated with the inversion of EM measurements.
To account for the effects of the model-error 
during the real-time inversion, we use the FlexIES.
We compare the results of FlexIES with the classical ensemble smoother and show that the combination of the two can detect multi-modality or local modes (minima) of the inverse problem.

The outline of this paper is as follows. In Section~(\ref{sec:formulation}), the problem description of real-time inversion of borehole electromagnetic (EM) measurements is discussed. Following that, the methodology of real-time inversion using iterative ensemble smoothers is explained in Section~(\ref{sec:methodology}). The real-time inversion of EM measurements is evaluated and assessed on two test cases, which are described with results in Section~(\ref{sec:test_cases}). The conclusions of this paper are listed in Section~(\ref{sec:conclusions}).

\section{Problem description} \label{sec:formulation}

Accurate knowledge of the subsurface properties is needed for performing optimal geosteering. 
Inferring the subsurface properties relies on many data sources including LWD, 
which are recorded from instruments located in the bottom-hole drilling assembly and are transmitted in real time during an operation.
The DeepEM logs have the most sensitivity and can provide the most reliable information about the subsurface geology. Therefore DeepEM logs are used to quantify the geology (e.g.~layer positions) and resistivities in the subsurface environment. Other petrophysical properties could also be implicitly inferred \citep{luo2015ensemble,nazainin2021SPWLA}.  
Mathematically, DeepEM signals can be represented as a vector function of the subsurface properties and can be written as follows:

\begin{equation}
		\mathbf{EM} = F\mathbf{(S,t)},
		\label{eq:physics}
\end{equation}
where $F$ is the function which maps subsurface properties $\mathbf{S}$ to electromagnetic $\mathbf{EM}$ signals for a given well trajectory $\mathbf{t}$. 
The components of $F$ can be understood as the magnetic component of the solution to Maxwell's equations evaluated at the points where the receivers are placed.
The equations  are solved for all the transmitter locations and frequencies.
We note that {\color{black} Maxwell's} equations also constitute a physics-based approximation of  $F$ and that Eq.~(\ref{eq:physics}) should be understood as the actual physical response and not a modeled one.

The reliable assessment of uncertainties and non-uniqueness during real-time inversion require thousands of function evaluations of the forward model within no more than a minute. 
In these situations, the use of PDE solvers for forward modeling may not be a good choice because of the computational cost.
Recently, it was shown that the DNN models could provide a good approximation of the electromagnetic geophysical logs \citep{Shahriari_deep_forward,alyaev2020modeling}.

\begin{figure}
\centering
	\includegraphics[width=0.4\textwidth]{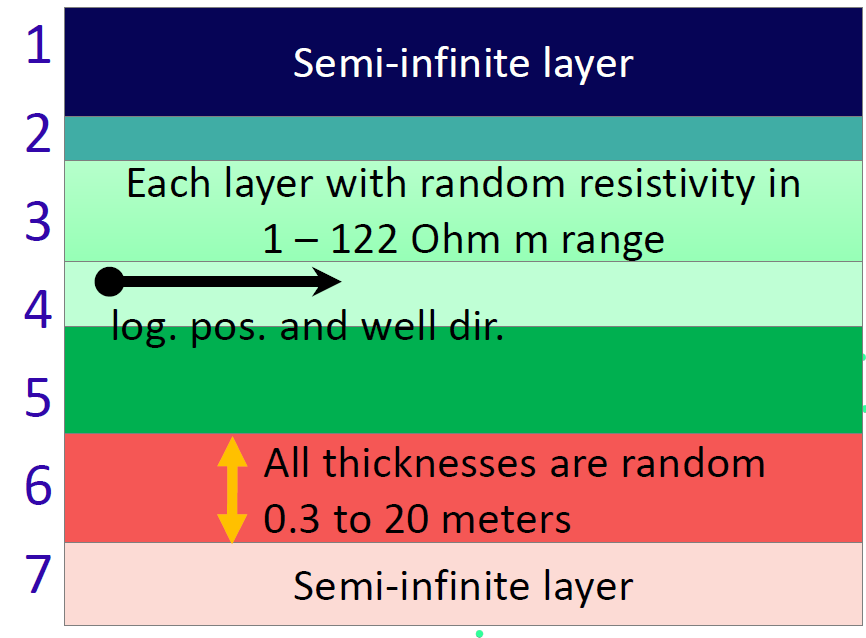}
\caption{Schematic diagram of inputs for DNN. Dot with arrow represents the measurement position. \citep{alyaev2020modeling}}
\label{dnn_input}
\end{figure}

\begin{figure}
    \centering
    \includegraphics[width=1\textwidth]{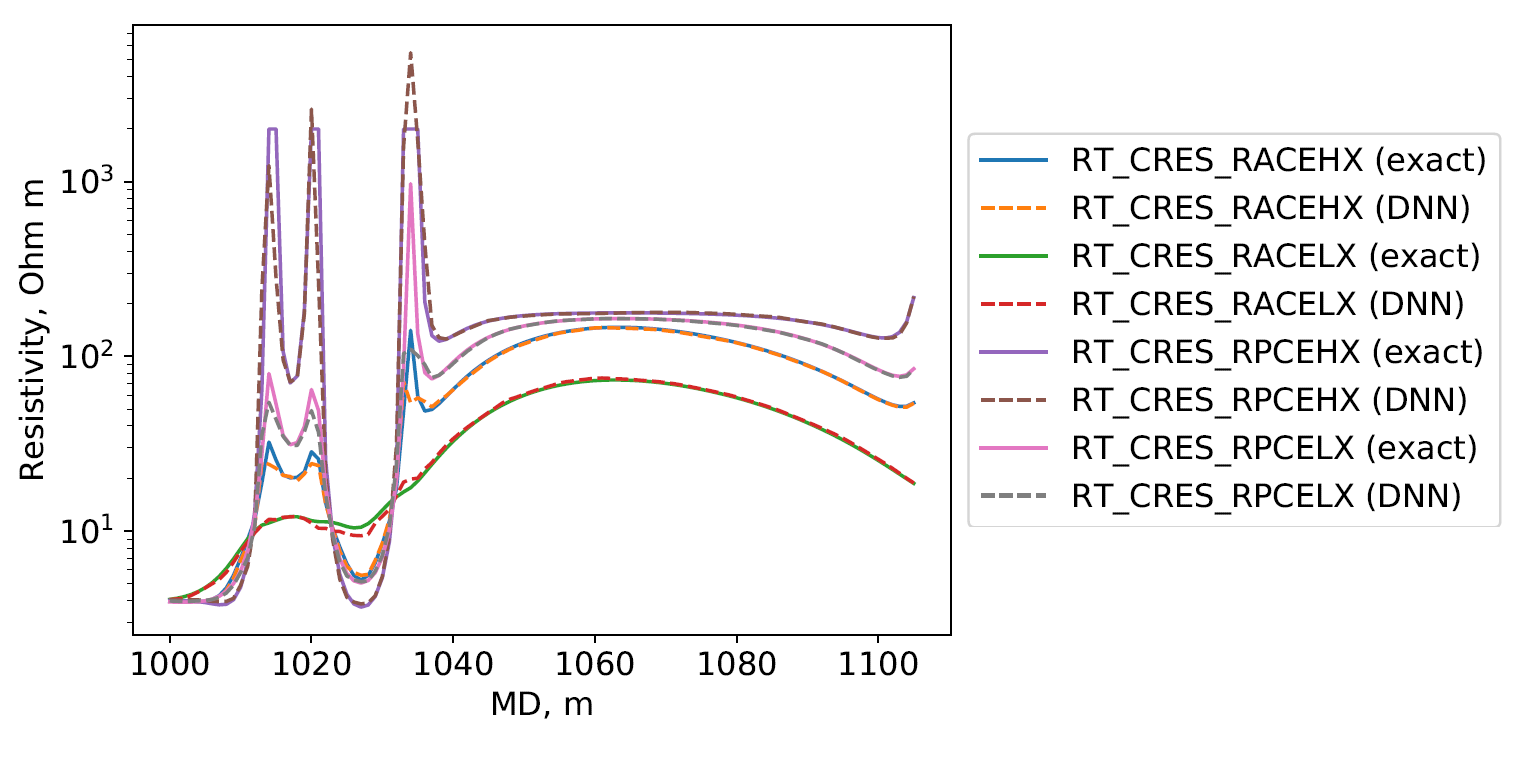}
    (a)
    
    \includegraphics[width=1\textwidth]{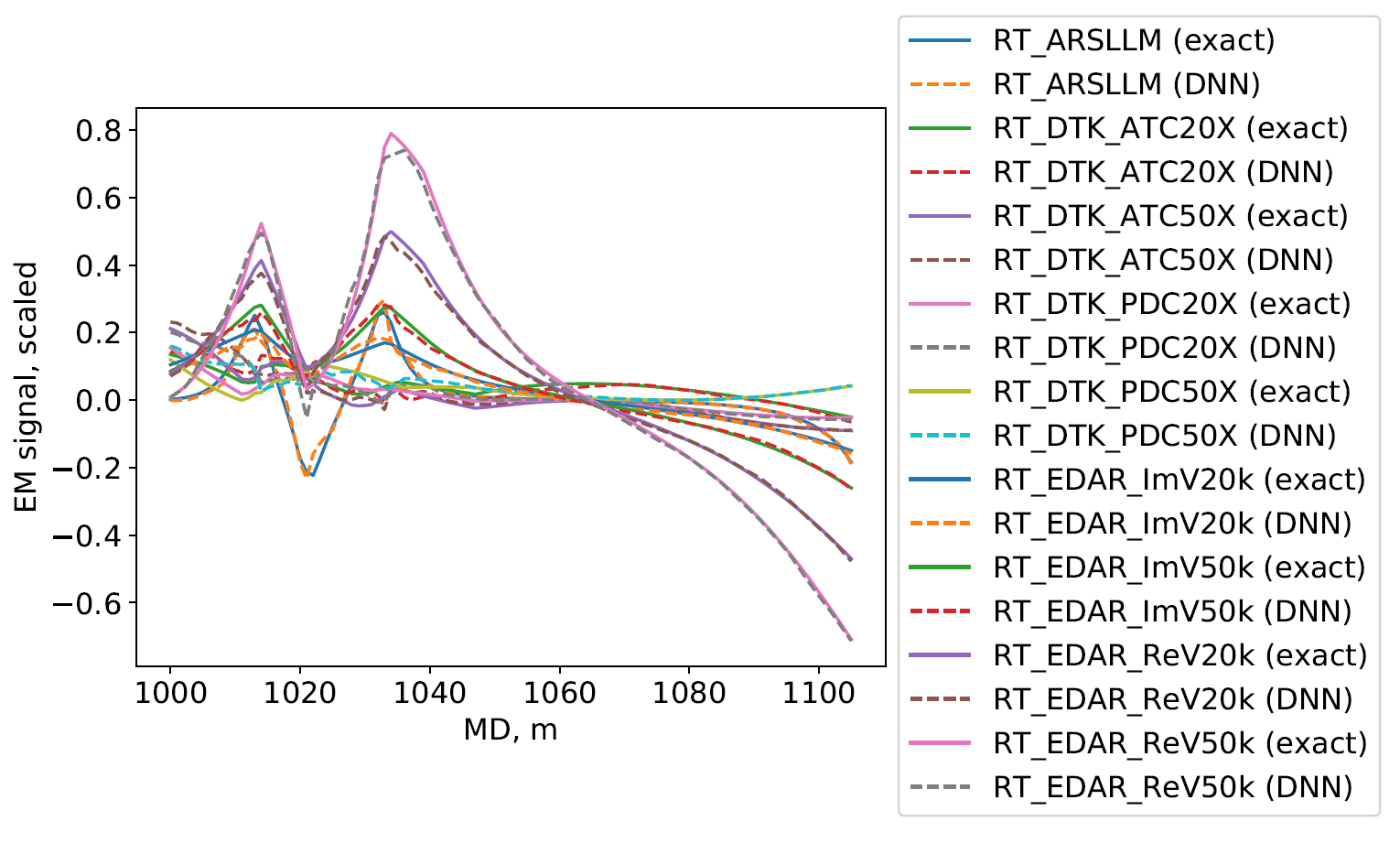}
    (b)

\caption{An example of EM measurements and their approximations using the DNN model from Eq.~(\ref{dnn_eq1}) as a function of the position within a well. (a) Four 'shallow' measurements converted to apparent resistsivities. (b) Nine 'deep' measurements scaled to non-dimensional units.}
    \label{fig:logs}
\end{figure}

In this work, a DNN is utilized as the forward model during the real-time probabilistic inversion. 
We use the trained DNN described in \citet{alyaev2020modeling}.
The DNN model is trained on the dataset containing around seventy thousand samples, which are created using a high-fidelity physics-based simulator provided by the tool vendor \citep{simulator2014}.
The input for each sample is a 1D-layered environment as shown in Figure~(\ref{dnn_input}). 
The high-fidelity simulator is a PDE solver for Maxwell's equations specialized for 1D-layered media. 
It computes the EM responses for all deep and ultra-deep tools from layer positions, resistivities, and the well inclination. 
The DNN model uses the subsurface properties $\mathbf{S}$ and produces the approximated electromagnetic signals $\mathbf{\widetilde{EM}}$ as the output as shown in Eq.~(\ref{dnn_eq1}) for a given well trajectory $\mathbf{t}$:
\begin{equation}
		\mathbf{\widetilde{EM}} = DNN(\mathbf{S,t}).
\label{dnn_eq1}		
\end{equation}
This approximation allows us to perform thousands of function evaluations within seconds 
during the probabilistic inversion allowing for reliable assessment of uncertainties in real-time.
Figure~(\ref{dnn_input}) shows a schematic diagram of the inputs to the DNN model.  
The model takes as an input the boundaries and resistivities of three layers above and below from the hosting layer to the logging instrument (total of seven resistivities and six boundary positions). The output of the utilized DNN model are approximations of electromagnetic measurements, (see Figure~(\ref{fig:logs})). These electromagnetic measurements are used to estimate a relative position of a wellbore within the geology along with resistivity of the layers. 

In this work, we solve the following inverse problem: estimate the distribution of layers' resistivities and thicknesses given the EM measurements. We use ensemble smoothers which have only access to the approximate DNN model for the EM measurements. The objective of this work is to address the challenges associated with the model imperfection and possible simulated measurement discrepancy (i.e.~model-error) during real-time probabilistic inversion. We further discuss the possible solutions to alleviate the multi-modality or local modes of this inverse problem when observed.

\section{Methodology for real-time probabilistic inversion} \label{sec:methodology}

Bayesian inversion allows us to quantify the non-uniqueness and uncertainties in the real-time estimation of the subsurface properties. For this purpose, we perform the real-time inversion using iterative ensemble smoothers due to their computational efficiency, flexibility for highly non-linear models and parallelizable nature of the algorithms. 
In this work, we use two different variants of the iterative ensemble smoothers: classical \citep{emerick2013} and flexible \citep{rammay2020flexible}.
Through the comparison, we evaluate the performance of real-time inversion in the absence and presence of the model-error along with the identification of the multi-modality and local modes of the problem. 
The implementation of the two variants of the iterative ensemble smoothers for real-time inversion of EM measurements while utilizing DNN as a forward model are shown in the subsections (\ref{subsec:ES}) and (\ref{subsec:FES}).

\subsection{Ensemble Smoother with Multiple Data Assimilation (ESMDA)}\label{subsec:ES}

In this work, {\color{black} we use ESMDA for real-time Bayesian inversion of borehole EM measurements. The goal of the inversion is to estimate the subsurface properties while utilizing the DNN proxy as an approximated forward model.} {\color{black} ESMDA is an iterative form of ensemble smoother in which the observed data is assimilated multiple times with an inflated covariance of the measurement error $\mathbf{C}_D$. The inflation coefficient/parameter $\alpha$ is usually equal to the number of data assimilations $N_a$ \citep{emerick2013}. In ESMDA, generally four to eight iterations are used as a termination criteria for the inversion/data assimilation \citep{emerick2013, rammay2019quantification}.} The first step is related to the computation of the $N_e$ number of the prior realizations of $N_m$ number of the subsurface properties $\mathbf{S}$ from a known or assumed statistical distribution. In this work, the prior realizations $\mathbf{M}_{prior} = [\mathbf{S}_{1} \; \mathbf{S}_{2} \; \mathbf{S}_{3} \; ...... \; \mathbf{S}_{N_e}]$  are sampled from a uniform distribution. In the next step, the realizations of the outputs $\mathbf{D}$ {(\color{black} ensemble of the EM signals)} are computed by passing the prior realizations to the DNN for a given well trajectory $\mathbf{t}$, i.e.,~$\mathbf{D} = DNN(\mathbf{M, t}) \in [\mathbf{\widetilde{EM}}_{1} \; \mathbf{\widetilde{EM}}_{2} \; \mathbf{\widetilde{EM}}_{3} \; ...... \; \mathbf{\widetilde{EM}}_{N_e}]$. 
{\color{black} The prior ensemble of the subsurface properties and electromagnetic outputs are used to estimate the posterior ensemble of subsurface properties $\mathbf{M}_{post}$ and outputs $\mathbf{D}_{post}$ using the ESMDA algorithm} for the observed electromagnetic measurements $\mathbf{EM}_{obs}$ of size $N_d$:
\begin{equation}
    \mathbf{M}_{post}, \mathbf{D}_{post} \gets {ESMDA}(\mathbf{M}_{prior}, \mathbf{EM}_{obs}, \mathbf{C}_D, DNN, \mathbf{t}),
    \label{alg:A1}
\end{equation}
The schematic of the ESMDA is shown in Figure~(\ref{esmda_diag}) and the detailed algorithm is described in Appendix \ref{sec:AppendixA}.

\subsection{Flexible iterative ensemble smoother (FlexIES)}\label{subsec:FES}

ESMDA can account for the measurement errors, however it is not flexible for handling model-errors and imperfect models. Even though a DNN model is trained to give minimal errors on the training dataset, it can contain model-errors at the regions which are not covered by the training dataset. In real-time inversion, ignoring model-errors could produce a bias in the estimate of the input parameters. \citet{rammay2020flexible} showed that FlexIES can be useful to reduce the bias in input parameters and prohibit to converge to the wrong solution. {\color{black} FlexIES is similar to the ESMDA in terms of the structure and steps of the algorithm. However, in FlexIES, the model-error ensemble is estimated by splitting of residual (data mismatch) into two parts. One part is used for parameter estimation/update and the second part is used to represent the model error. The initial value of the split parameter $s_p$ is computed based on the ratio of norm of mean residual $\sigma_m$ (mean deviation from observed data) and norm of maximum residual $\sigma_{max}$ (maximum absolute deviation from observed data). During the inversion this split parameter is updated based on the ratio of norm of mean residuals obtained in current and previous iterations, which results in an updated model-error ensemble. If the split parameter approaches to 1.0, the parameter update is stopped and the remaining residual is considered as model-error. However, the termination criteria of FlexIES is similar to ESMDA. It is important to note that, if the forward model is perfect and split parameter approaches to 1.0, FlexIES sometimes fails to reduce the data mismatch.
This could be attributed to either a misspecified/misinformed prior or the inverse problem is multi-modal \citep{rammay2020flexible}.} Therefore to reduce and quantify the model-error effect along with multi-modality, we use FlexIES algorithm
\begin{equation}
    \mathbf{M}_{post}, \mathbf{D}_{post} \gets   FlexIES(\mathbf{M}_{prior}, \mathbf{EM}_{obs}, \mathbf{C}_D, DNN, \mathbf{t}).
    \label{alg:A2}
\end{equation}
The steps of the inversion of the {\color{black} EM measurements using DNN model and FlexIES} are shown in Appendix \ref{sec:AppendixA}. Practically, we are uncertain about the models' perfection or imperfection. In other words, we are uncertain about the extent of the model-error. \citep{rammay2020flexible} showed that FlexIES can be useful under these situations because it uses the data misfit/residual to account for model-error. Moveover, by design, and as illustrated in \citep{rammay2020flexible}, FlexIES can still be used for inverting perfect models  where we have no model-error. {\color{black} Generally in FlexIES, eight number of iterations can be used as a termination criteria of the inversion. However, a more robust way to verify is by checking if the split parameter approaches to one \citep{rammay2020flexible}. In the cases where the split parameter is significantly less than one, then it is recommended to increase the number of iterations/data assimilations cycles.}

\begin{figure}[H]
\begin{center}
\includegraphics[width=1\linewidth]{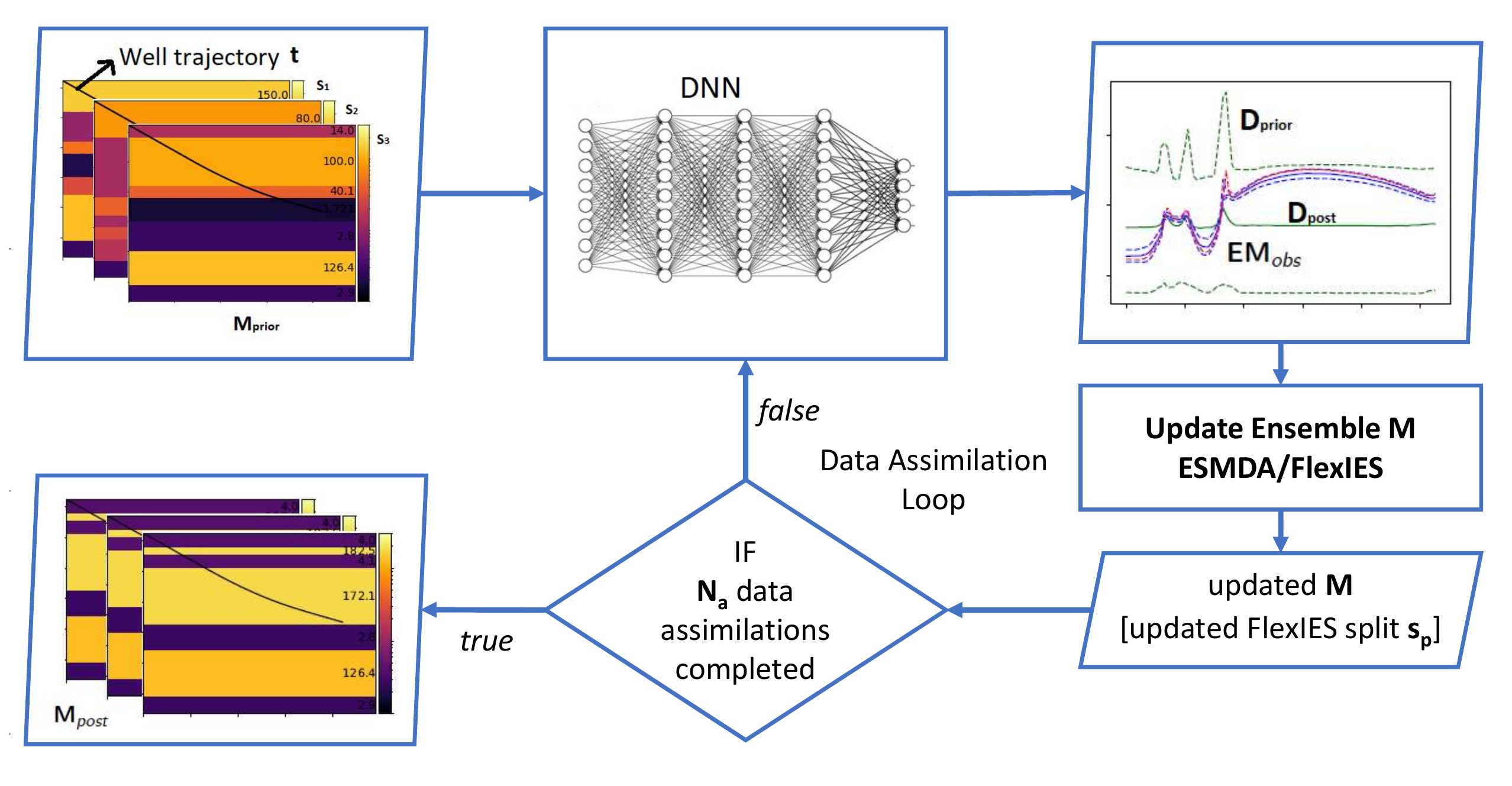}
\end{center}
\caption{Schematic diagram of ESMDA and FlexIES algorithms for real-time inversion of EM measurements by utilizing DNN as a forward model.}
\label{esmda_diag}
\end{figure}

\section{Numerical results}  \label{sec:test_cases}

We test the presented probabilistic inversion methods to estimate the resistivities and boundary locations in a layered environment using EM measurements. To simulate the EM logs during the inversion, we only have access to a pre-trained DNN which contains model-errors, as described in Section \ref{sec:formulation}. 

\begin{figure}[!ht]
\begin{center}
	\includegraphics[scale=0.5]{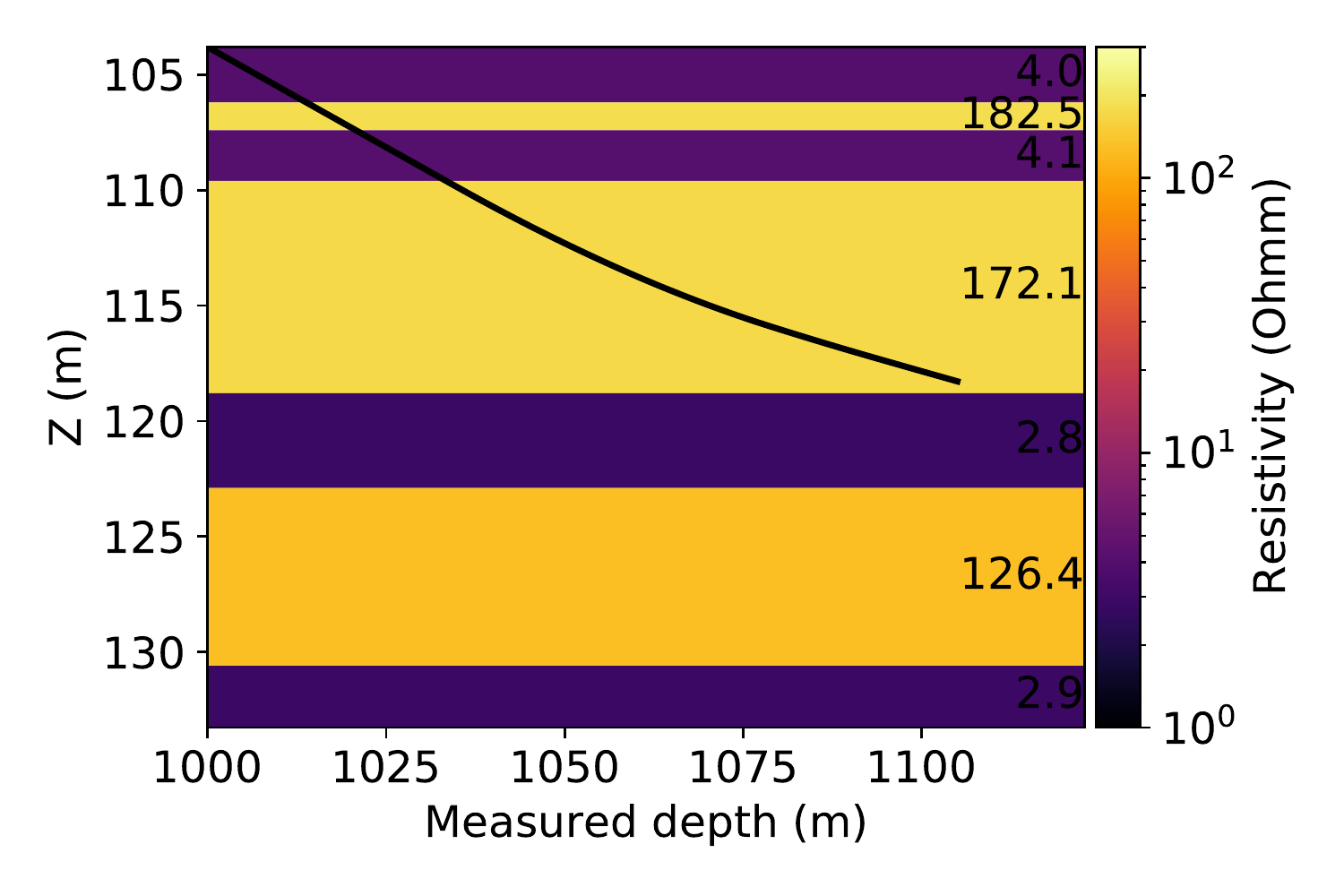}
\end{center}	
\caption{The reference model/synthetic truth of the layers resistivities and thicknesses (boundary positions) along with well trajectory in measured depth vs true vertical depth coordinates.}
\label{ref_traj}
\end{figure}

We present test cases in order of the complexity growth. In the first test case, layer resistivities are estimated by the {\color{black} inversion of EM measurements using DNN as a forward model}, while the boundary positions remain fixed. 
This test case allows us to evaluate the performance of the real-time inversion of EM measurements in the absence of pronounced multi-modality, where an ensemble method can avoid local minima \citep{nazainin2021SPWLA}. 
In the second test case, the top boundary location, layers' thicknesses, and resitivities are jointly estimated. We note that such a setup could allow the multiple correct solutions, e.g.~when two consecutive layers have same resistivity. 
From this test case we observe the performance of the {\color{black} real-time inversion of EM measurements utilizing DNN as a forward model}, in the presence of local minima or multiple local modes in the posterior. In both test cases, the observed data of the EM measurements are generated from the reference 1D-layered earth model shown in Figure~(\ref{ref_traj}) using the full-physics simulator (high-fidelity model), which mimics a section of a real operation in the Barents Sea presented in \cite{larsen2015extra}.

In order to explore the effect of model-error, the inversion is performed using both the standard ESMDA (Algo.~\ref{alg:A1}) and the FlexIES algorithms (Algo.~\ref{alg:A2}). 
{\color{black} We found that an ensemble of $1000$ members and $8$ iterations are sufficient and thus that setup is used for the inversion of all test cases.}   
The associated computational time is around 80 seconds for 1000 ensemble members and 8 iterations (i.e.~9000 function evaluation) for both algorithms (\ref{alg:A1}) and (\ref{alg:A2}), which allows the inversion to be performed in real-time geosteering. Furthermore, the inversion is performed using two procedures in both cases. 
In the first procedure, we consider that the DNN is a perfect model, i.e. the reference solution is generated from the same DNN model. 
This procedure allows us to evaluate the inversion in the absence of model-error. 
The second procedure is the realistic real-time inversion, where the observed data is obtained from the observed EM measurements, which is generated from a vendor-provided high-fidelity simulator.

\subsection{Estimation of layers resistivities}
\label{sec:case1}

In this test case, we solve a simple problem, where we only estimate layers resistivities by inverting {\color{black} the EM measurements using DNN as a forward model.}
Comparing the classical and the FlexIES, we identify the influence of model-error on the inversion.
If there is no model-error, we expect both algorithms to give the same results.
The measurement errors in the observed data are set to be negligible (i.e.~order of magnitude $10^{-9}$). 
In this case, the inversions are performed using the two procedures, i.e.,~considering DNN model as a perfect and an imperfect model. 
In both procedures, the prior realizations of the layers' {\color{black} log10} resistivities are sampled from the uniform distribution $U \sim [0, \quad 2.34]$. 

\subsubsection{DNN as a perfect model}

\begin{figure}[H]
\begin{center}

 \hspace{-0.5in}
    \begin{subfigure}[normal]{0.4\textwidth}
	\includegraphics[scale=0.5]{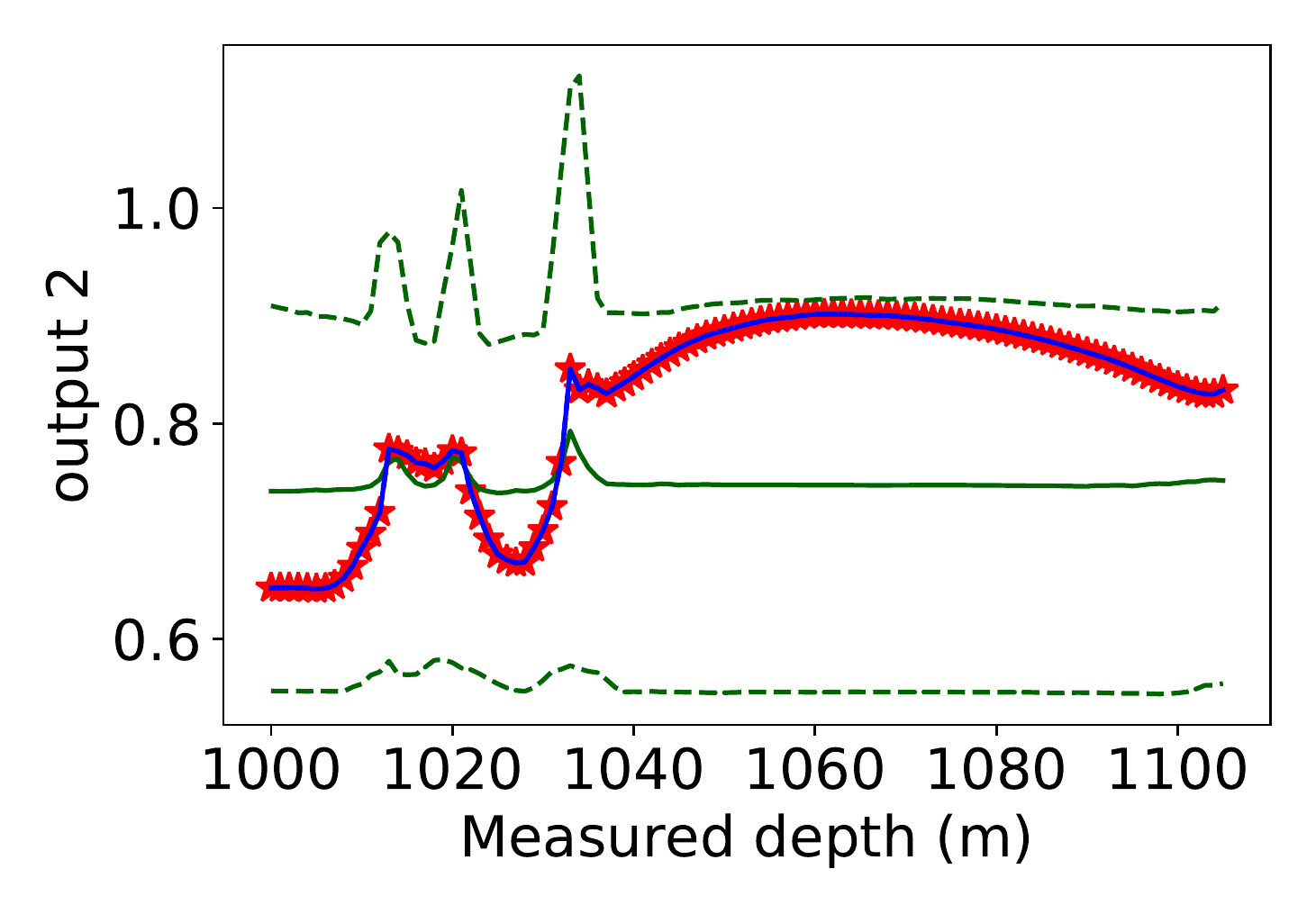}	
	\end{subfigure}
 \hspace{0.5in}
	\begin{subfigure}[normal]{0.4\textwidth}
	\includegraphics[scale=0.5]{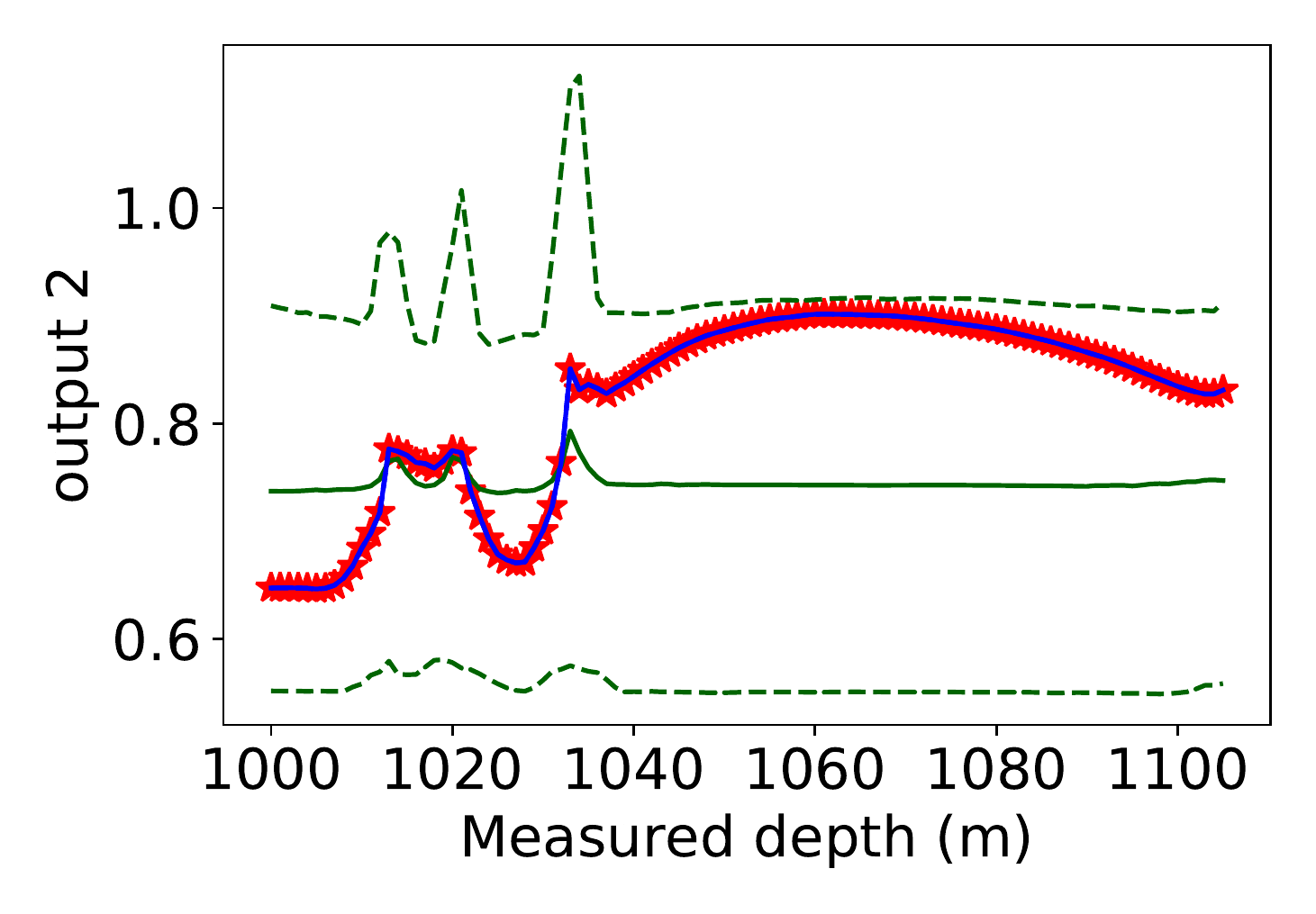}	
	\end{subfigure}

 \hspace{-0.5in}
    \begin{subfigure}[normal]{0.4\textwidth}
	\includegraphics[scale=0.5]{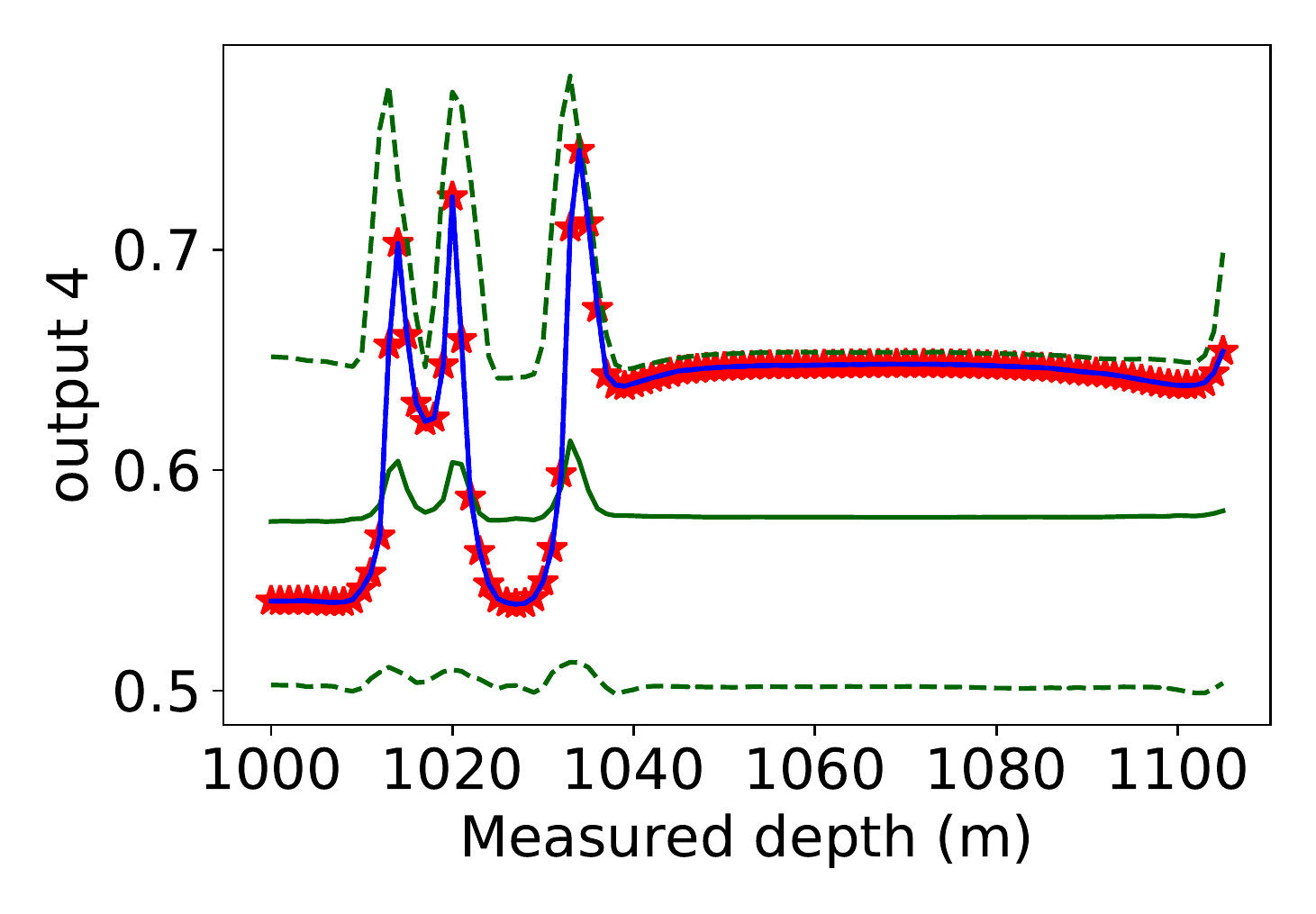}	
	\end{subfigure}
 \hspace{0.5in}
	\begin{subfigure}[normal]{0.4\textwidth}
	\includegraphics[scale=0.5]{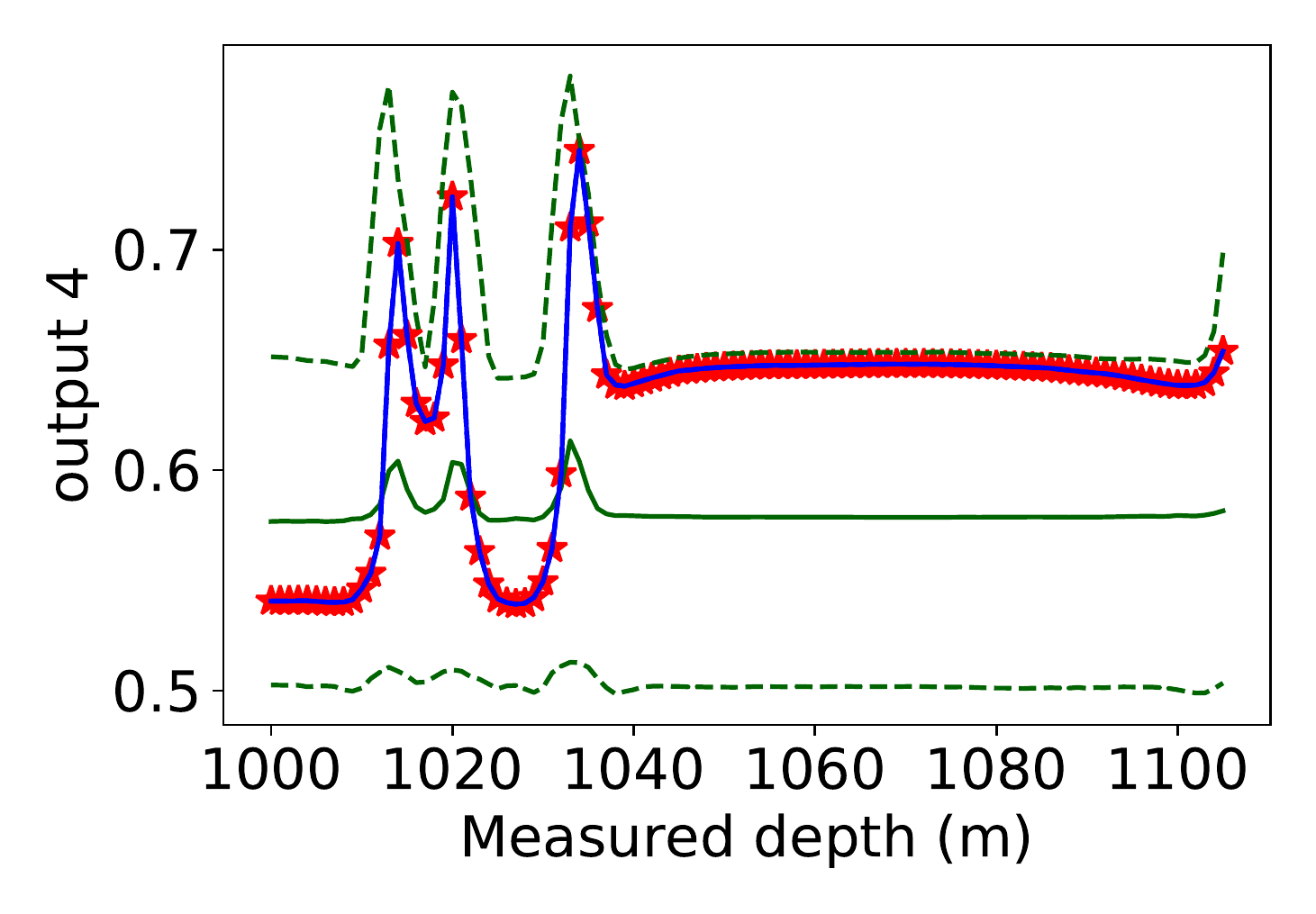}	
	\end{subfigure}

 \hspace{-0.5in}
    \begin{subfigure}[normal]{0.4\textwidth}
	\includegraphics[scale=0.5]{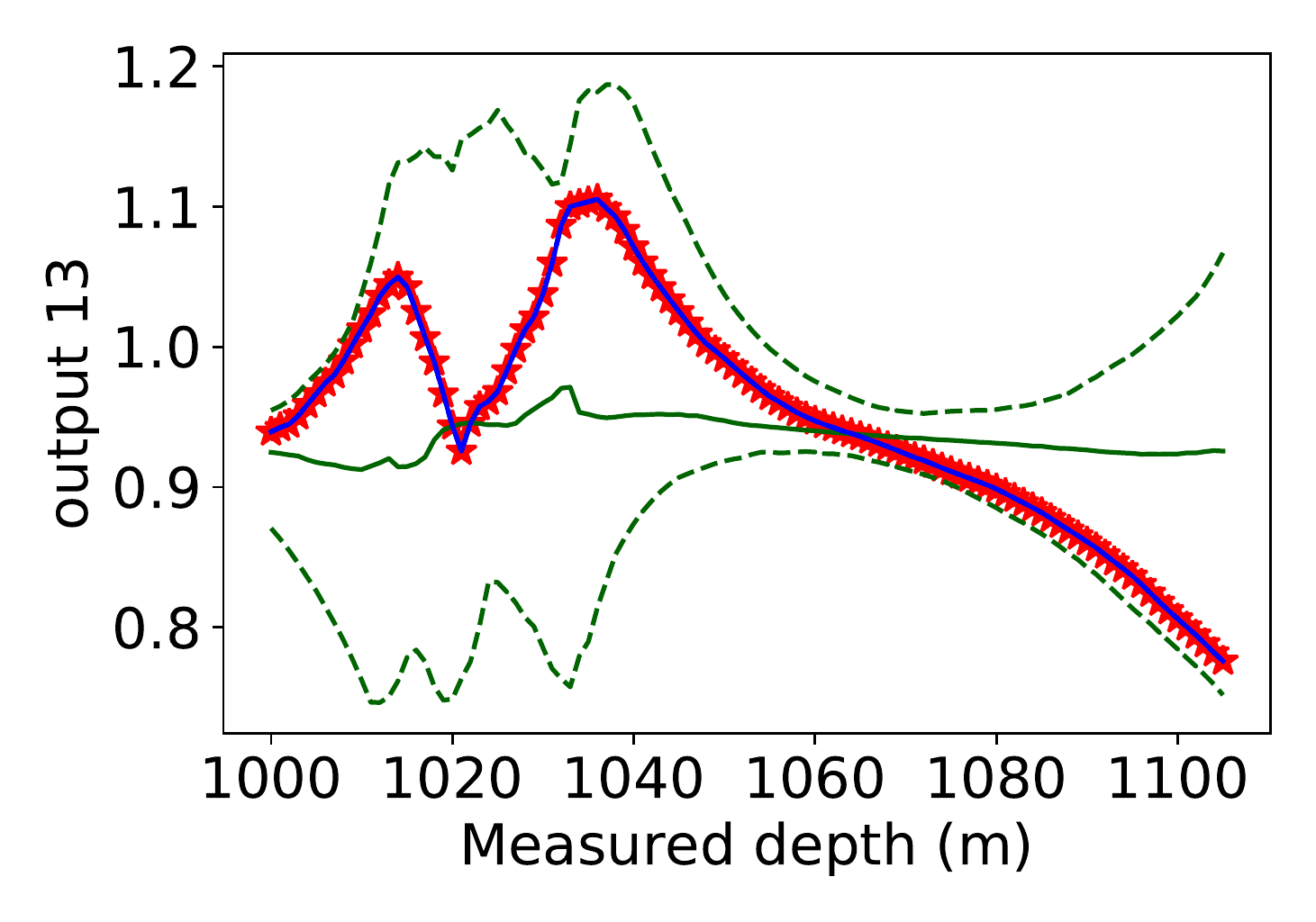}	
	\end{subfigure}
 \hspace{0.5in}
	\begin{subfigure}[normal]{0.4\textwidth}
	\includegraphics[scale=0.5]{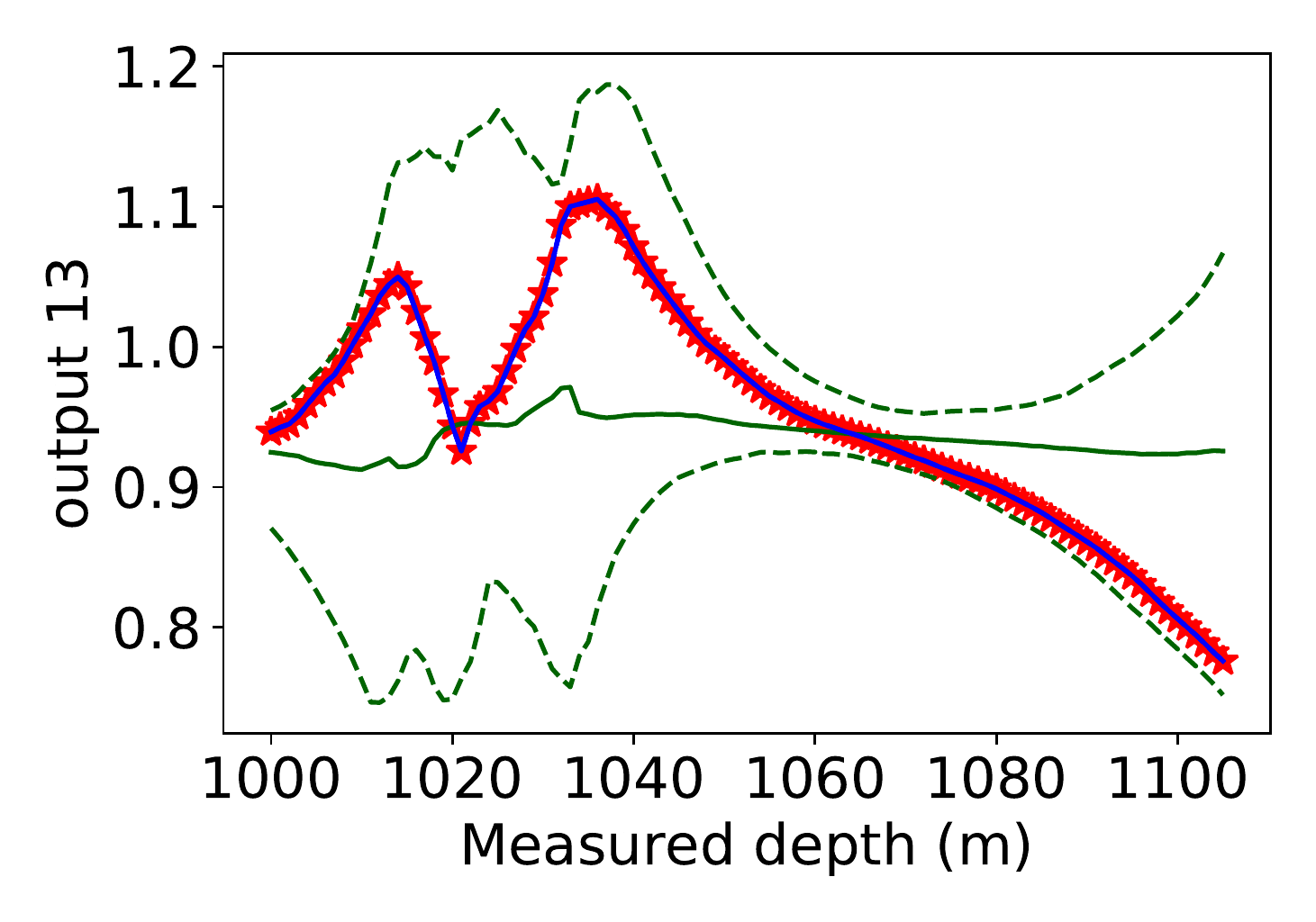}	
	\end{subfigure}

 \hspace{-0.1in}
		 ESMDA
 \hspace{2.5in}
 		 FlexIES
 \end{center}

\caption{Prior and posterior distribution of EM outputs considering DNN as a perfect forward model. Green and blue lines show ensemble approximation of the prior and posterior distribution respectively and red stars show observed EM measurements. Solid green and solid blue lines show 50th percentile ($p50$) and dashed green and dashed blue lines show $99\%$ confidence interval respectively of the prior and posterior distribution respectively. The posterior distribution appears as the point estimate therefore solid blue lines overlaps dashed blue lines. The sub-figures in the first column show results obtained from the ESMDA algorithm and the sub-figures in the second column show results from the FlexIES algorithm.}
\label{post_outputs_1_1}
\end{figure}

In the first procedure, we consider the DNN model as a perfect model, i.e.,~we generate the reference/observed data of electromagnetic measurements from the same DNN model under consideration. 
We have negligible measurement error in the observations, so we don't have any source of uncertainty in this procedure.
We expect that this problem has no multi-modality, thus we expect an exact match of the output of the DNN model to the observed data as shown in Figure~(\ref{post_outputs_1_1}). 
The posterior distribution of the measurements converges to a single point for both classical and flexible algorithms, which also coincides with the exact data.
This is due to the fact that we are considering the DNN model as a perfect model and in the absence of measurement error and model error, we obtain an exact match of the data.

\begin{figure}[H]
\begin{center}

 \hspace{-0.5in}
 	\begin{subfigure}[normla]{0.5\textwidth}
		\begin{tabular}{ccc}
		\hspace{-0.2in}
		\includegraphics[scale=0.35]{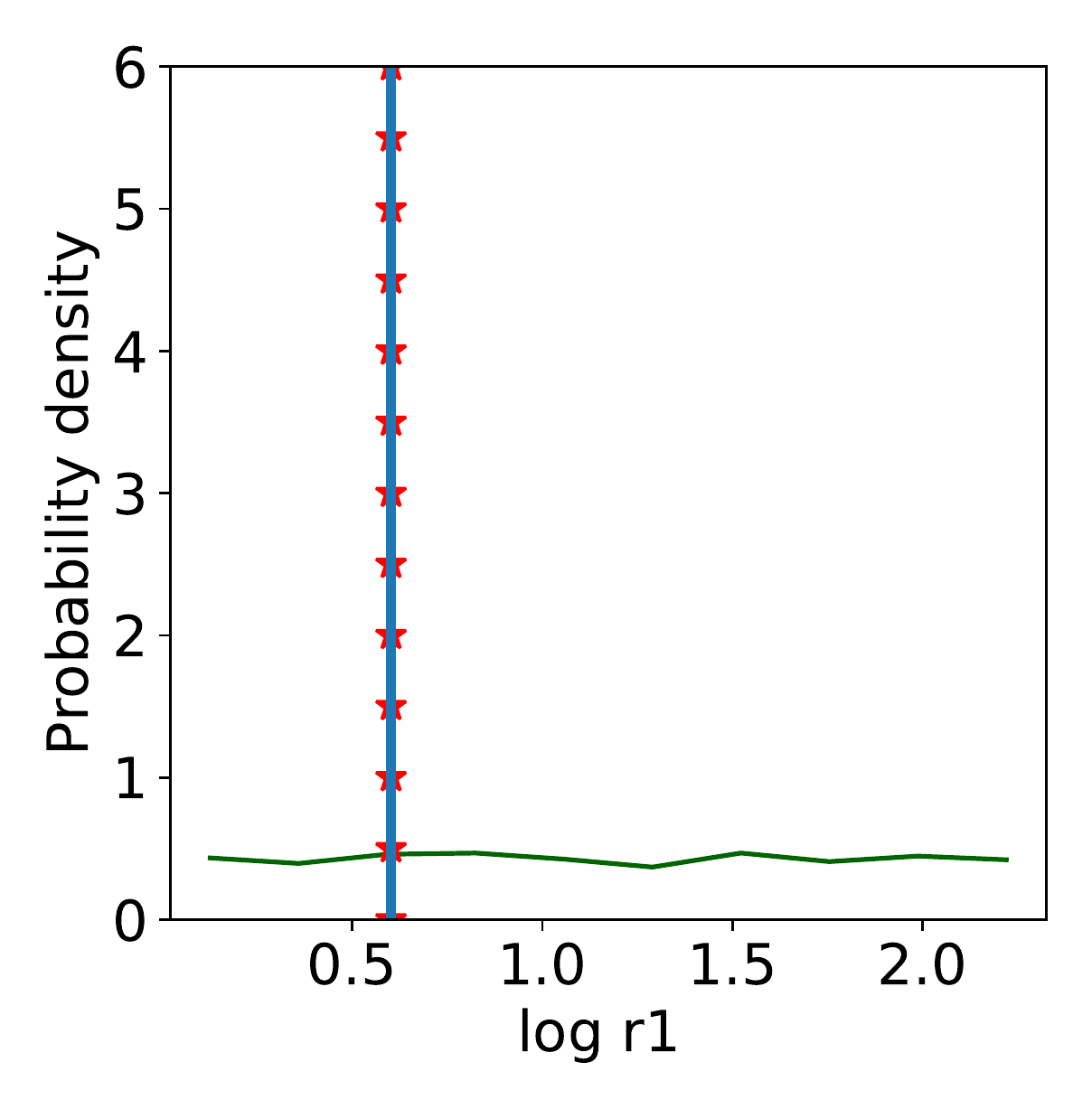} &
		\hspace{-0.2in}
		\includegraphics[scale=0.35]{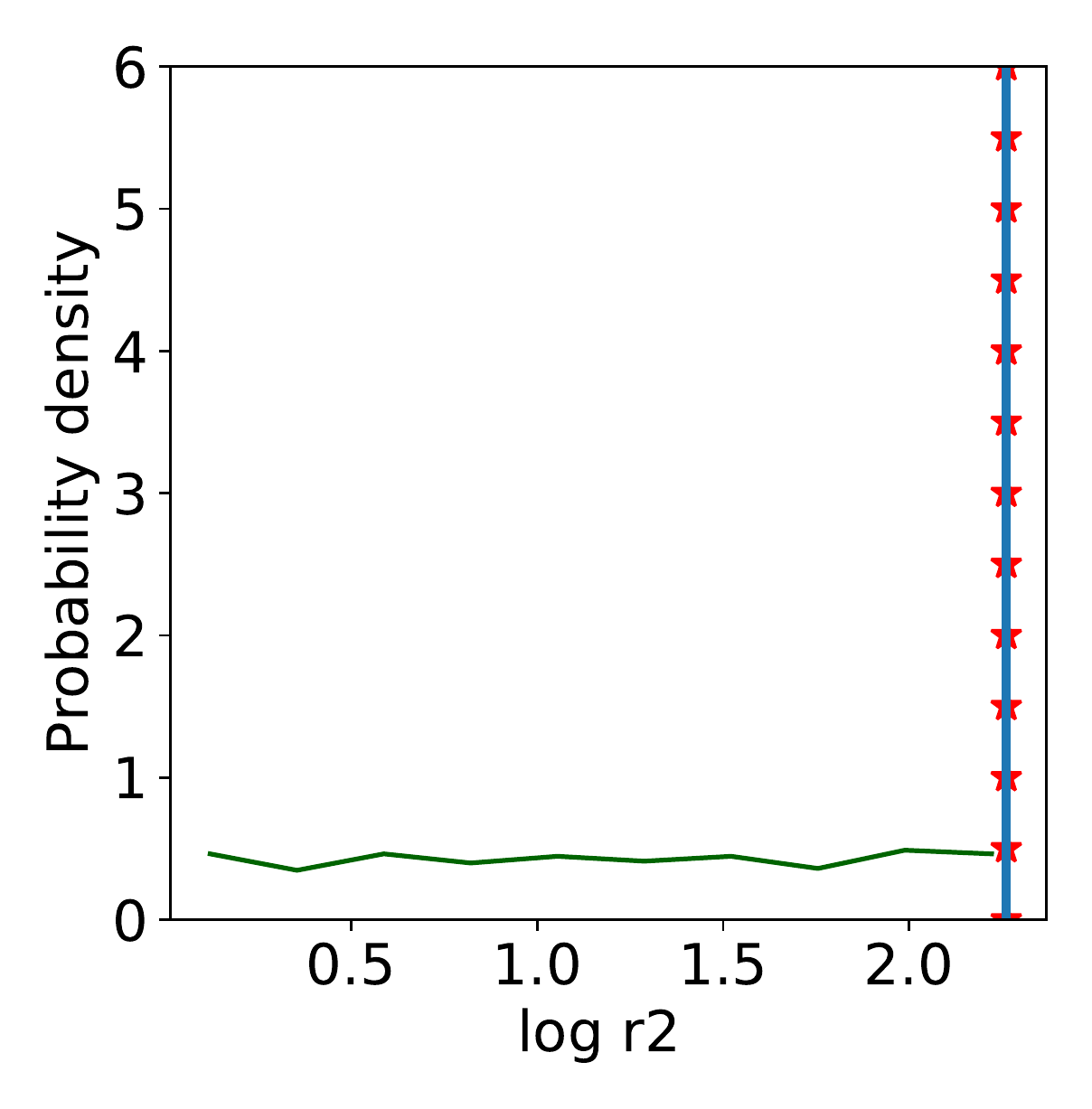} \\ 
		\end{tabular}
	\end{subfigure}
 \hspace{0.2in}	
	\begin{subfigure}[normla]{0.5\textwidth}
	   \begin{tabular}{cc}
		\includegraphics[scale=0.35]{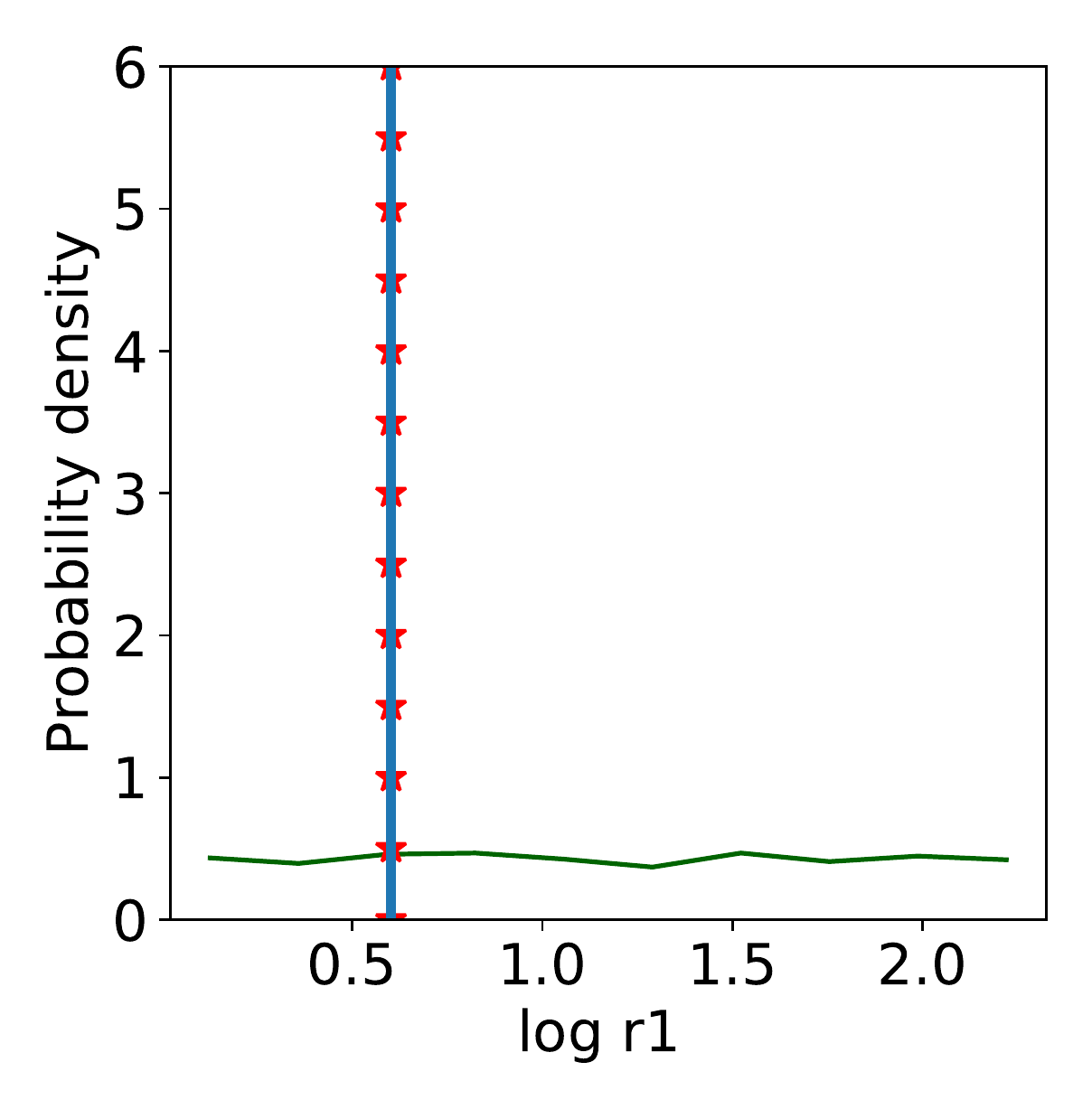} &
		\hspace{-0.2in}
		\includegraphics[scale=0.35]{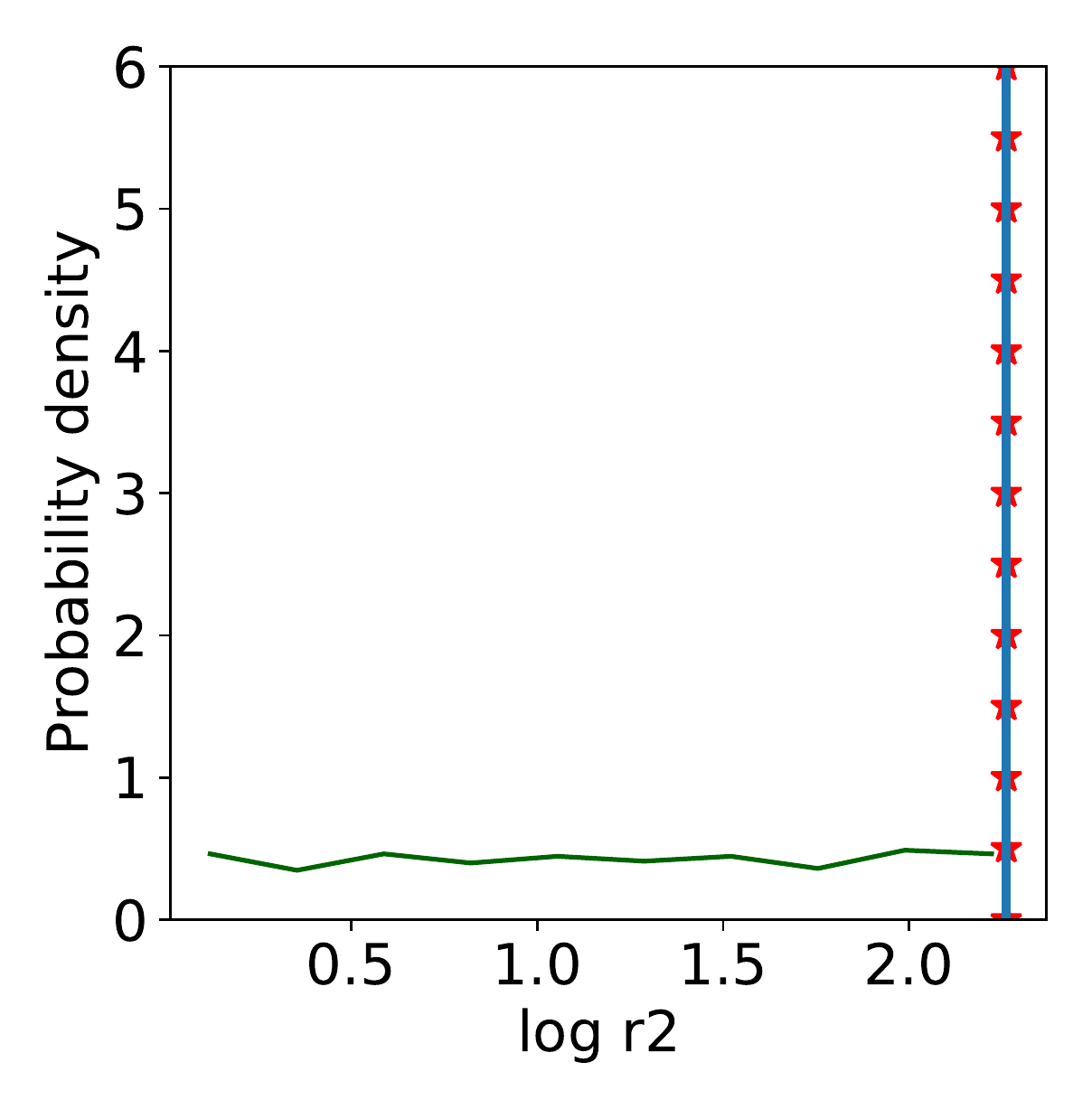} \\
		\end{tabular}
	\end{subfigure}

 \hspace{-0.5in}
 	\begin{subfigure}[normla]{0.5\textwidth}
		\begin{tabular}{ccc}
		\hspace{-0.2in}
		\includegraphics[scale=0.35]{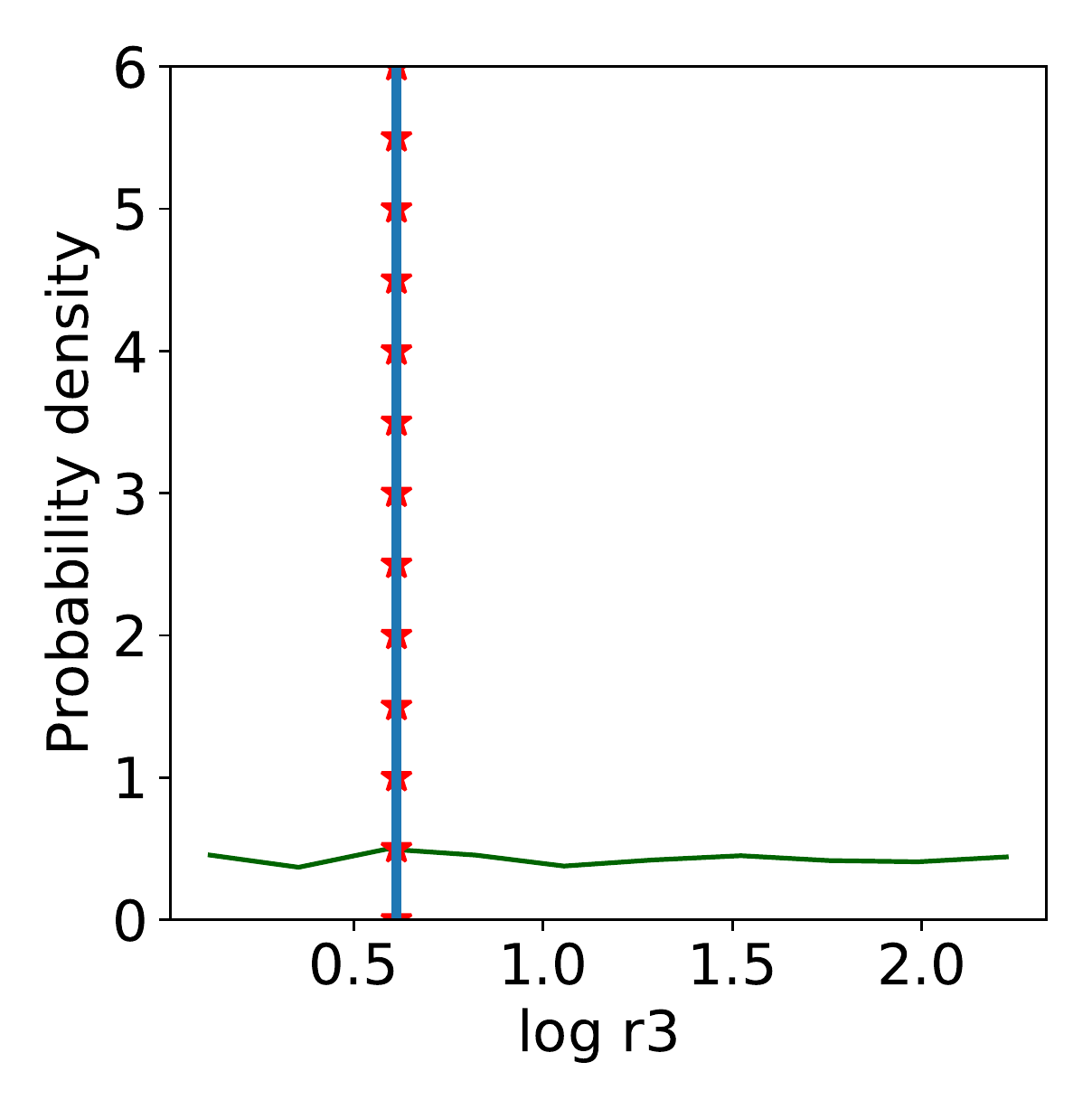} &
		\hspace{-0.2in}
		\includegraphics[scale=0.35]{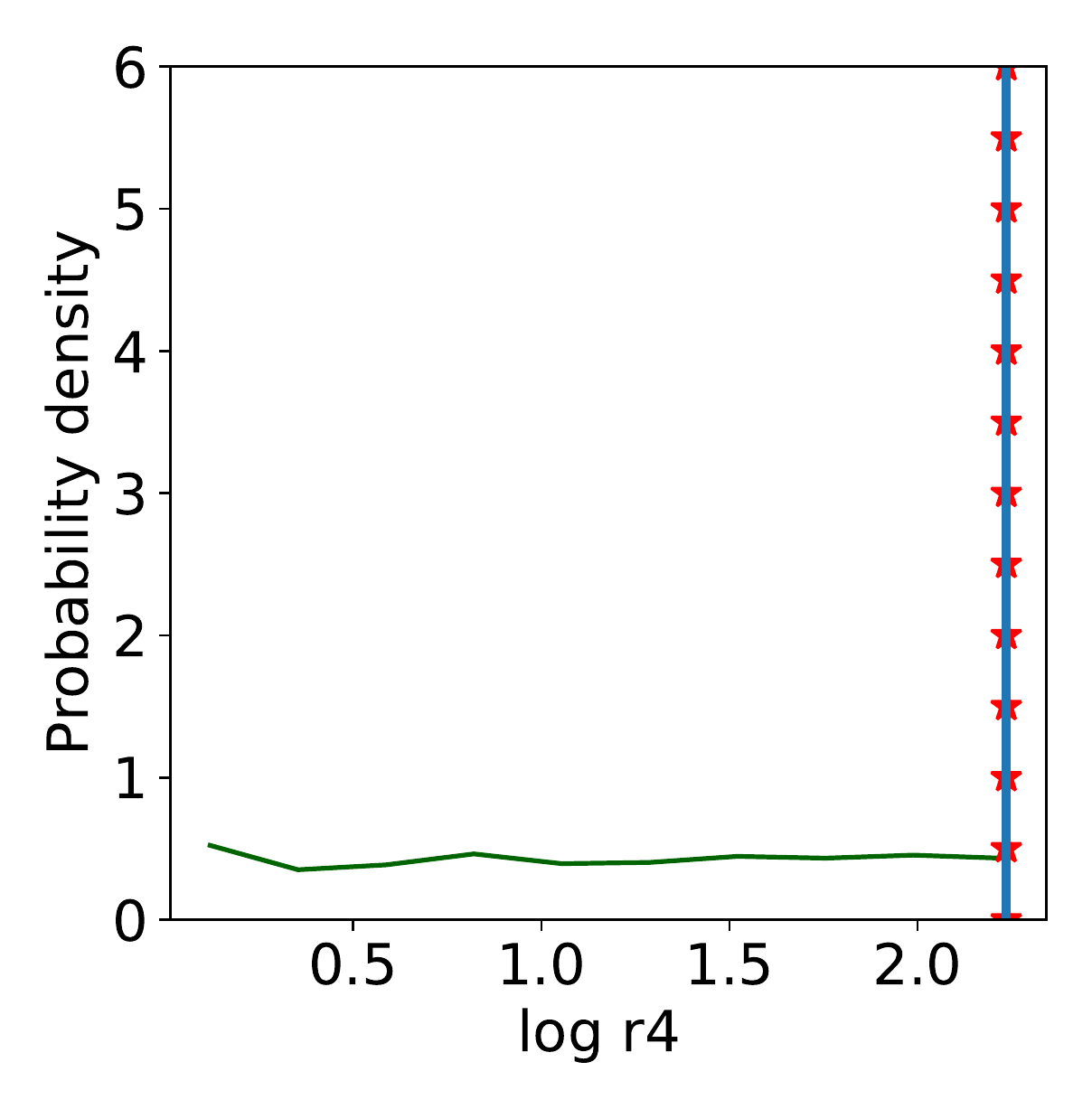} \\ 
		\end{tabular}
	\end{subfigure}
 \hspace{0.2in}	
	\begin{subfigure}[normla]{0.5\textwidth}
	   \begin{tabular}{cc}
		\includegraphics[scale=0.35]{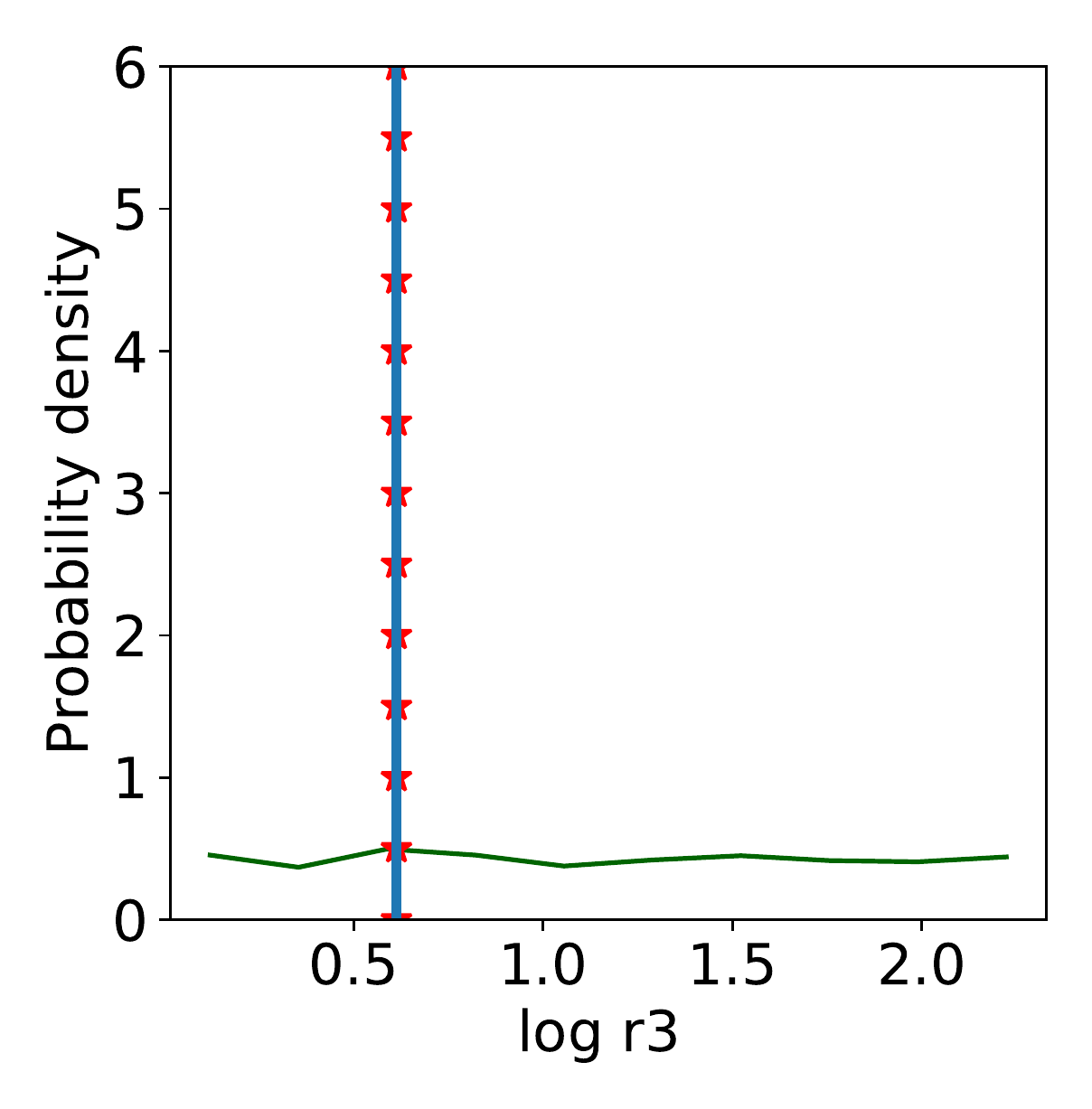} &
		\hspace{-0.2in}
		\includegraphics[scale=0.35]{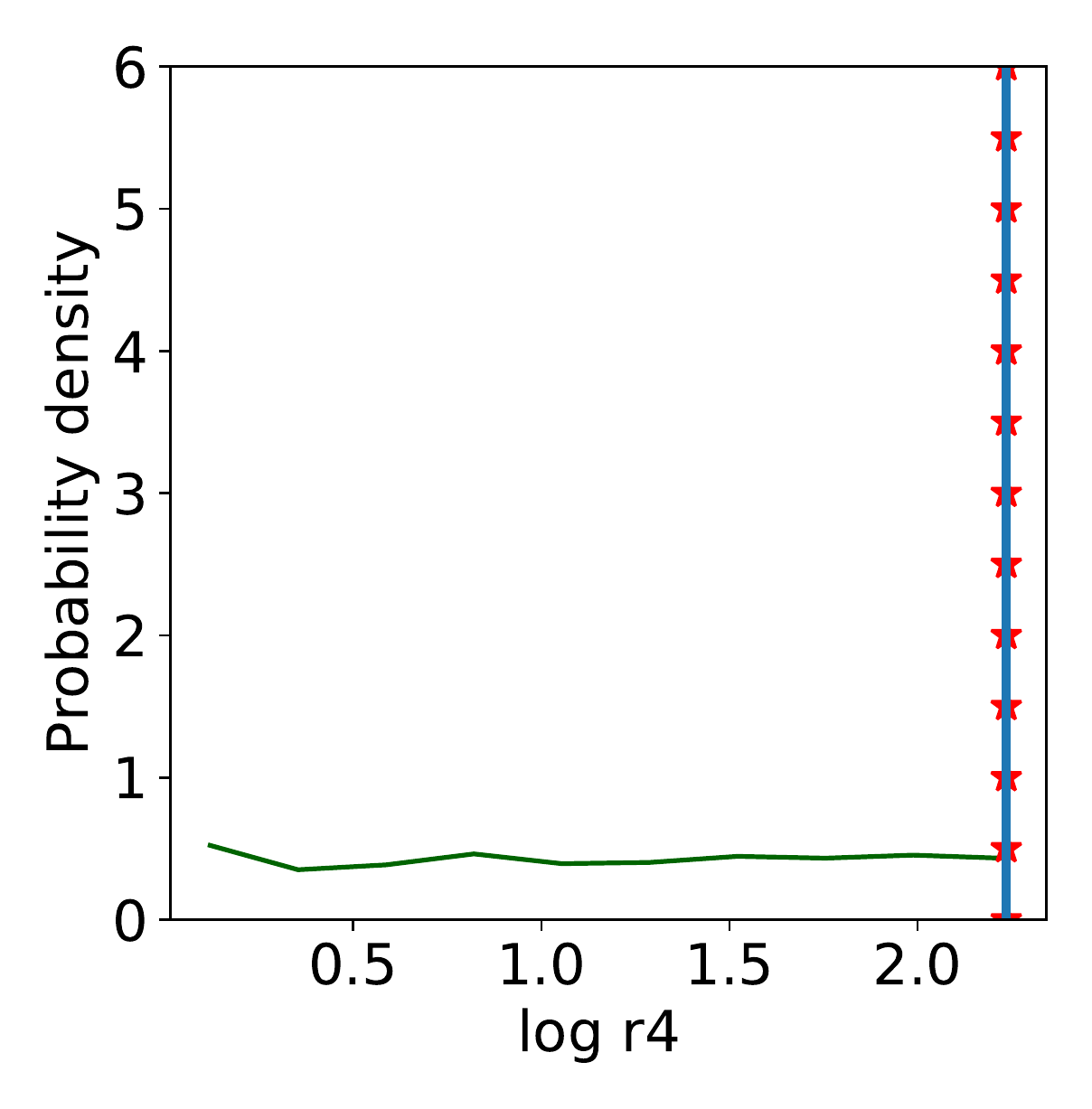} \\
		\end{tabular}
	\end{subfigure}

 \hspace{-0.5in}
 	\begin{subfigure}[normla]{0.5\textwidth}
		\begin{tabular}{ccc}
		\hspace{-0.2in}
		\includegraphics[scale=0.35]{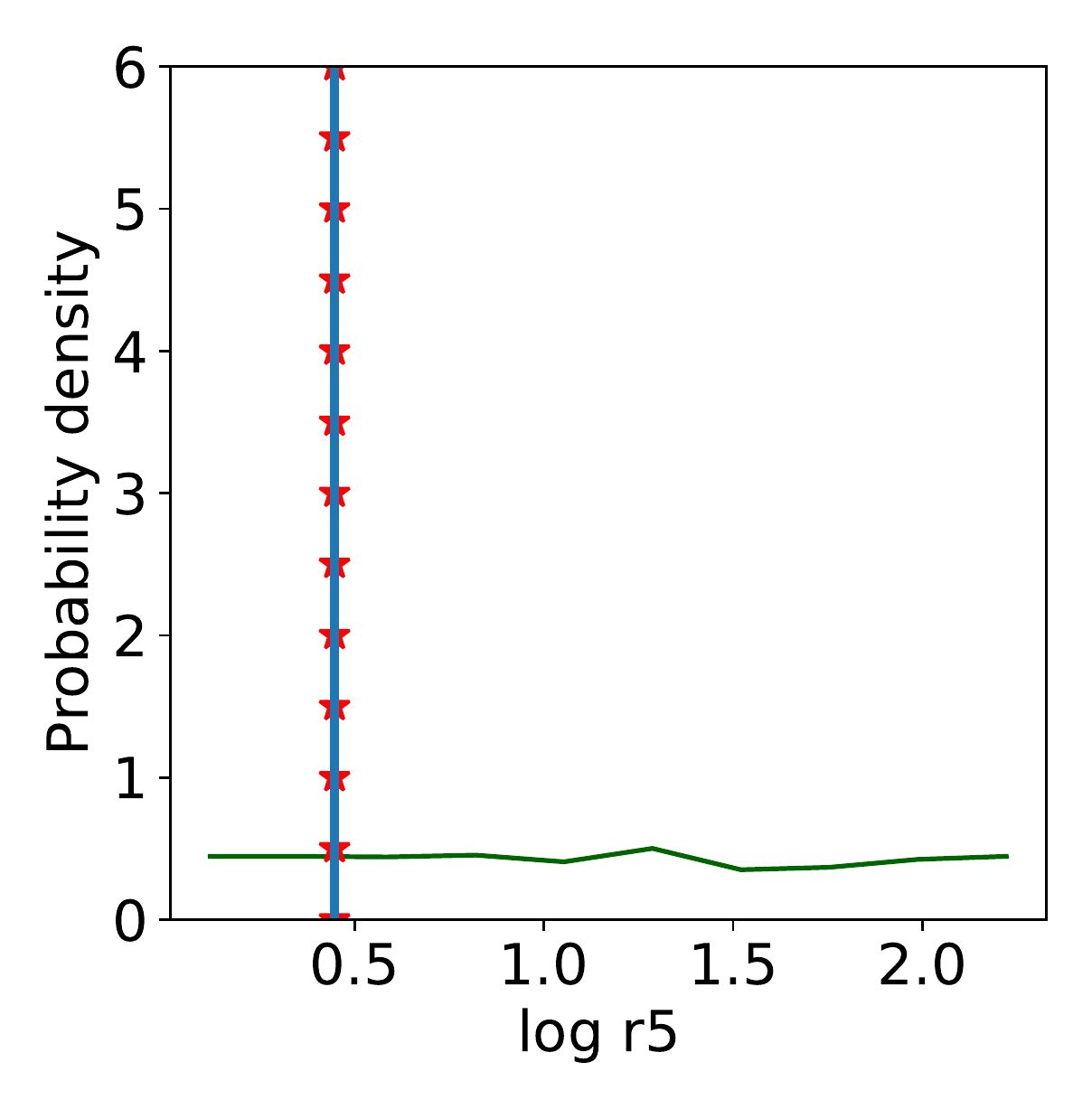} &
		\hspace{-0.2in}
		\includegraphics[scale=0.35]{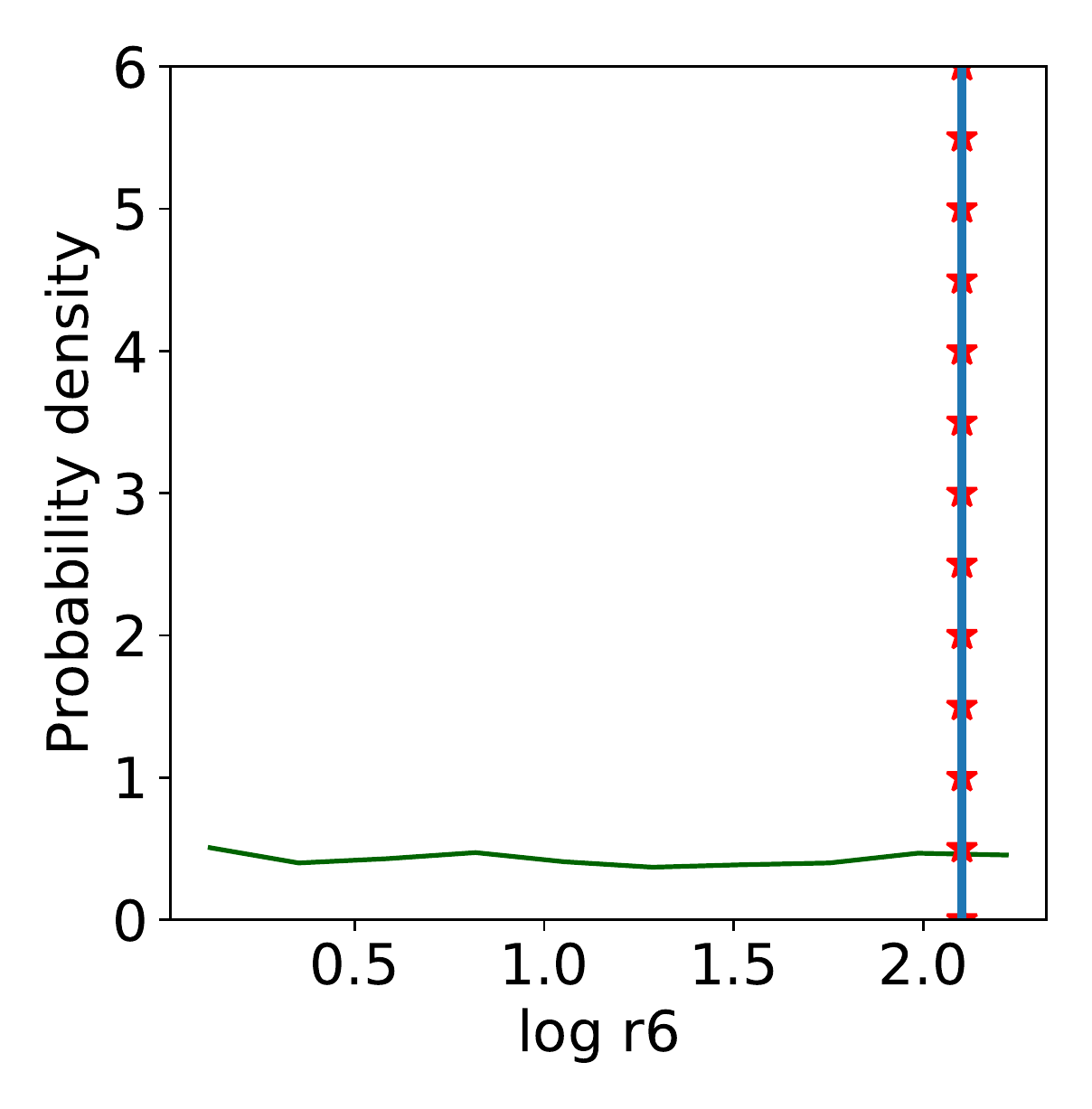} \\ 
		\end{tabular}
	\end{subfigure}
 \hspace{0.2in}	
	\begin{subfigure}[normla]{0.5\textwidth}
	   \begin{tabular}{cc}
		\includegraphics[scale=0.35]{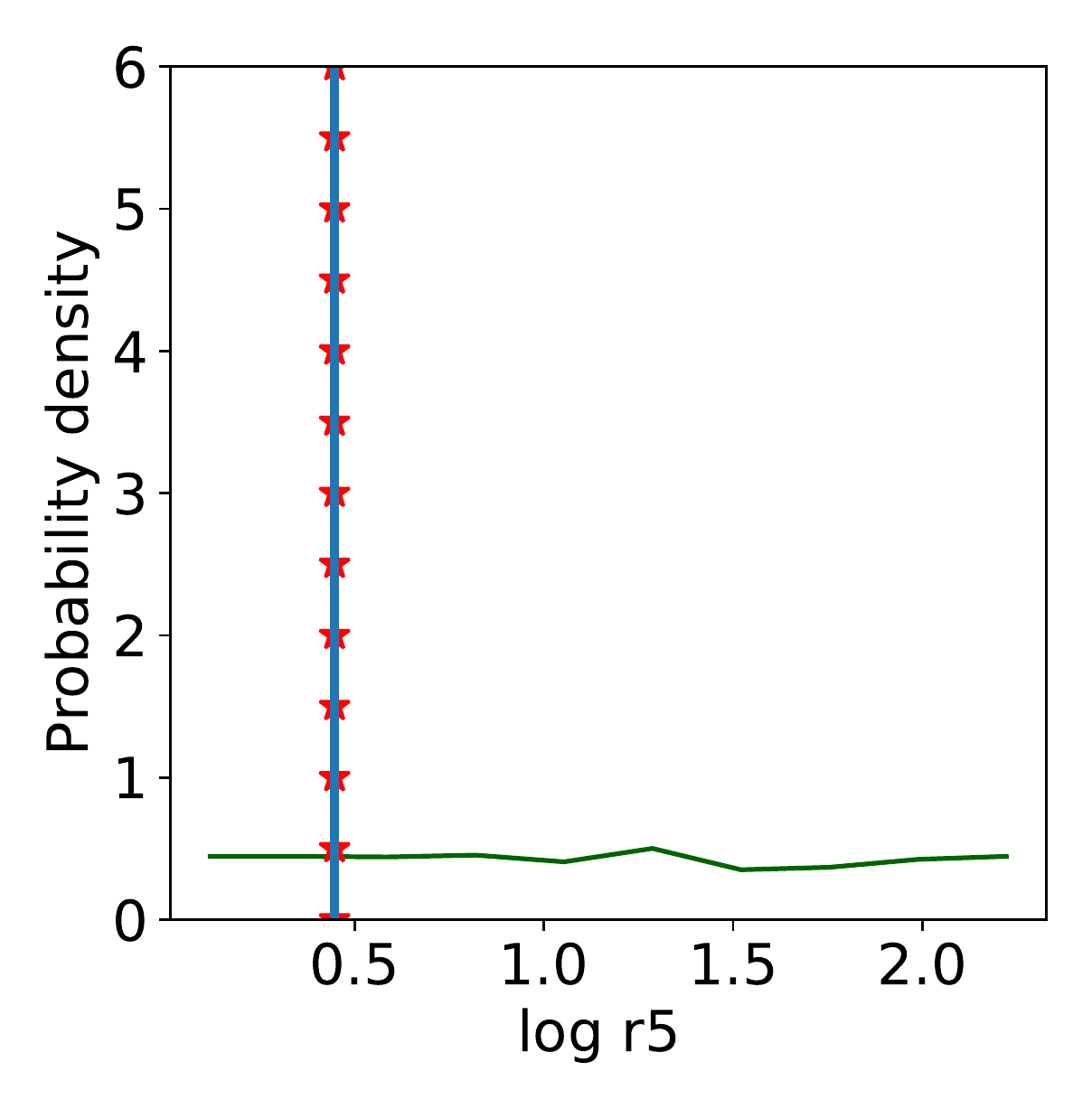} &
		\hspace{-0.2in}
		\includegraphics[scale=0.35]{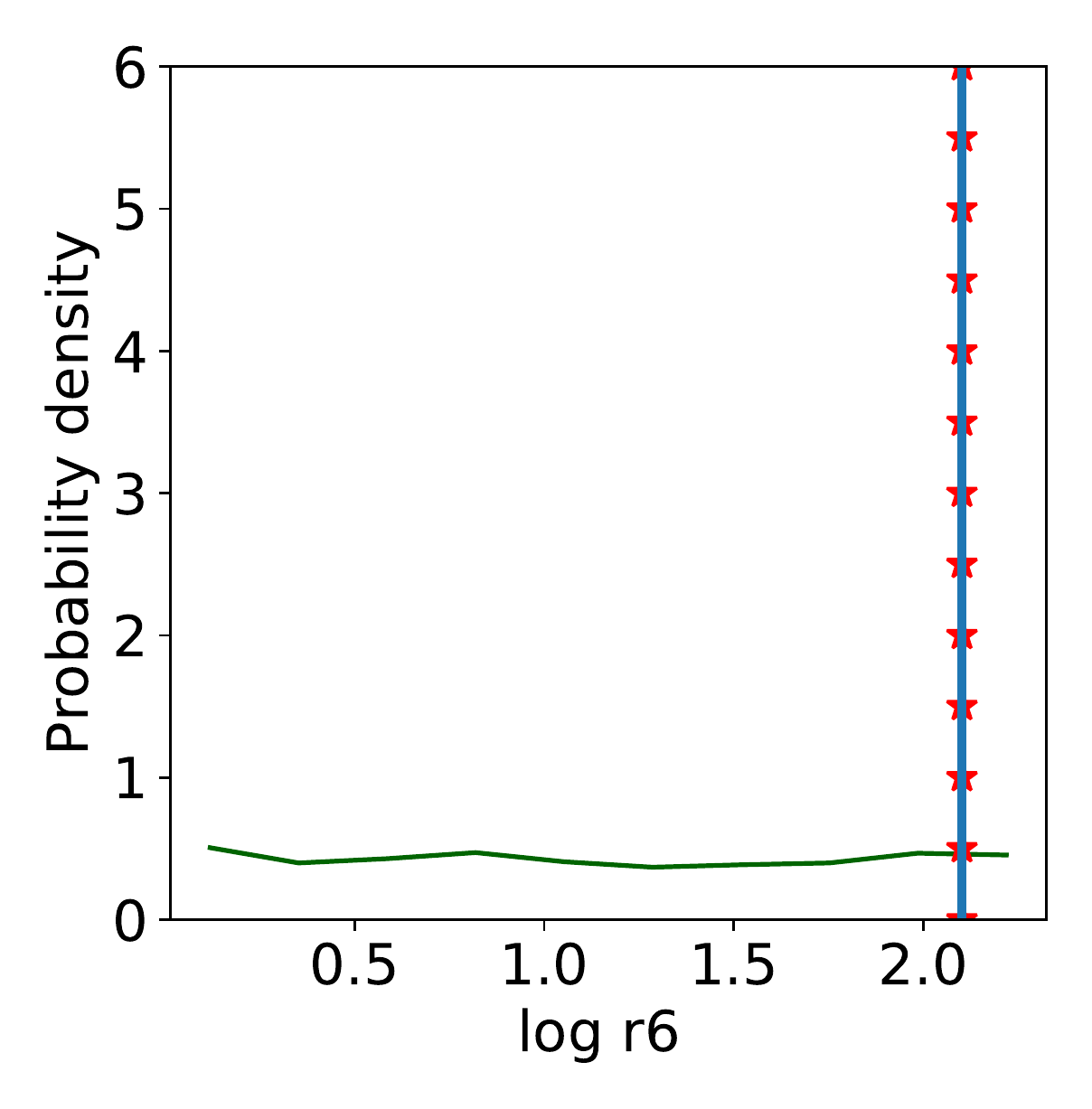} \\
		\end{tabular}
	\end{subfigure}
		
  \hspace{-0.1in}
		 ESMDA
 \hspace{2.5in}
 		 FlexIES

 \end{center}

\caption{Prior and posterior distribution of layers resistivities considering DNN as a perfect forward model. Green and blue lines show ensemble approximation of the prior and posterior distribution respectively and red stars show reference truth. Solid blue lines show $p50$ of the posterior distribution. The posterior distributions appear as the point estimate. The sub-figures in the first and second columns show results obtained from the ESMDA algorithm and the sub-figures in the third and fourth columns show results from the FlexIES algorithm.}
\label{post_inputs_1_1}
\end{figure}

Figure~(\ref{post_inputs_1_1}) shows the estimation results of the log10 resistivities of the layers. We observe that both algorithms successfully recover/estimate the exact/reference resistivity.
Furthermore, the estimated posterior appears as the exact point estimate in both cases. 
This confirms that FlexIES can estimate the parameter exactly if there is no uncertainty in terms of model-error and measurement error for a well-posed problem.

\subsubsection{Realistic inversion in presence of model-errors}

In the second procedure, we take the reference data from the high-fidelity log simulator provided by the tool vendor \cite{simulator2014}. {\color{black} The simulator produces the EM signals using layers properties and well trajectory by solving Maxwell's equations.}
Thus, the DNN approximation {\color{black} could contain the model-errors, and therefore this test depicts the more realistic scenario for real-time inversion of EM measurements.} 
In this procedure, we still have negligible measurement error in the observations, so the model-error is the main source of estimation uncertainty. 

\begin{figure}[H]
\begin{center}

 \hspace{-0.5in}
    \begin{subfigure}[normal]{0.4\textwidth}
	\includegraphics[scale=0.5]{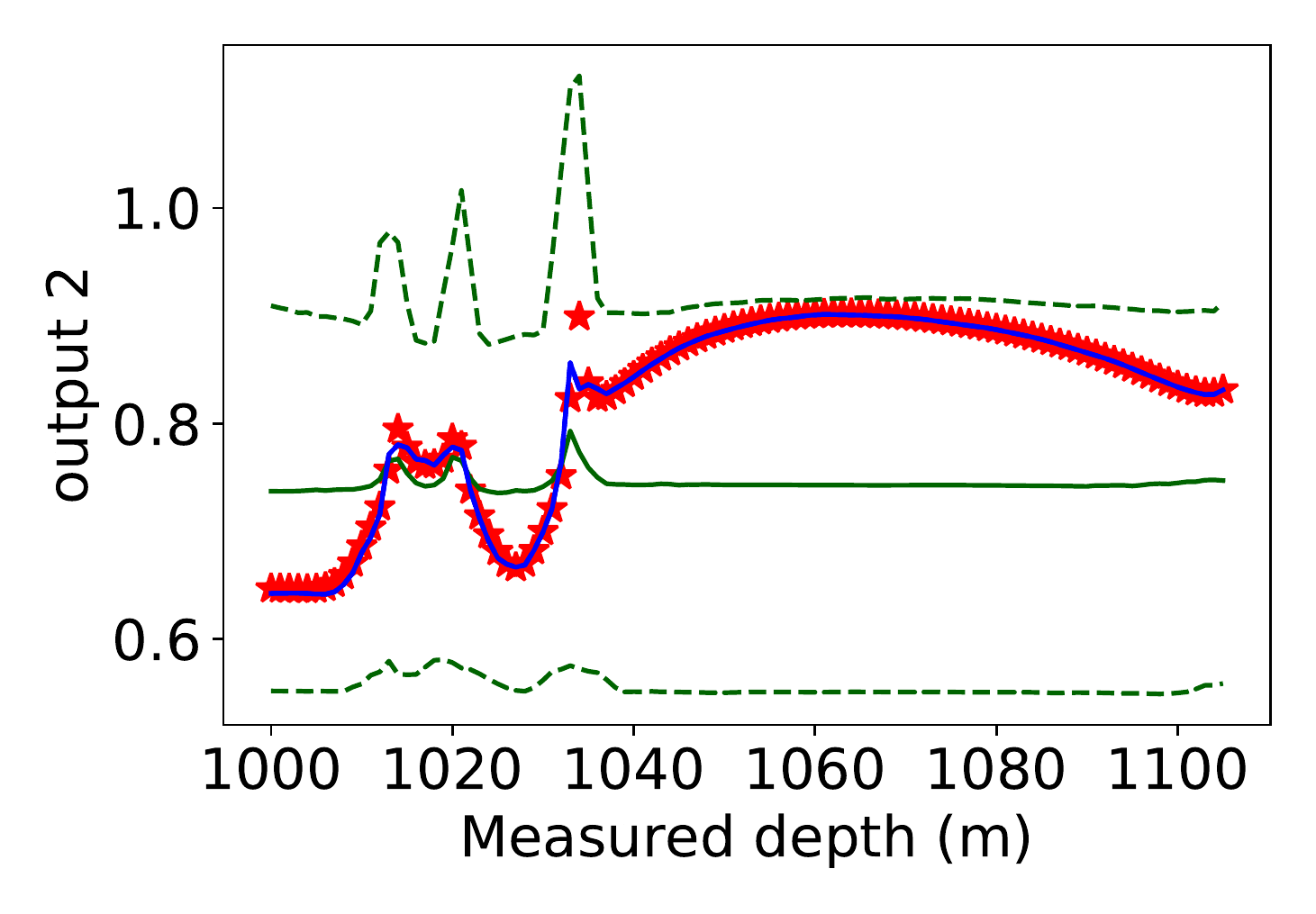}	
	\end{subfigure}
 \hspace{0.5in}
	\begin{subfigure}[normal]{0.4\textwidth}
	\includegraphics[scale=0.5]{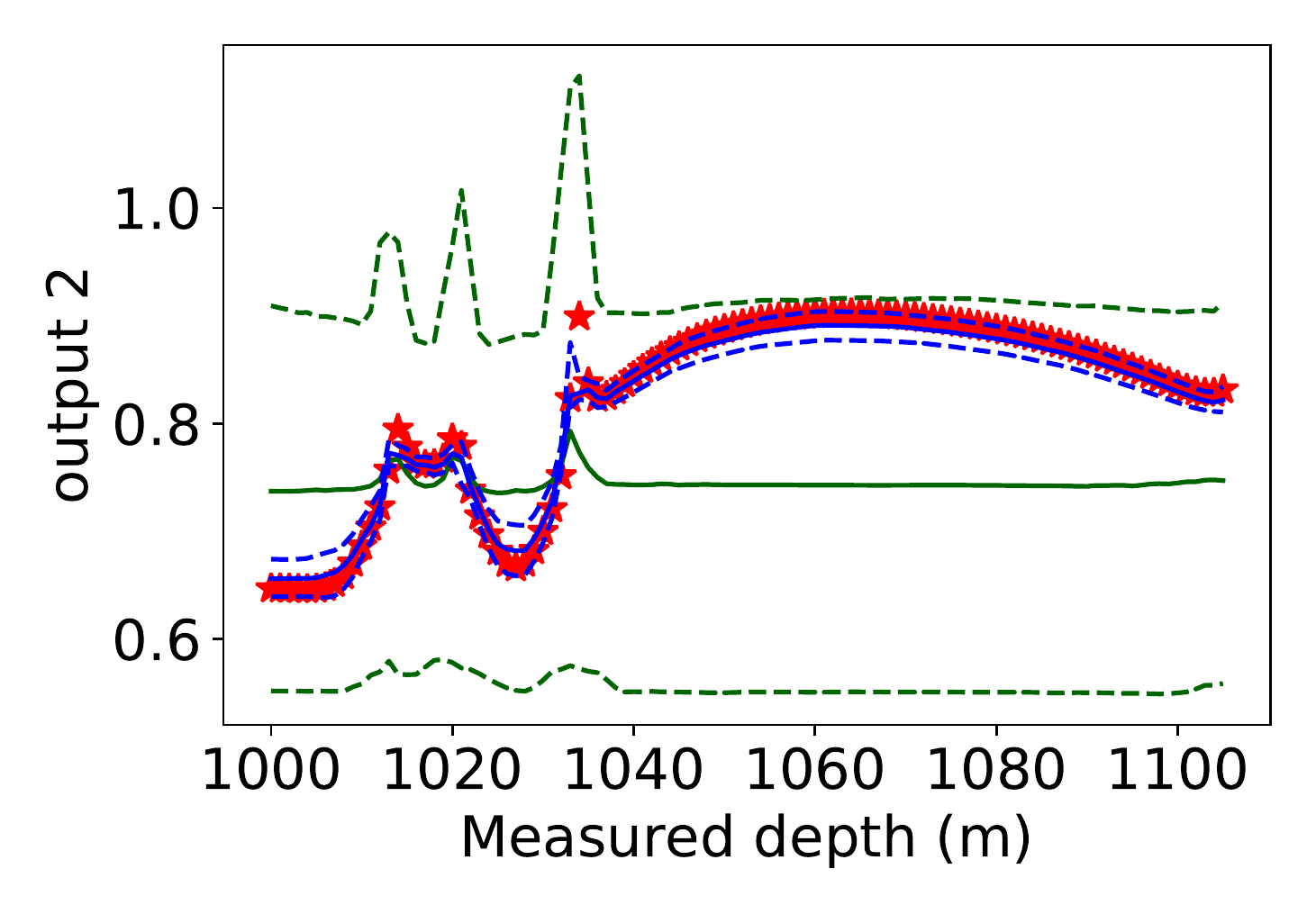}	
	\end{subfigure}

 \hspace{-0.5in}
    \begin{subfigure}[normal]{0.4\textwidth}
	\includegraphics[scale=0.5]{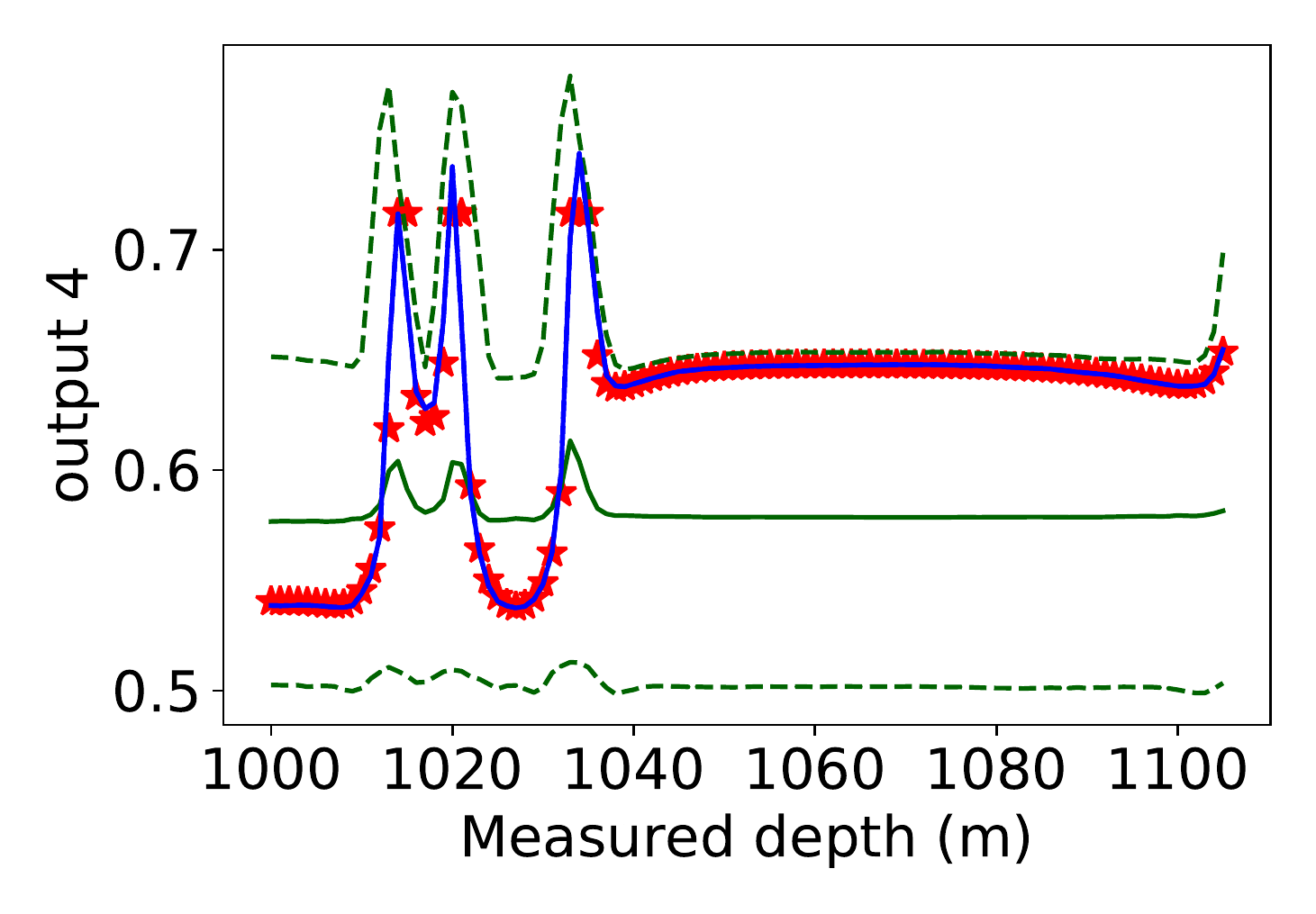}	
	\end{subfigure}
 \hspace{0.5in}
	\begin{subfigure}[normal]{0.4\textwidth}
	\includegraphics[scale=0.5]{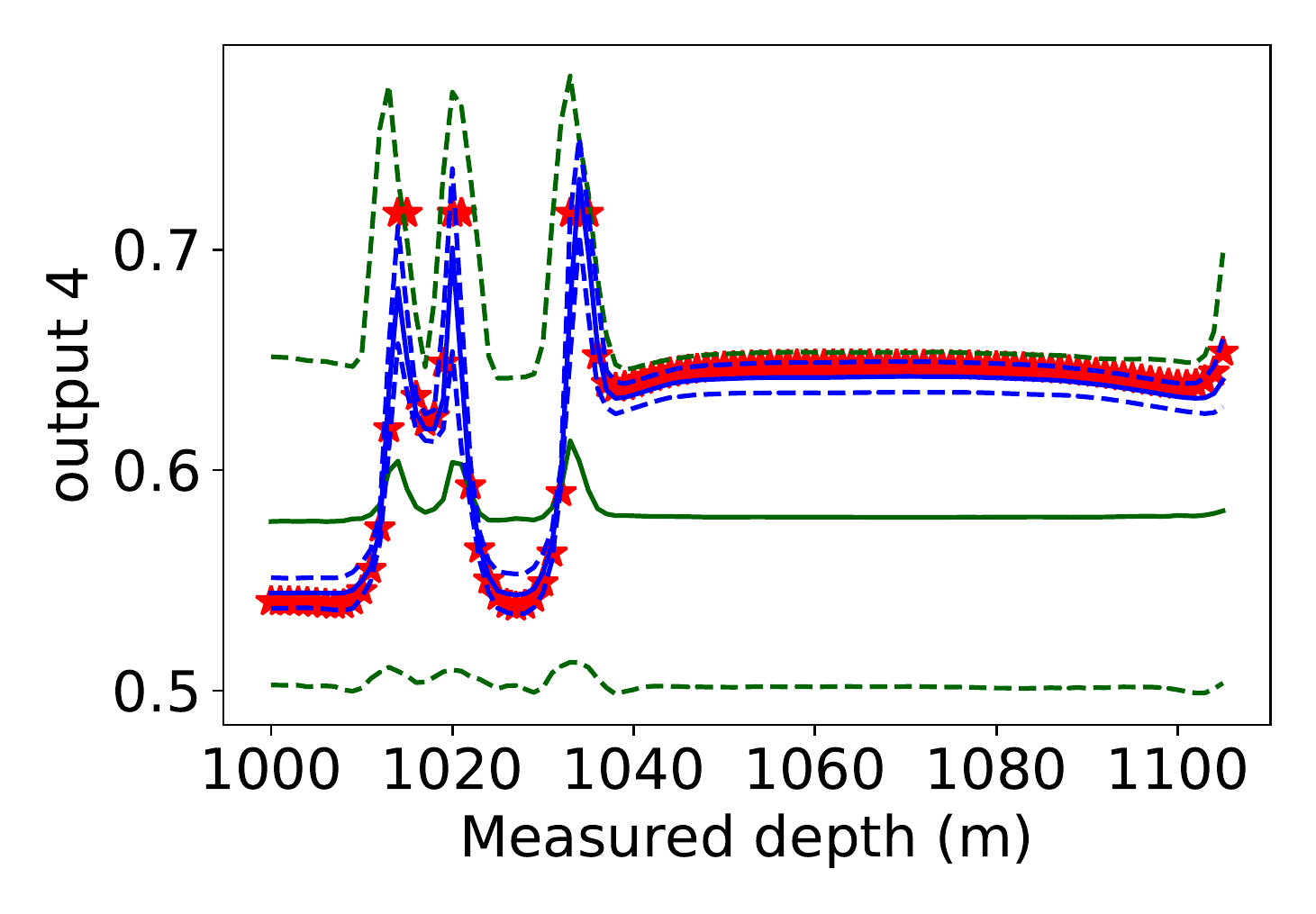}	
	\end{subfigure}

 \hspace{-0.5in}
    \begin{subfigure}[normal]{0.4\textwidth}
	\includegraphics[scale=0.5]{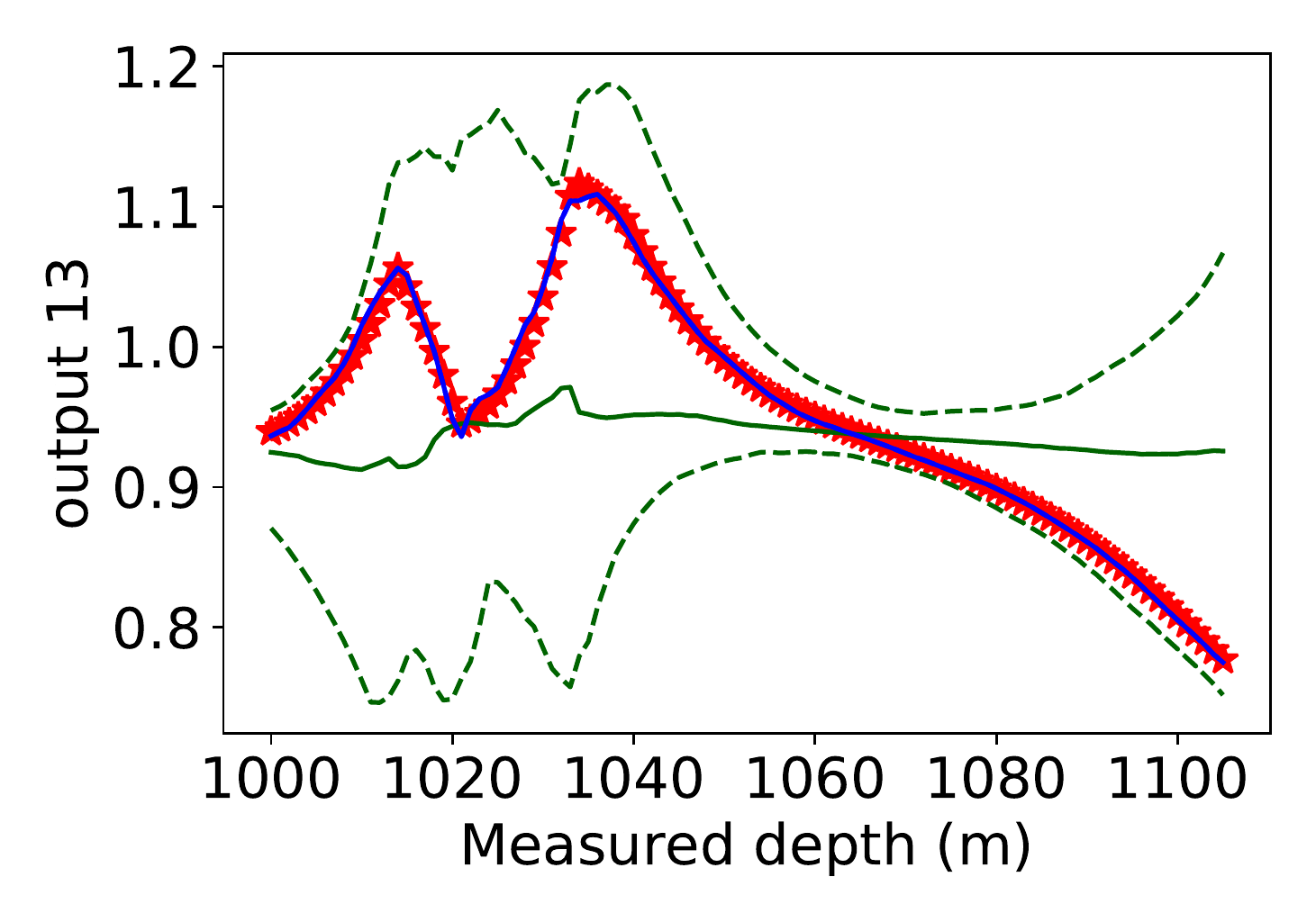}	
	\end{subfigure}
 \hspace{0.5in}
	\begin{subfigure}[normal]{0.4\textwidth}
	\includegraphics[scale=0.5]{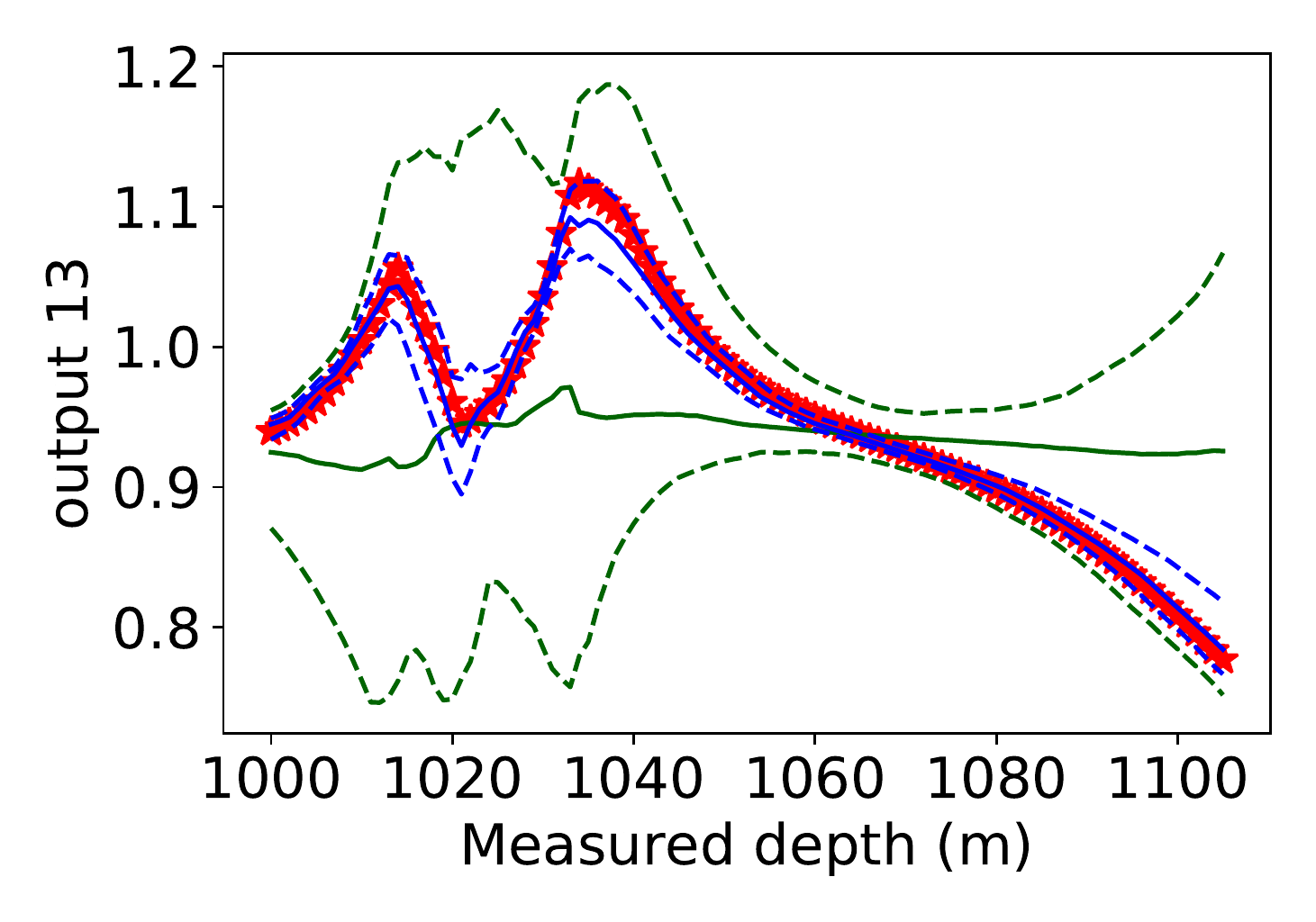}	
	\end{subfigure}

 \hspace{-0.1in}
		 ESMDA
 \hspace{2.5in}
 		 FlexIES
 \end{center}

\caption{Prior and posterior distribution of EM outputs in the presence of model-errors. Green and blue lines show ensemble approximation of the prior and posterior distribution respectively and red stars show observed EM measurements. Solid green and solid blue lines show $p50$ and dashed green and dashed blue lines show $99\%$ confidence interval respectively of the prior and posterior distribution respectively. The sub-figures in the first column show results obtained from the ESMDA algorithm and the sub-figures in the second column show results from the FlexIES algorithm. In first column of the sub-figures posterior distribution appears as the point estimate therefore solid blue lines overlaps dashed blue lines.}
\label{post_outputs_2_1}
\end{figure}

Classical ESMDA achieves a reasonable data match also in the presence of model-error as shown in Figure~(\ref{post_outputs_2_1}). However the posterior distribution of predicted measurements appears as a point estimate, thus showing no sign of uncertainty since ESMDA does not account for the model-error effect. 
On the other hand, FlexIES can take into account the model-error during real-time inversion.
This results in a broader posterior distribution that also covers the part of the data where the classical method had mismatches as shown in Figure~(\ref{post_outputs_2_1}). We note that the predicted ensemble mean using the FlexIES stays close to the estimates from the classical method. 

\begin{figure}[H]
\begin{center}

 \hspace{-0.3in}
 	\begin{subfigure}[normla]{0.5\textwidth}
		\begin{tabular}{ccc}
		\hspace{-0.2in}
		\includegraphics[scale=0.35]{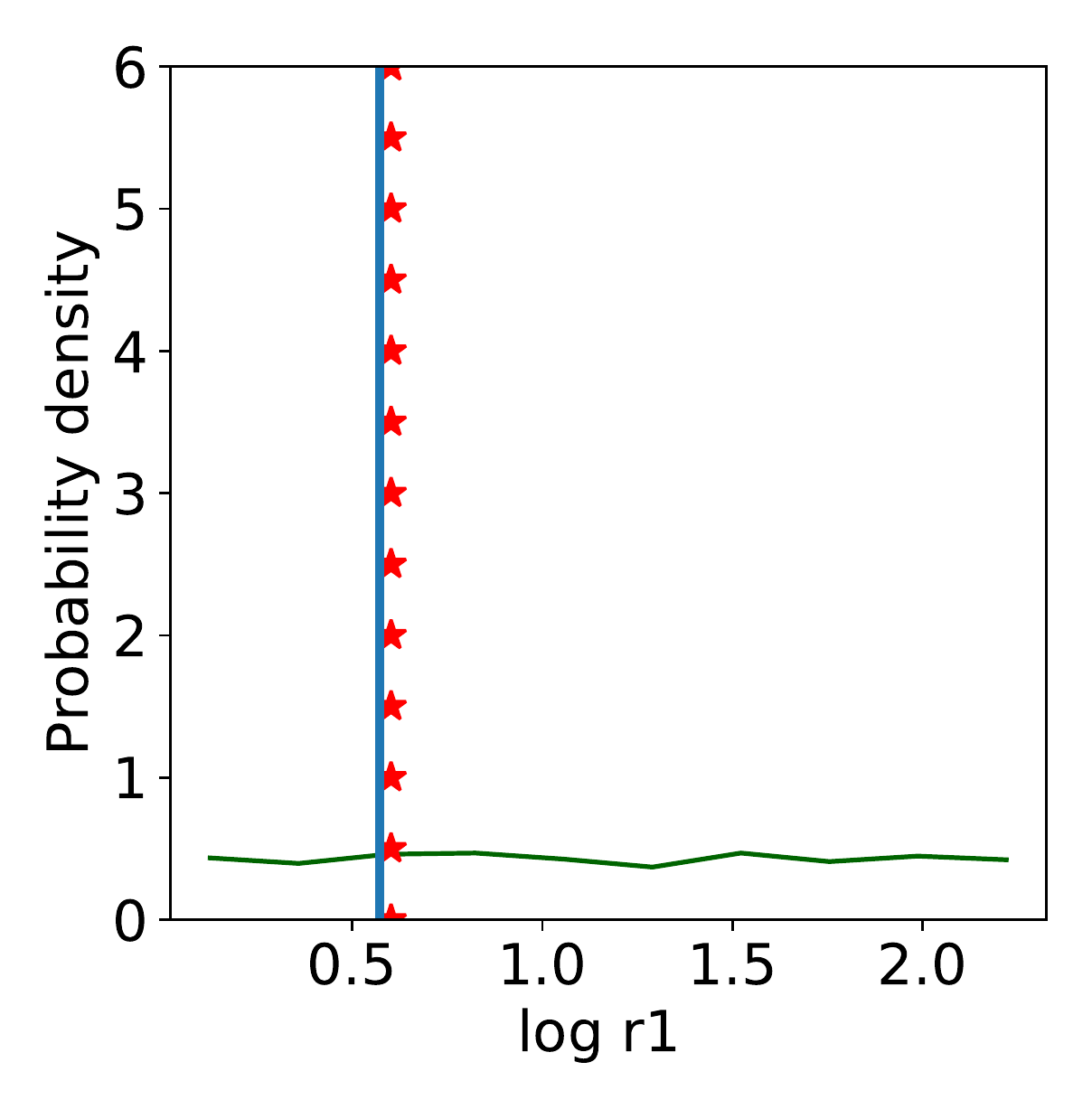} &
		\hspace{-0.2in}
		\includegraphics[scale=0.35]{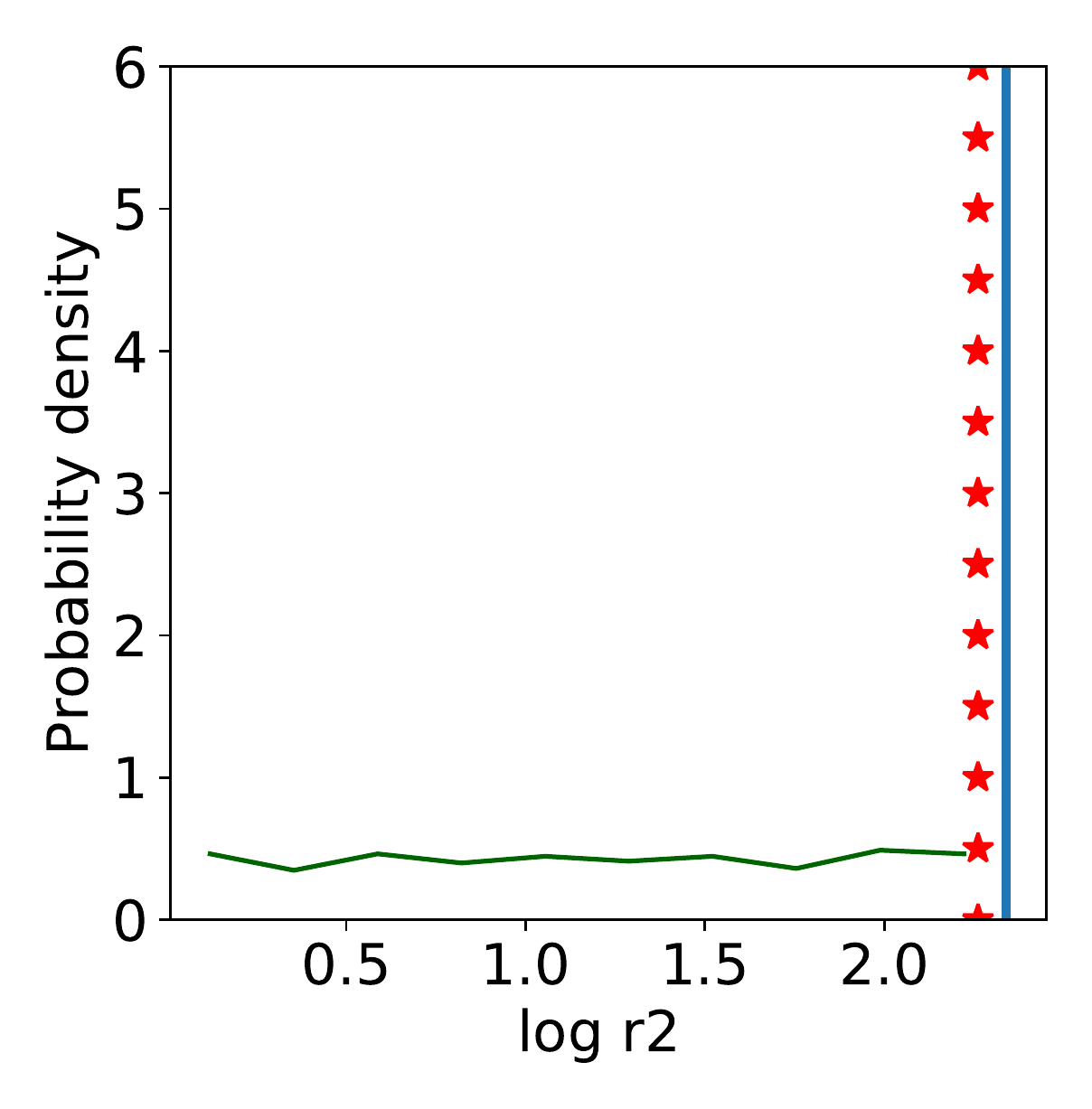} \\ 
		\end{tabular}
	\end{subfigure}
 \hspace{0.15in}	
	\begin{subfigure}[normla]{0.5\textwidth}
	   \begin{tabular}{cc}
		\includegraphics[scale=0.35]{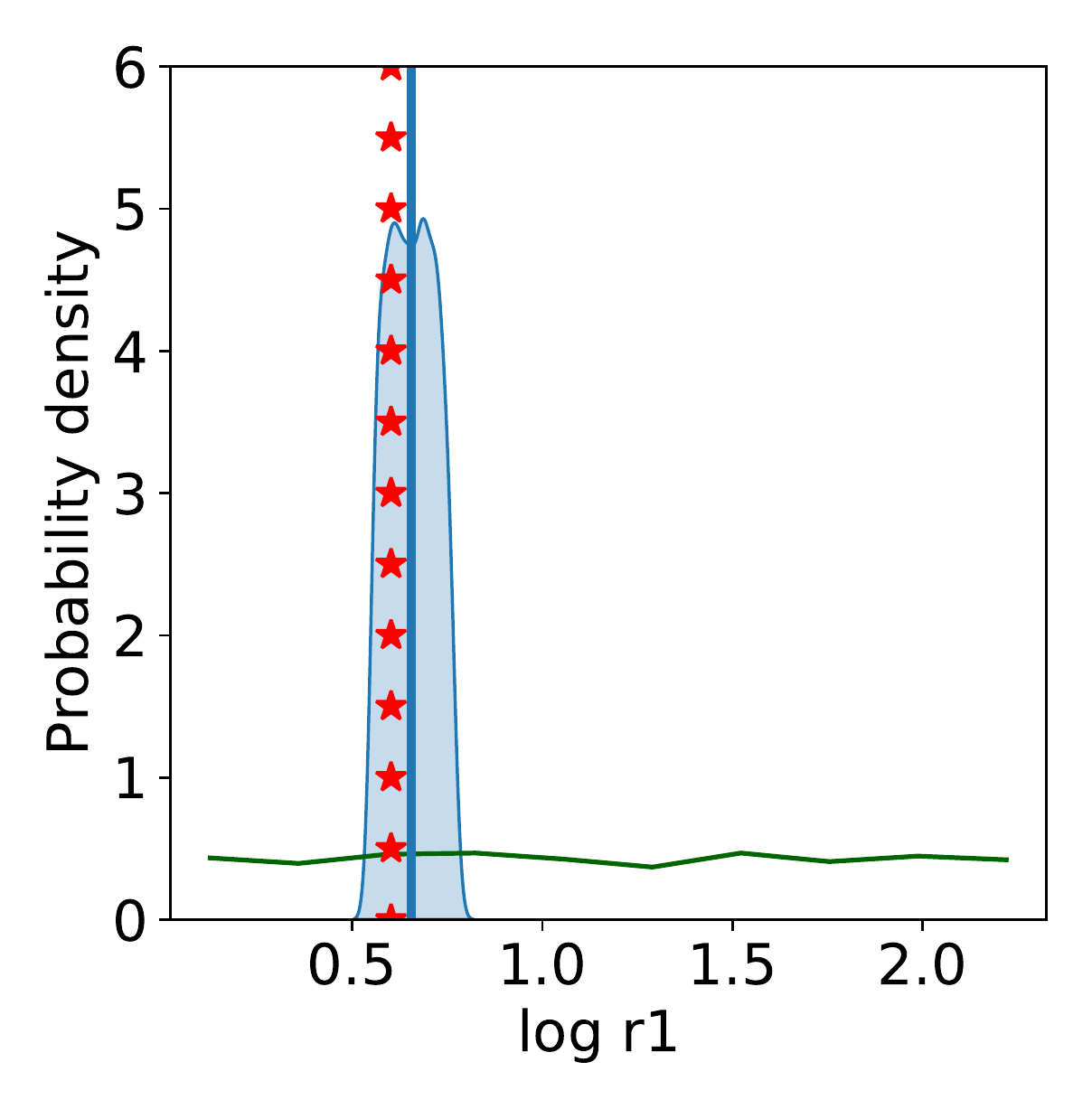} &
		\hspace{-0.2in}
		\includegraphics[scale=0.35]{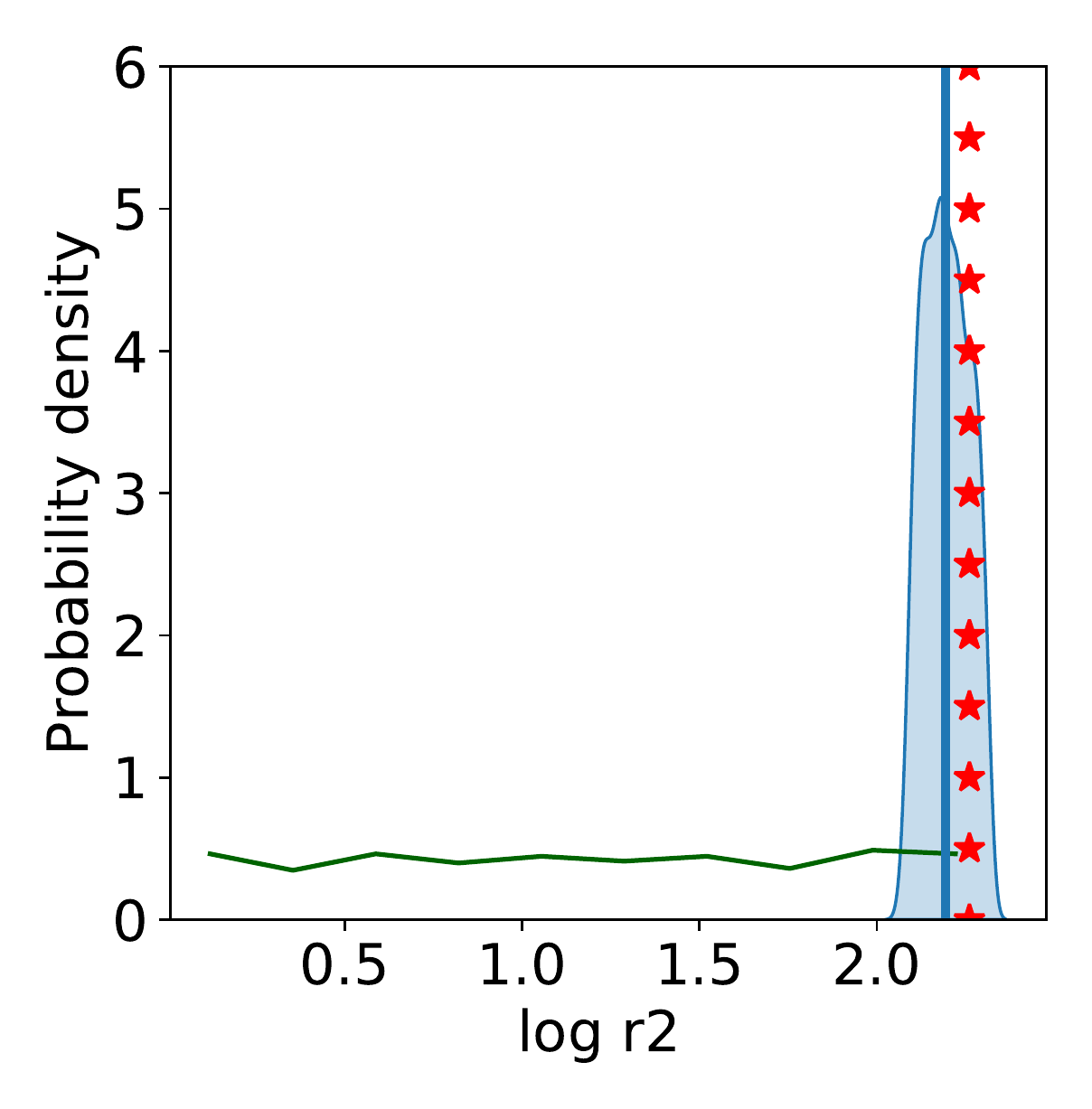} \\
		\end{tabular}
	\end{subfigure}

 \hspace{-0.3in}
 	\begin{subfigure}[normla]{0.5\textwidth}
		\begin{tabular}{ccc}
		\hspace{-0.2in}
		\includegraphics[scale=0.35]{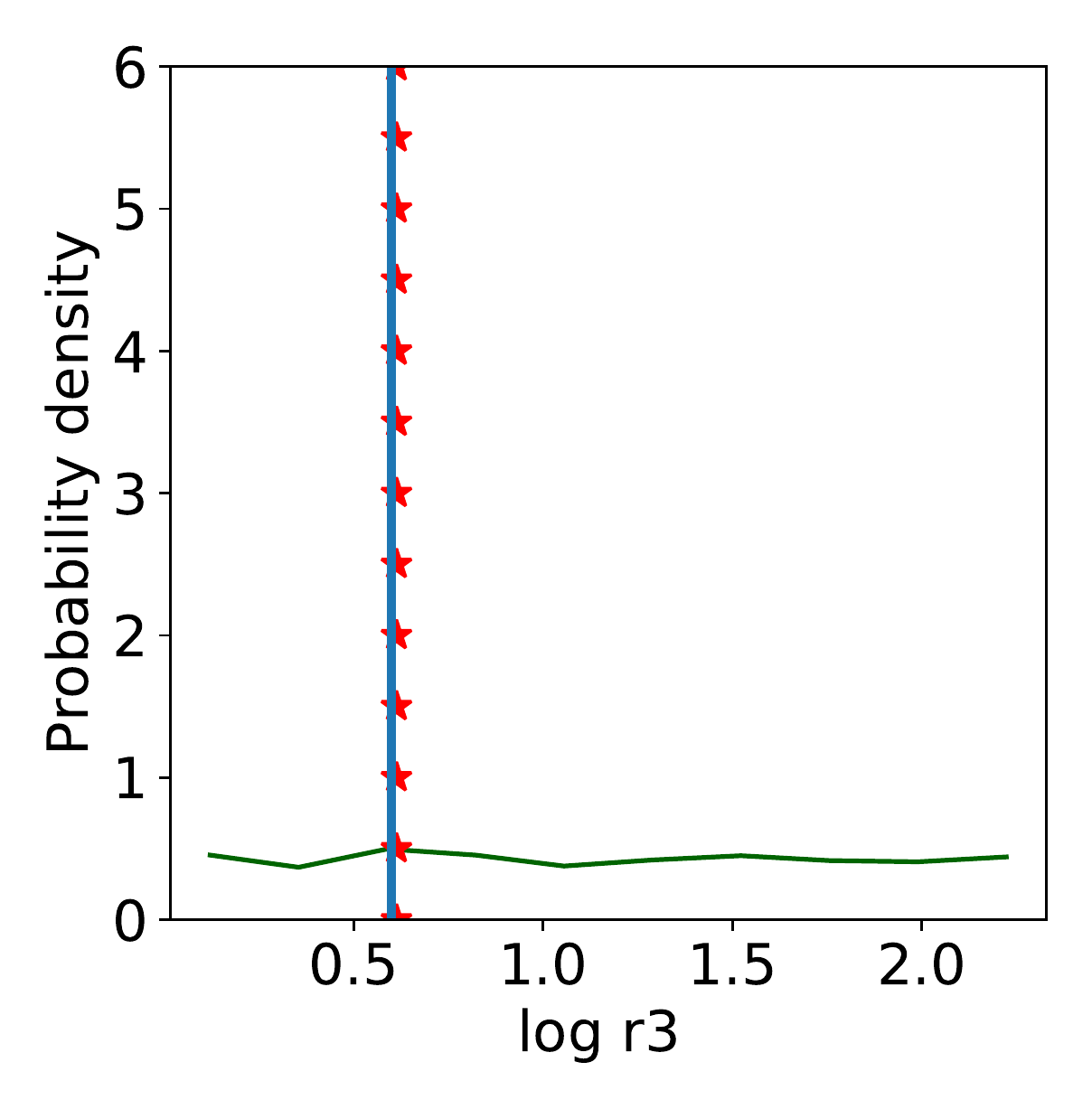} &
		\hspace{-0.2in}
		\includegraphics[scale=0.35]{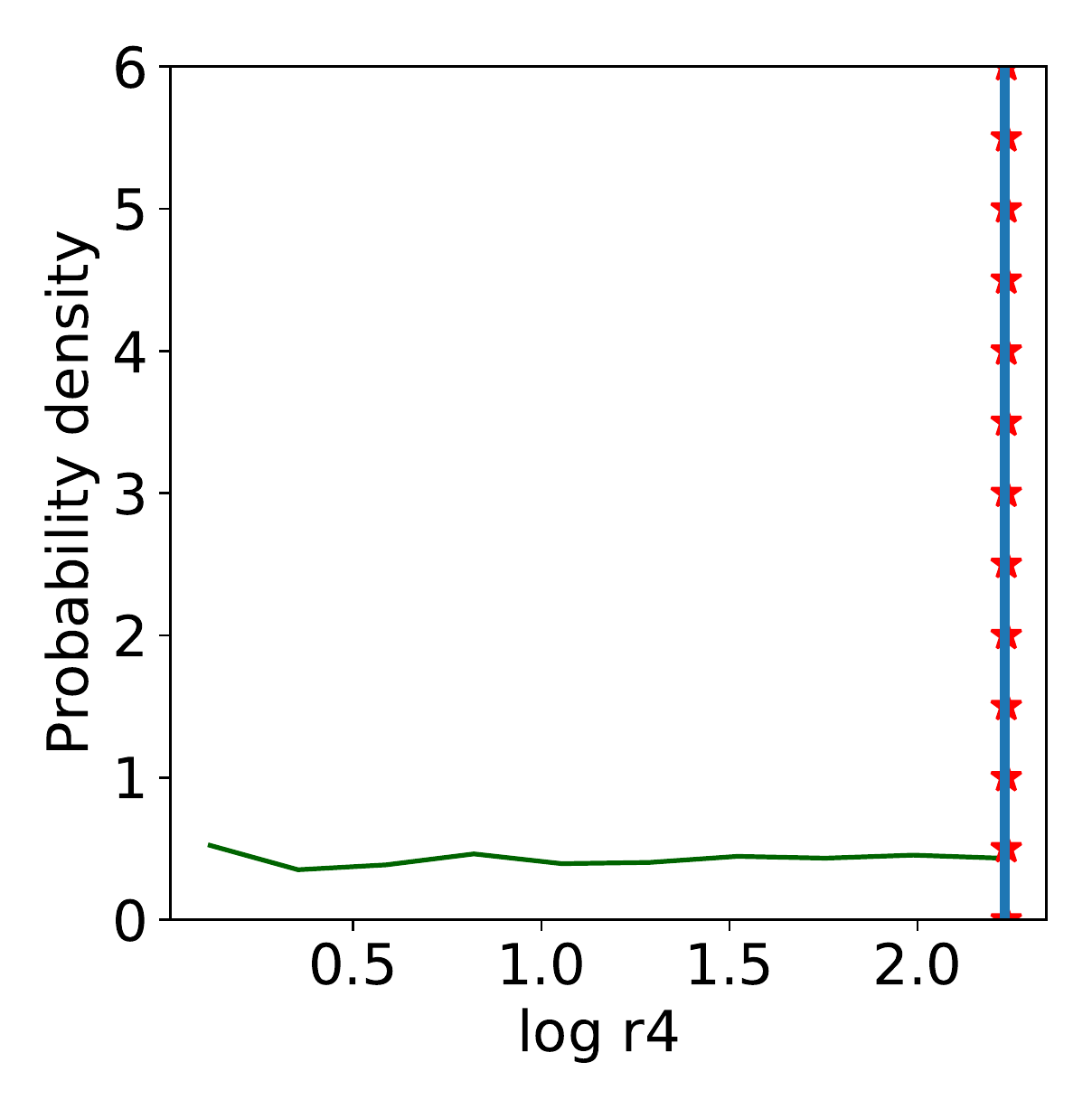} \\ 
		\end{tabular}
	\end{subfigure}
 \hspace{0.000001in}	
	\begin{subfigure}[normla]{0.5\textwidth}
	   \begin{tabular}{cc}
		\includegraphics[scale=0.35]{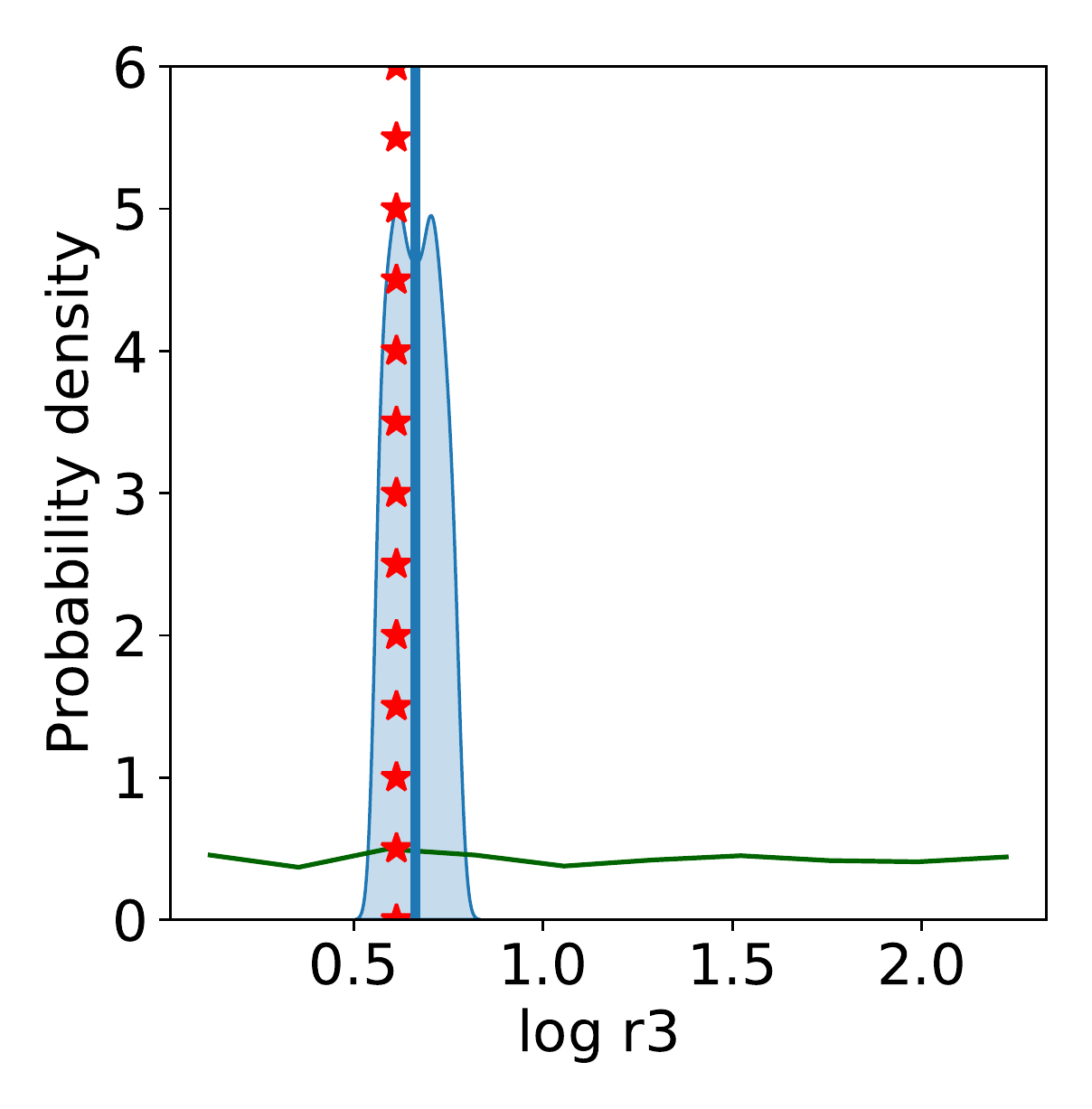} &
		\hspace{-0.2in}
		\includegraphics[scale=0.35]{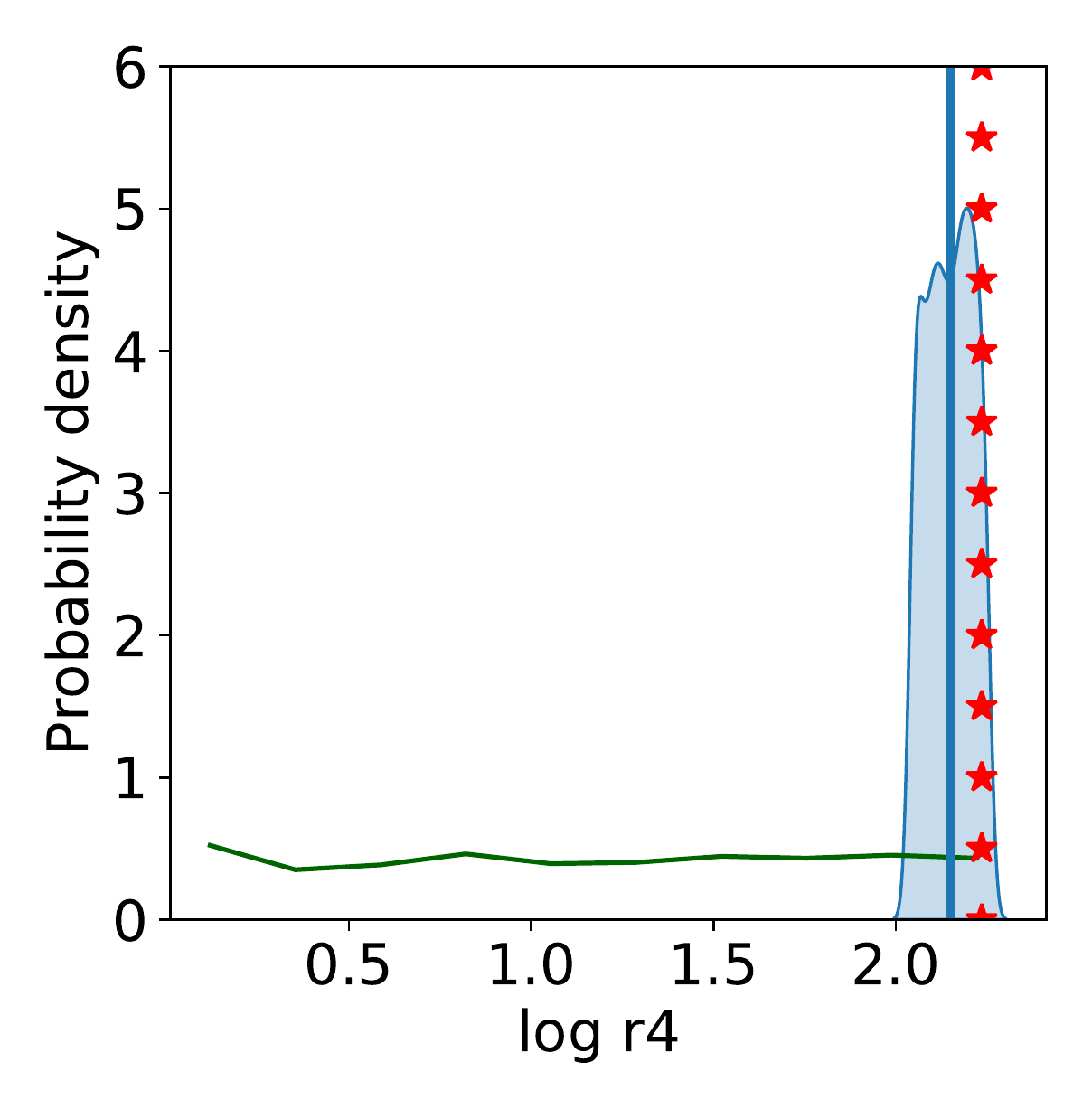} \\
		\end{tabular}
	\end{subfigure}

 \hspace{-0.3in}
 	\begin{subfigure}[normla]{0.5\textwidth}
		\begin{tabular}{ccc}
		\hspace{-0.2in}
		\includegraphics[scale=0.35]{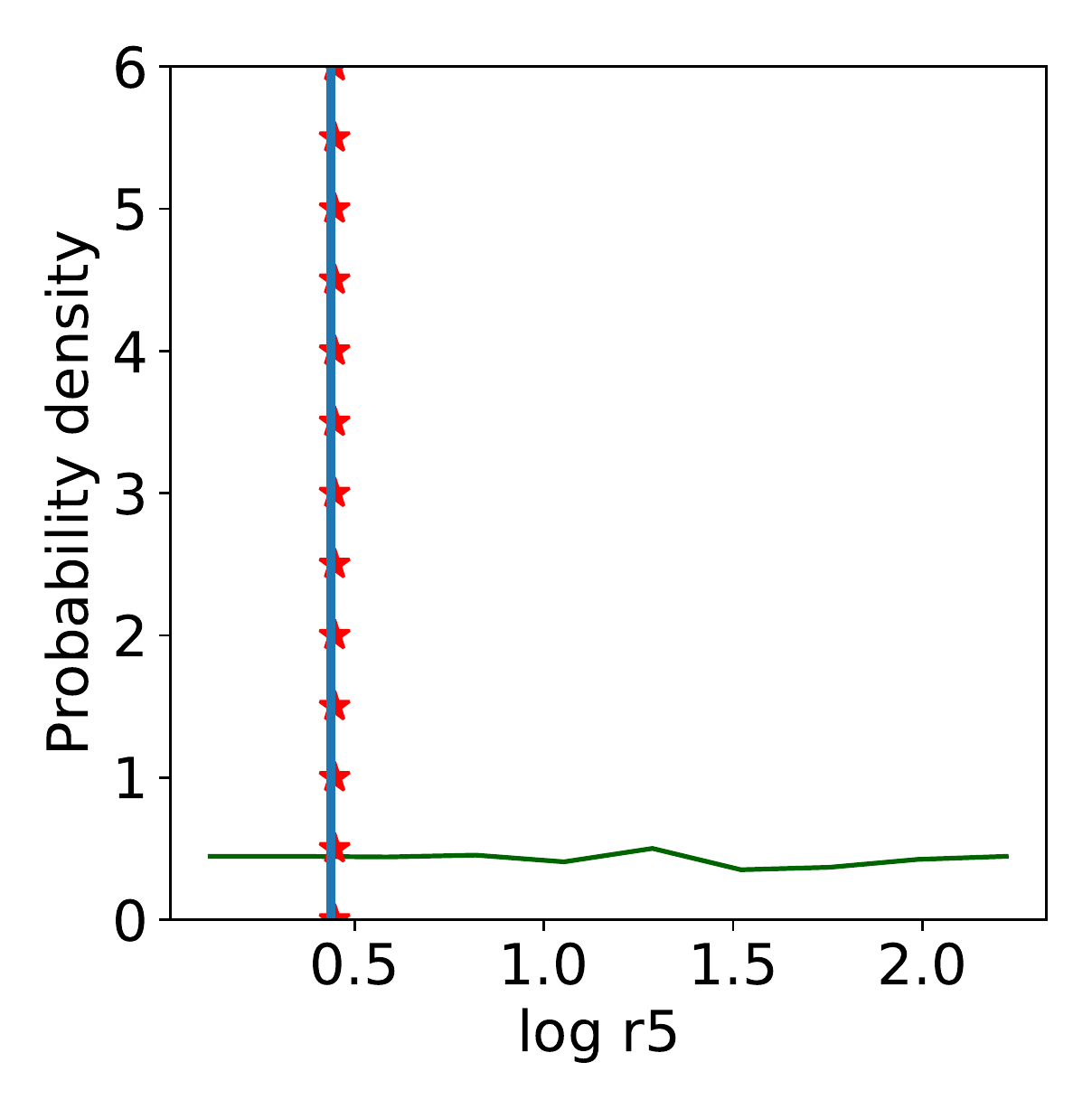} &
		\hspace{-0.2in}
		\includegraphics[scale=0.35]{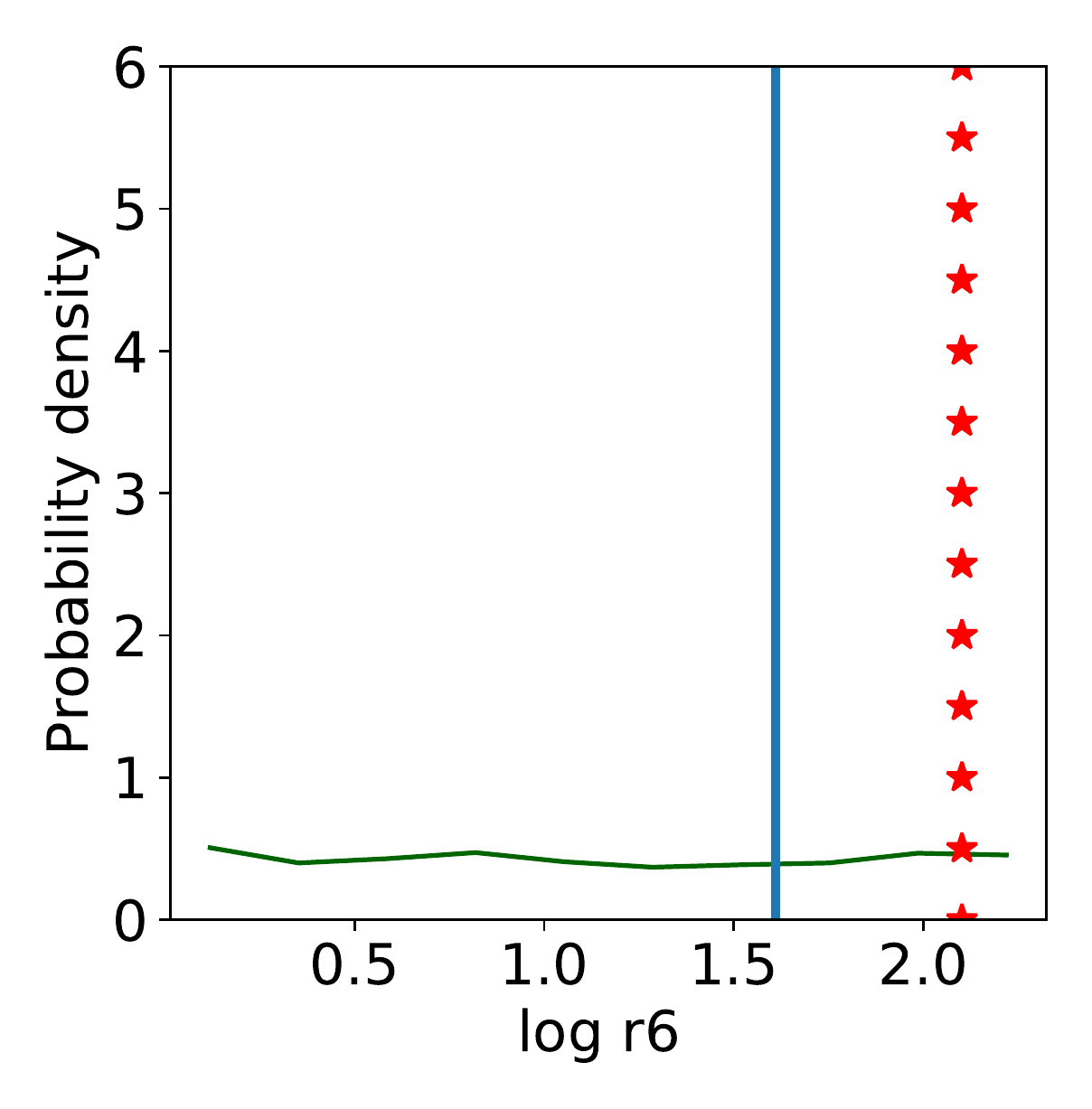} \\ 
		\end{tabular}
	\end{subfigure}
 \hspace{0.000001in}	
	\begin{subfigure}[normla]{0.5\textwidth}
	   \begin{tabular}{cc}
		\includegraphics[scale=0.35]{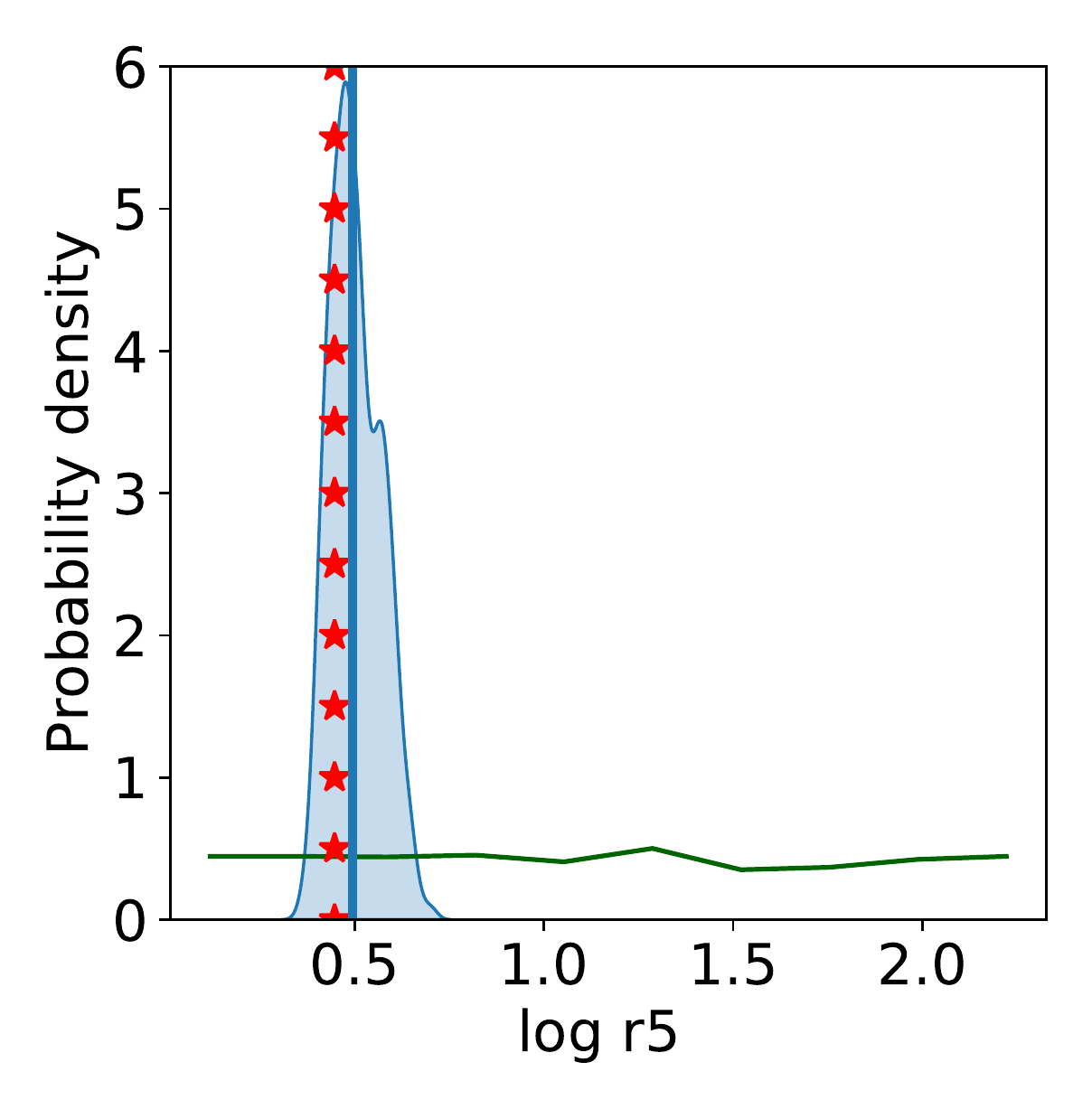} &
		\hspace{-0.2in}
		\includegraphics[scale=0.35]{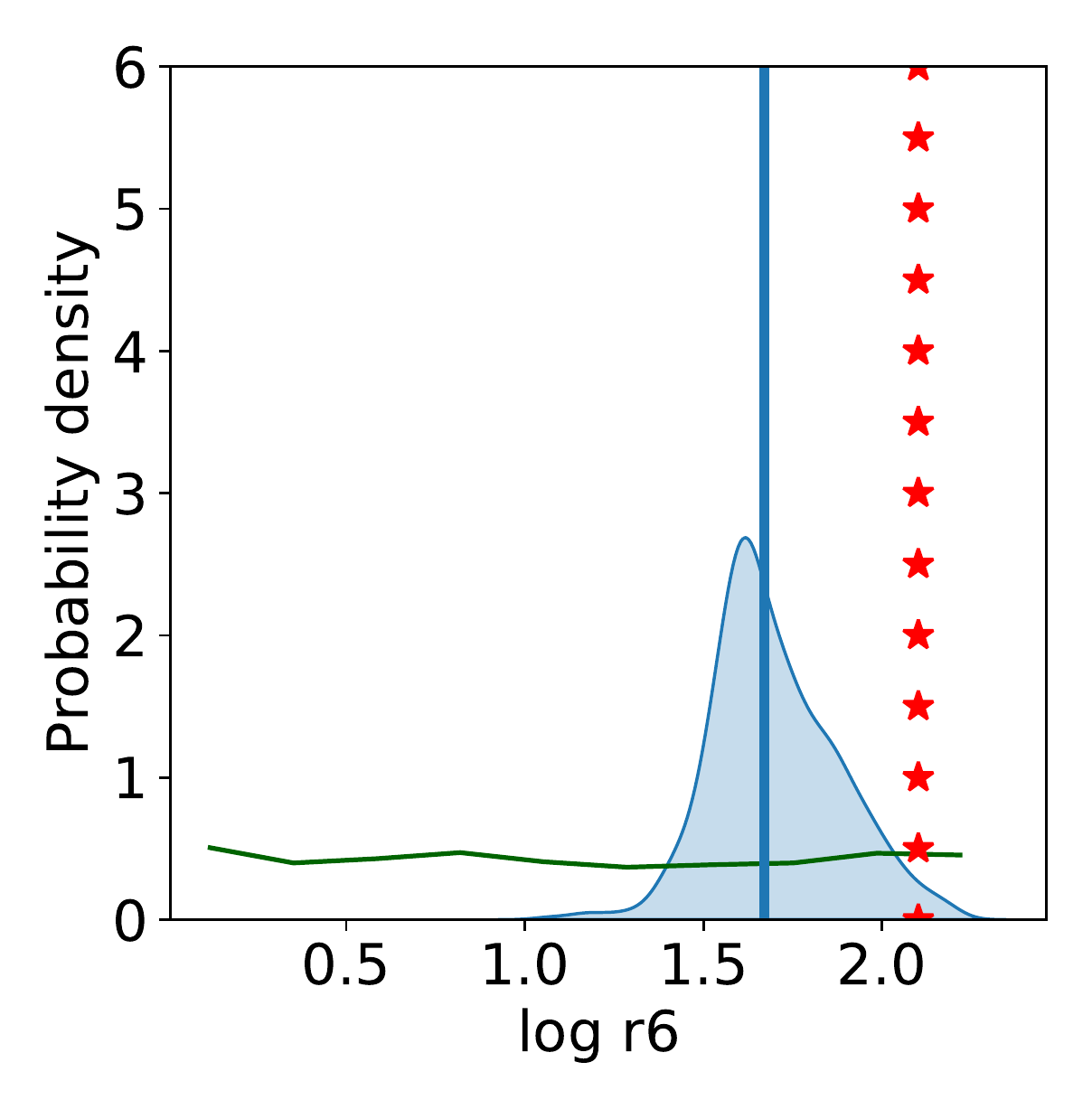} \\
		\end{tabular}
	\end{subfigure}
		
  \hspace{-0.1in}
		 ESMDA
 \hspace{2.5in}
 		 FlexIES
 \end{center}

\caption{Prior and posterior distribution of layers resistivities in the presence of model-errors. Green and blue lines show ensemble approximation of the prior and posterior distribution respectively and red stars show reference truth. Solid blue lines show $p50$ of the posterior distribution. The sub-figures in the first and second columns show results obtained from the ESMDA algorithm and the sub-figures in the third and fourth columns show results from the FlexIES algorithm. The posterior distributions appear as the point estimate in the first and second columns of the sub-figures.}
\label{post_inputs_2_1}
\end{figure}

Figure~(\ref{post_inputs_2_1}) shows the estimation results of the log resistivities in the presence of model-error using the classical and the flexible algorithms. 
The estimated log resistivities of the layers using classical ESMDA are biased due to the model-error. 
Similar to deterministic inversion, for the part of the model where more data are available, the bias is low (log r3, log r4, log r5). The sixth layer resistivity (log r6) significantly deviates from the reference value, which shows very large bias. Since FlexIES can take into account the model-error during the  inversion, 
the resulting posterior distributions of the parameters is more reliable and the posterior distribution of the resistivities cover the reference truth.

\begin{figure}[H]
\begin{center}

 \hspace{-0.5in}
    \begin{subfigure}[normal]{0.4\textwidth}
	\includegraphics[scale=0.5]{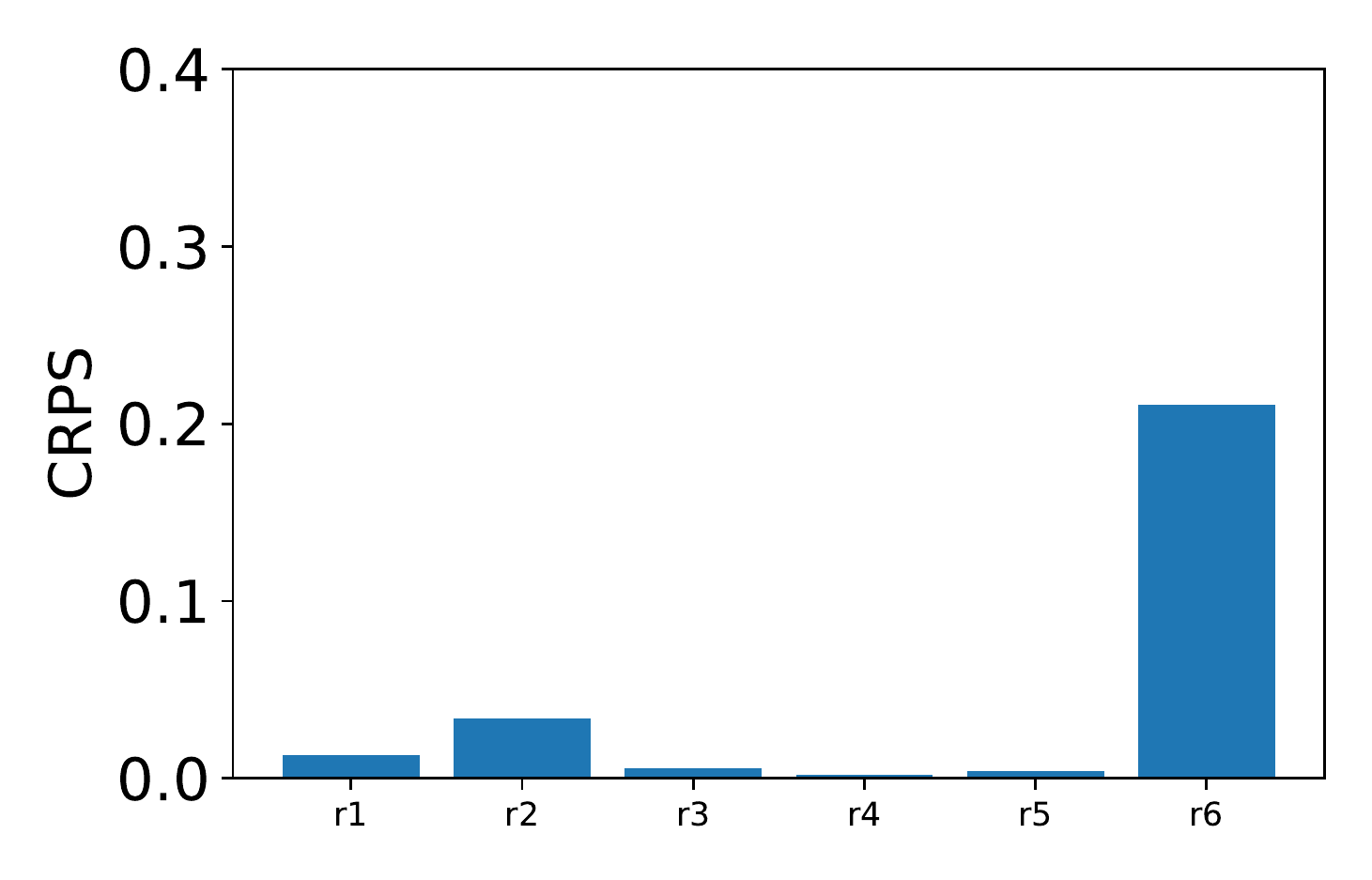}	
	\end{subfigure}
 \hspace{0.5in}
	\begin{subfigure}[normal]{0.4\textwidth}
	\includegraphics[scale=0.5]{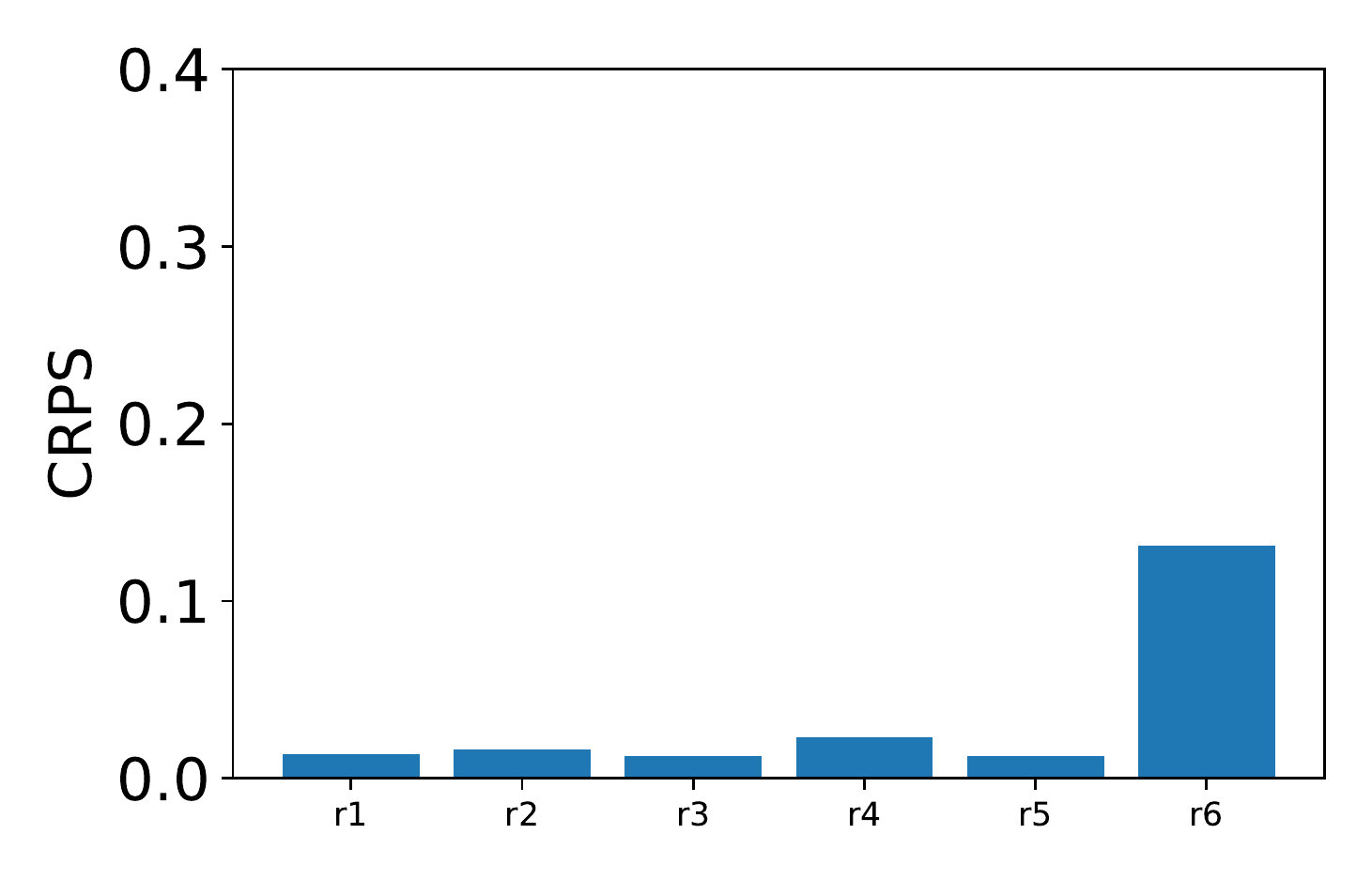}	
	\end{subfigure}

  \hspace{-0.1in}
		 ESMDA
 \hspace{2.5in}
 		 FlexIES
 \end{center}

\caption{Continuous ranked probability score (CRPS) of the layers' log-resistivities obtained from inversion of EM measurements using ESMDA and FlexIES in the presence of model-errors. Lower values are better.}
\label{CRPS_1}
\end{figure}

We further use the continuous ranked probability score (CRPS) in order to quantify the quality of the inversion results. CRPS is a useful metric for the assessment of the probabilistic inversion as it quantifies both precision and accuracy by comparing cumulative density function/distribution (CDF) of the posterior distribution to the observed data or reference truth \citep{hersbach2000decomposition}. The mathematical description of CRPS is listed in the Appendix \ref{sec:AppendixB}. Higher values of CRPS indicate less accurate or less reliable estimation and vice versa. Figure~(\ref{CRPS_1}) shows the comparison of the inversion results for the estimation of log resistivities of the layers. The CRPS of the sixth layer improves significantly using FlexIES as compared to ESMDA due to the inclusion of the model-error during inversion. FlexIES also improves the CRPS of the log resistivities of the first and the second layer as compared to ESMDA. However, ESMDA shows lower CRPS for the third, fourth and fifth layers resistivities, as these parameters are not affected by the model-error and are very close to the reference truth.

\subsection{Joint estimation of layers thicknesses and resistivities}

In this test case, the top boundary location, layers thicknesses and resistivities are jointly estimated by inverting EM measurements in real-time using Algorithms (\ref{alg:A1}) and (\ref{alg:A2}). 
For convenience of the representation, we replace the position of the top boundary by a virtual 'thickness' of the overburden layer defined as distance from 100 m (datum) depth to the top boundary. 
To isolate the influence of model-errors, we make the measurement errors in the observed data negligible (i.e.,~order of magnitude $10^{-9}$). 
Similar to the first case, the inversions are performed using two procedures, i.e.,~considering DNN model as a perfect and an imperfect model. 
Initially the prior realizations of the layers thicknesses (in meter) and log10 resistivities are sampled from the uniform distribution $U \sim [0.3, \quad 20]$ and $U \sim [0, \quad 2.34]$, respectively. 
However, we couldn't get an exact data match even in the absence of model-error.
Thus in the first example of this section, we consider a more narrow (informed) prior, which makes this problem well-posed. This example is useful to evaluate the performance of joint inversion in the presence of the model-error.
In the second example, we use the full uninformed prior.
{\color{black} We show that for the latter case, the proposed methodology can be used to identify and alleviate the multi-modality/local modes of the real-time inversion, by preserving the uncertainty in the posterior.}

\subsubsection{Joint probabilistic inversion for a well-posed problem}

In this section, we consider a well-posed problem of joint inversion of bed boundary positions/layers thicknesses and resistivities simultaneously.
Well-posedness here means that in the absence of model-error, the iterative ensemble smoothers {\color{black} converges to the correct solutions.} {\color{black} To achieve well-posedness, we select an informed uniform prior by trial and error approach and select the following min-max ranges for the six layers thicknesses}:
$min = [2,0.3,0.3,7,2,4]$ meter and $max=[10,4,4,11,7,10]$ meter.
The real-time inversions are performed in the presence of the model-error, i.e., we take the reference/observed data of electromagnetic measurements from the high-fidelity log simulator. The prior realizations of the layers thickness are sampled from the informed uniform distribution $U \sim [(2,0.3,0.3,7,2,4), \quad (10,4,4,11,7,10)]$.

\begin{figure}[H]
\begin{center}

 \hspace{-0.5in}
    \begin{subfigure}[normal]{0.4\textwidth}
	\includegraphics[scale=0.5]{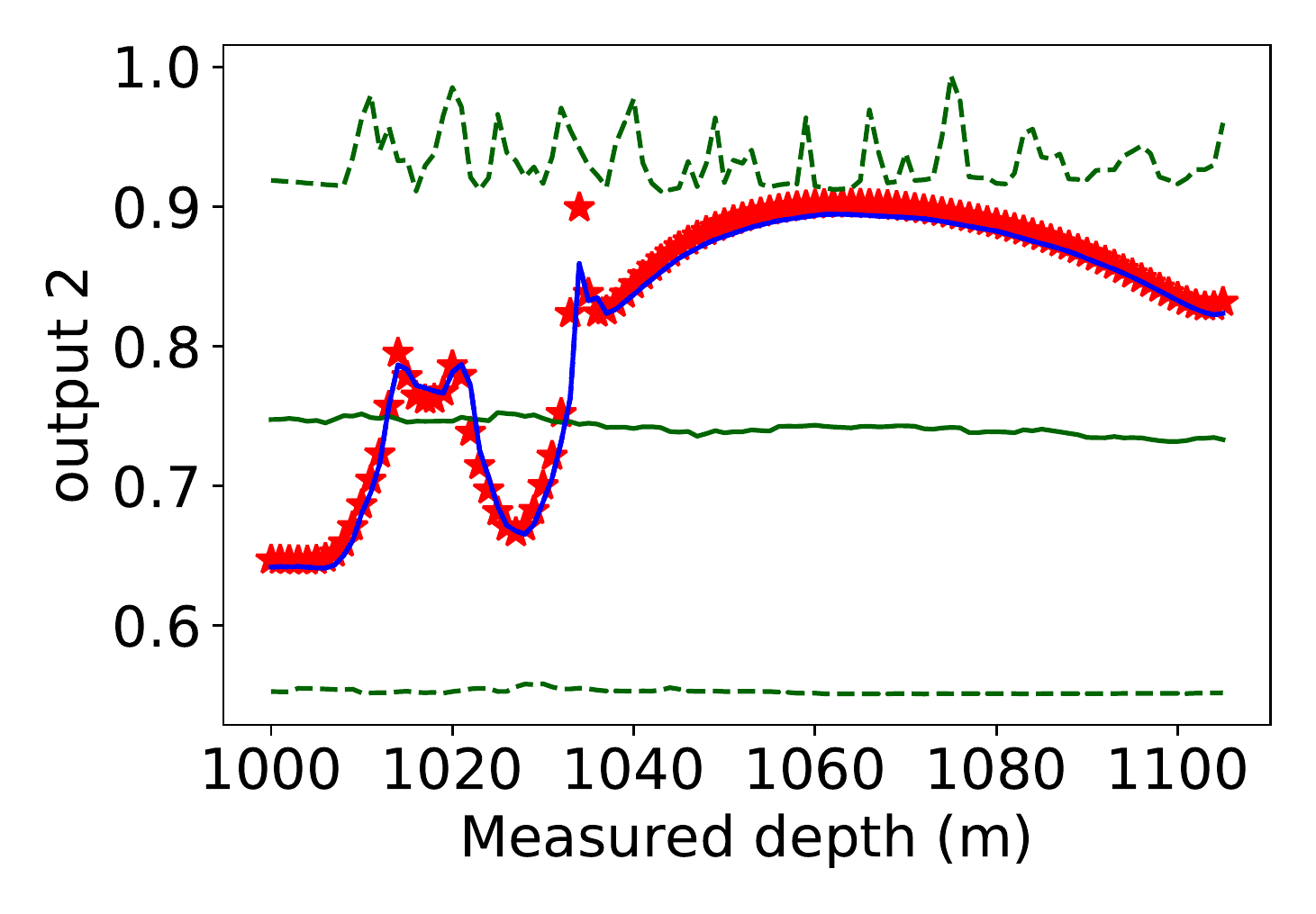}	
	\end{subfigure}
 \hspace{0.5in}
	\begin{subfigure}[normal]{0.4\textwidth}
	\includegraphics[scale=0.5]{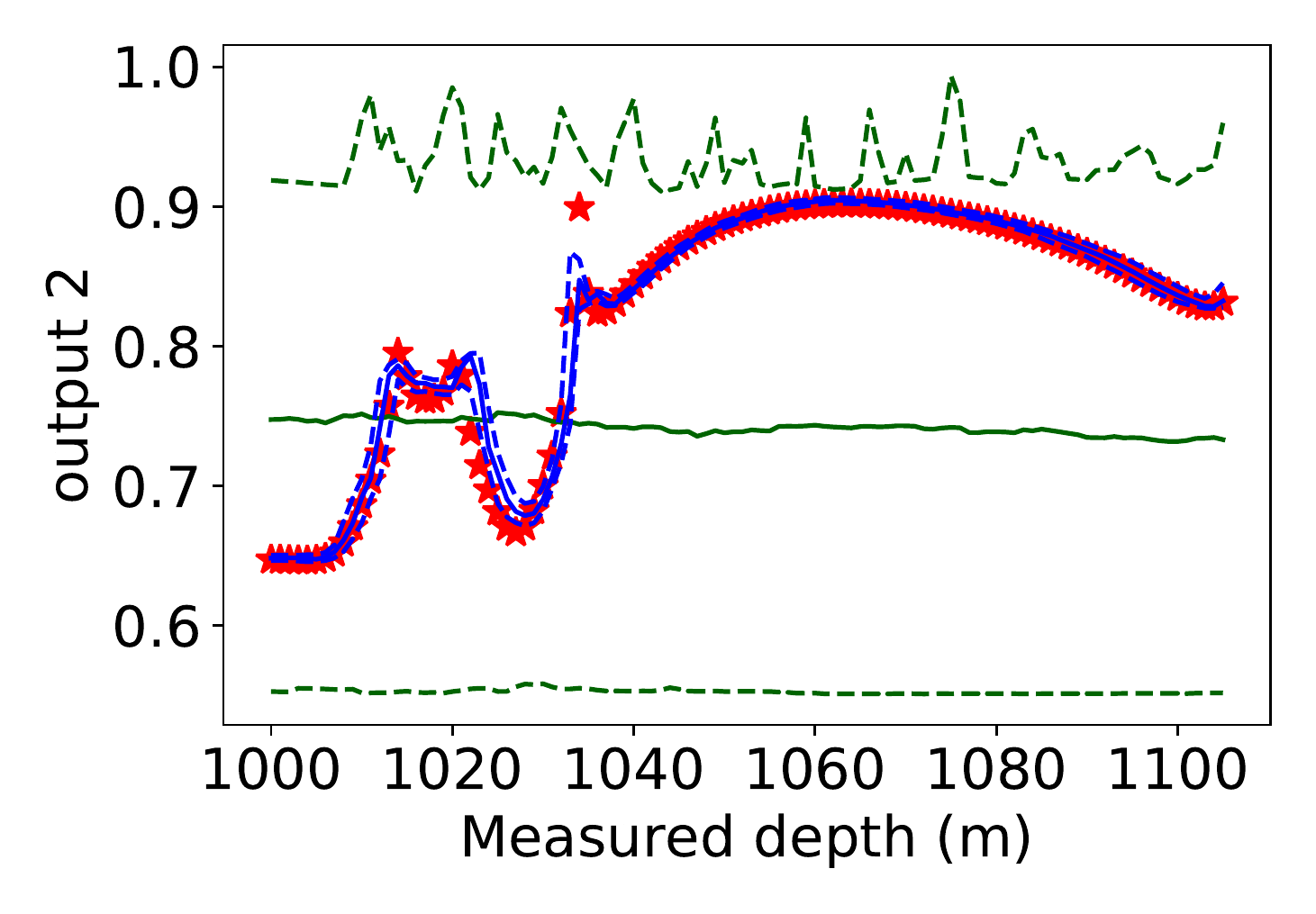}	
	\end{subfigure}

 \hspace{-0.5in}
    \begin{subfigure}[normal]{0.4\textwidth}
	\includegraphics[scale=0.5]{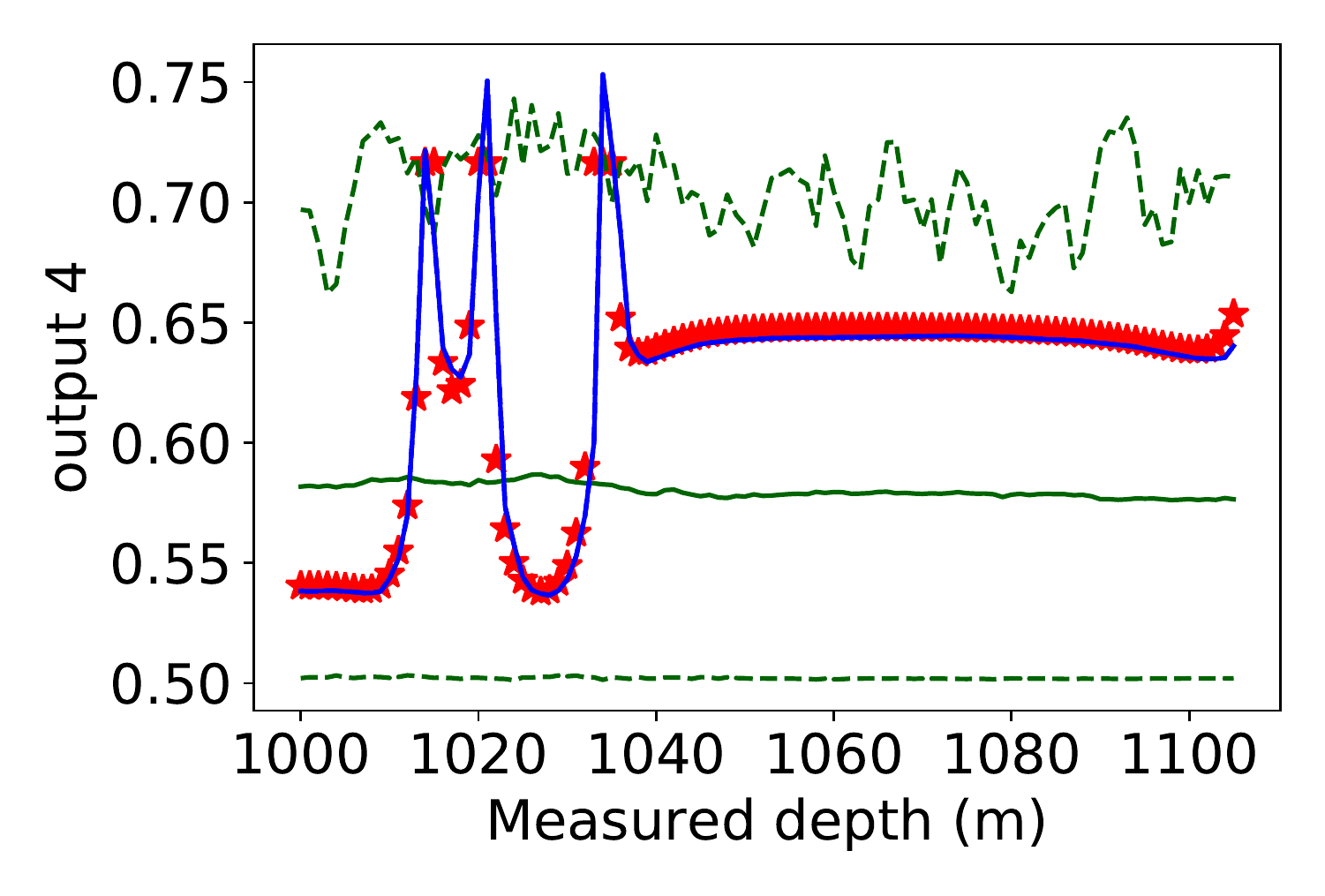}	
	\end{subfigure}
 \hspace{0.5in}
	\begin{subfigure}[normal]{0.4\textwidth}
	\includegraphics[scale=0.5]{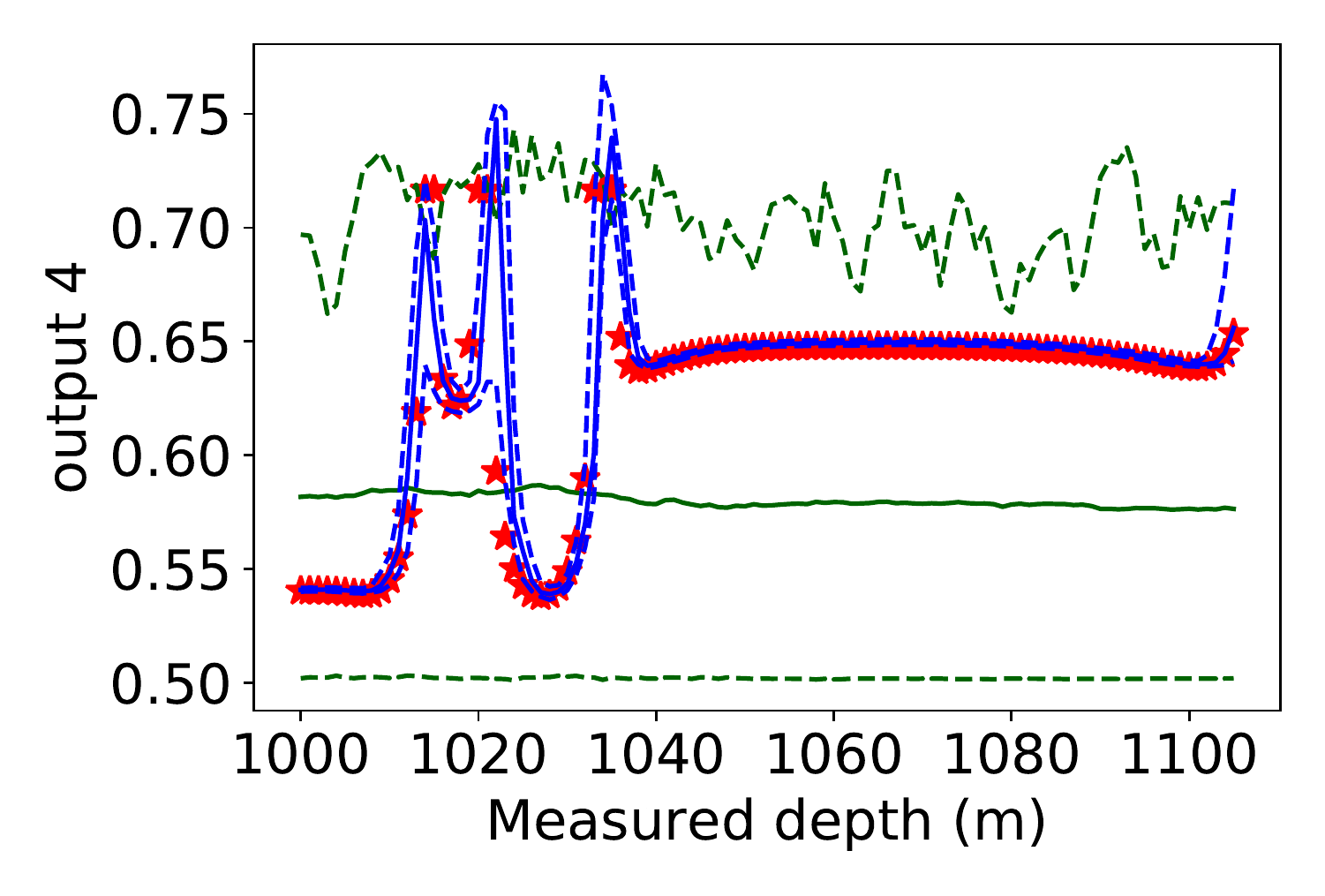}	
	\end{subfigure}

 \hspace{-0.5in}
    \begin{subfigure}[normal]{0.4\textwidth}
	\includegraphics[scale=0.5]{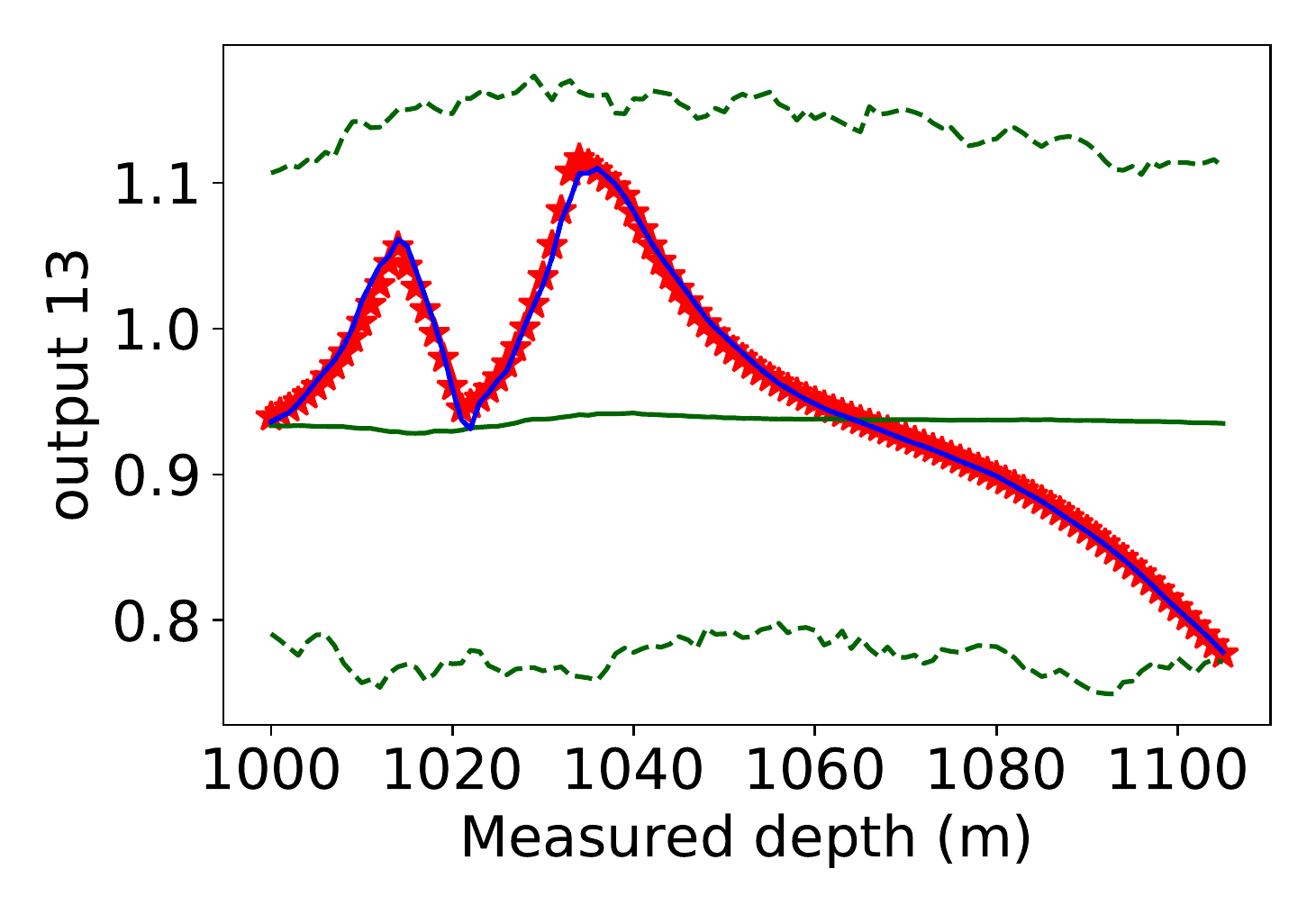}	
	\end{subfigure}
 \hspace{0.5in}
	\begin{subfigure}[normal]{0.4\textwidth}
	\includegraphics[scale=0.5]{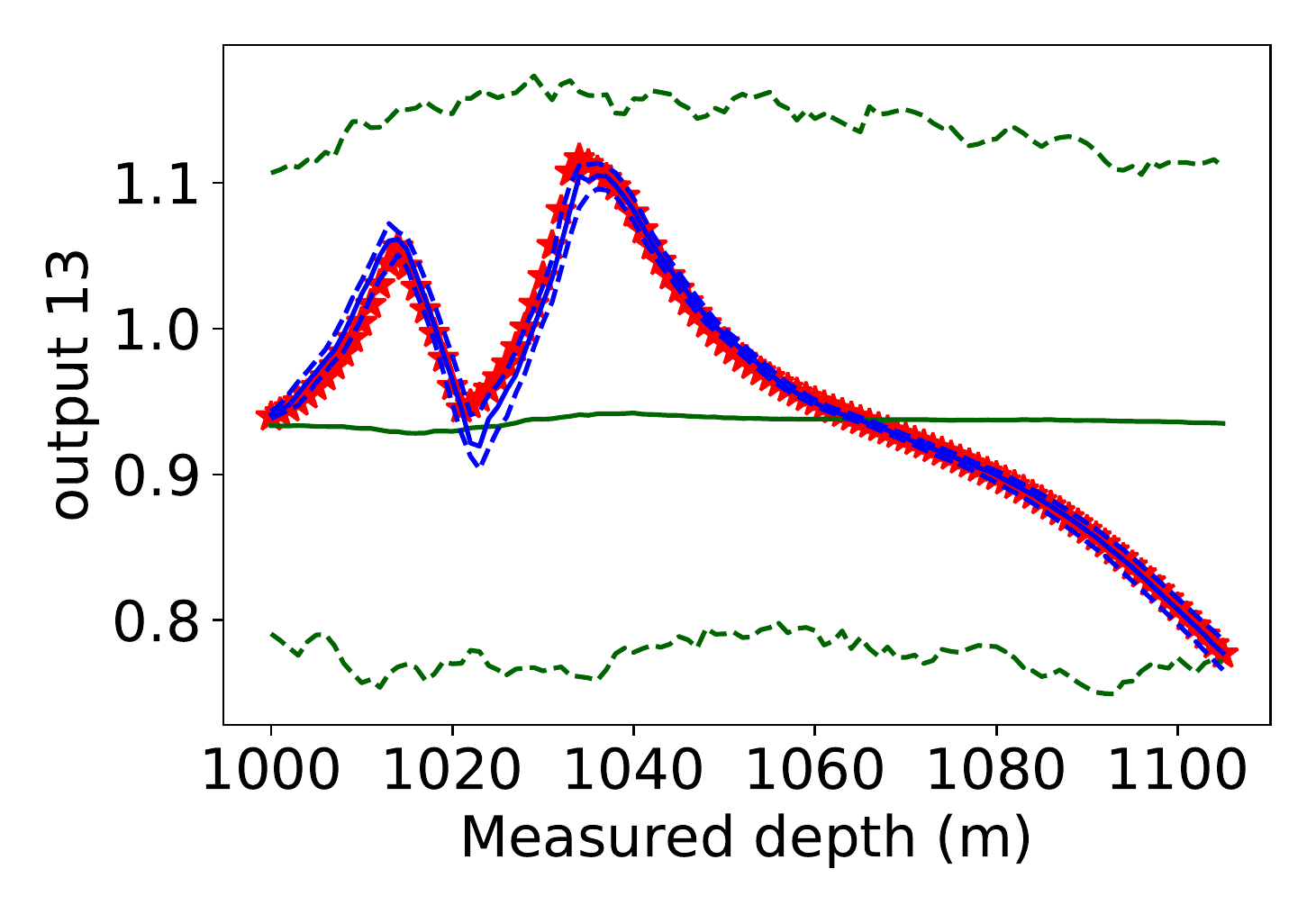}	
	\end{subfigure}

 \hspace{-0.1in}
		 ESMDA
 \hspace{2.5in}
 		 FlexIES
 \end{center}

\caption{Prior and posterior distribution of EM outputs for joint inversion in presence of model-errors as functions of measured depth. Green and blue lines show ensemble approximation of the prior and posterior distribution respectively and red stars show observed EM measurements. Solid green and solid blue lines show $p50$ and dashed green and dashed blue lines show $99\%$ confidence interval respectively of the prior and posterior distribution respectively. The sub-figures in the first column show results obtained from the ESMDA algorithm and the sub-figures in the second column show results from the FlexIES algorithm. In first column of the sub-figures posterior distribution appears as the point estimate therefore solid blue lines overlaps dashed blue lines.}
\label{post_outputs_4_1}
\end{figure}

\begin{figure}[H]
\begin{center}

\hspace{-0.5in}
 	\begin{subfigure}[normla]{0.5\textwidth}
		\begin{tabular}{ccc}
		\hspace{-0.2in}
		\includegraphics[scale=0.35]{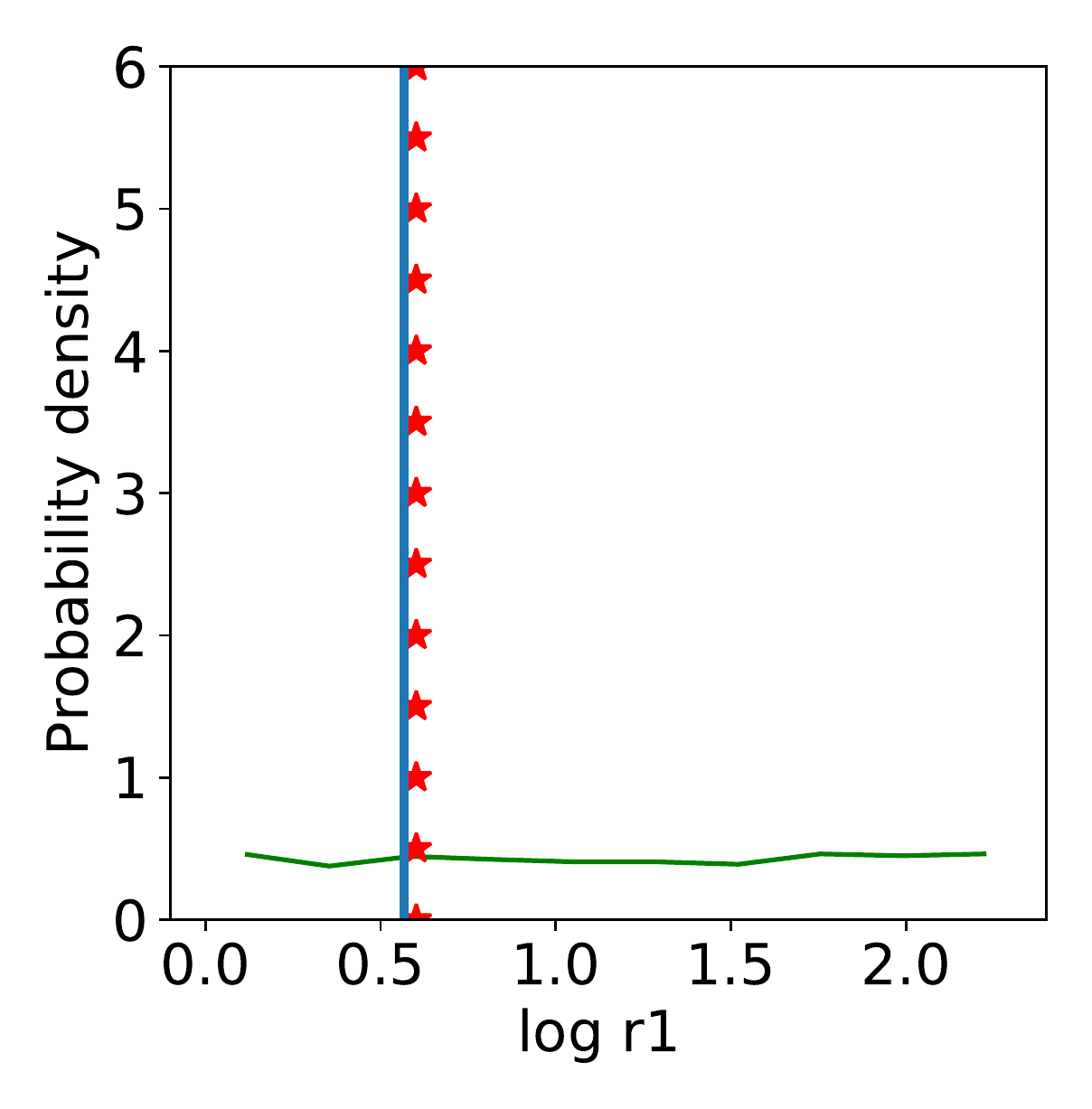} &
		\hspace{-0.2in}
		\includegraphics[scale=0.35]{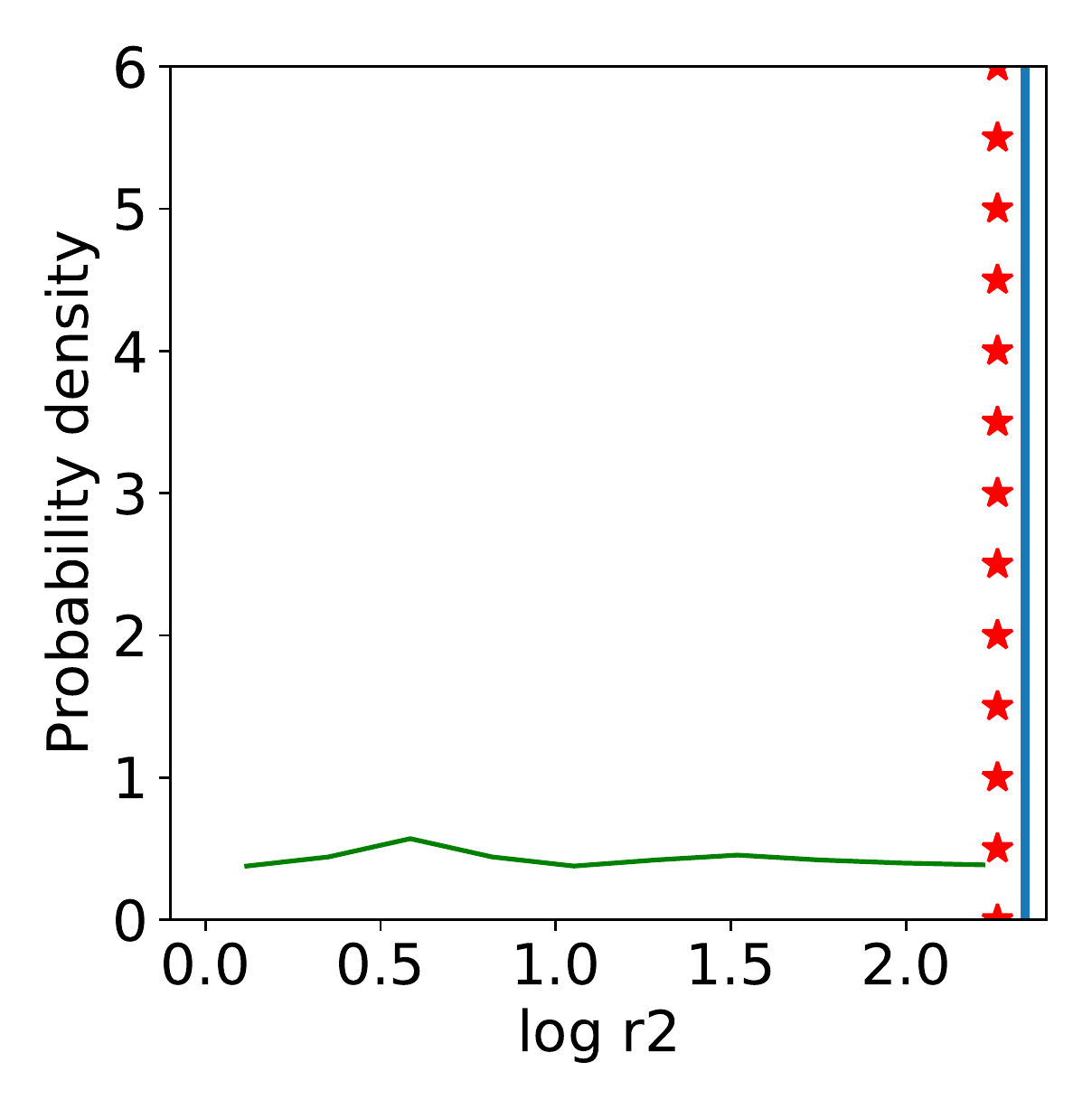} \\ 
		\end{tabular}
	\end{subfigure}
 \hspace{0.2in}	
	\begin{subfigure}[normla]{0.5\textwidth}
	   \begin{tabular}{cc}
		\includegraphics[scale=0.35]{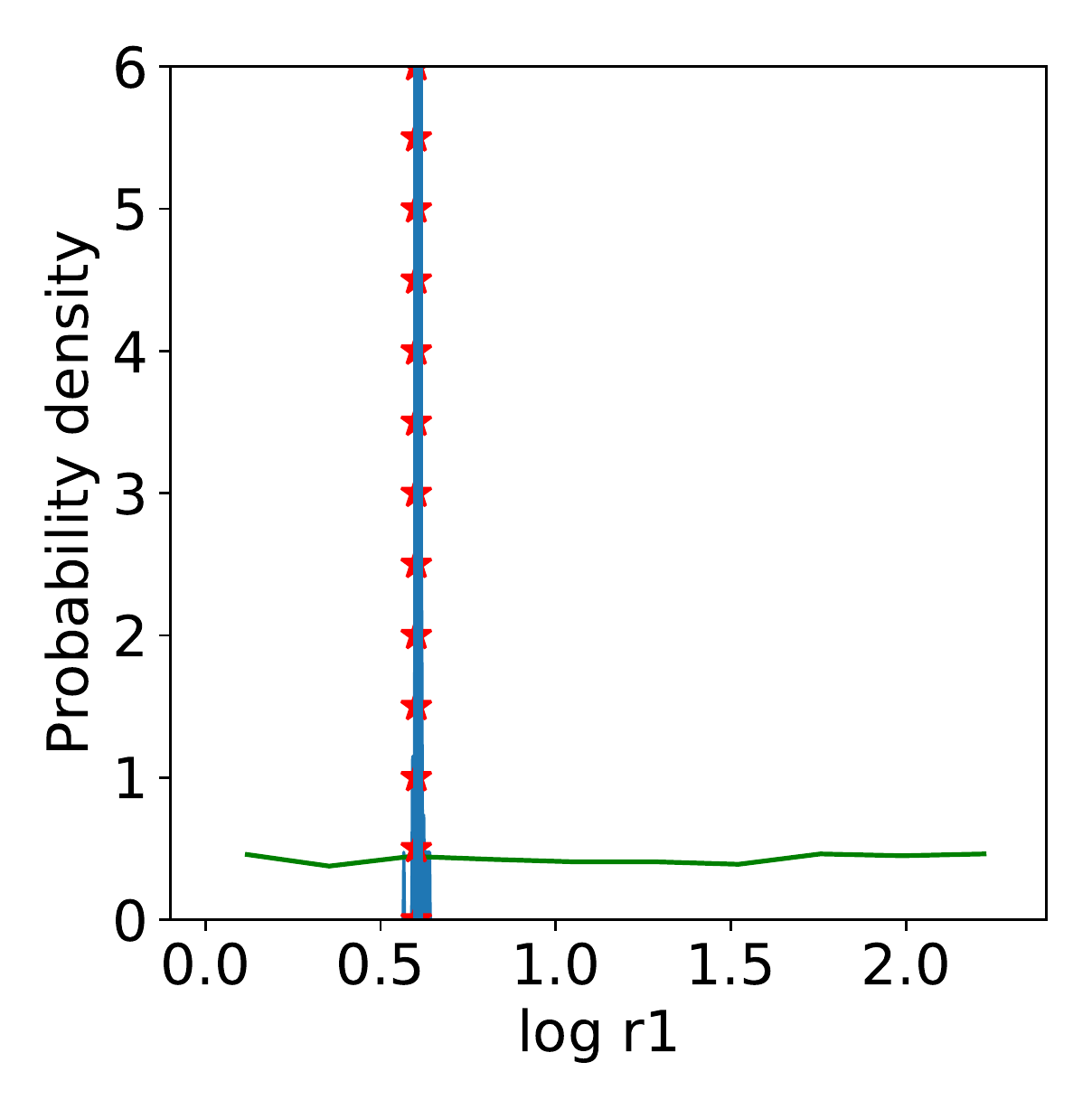} &
		\hspace{-0.2in}
		\includegraphics[scale=0.35]{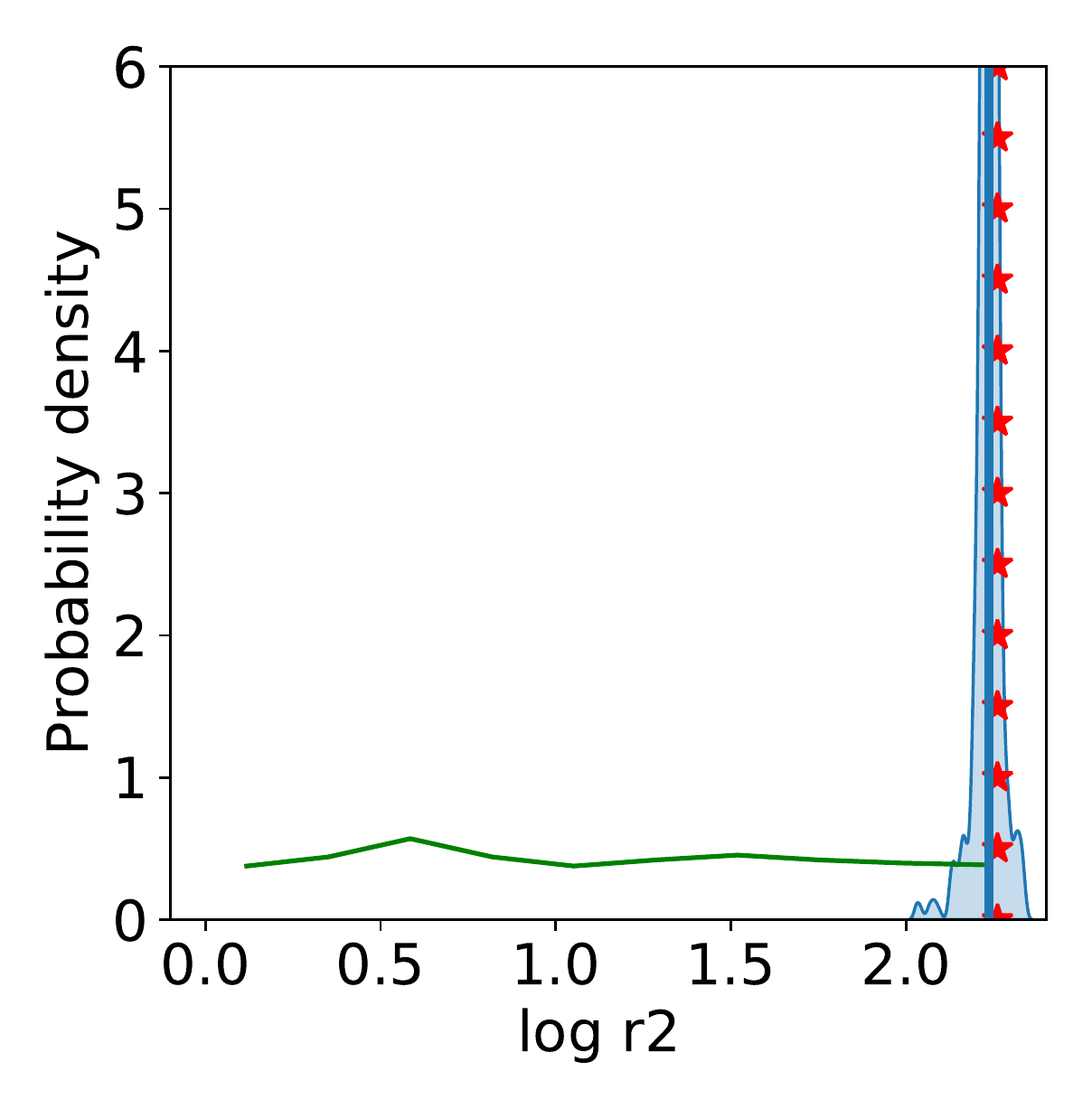} \\
		\end{tabular}
	\end{subfigure}

 \hspace{-0.5in}
 	\begin{subfigure}[normla]{0.5\textwidth}
		\begin{tabular}{ccc}
		\hspace{-0.2in}
		\includegraphics[scale=0.35]{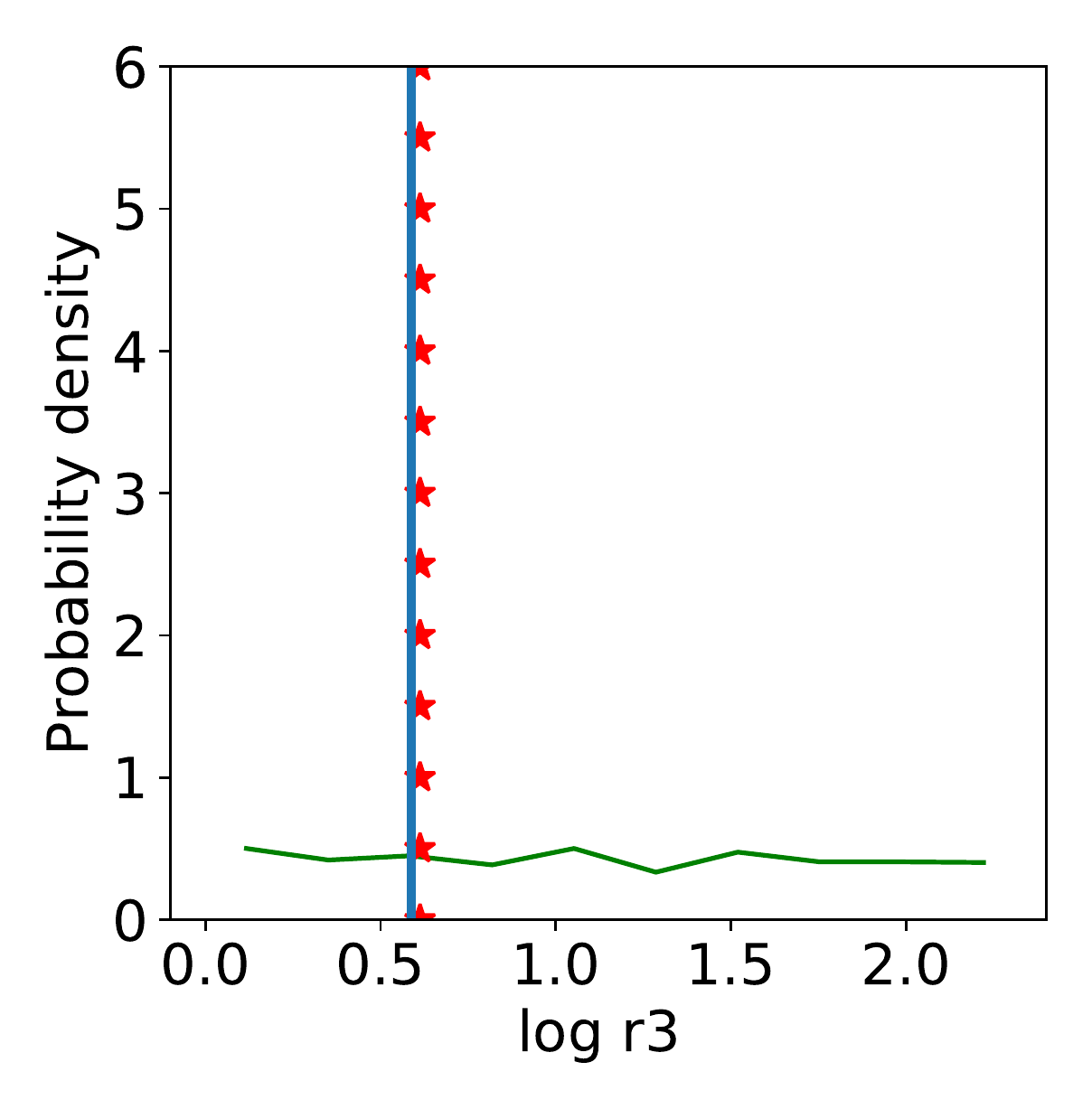} &
		\hspace{-0.2in}
		\includegraphics[scale=0.35]{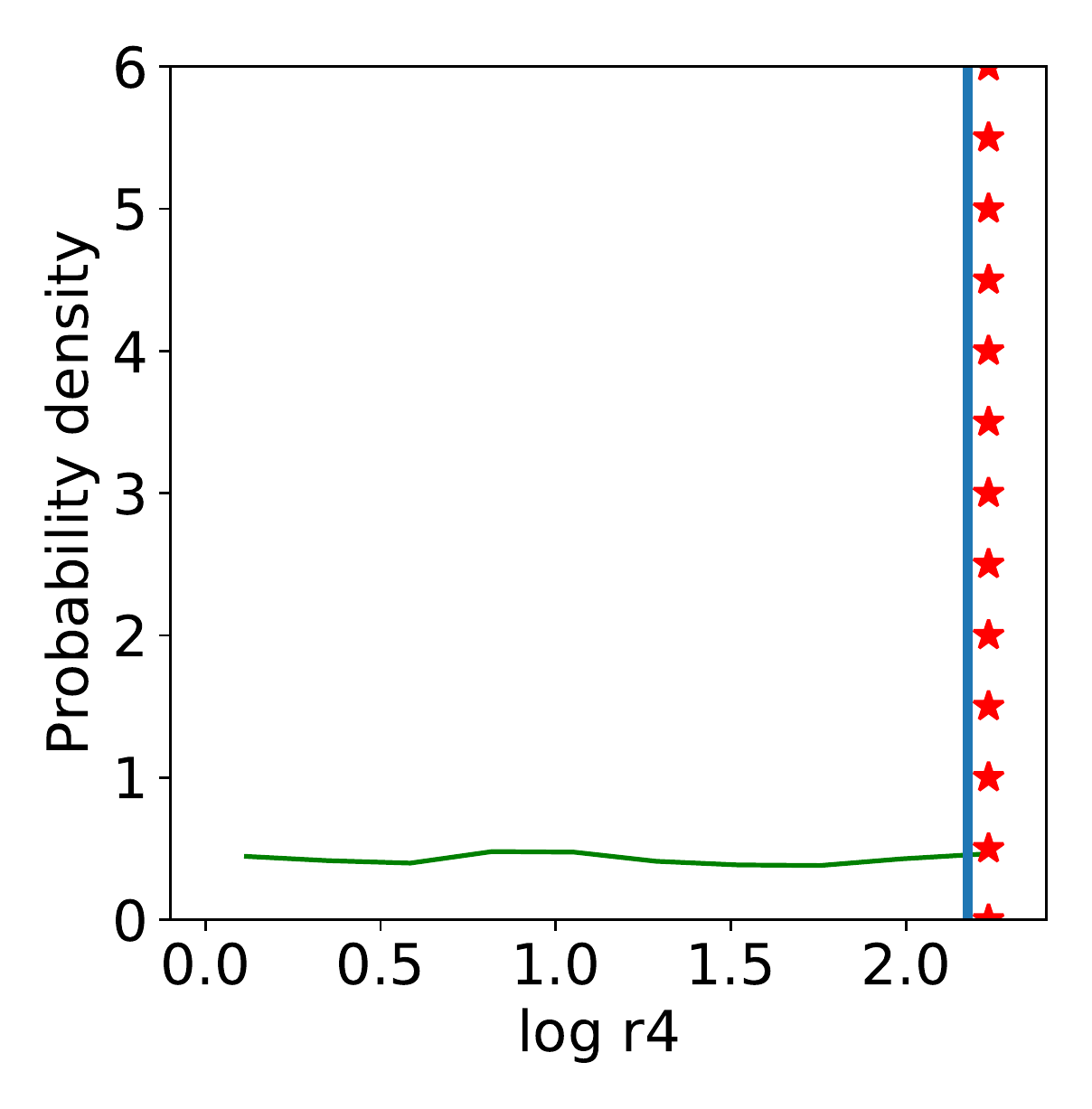} \\ 
		\end{tabular}
	\end{subfigure}
 \hspace{0.2in}	
	\begin{subfigure}[normla]{0.5\textwidth}
	   \begin{tabular}{cc}
		\includegraphics[scale=0.35]{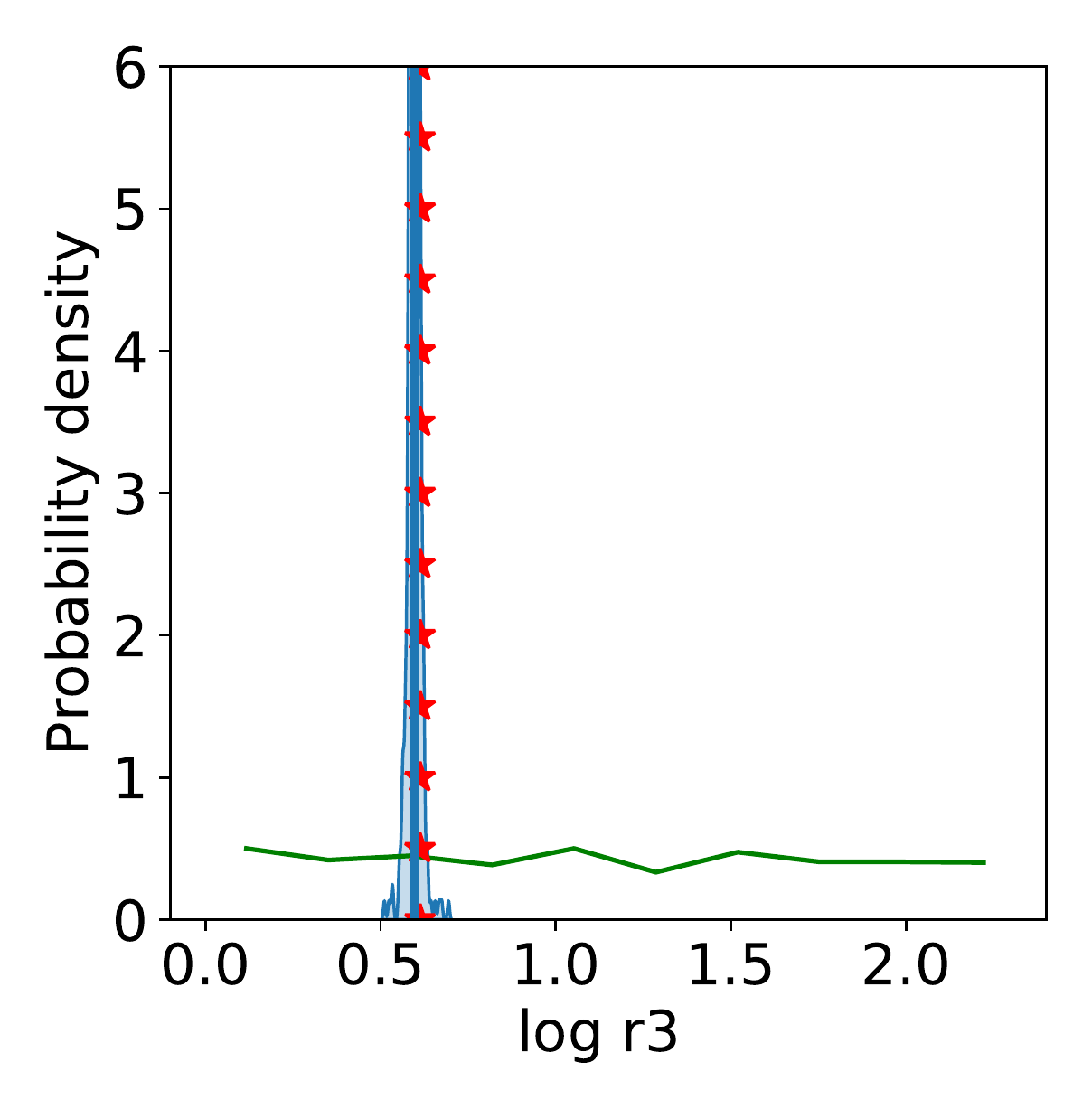} &
		\hspace{-0.2in}
		\includegraphics[scale=0.35]{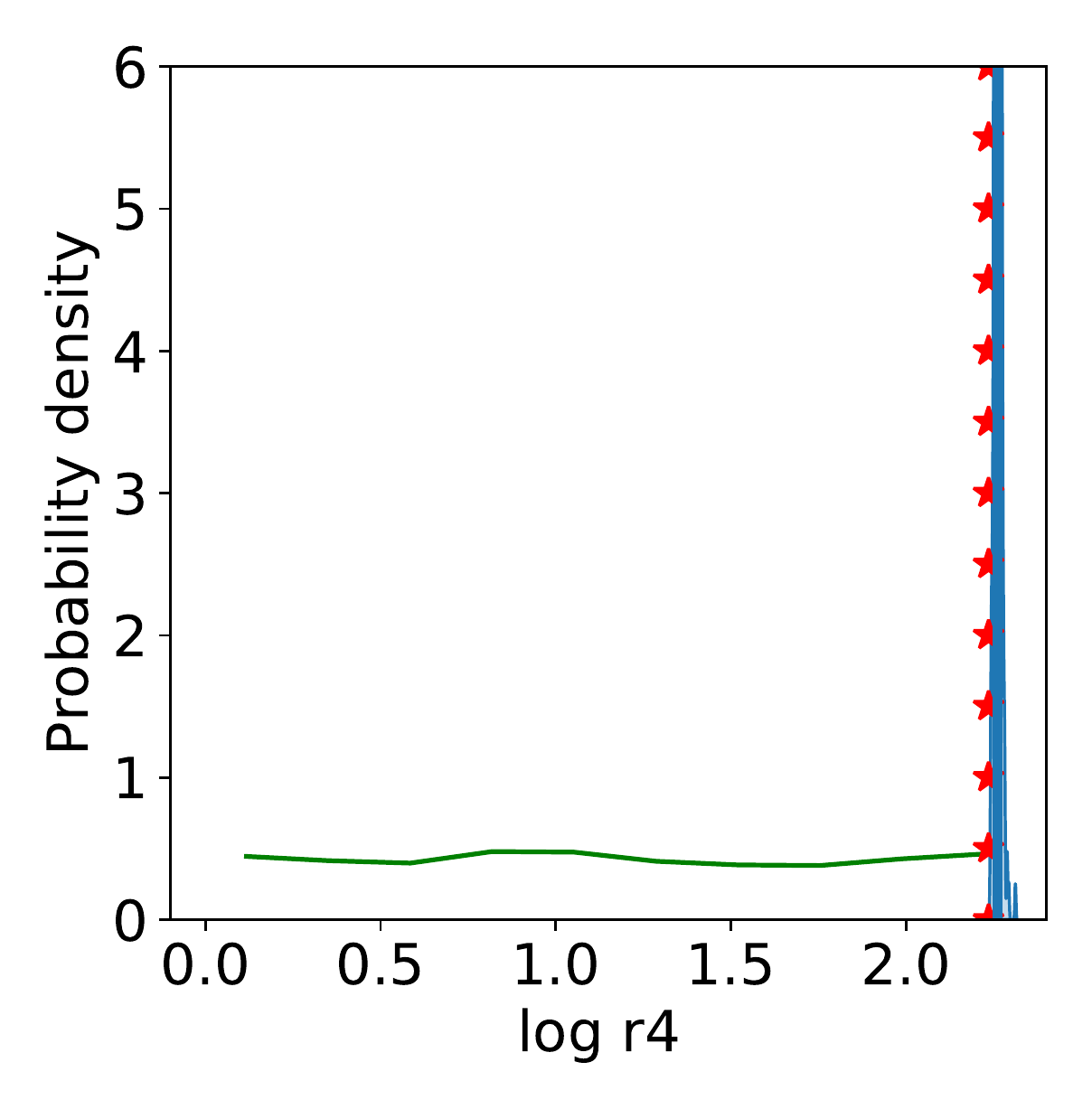} \\
		\end{tabular}
	\end{subfigure}

 \hspace{-0.5in}
 	\begin{subfigure}[normla]{0.5\textwidth}
		\begin{tabular}{ccc}
		\hspace{-0.2in}
		\includegraphics[scale=0.35]{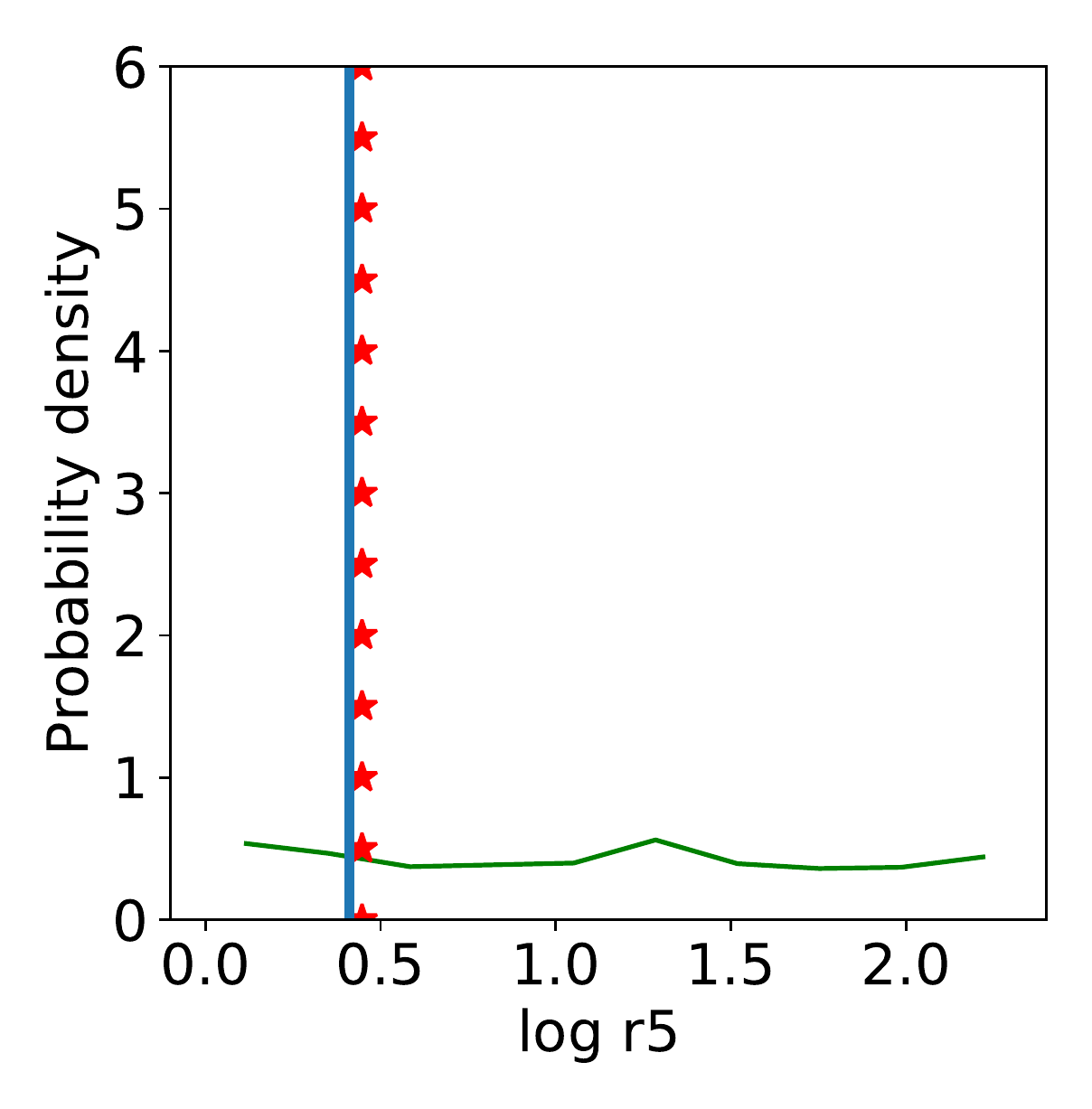} &
		\hspace{-0.2in}
		\includegraphics[scale=0.35]{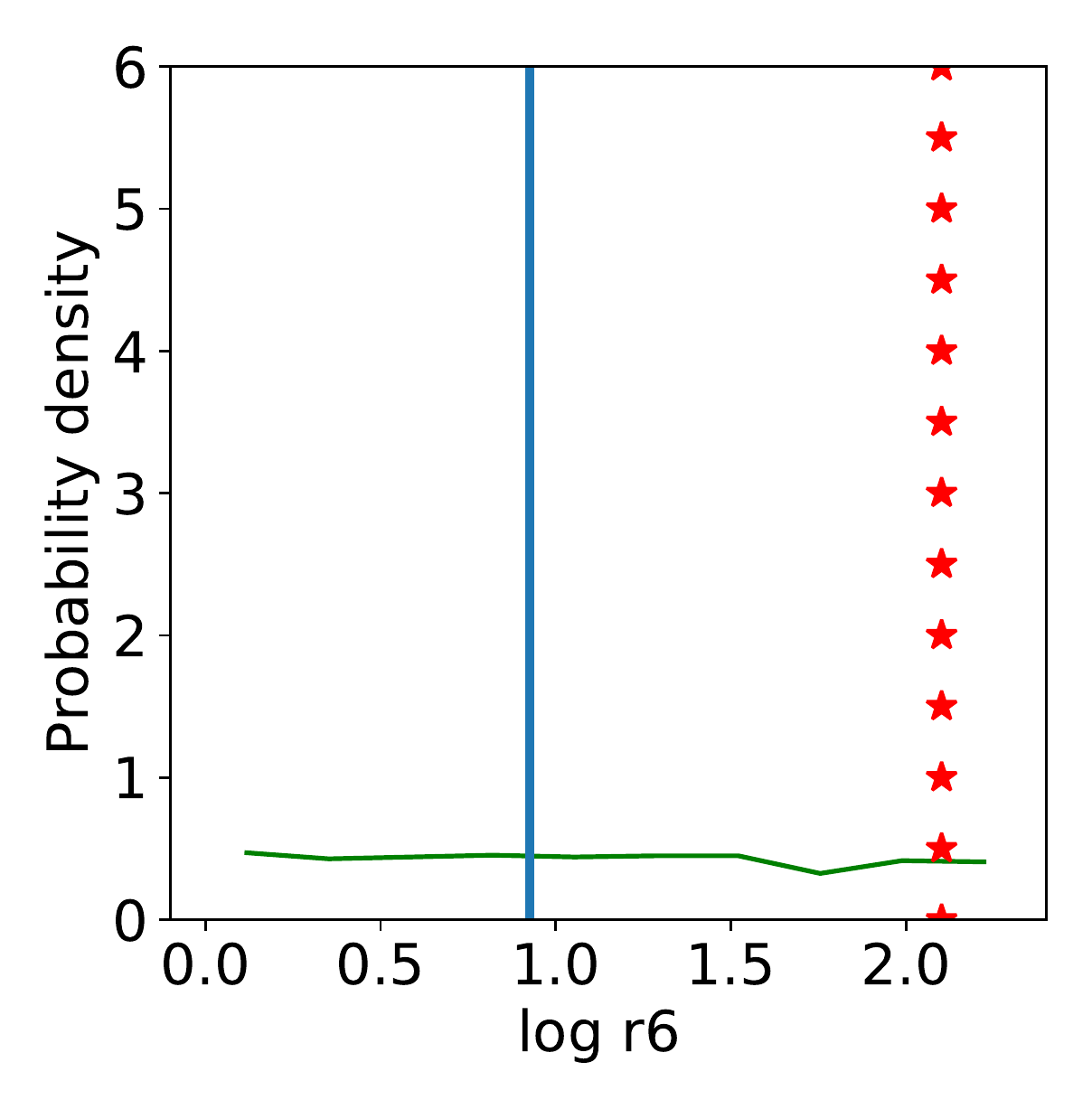} \\ 
		\end{tabular}
	\end{subfigure}
 \hspace{0.2in}	
	\begin{subfigure}[normla]{0.5\textwidth}
	   \begin{tabular}{cc}
		\includegraphics[scale=0.35]{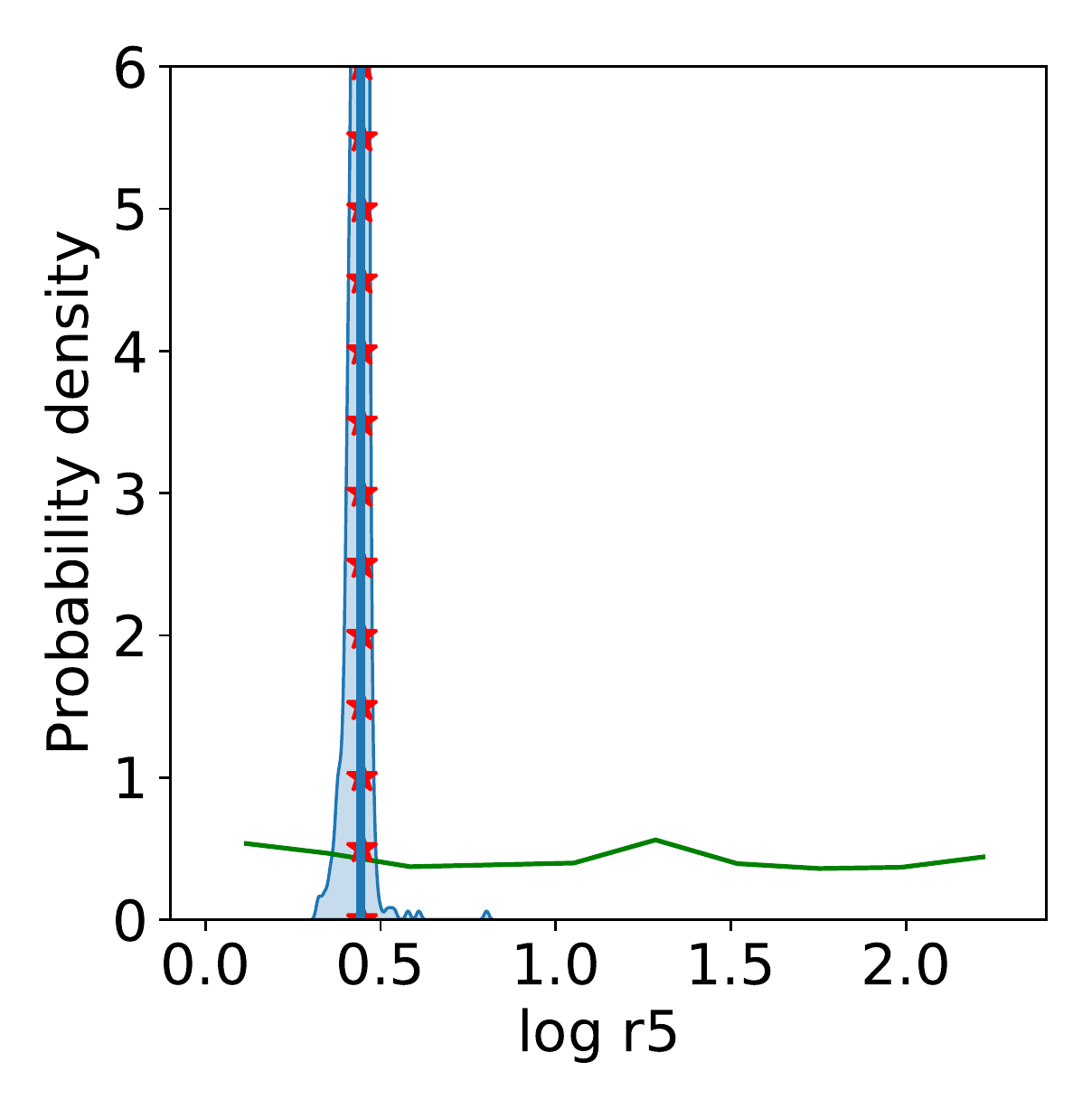} &
		\hspace{-0.2in}
		\includegraphics[scale=0.35]{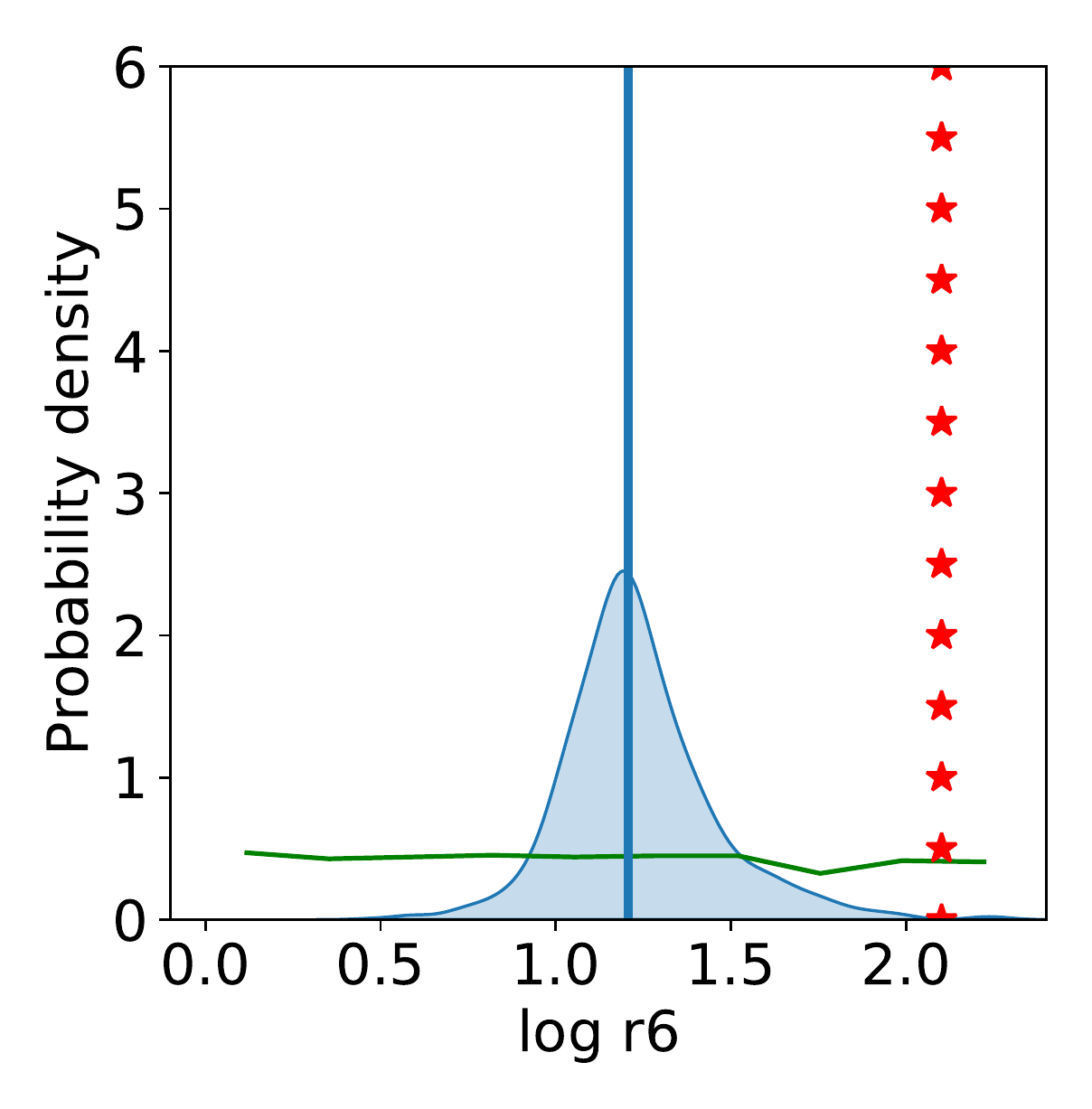} \\
		\end{tabular}
	\end{subfigure}
		
  \hspace{-0.1in}
		 ESMDA
 \hspace{2.5in}
 		 FlexIES
 \end{center}

\caption{Prior and posterior distribution of layers resistivity for joint inversion in presence of model-errors. Green and blue lines show ensemble approximation of the prior and posterior distribution respectively and red stars show reference truth. Solid blue lines show $p50$ of the posterior distribution. The sub-figures in the first and second columns show results obtained from the ESMDA algorithm and the sub-figures in the third and fourth columns show results from the FlexIES algorithm. The posterior distributions appear as the point estimate in the first and second columns of the sub-figures.}
\label{post_outputs_4_2}
\end{figure}

\begin{figure}[H]
\begin{center}

\hspace{-0.5in}
 	\begin{subfigure}[normla]{0.5\textwidth}
		\begin{tabular}{ccc}
		\hspace{-0.2in}
		\includegraphics[scale=0.35]{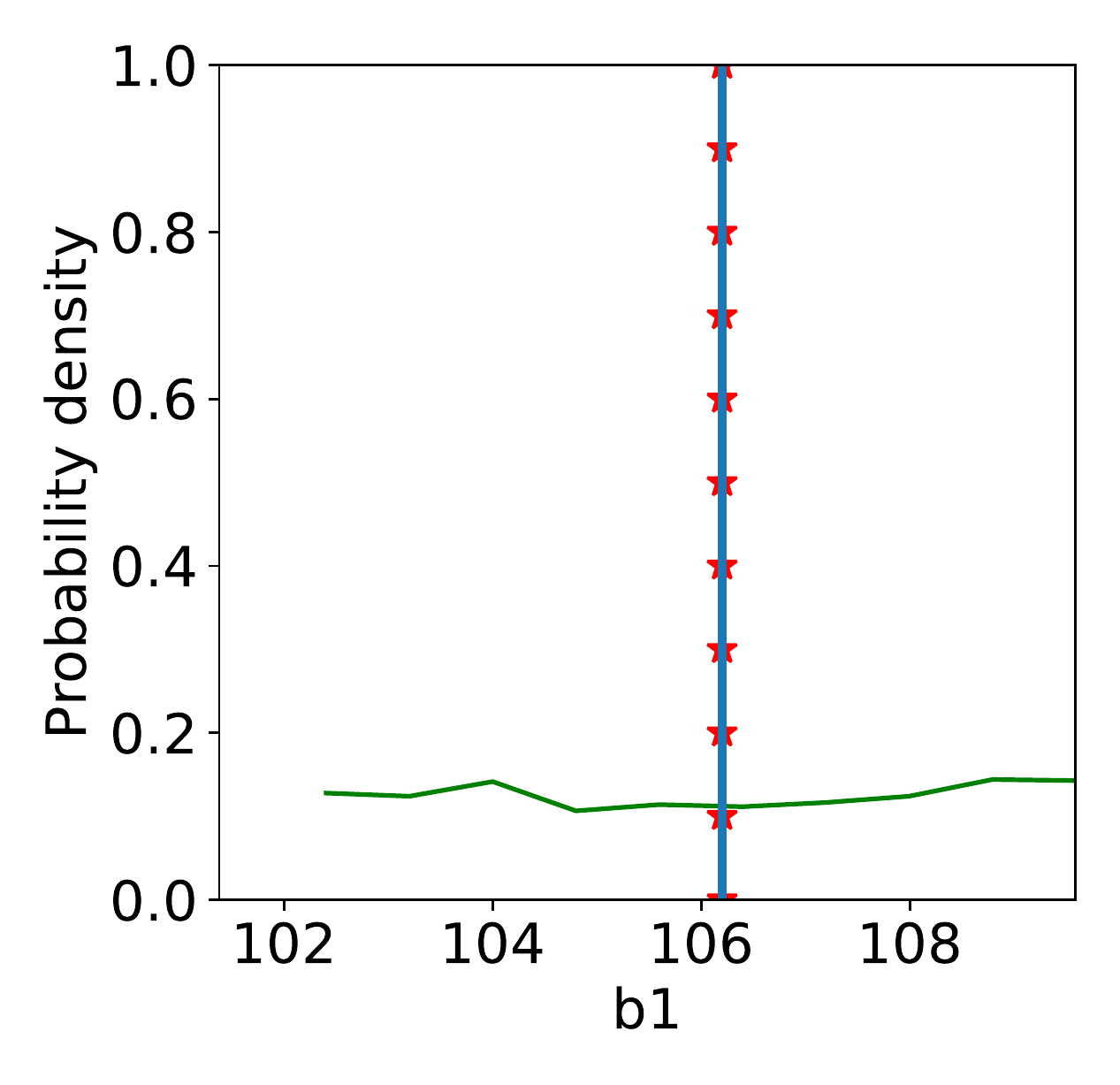} &
		\hspace{-0.2in}
		\includegraphics[scale=0.35]{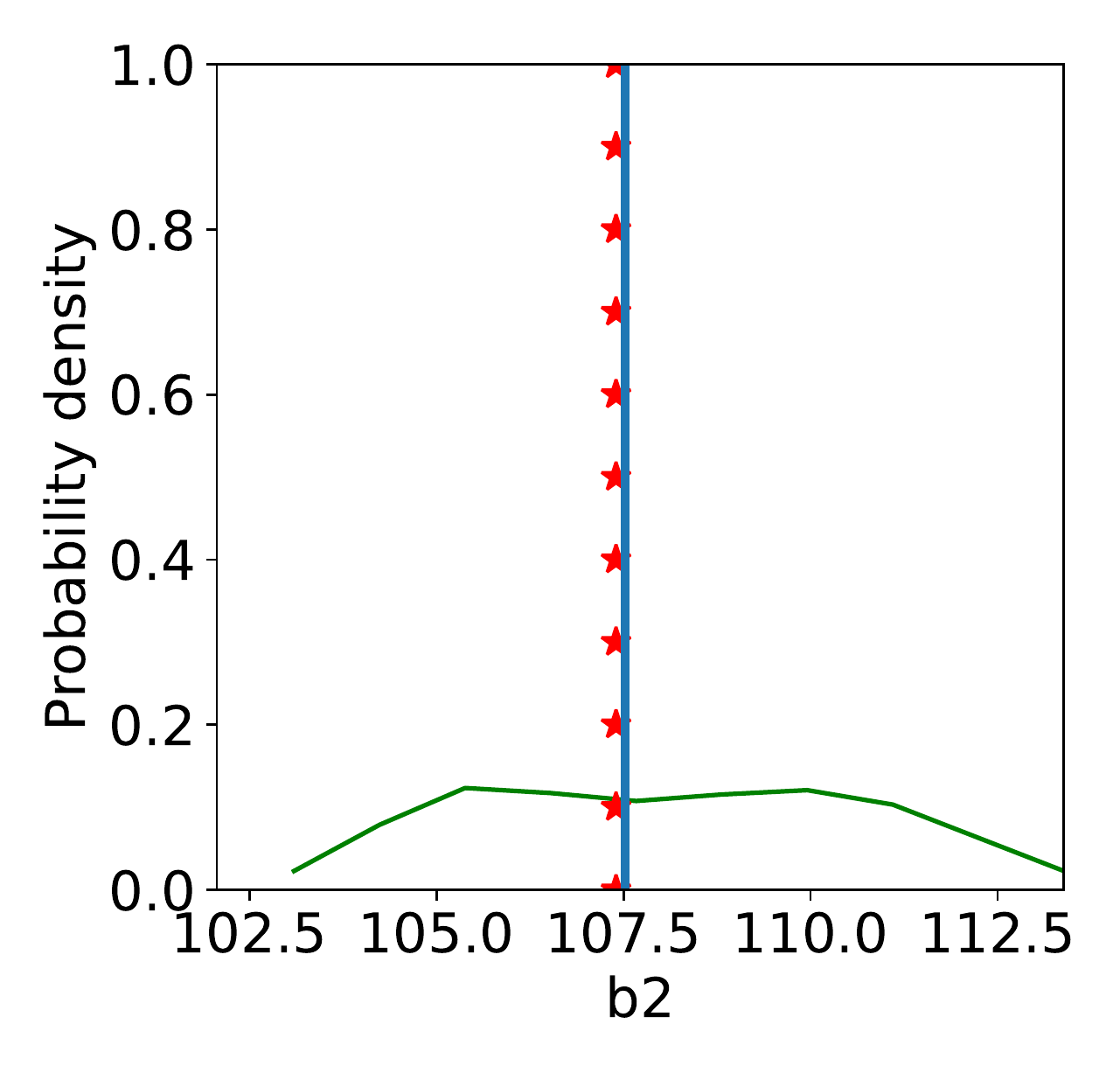} \\ 
		\end{tabular}
	\end{subfigure}
 \hspace{0.2in}	
	\begin{subfigure}[normla]{0.5\textwidth}
	   \begin{tabular}{cc}
		\includegraphics[scale=0.35]{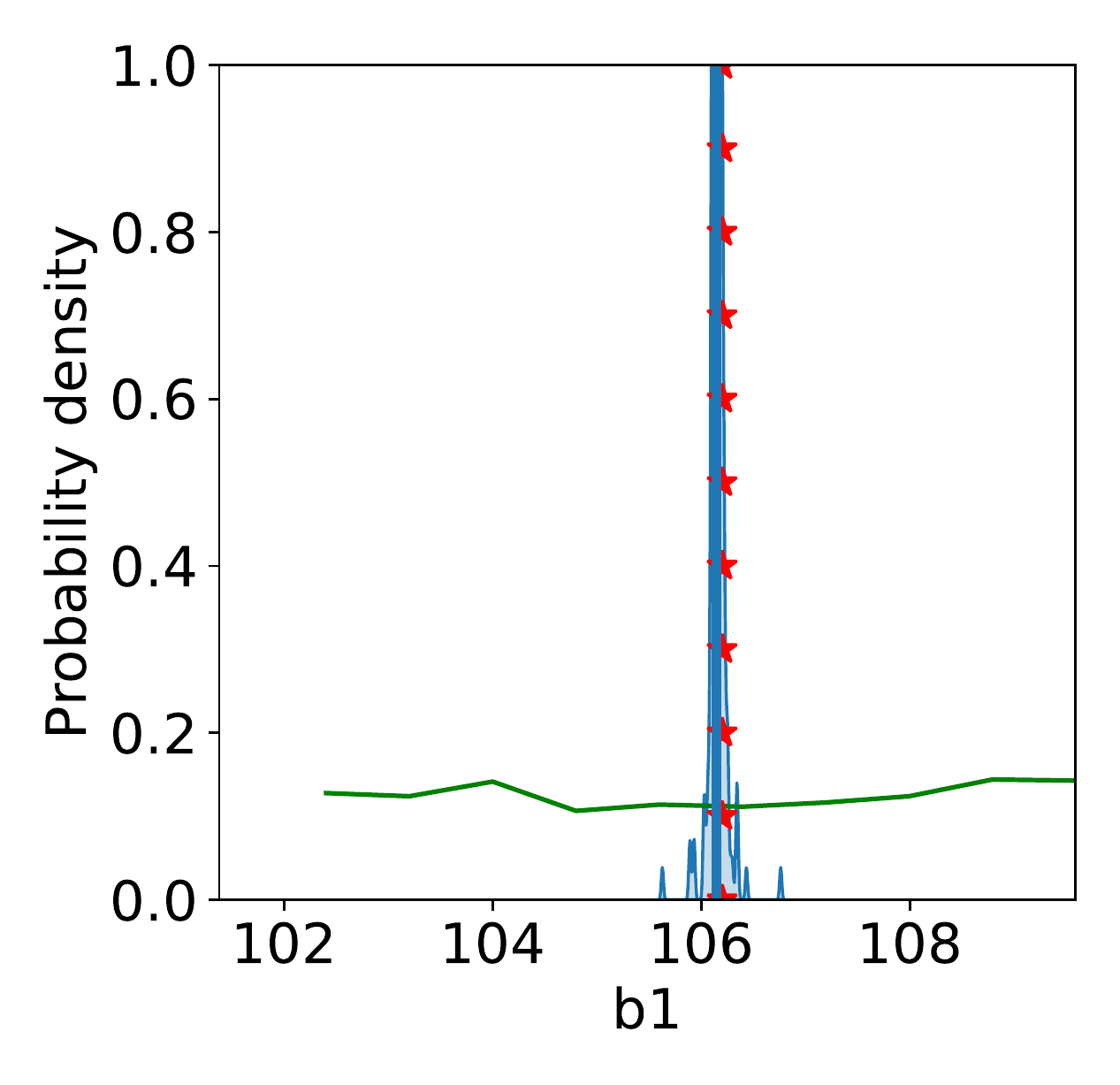} &
		\hspace{-0.2in}
		\includegraphics[scale=0.35]{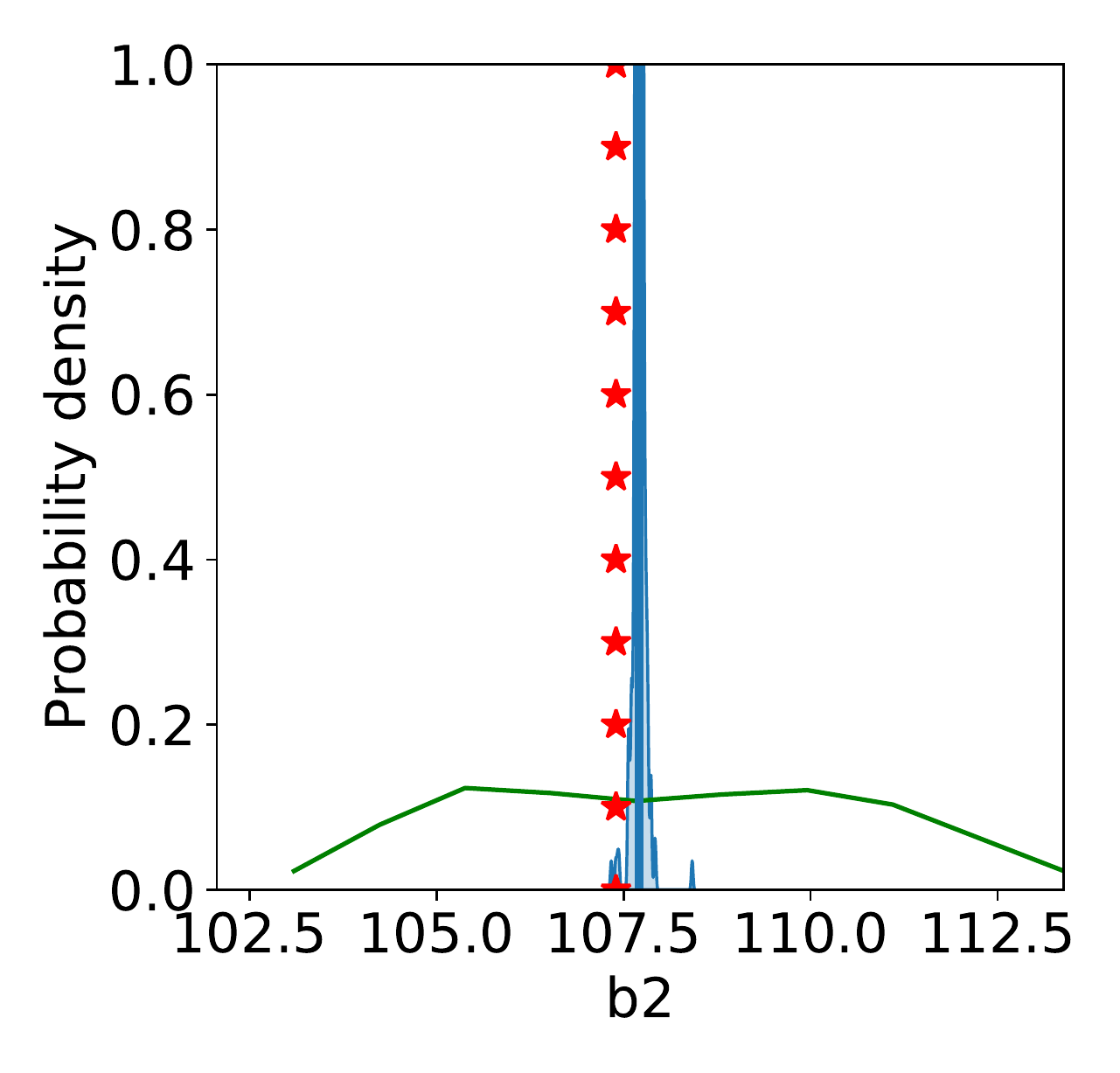} \\
		\end{tabular}
	\end{subfigure}

 \hspace{-0.5in}
 	\begin{subfigure}[normla]{0.5\textwidth}
		\begin{tabular}{ccc}
		\hspace{-0.2in}
		\includegraphics[scale=0.35]{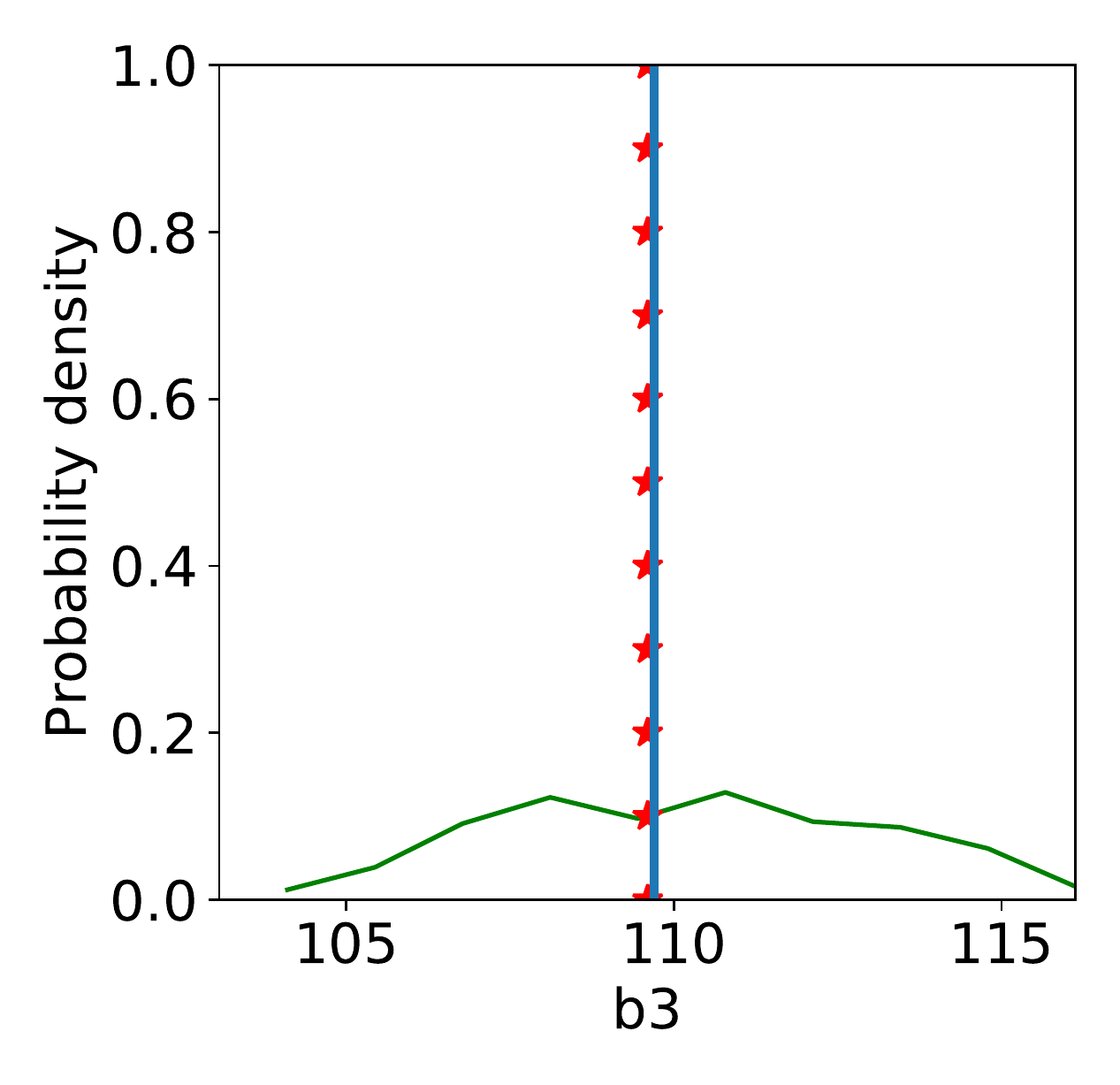} &
		\hspace{-0.2in}
		\includegraphics[scale=0.35]{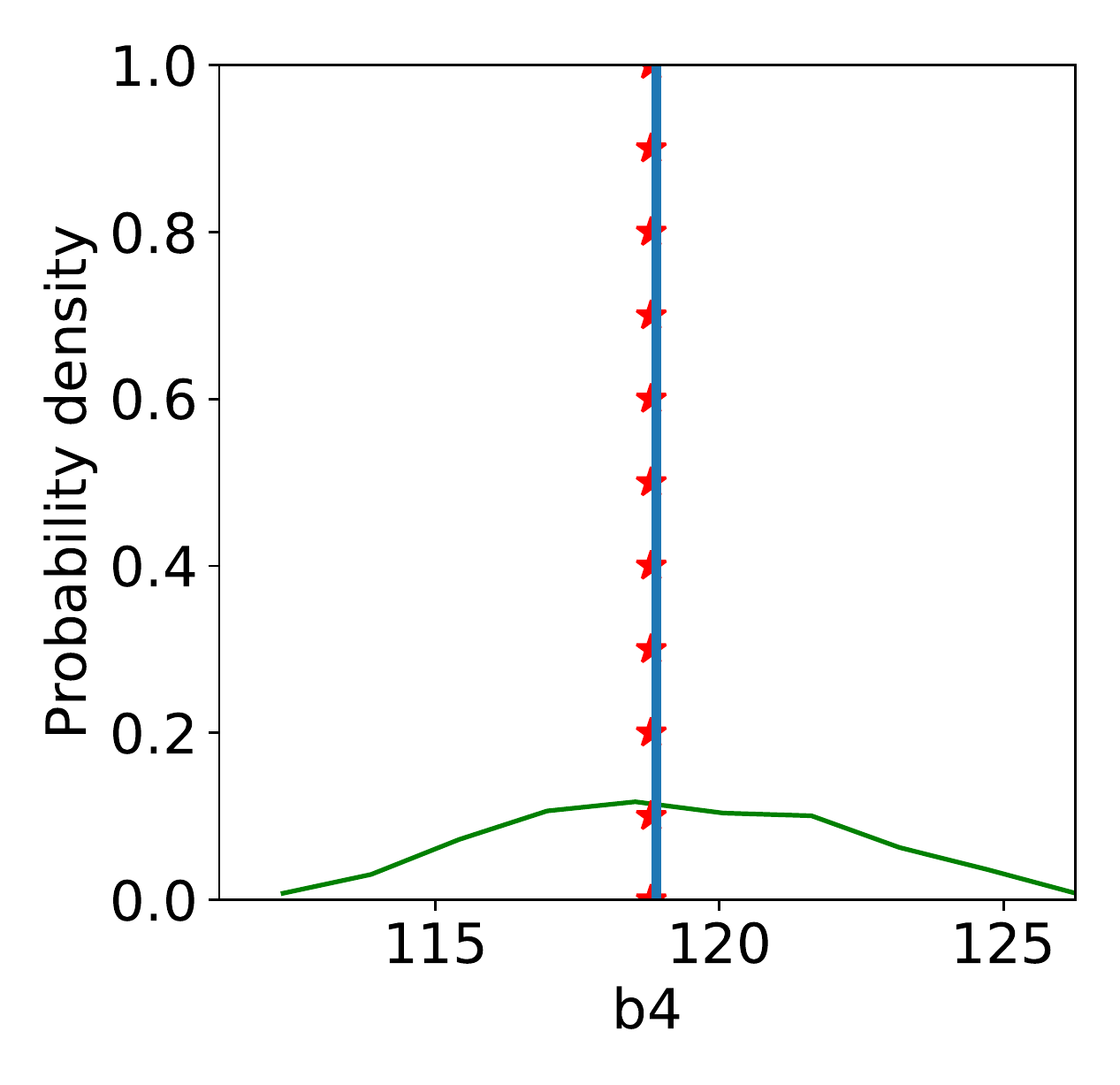} \\ 
		\end{tabular}
	\end{subfigure}
 \hspace{0.2in}	
	\begin{subfigure}[normla]{0.5\textwidth}
	   \begin{tabular}{cc}
		\includegraphics[scale=0.35]{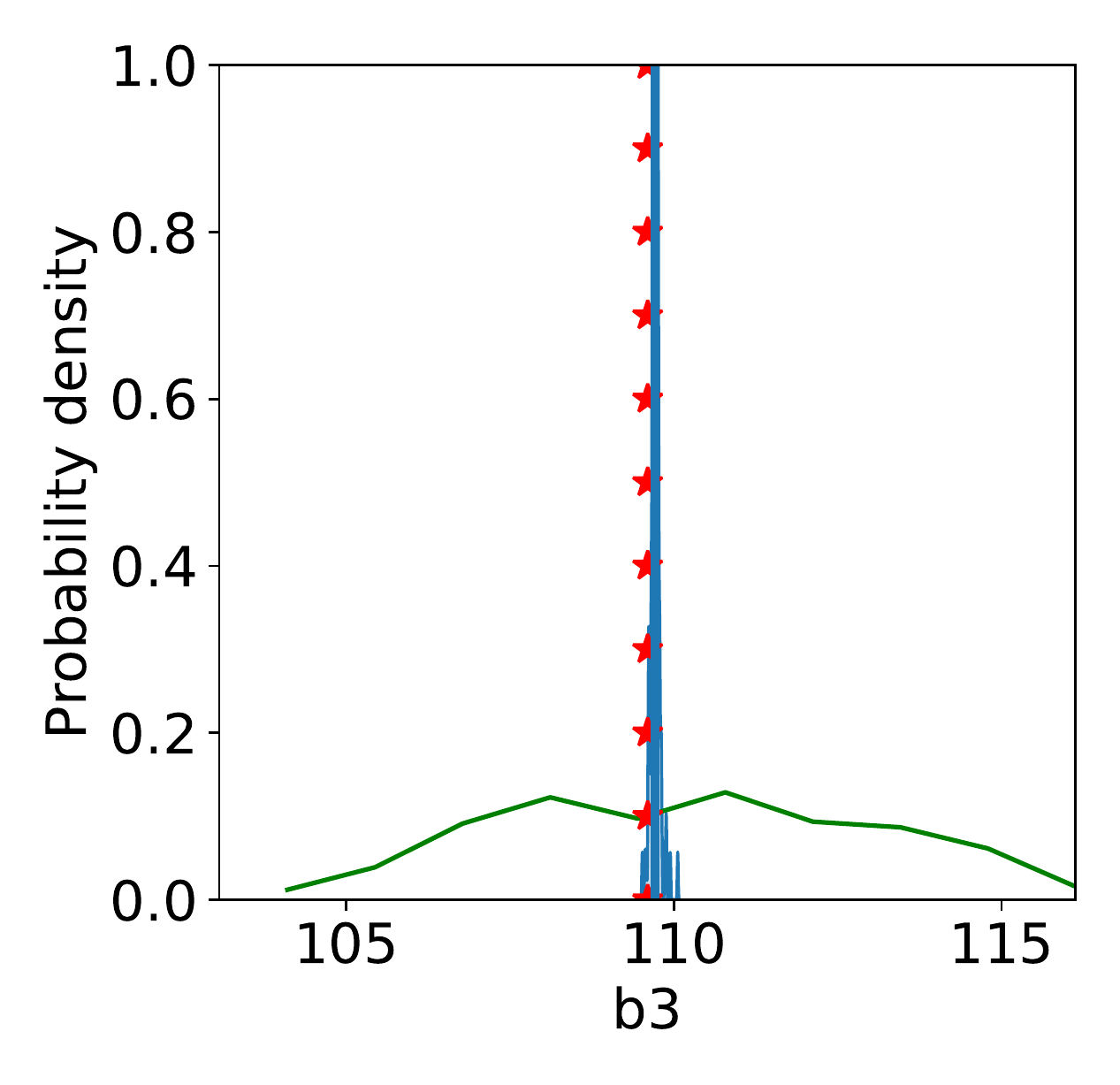} &
		\hspace{-0.2in}
		\includegraphics[scale=0.35]{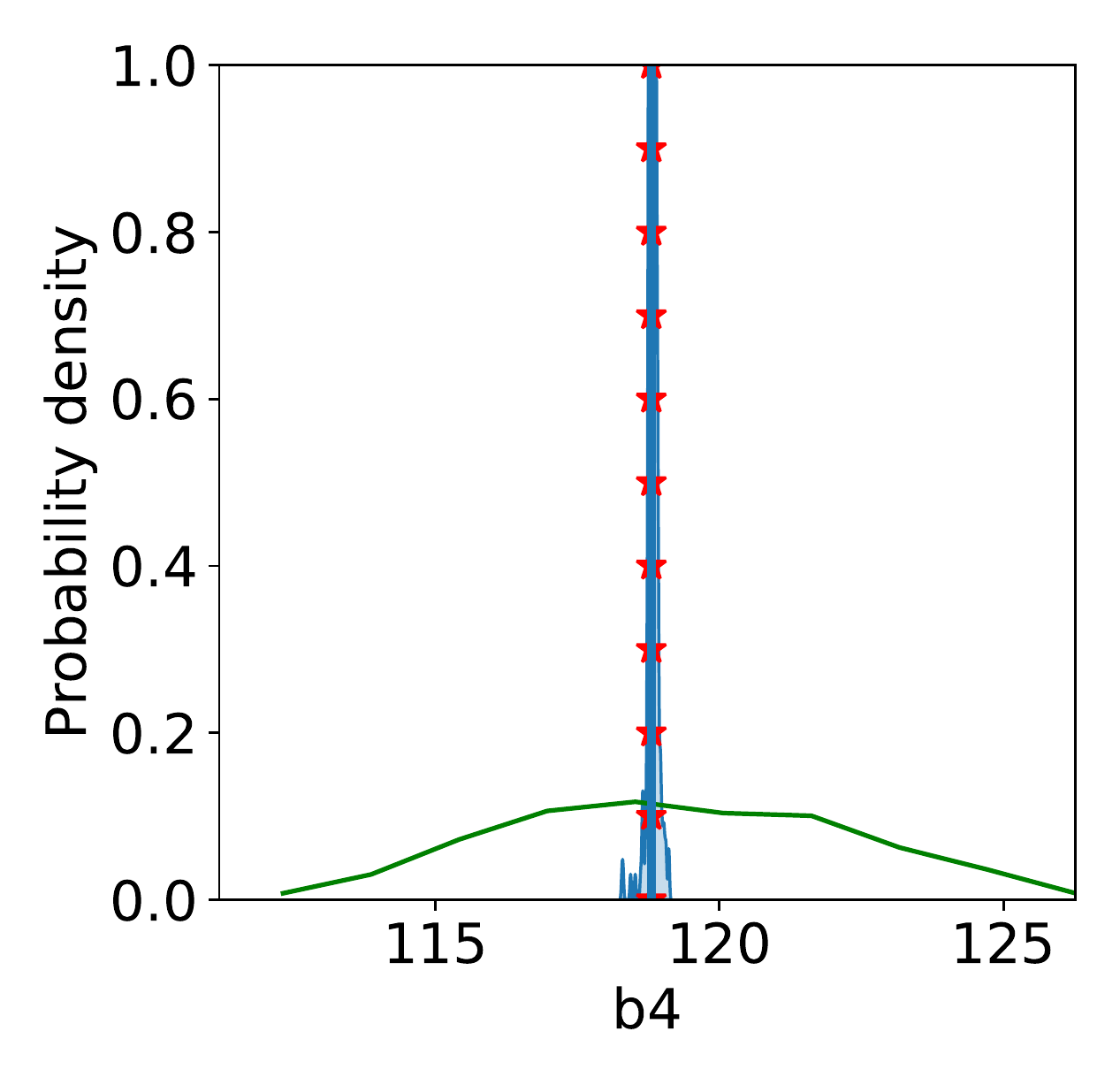} \\
		\end{tabular}
	\end{subfigure}

 \hspace{-0.5in}
 	\begin{subfigure}[normla]{0.5\textwidth}
		\begin{tabular}{ccc}
		\hspace{-0.2in}
		\includegraphics[scale=0.35]{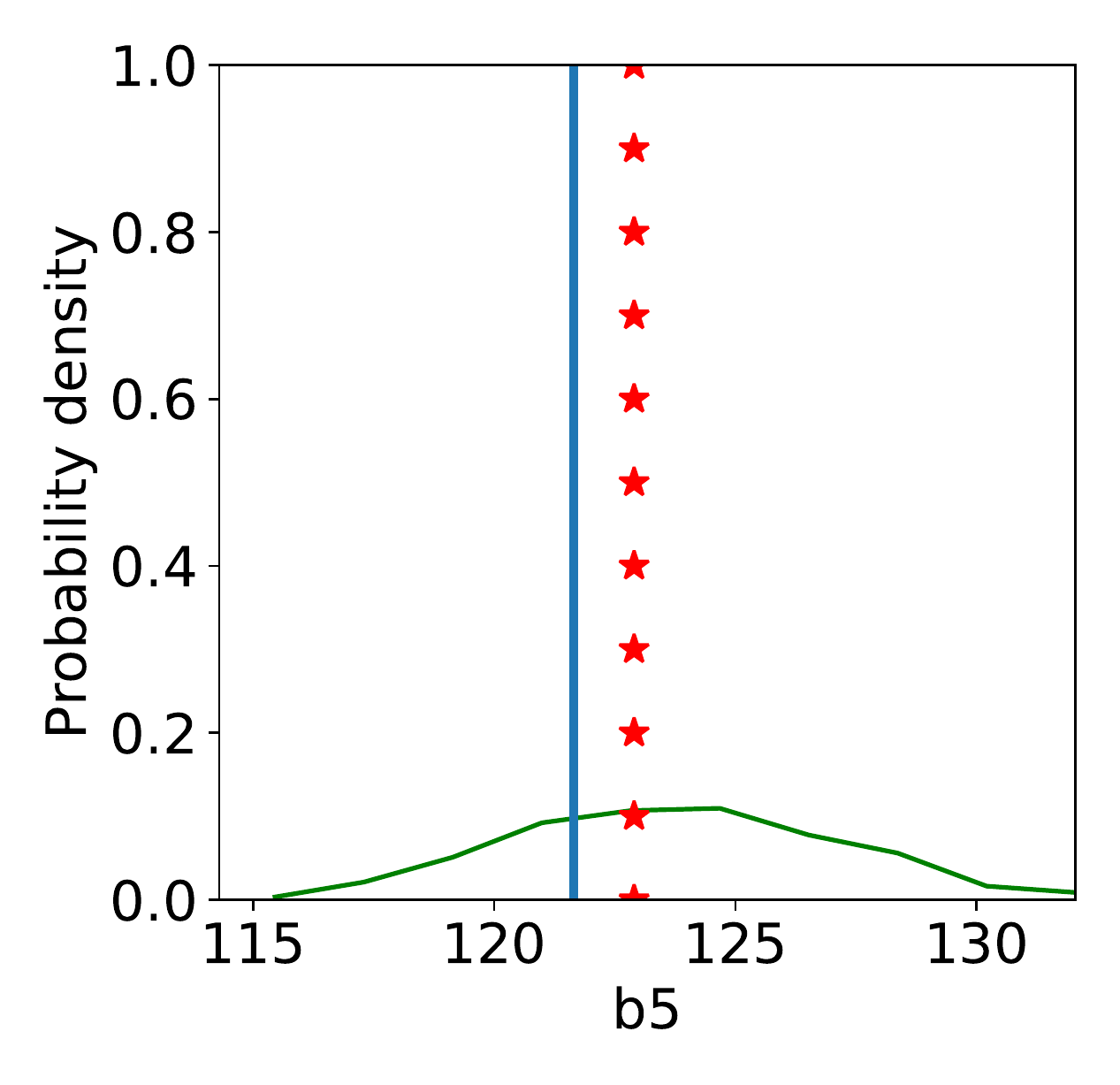} &
		\hspace{-0.2in}
		\includegraphics[scale=0.35]{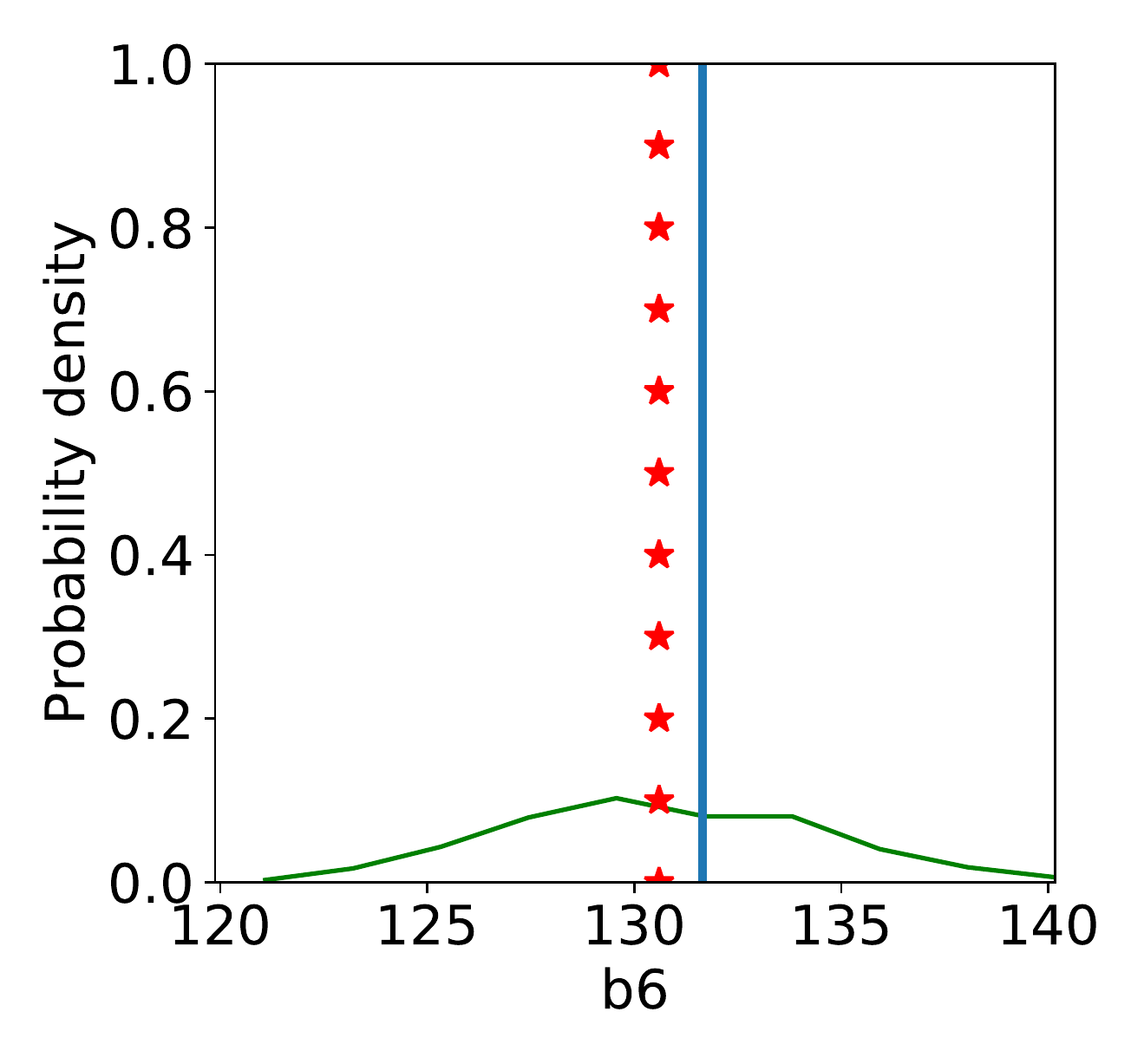} \\ 
		\end{tabular}
	\end{subfigure}
 \hspace{0.2in}	
	\begin{subfigure}[normla]{0.5\textwidth}
	   \begin{tabular}{cc}
		\includegraphics[scale=0.35]{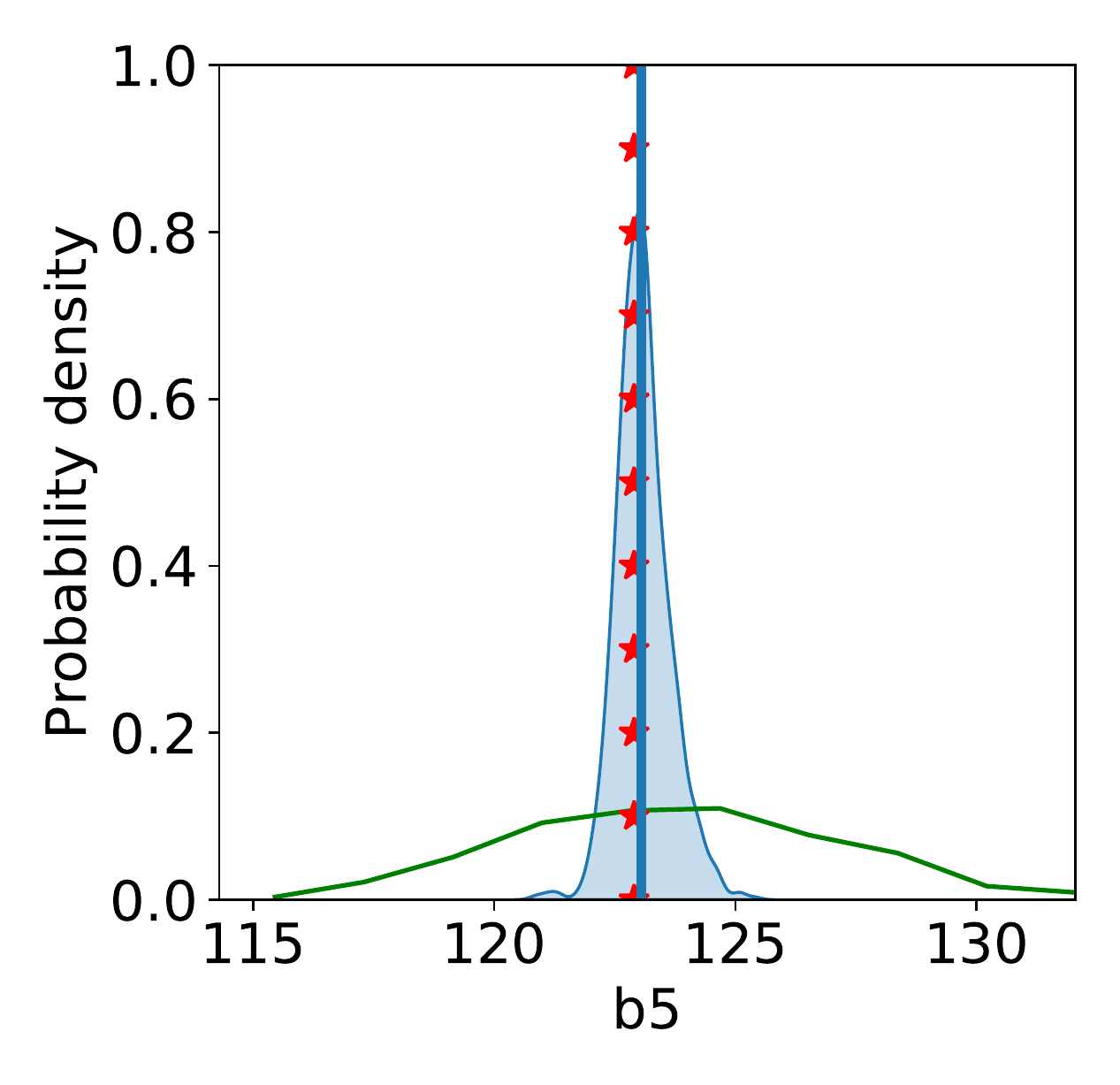} &
		\hspace{-0.2in}
		\includegraphics[scale=0.35]{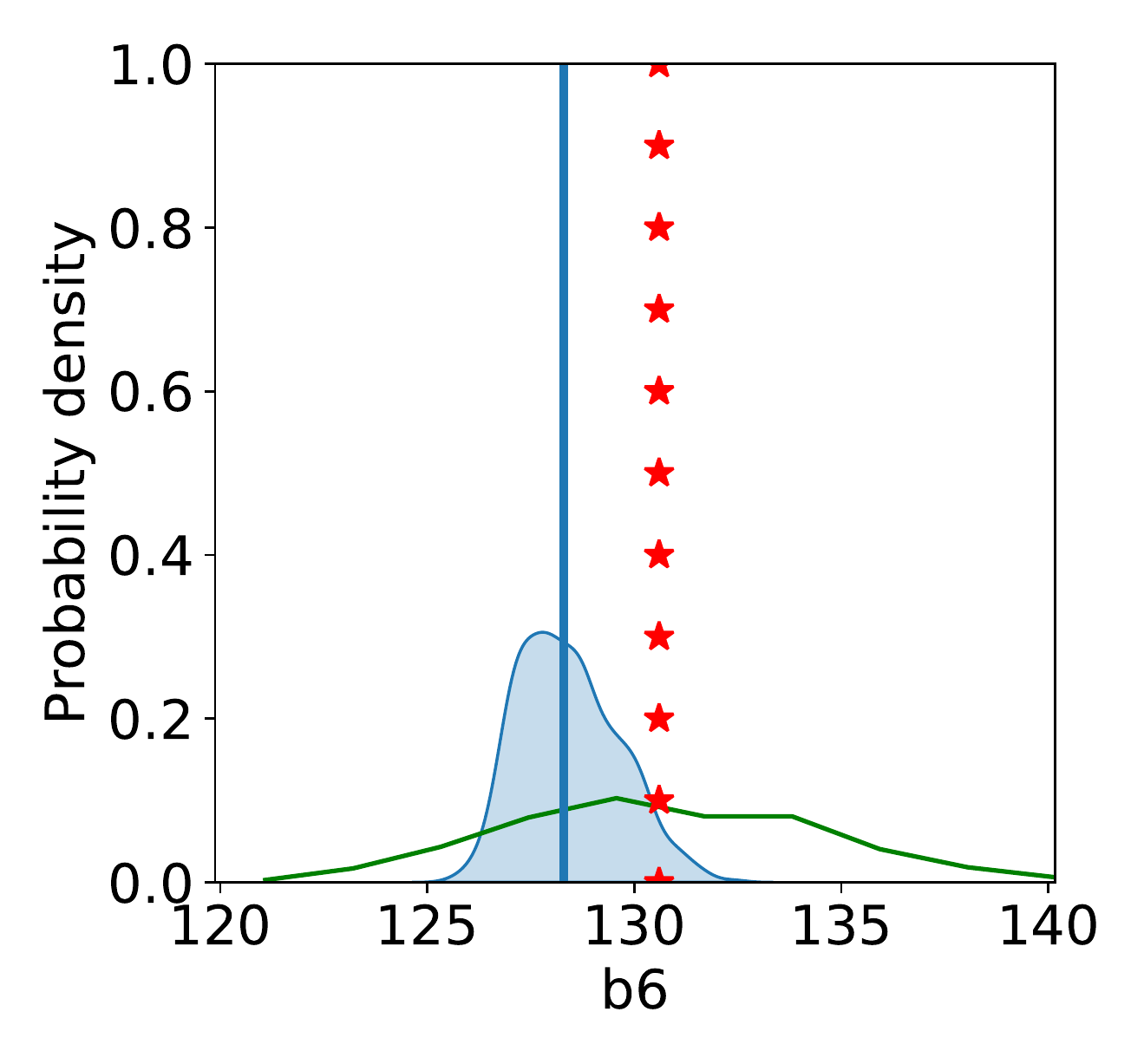} \\
		\end{tabular}
	\end{subfigure}
		
  \hspace{-0.1in}
		 ESMDA
 \hspace{2.5in}
 		 FlexIES
 \end{center}

\caption{Prior and posterior distribution of layer boundary positions (true depth converted from thicknesses) for joint inversion in presence of model-errors. Green and blue lines show ensemble approximation of the prior and posterior distribution respectively and red stars show reference truth. Solid blue lines show $p50$ of the posterior distribution. The sub-figures in the first and second columns show results obtained from the ESMDA algorithm and the sub-figures in the third and fourth columns show results from the FlexIES algorithm. The posterior distributions appear as the point estimate in the first and second columns of the sub-figures.}
\label{post_outputs_4_3}
\end{figure}

For this well-posed problem, we get a reasonable match of the observed data (Figure~(\ref{post_outputs_4_1})) using both the classical ESMDA and FlexIES algorithms despite the presence of model-error. 
However, the estimated log-resistivities and thicknesses of the layers are biased for the classical ESMDA.
That is, the posterior distribution does not cover the true solution for many of the estimated parameters, 
see Figures~(\ref{post_outputs_4_2}) and (\ref{post_outputs_4_3}).
Admittedly, resistivities and thicknesses of the first four layers are very close to the reference truth, thus showing negligible biases for the part of the model best covered by the measurement data.
The resistivities of the sixth layer and thicknesses of fifth and sixth layers show large bias. 
FlexIES treats model-error directly, which results 
in minimal biases, if any, in estimation of the layer resistivities and thicknesses (bed boundary positions) as shown in Figures~(\ref{post_outputs_4_2}) and (\ref{post_outputs_4_3}). 
For example, the estimated resistivity of the sixth layer and thicknesses of the fifth and the sixth layer are more reliable because the corresponding posterior distribution covers the reference truth.

\begin{figure}[H]
\begin{center}

 \hspace{-0.5in}
    \begin{subfigure}[normal]{0.4\textwidth}
	\includegraphics[scale=0.5]{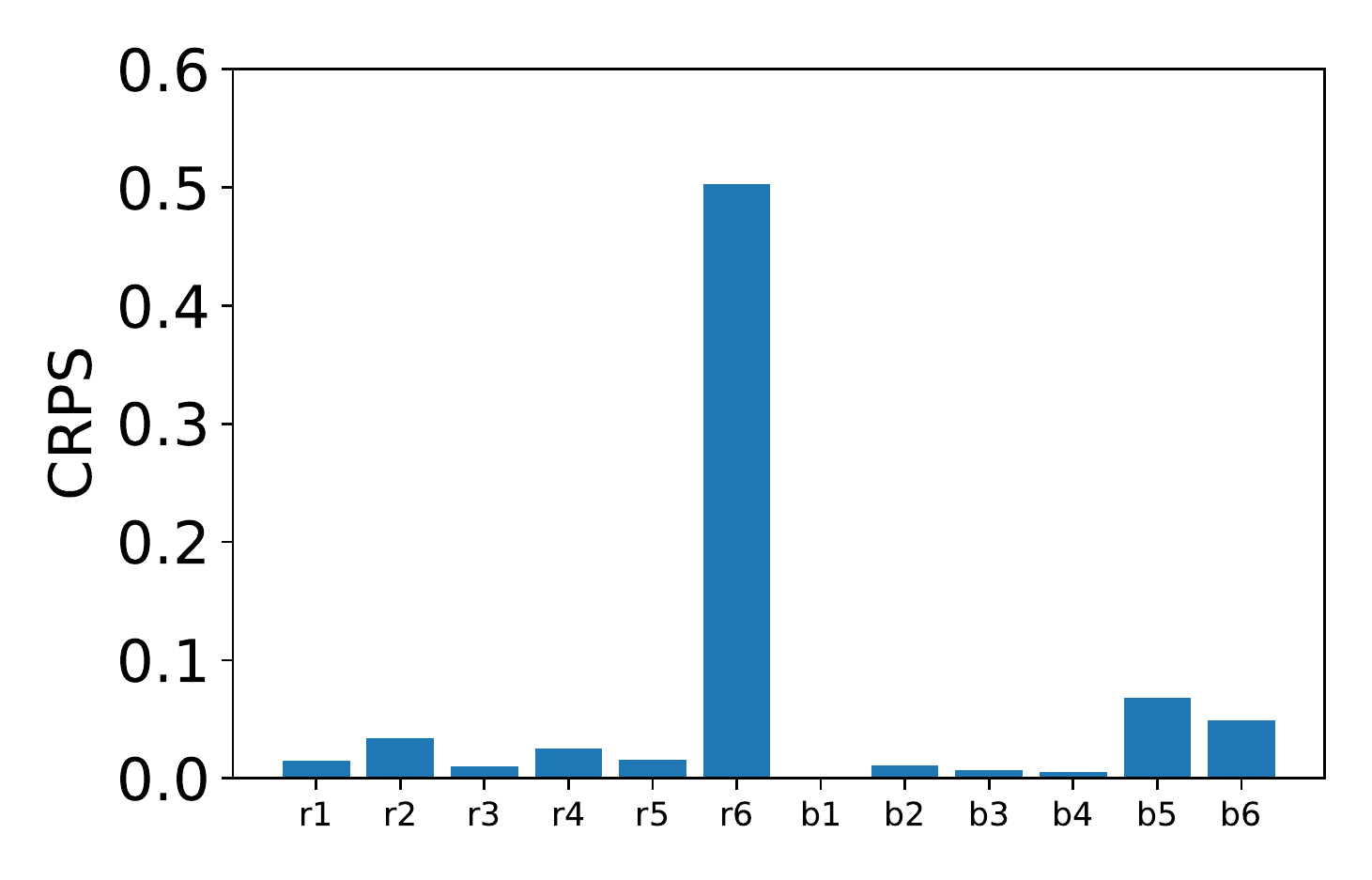}	
	\end{subfigure}
 \hspace{0.5in}
	\begin{subfigure}[normal]{0.4\textwidth}
	\includegraphics[scale=0.5]{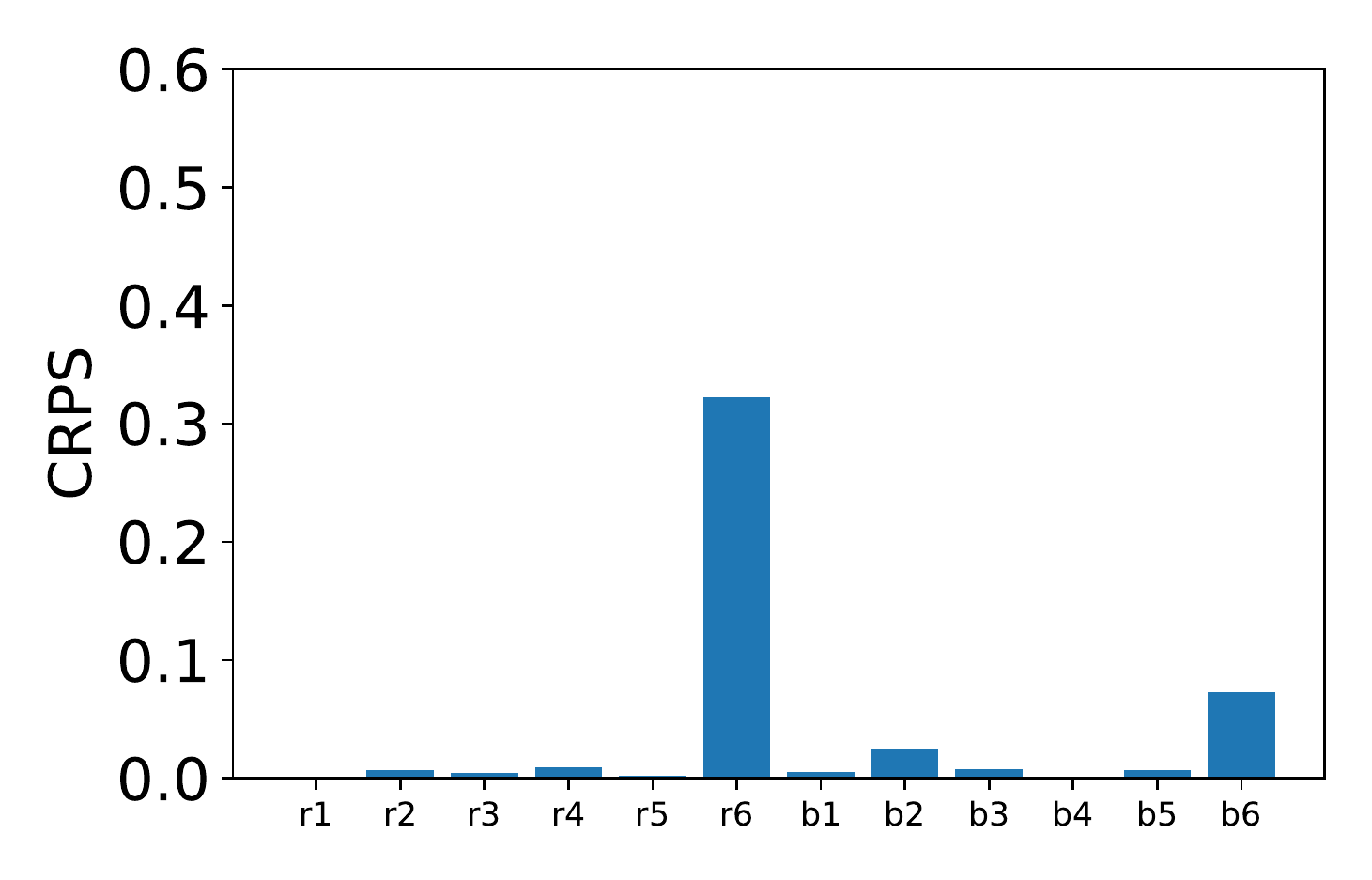}	
	\end{subfigure}

  \hspace{-0.1in}
		 ESMDA
 \hspace{2.5in}
 		 FlexIES
 \end{center}
\caption{Continuous ranked probability score (CRPS) of the log-resistivities and boundary positions of the layers obtained from joint inversion in the presence of model-errors using ESMDA and FlexIES. Lower values are better.}
\label{CRPS_2} 
\end{figure}

Figure~(\ref{CRPS_2}) shows the comparison of the inversion results for the estimation of log resistivities and thicknesses (bed boundary positions) of the layers. The CRPS of the estimated resistivities and thicknesses improves significantly using FlexIES as compared to ESMDA, since FlexIES reduces the estimation bias by capturing the model-error effects during real-time inversion.

\subsubsection{Probabilistic alleviation of multi-modality in case of a wide (uninformed) prior}

In many practical problems, one could not always get an informed prior needed for well-posedness of the inverse problem. 
We explore that FlexIES could be also useful in these scenarios. 
In order to illustrate the problem of multi-modality, we perform inversion by considering DNN model as a perfect model. 
We use the wide (uninformed) prior of the layers resistivities and thicknesses, where all layers can vary from  $U \sim [0.3, \quad 20]$ meter in thickness.

\begin{figure}[H]
\begin{center}

 \hspace{-0.5in}
    \begin{subfigure}[normal]{0.4\textwidth}
	\includegraphics[scale=0.5]{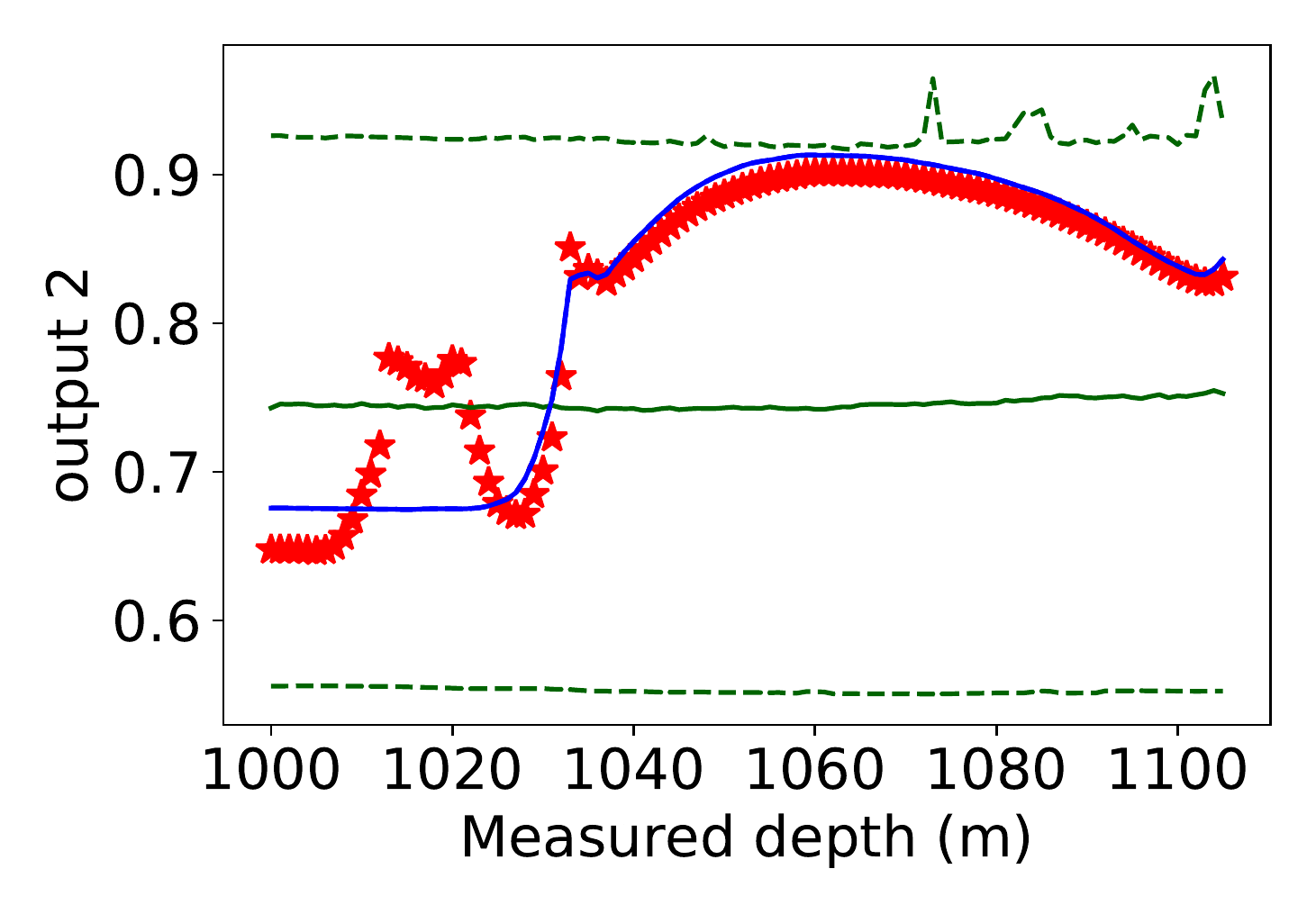}	
	\end{subfigure}
 \hspace{0.5in}
	\begin{subfigure}[normal]{0.4\textwidth}
	\includegraphics[scale=0.5]{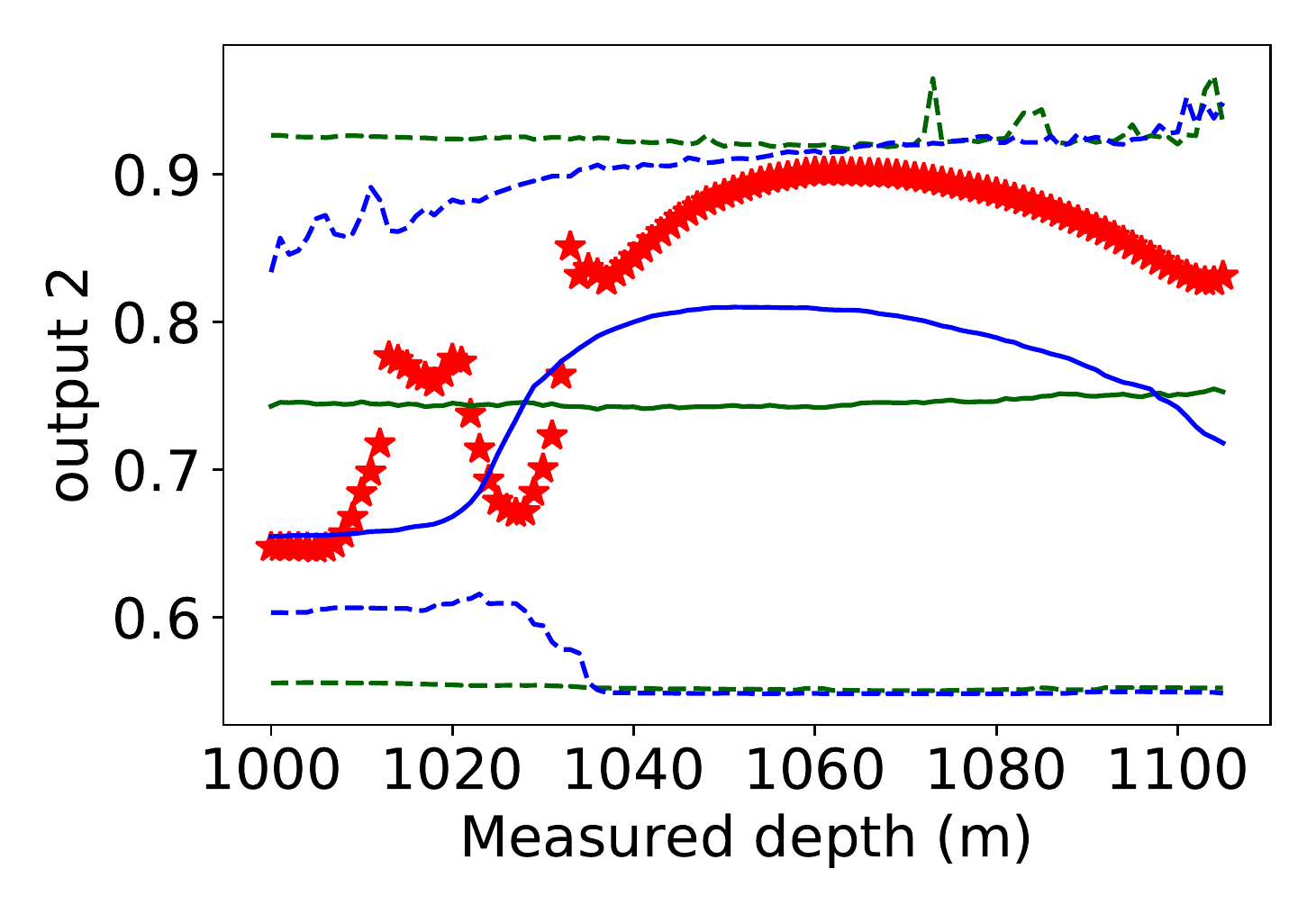}	
	\end{subfigure}

 \hspace{-0.5in}
    \begin{subfigure}[normal]{0.4\textwidth}
	\includegraphics[scale=0.5]{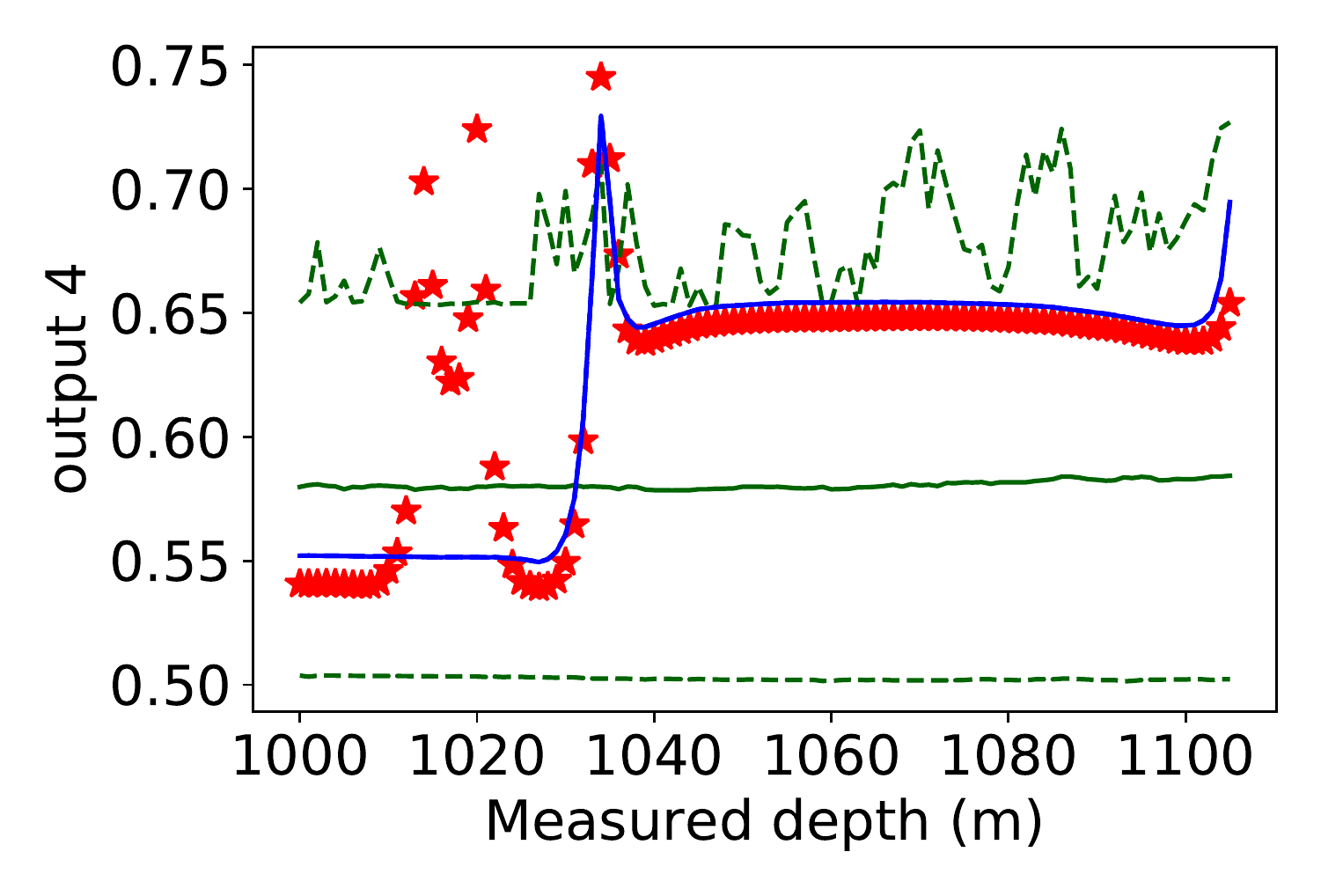}	
	\end{subfigure}
 \hspace{0.5in}
	\begin{subfigure}[normal]{0.4\textwidth}
	\includegraphics[scale=0.5]{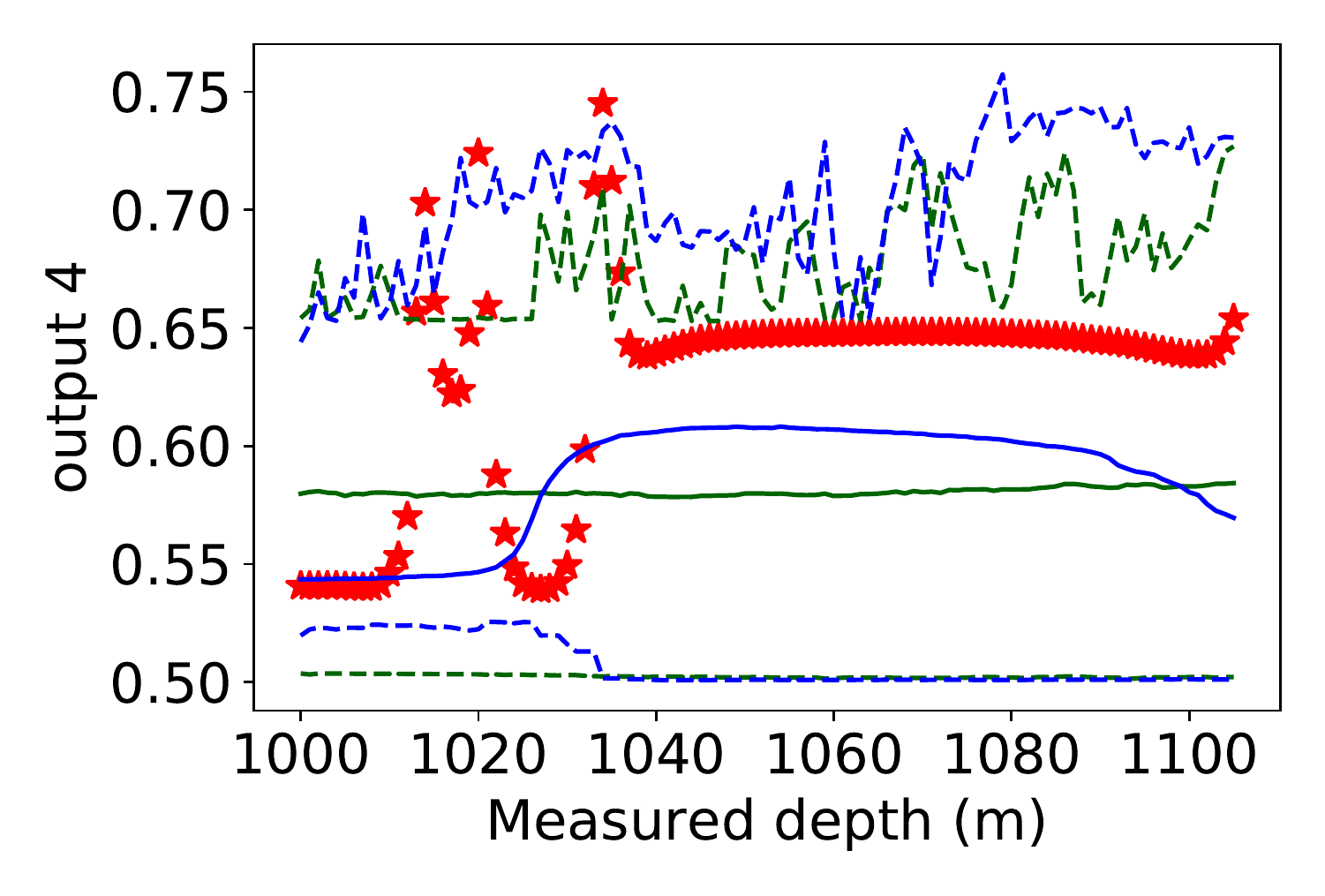}	
	\end{subfigure}

 \hspace{-0.5in}
    \begin{subfigure}[normal]{0.4\textwidth}
	\includegraphics[scale=0.5]{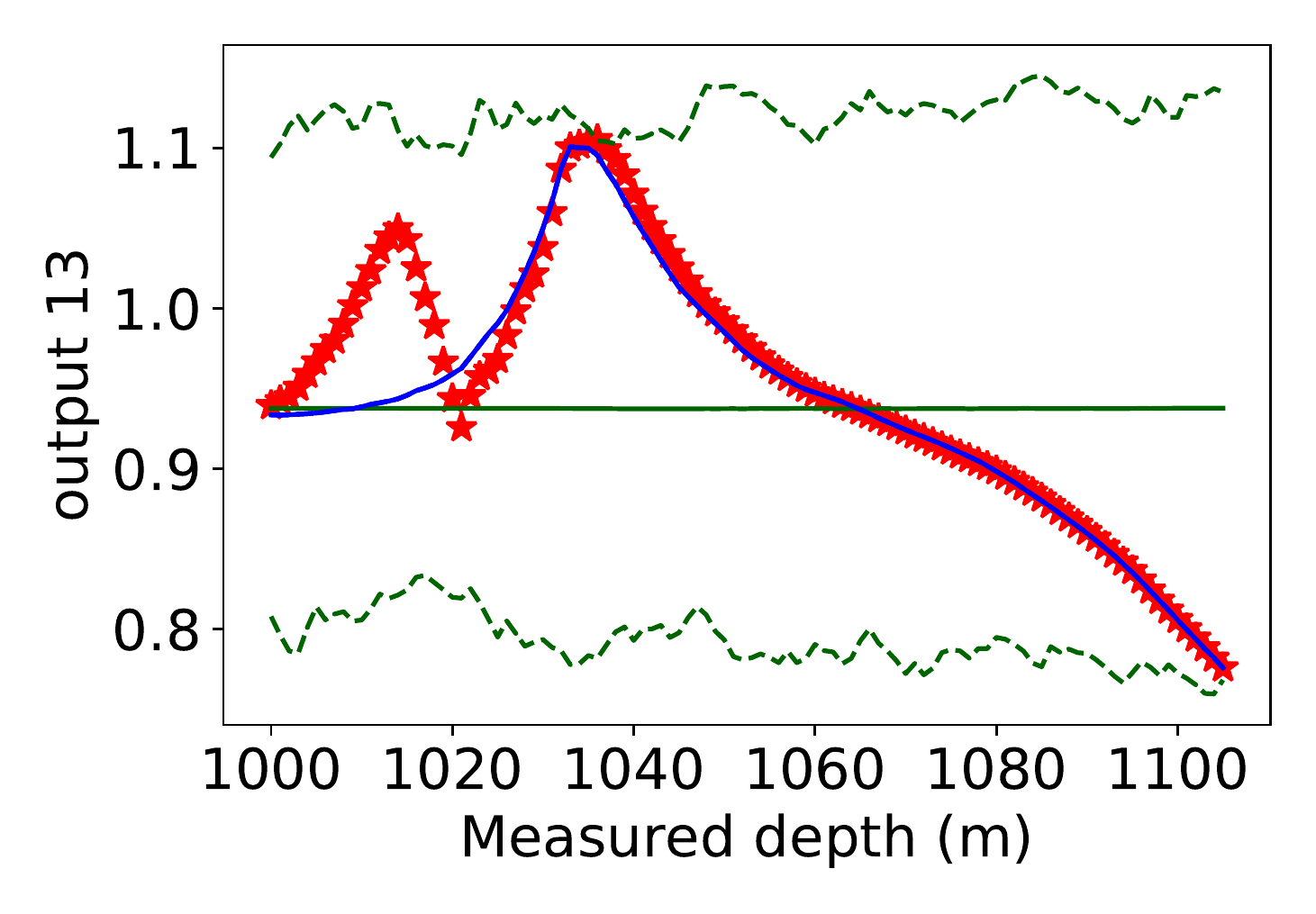}	
	\end{subfigure}
 \hspace{0.5in}
	\begin{subfigure}[normal]{0.4\textwidth}
	\includegraphics[scale=0.5]{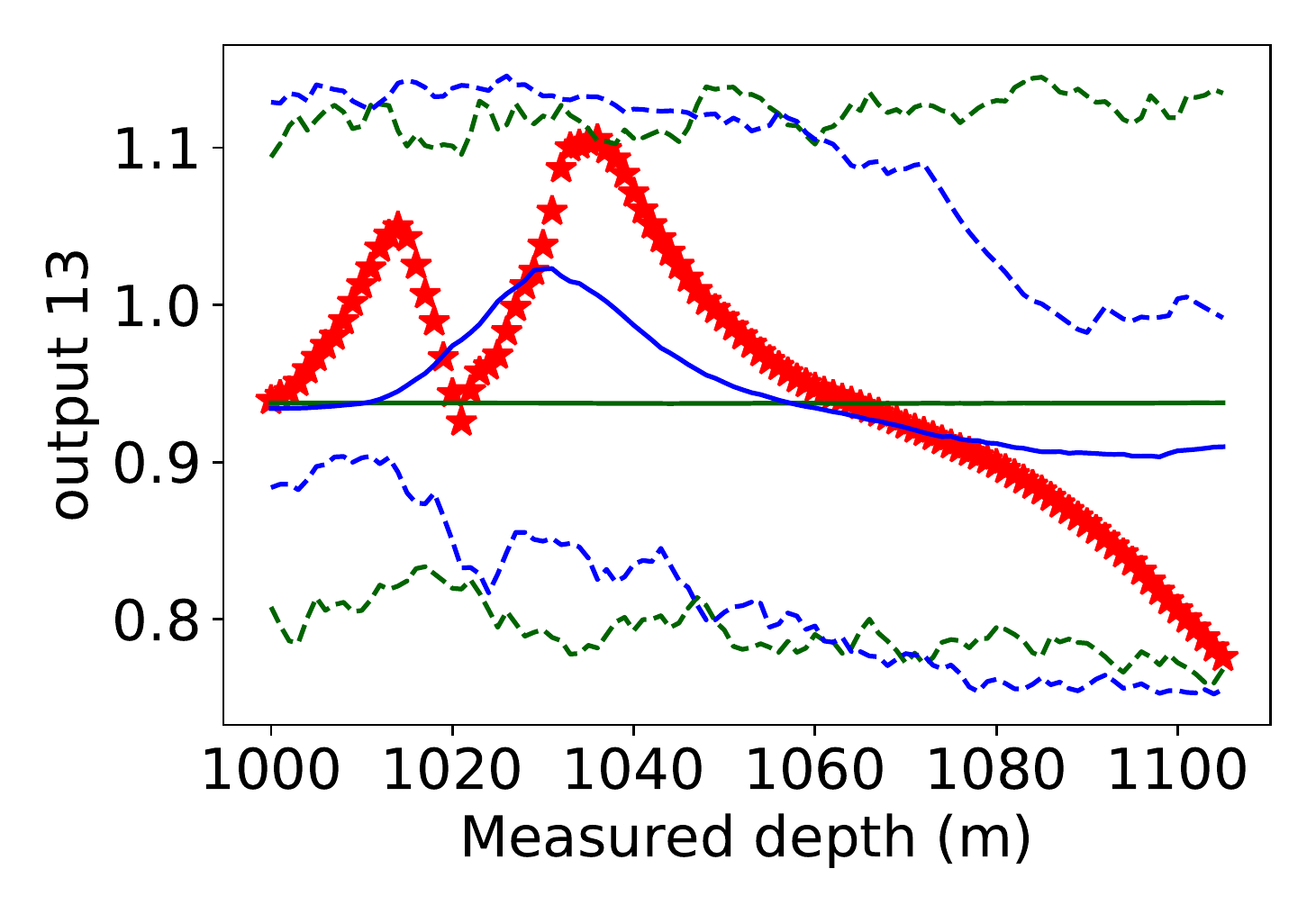}	
	\end{subfigure}

 \hspace{-0.1in}
		 ESMDA
 \hspace{2.5in}
 		 FlexIES
 \end{center}

\caption{Prior and posterior distribution of EM outputs as functions of measured depth in case of multi-modality and uninformed (wide) prior. Green and blue lines show ensemble approximation of the prior and posterior distribution respectively and red stars show observed EM measurements. Solid green and solid blue lines show $p50$ and dashed green and dashed blue lines show $99\%$ confidence interval respectively of the prior and posterior distribution respectively. The sub-figures in the first column show results obtained from the ESMDA algorithm and the sub-figures in the second column show results from the FlexIES algorithm. In first column of the sub-figures posterior distribution appears as the point estimate therefore solid blue lines overlaps dashed blue lines.}
\label{post_outputs_3_1}
\end{figure}

\begin{figure}[H]
\begin{center}

 \hspace{-0.5in}
 	\begin{subfigure}[normla]{0.5\textwidth}
		\begin{tabular}{ccc}
		\hspace{-0.2in}
		\includegraphics[scale=0.35]{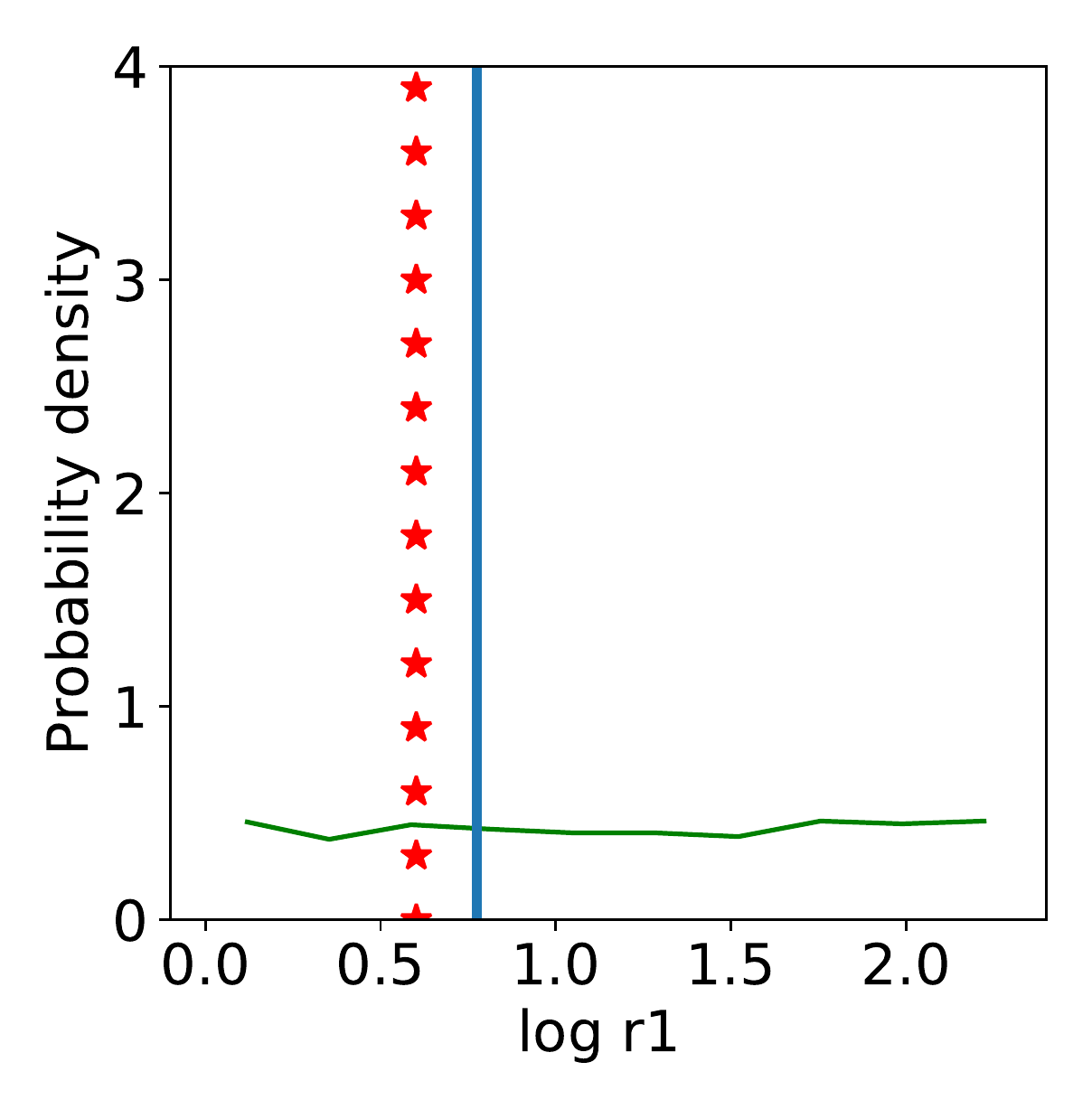} &
		\hspace{-0.2in}
		\includegraphics[scale=0.35]{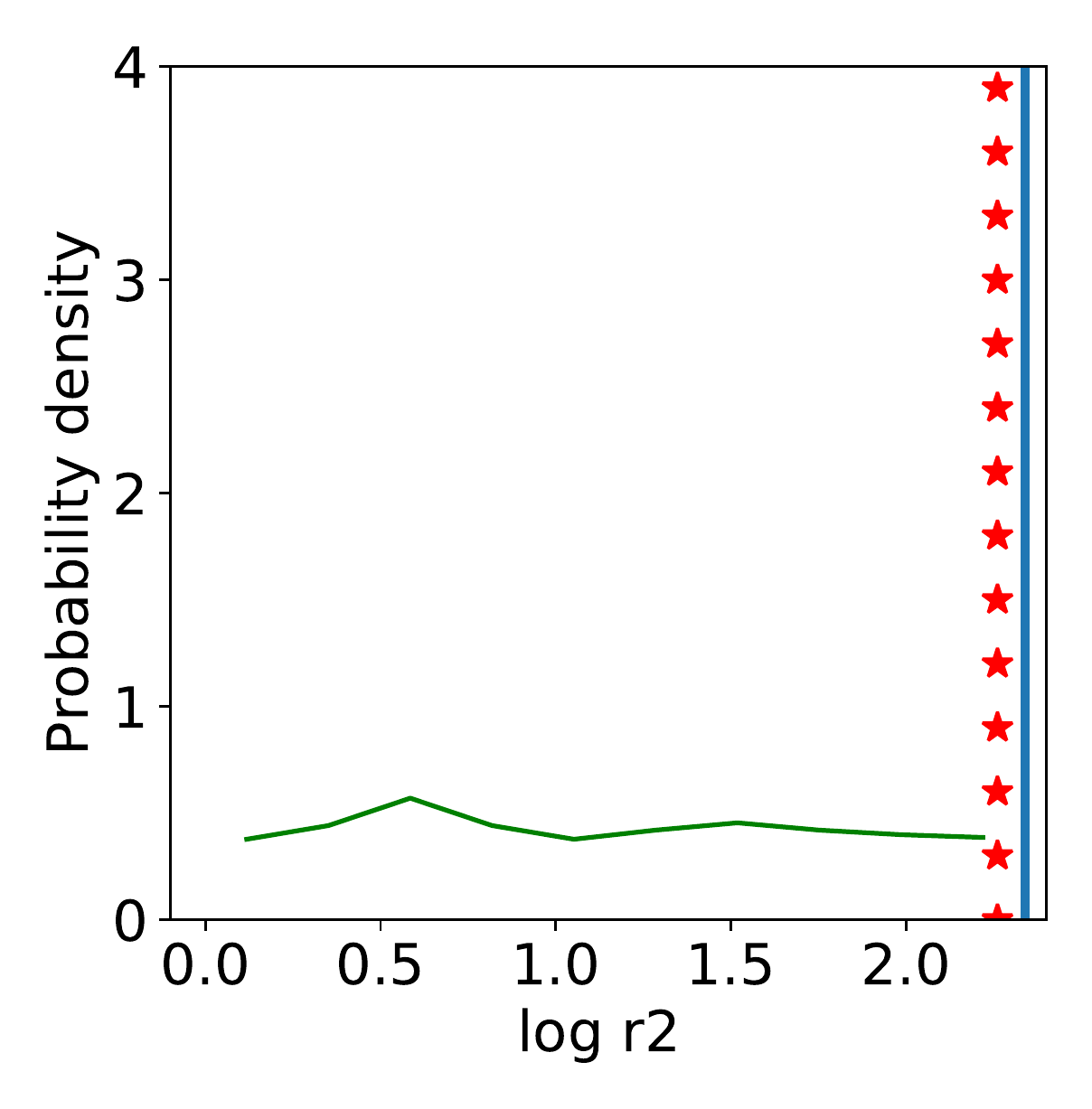} \\ 
		\end{tabular}
	\end{subfigure}
 \hspace{0.2in}	
	\begin{subfigure}[normla]{0.5\textwidth}
	   \begin{tabular}{cc}
		\includegraphics[scale=0.35]{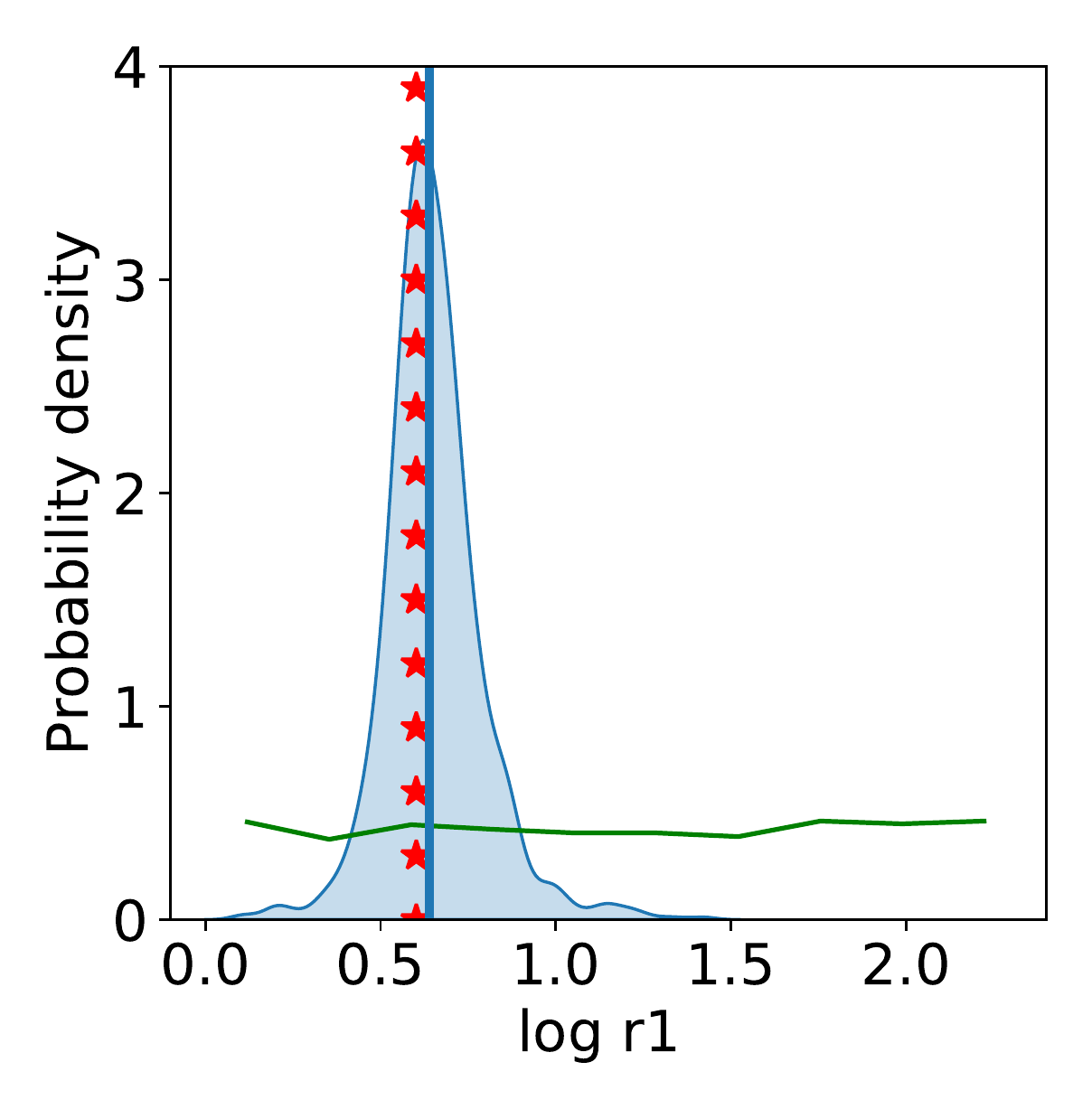} &
		\hspace{-0.2in}
		\includegraphics[scale=0.35]{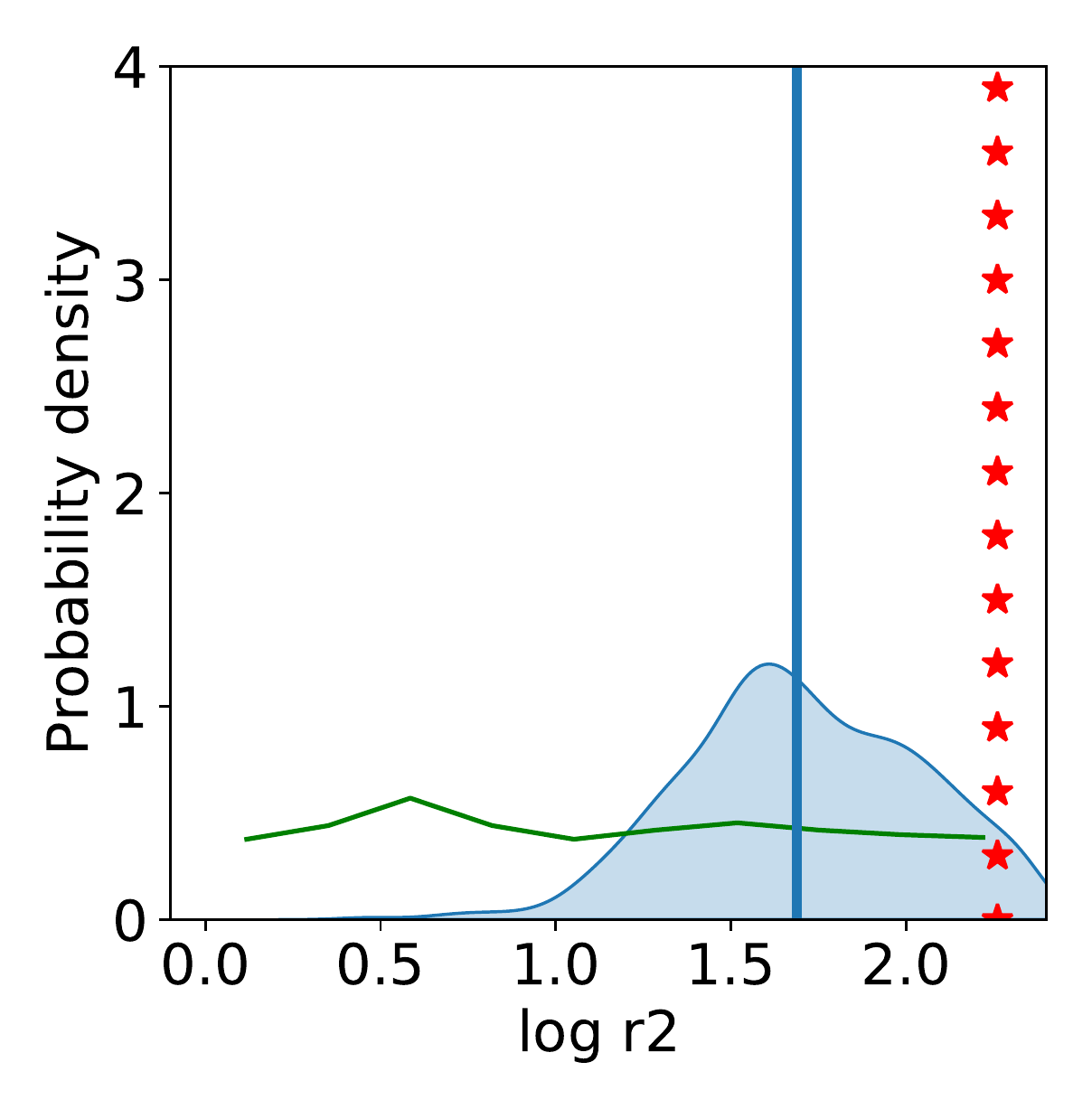} \\
		\end{tabular}
	\end{subfigure}

 \hspace{-0.5in}
 	\begin{subfigure}[normla]{0.5\textwidth}
		\begin{tabular}{ccc}
		\hspace{-0.2in}
		\includegraphics[scale=0.35]{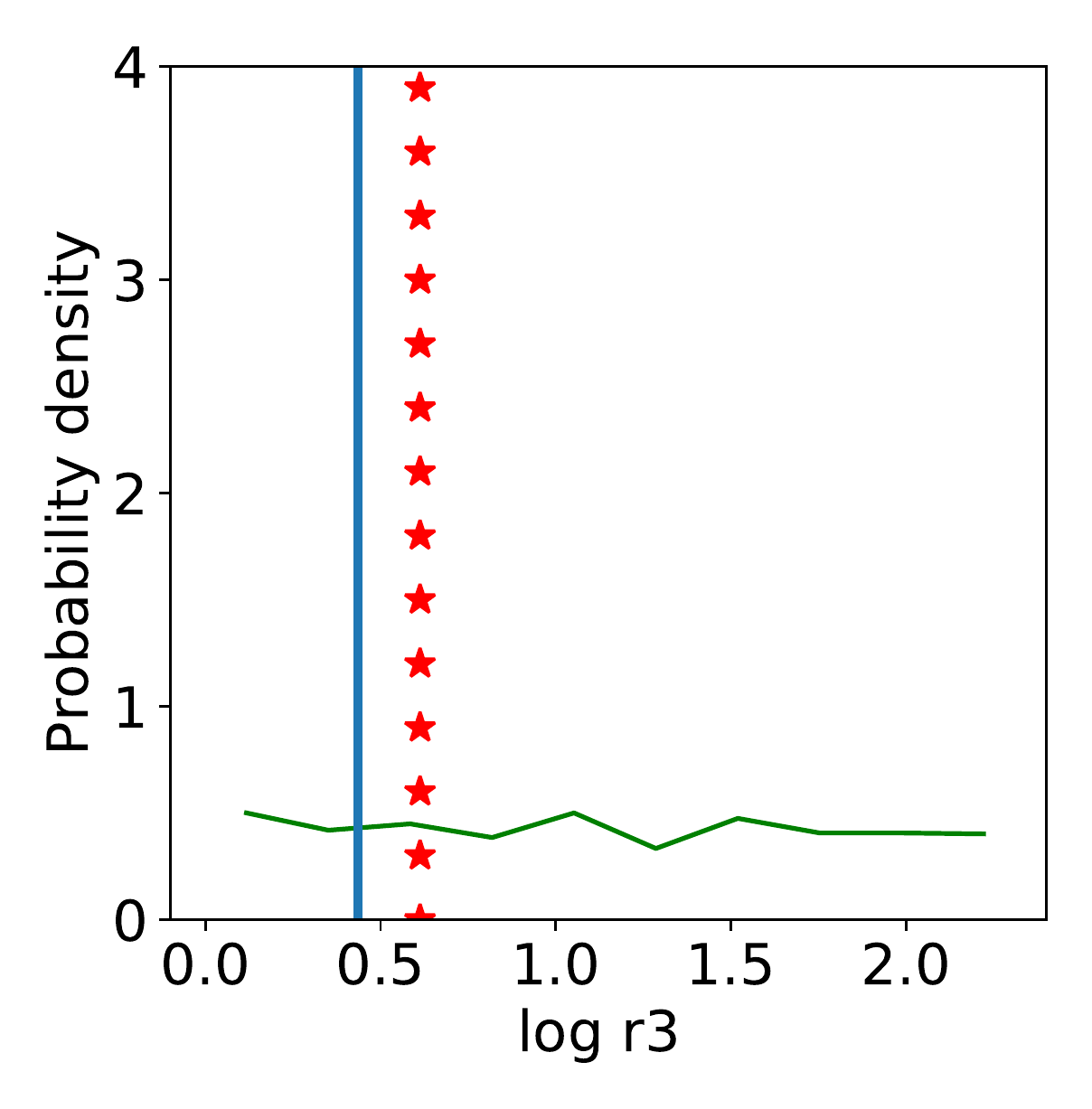} &
		\hspace{-0.2in}
		\includegraphics[scale=0.35]{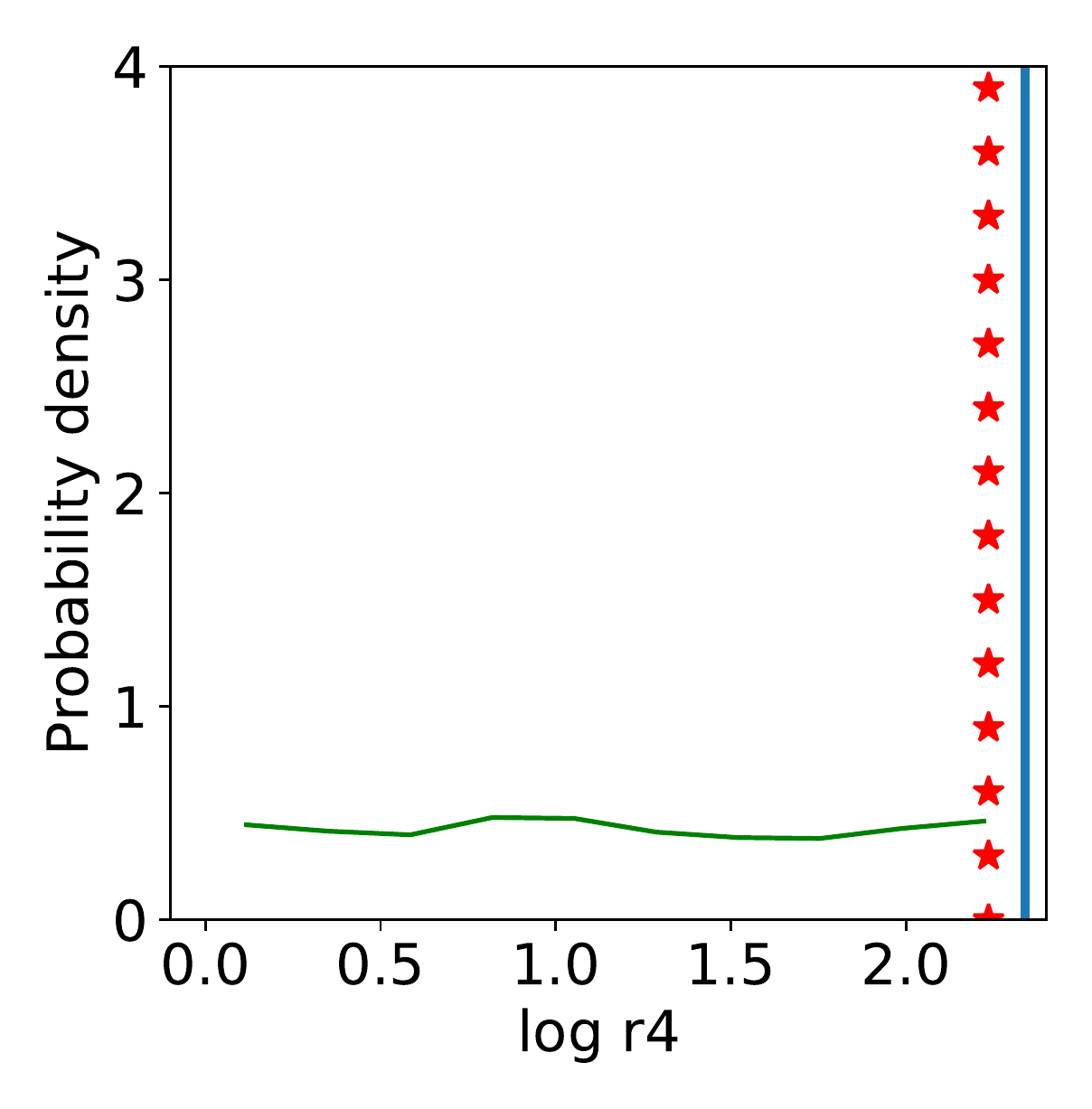} \\ 
		\end{tabular}
	\end{subfigure}
 \hspace{0.2in}	
	\begin{subfigure}[normla]{0.5\textwidth}
	   \begin{tabular}{cc}
		\includegraphics[scale=0.35]{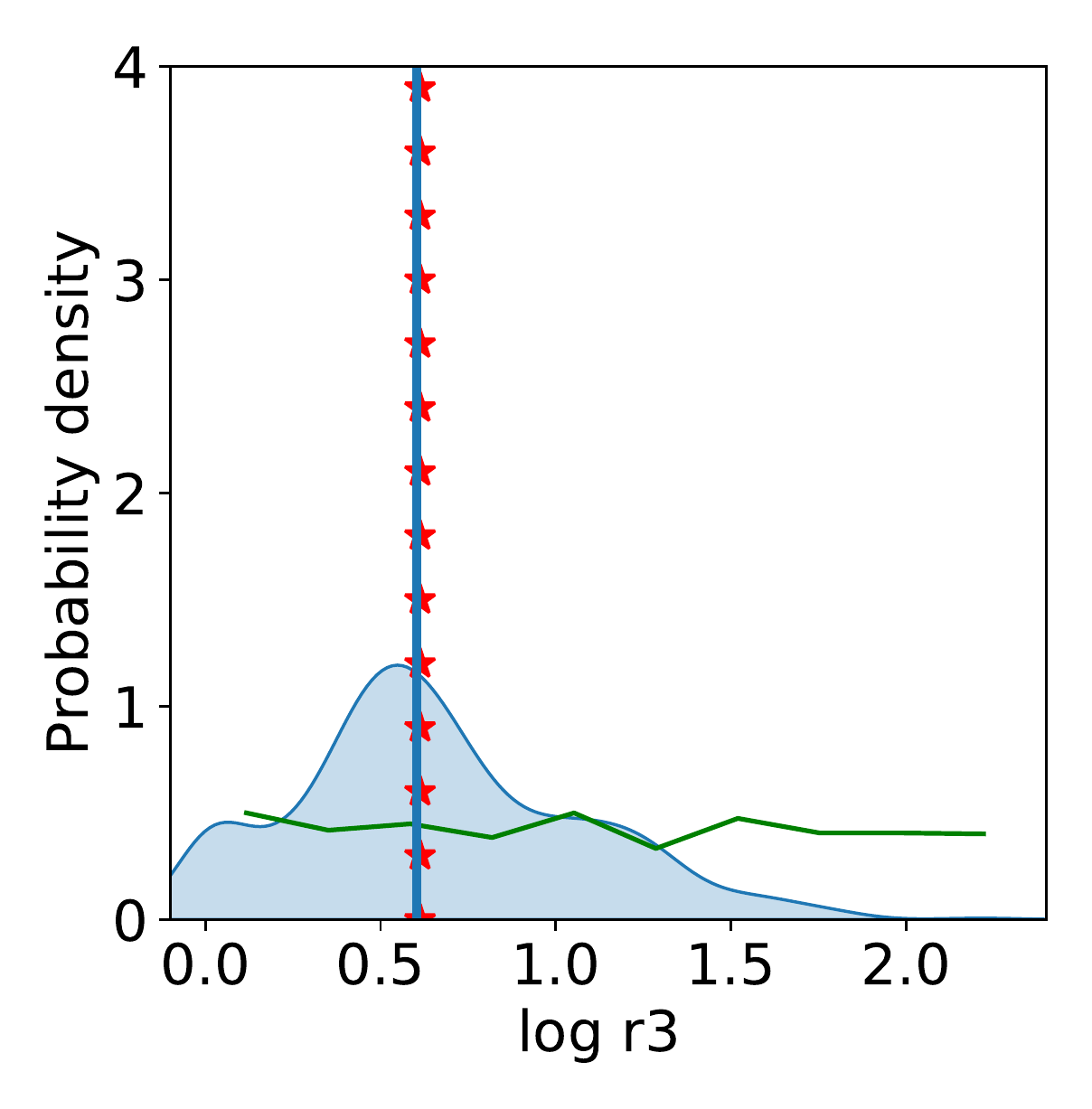} &
		\hspace{-0.2in}
		\includegraphics[scale=0.35]{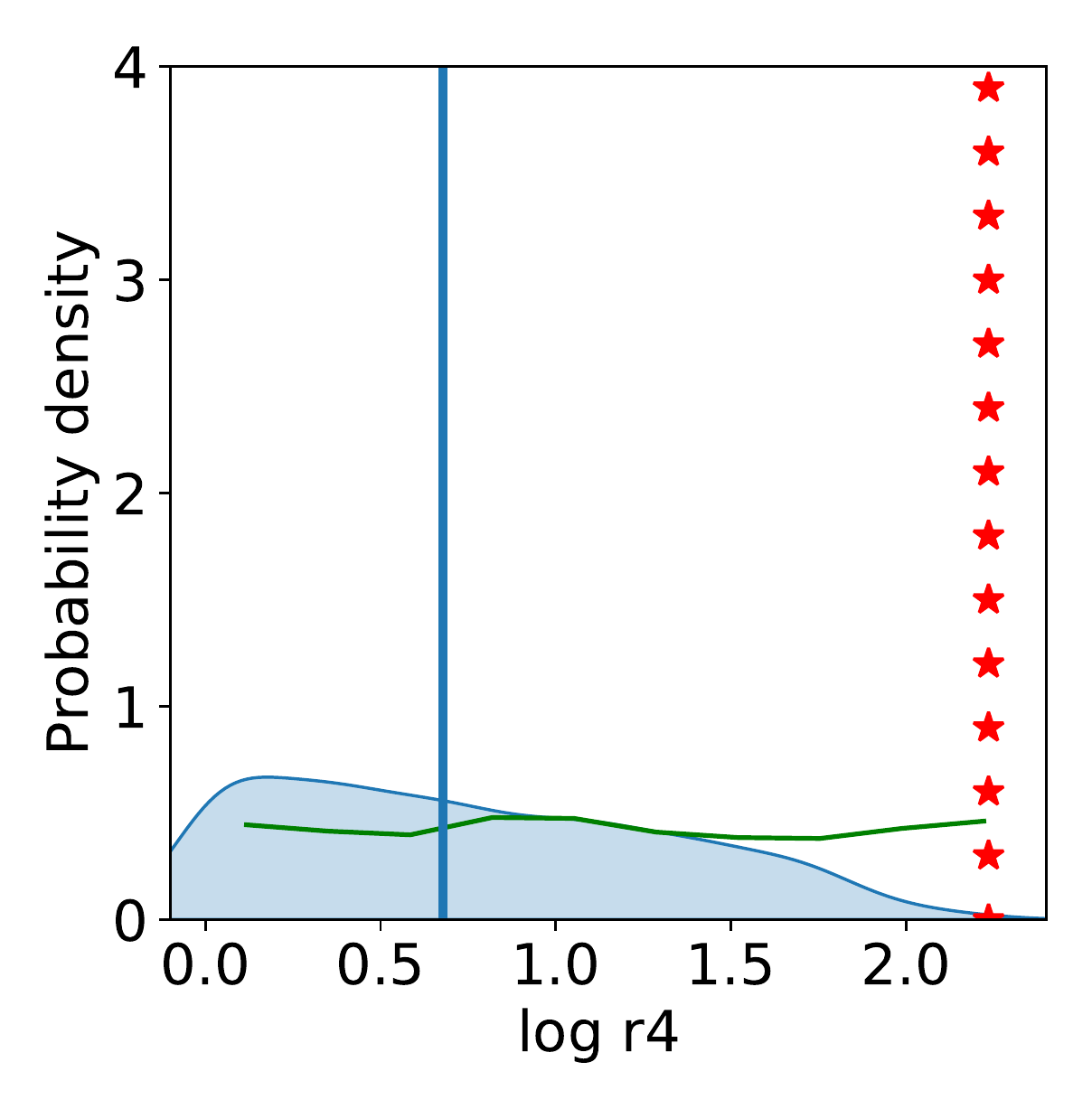} \\
		\end{tabular}
	\end{subfigure}

 \hspace{-0.5in}
 	\begin{subfigure}[normla]{0.5\textwidth}
		\begin{tabular}{ccc}
		\hspace{-0.2in}
		\includegraphics[scale=0.35]{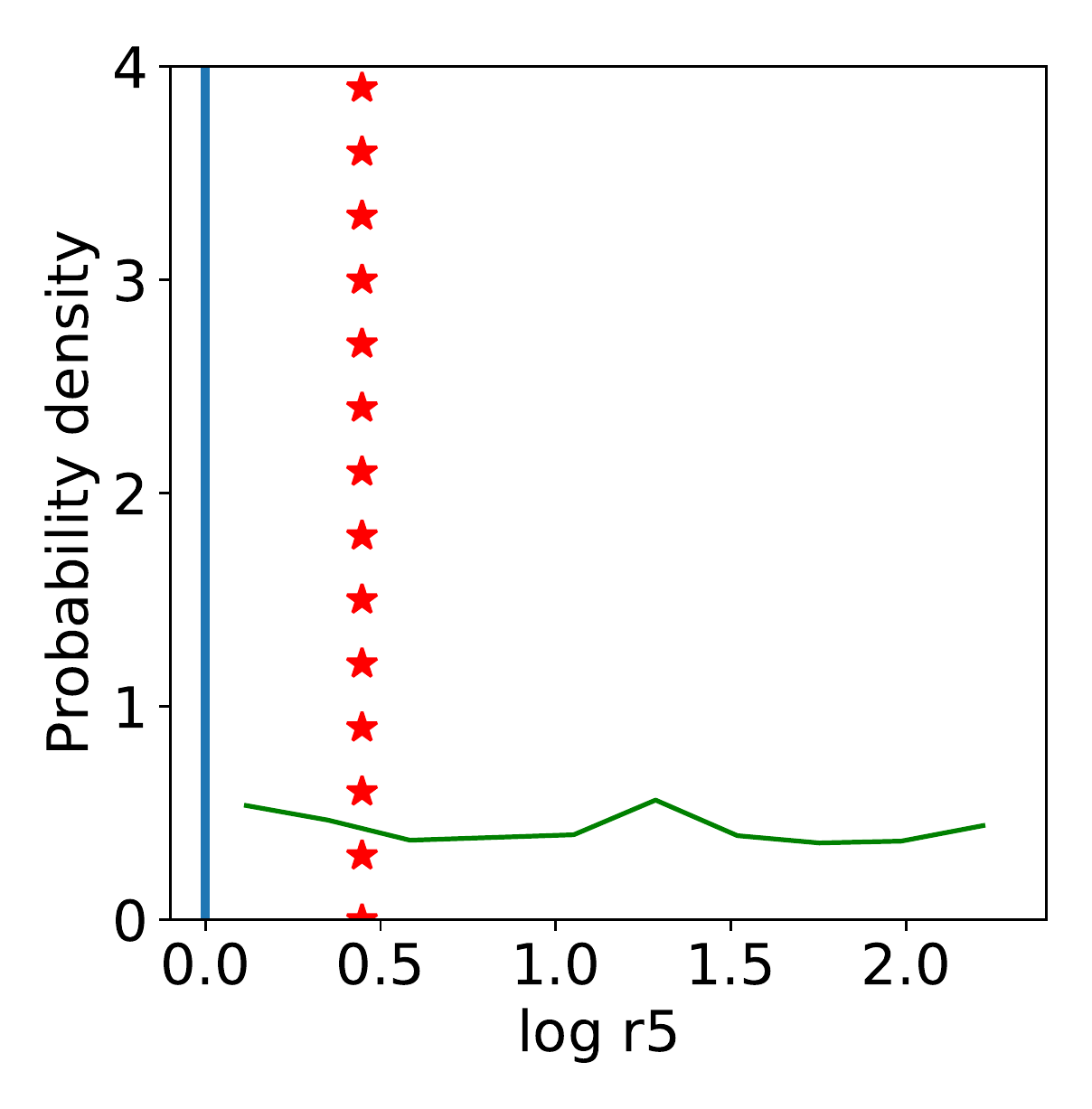} &
		\hspace{-0.2in}
		\includegraphics[scale=0.35]{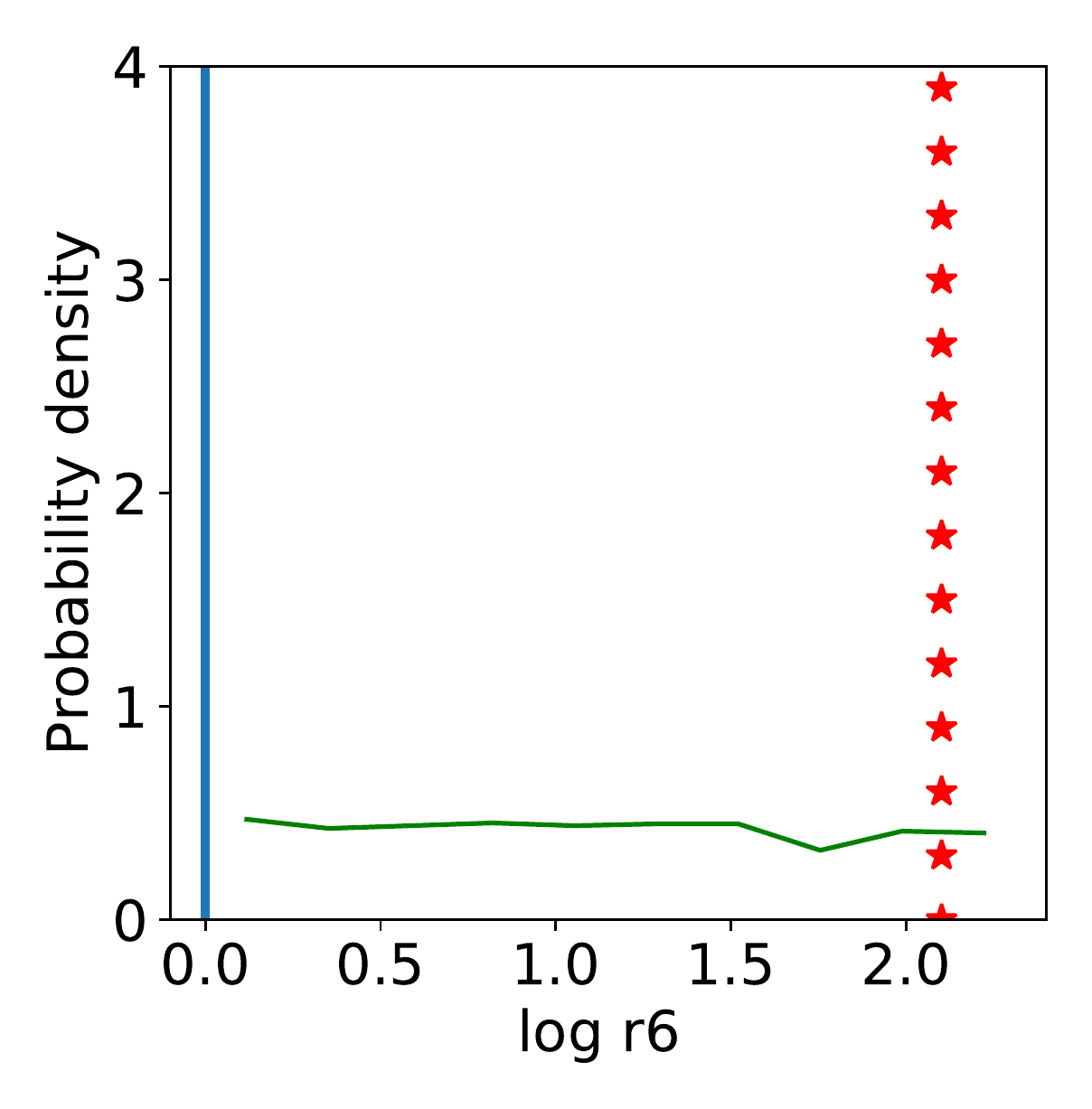} \\ 
		\end{tabular}
	\end{subfigure}
 \hspace{0.2in}	
	\begin{subfigure}[normla]{0.5\textwidth}
	   \begin{tabular}{cc}
		\includegraphics[scale=0.35]{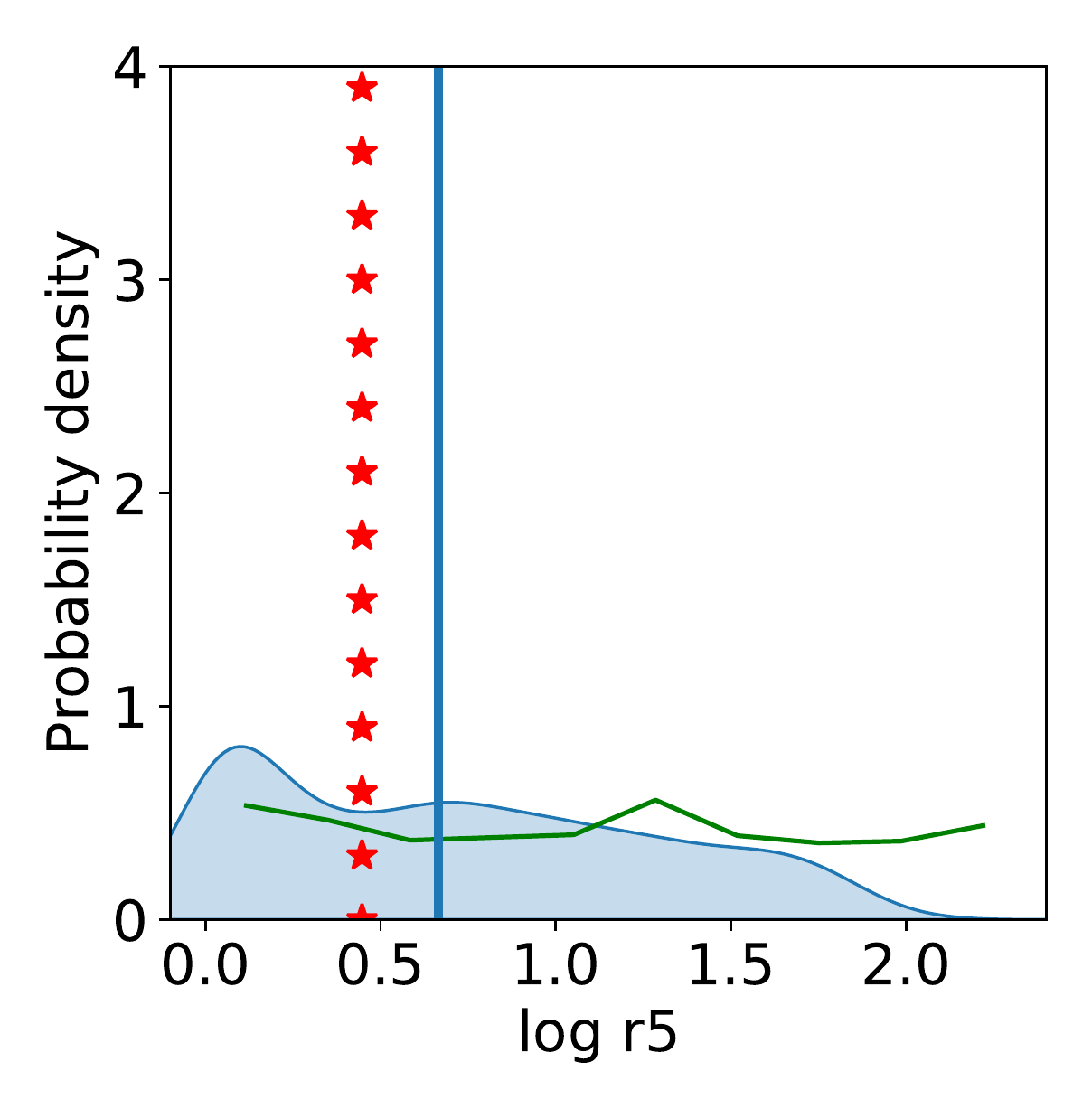} &
		\hspace{-0.2in}
		\includegraphics[scale=0.35]{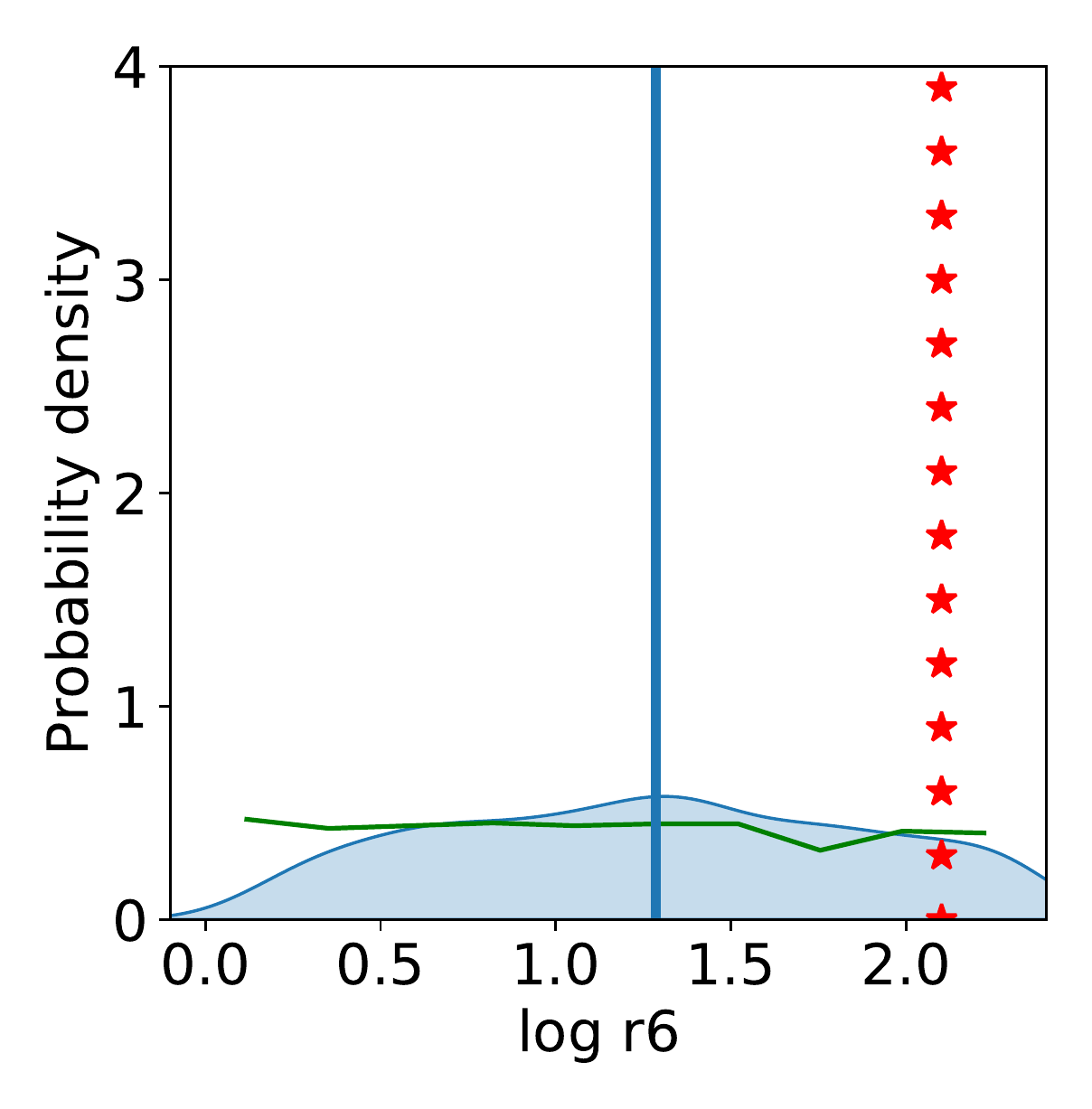} \\
		\end{tabular}
	\end{subfigure}
		
  \hspace{-0.1in}
		 ESMDA
 \hspace{2.5in}
 		 FlexIES
 \end{center}

\caption{Prior and posterior distribution of layers resistivities in case of multi-modality and uninformed (wide) prior. Green and blue lines show ensemble approximation of the prior and posterior distribution respectively and red stars show reference truth. Solid blue lines show $p50$ of the posterior distribution. The sub-figures in the first and second columns show results obtained from the ESMDA algorithm and the sub-figures in the third and fourth columns show results from the FlexIES algorithm. The posterior distributions appear as the point estimate in the first and second columns of the sub-figures.}
\label{post_outputs_3_2}
\end{figure}

\begin{figure}[H]
\begin{center}

 \hspace{-0.5in}
 	\begin{subfigure}[normla]{0.5\textwidth}
		\begin{tabular}{ccc}
		\hspace{-0.2in}
		\includegraphics[scale=0.35]{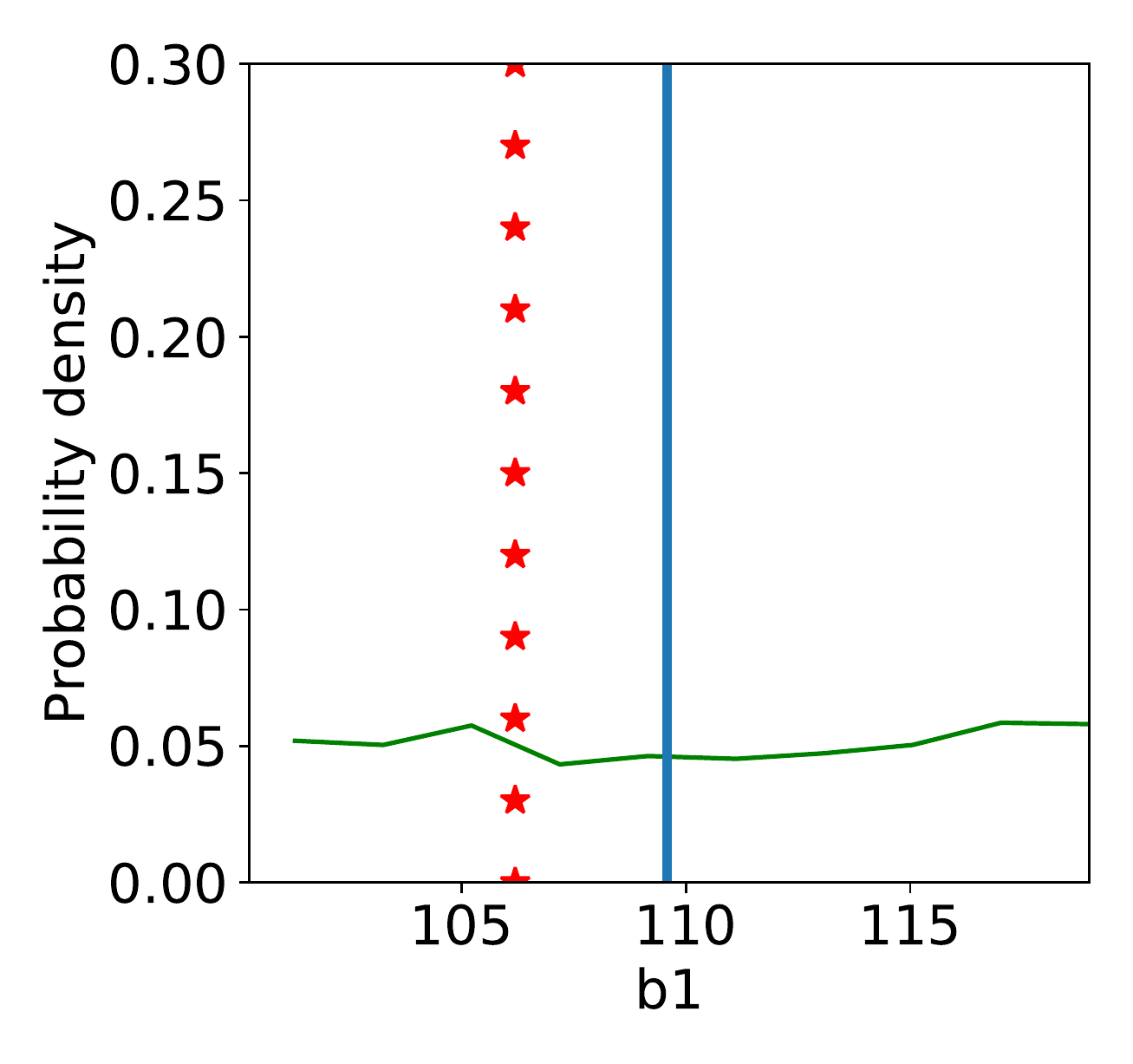} &
		\hspace{-0.2in}
		\includegraphics[scale=0.35]{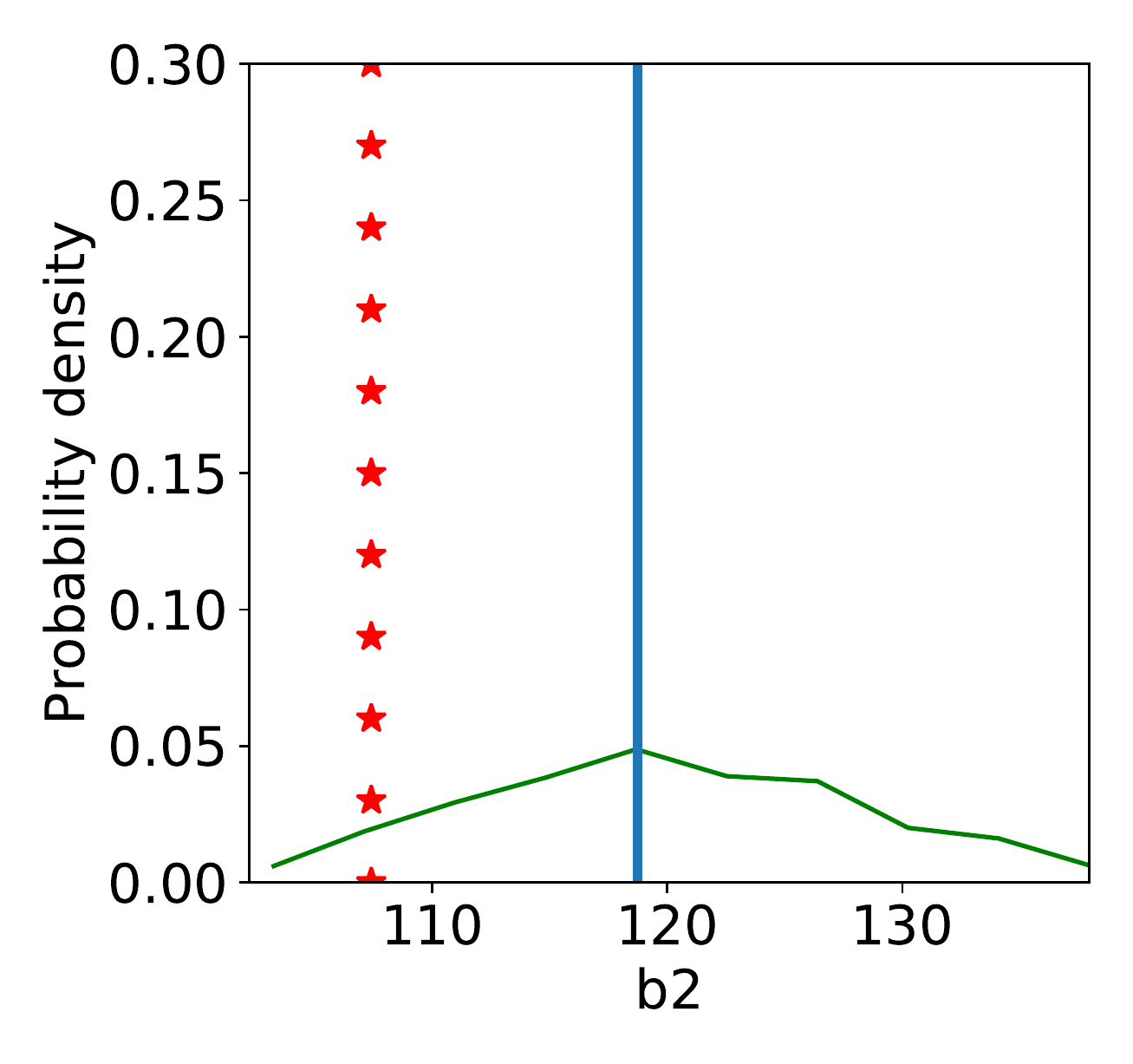} \\ 
		\end{tabular}
	\end{subfigure}
 \hspace{0.2in}	
	\begin{subfigure}[normla]{0.5\textwidth}
	   \begin{tabular}{cc}
		\includegraphics[scale=0.35]{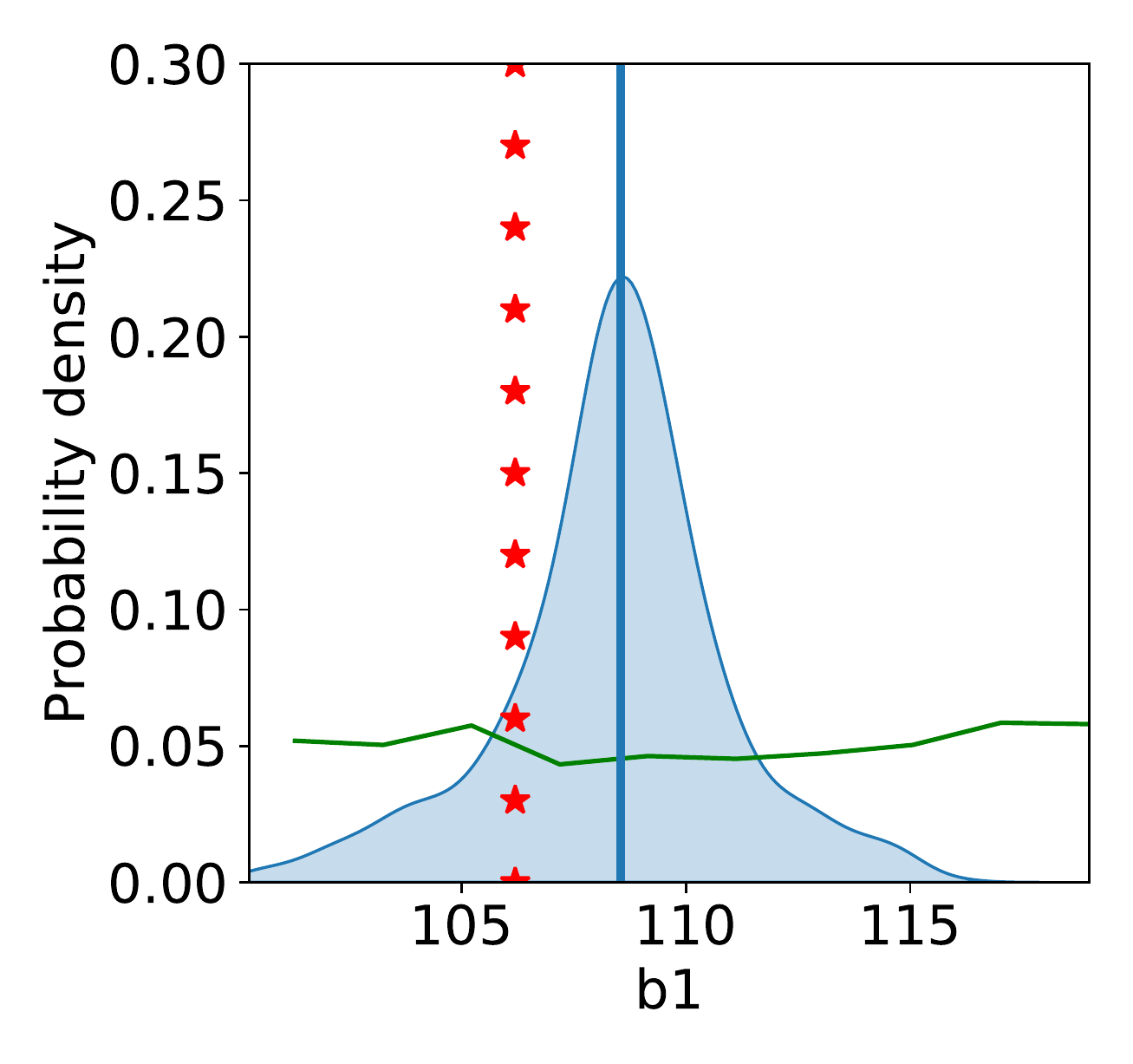} &
		\hspace{-0.2in}
		\includegraphics[scale=0.35]{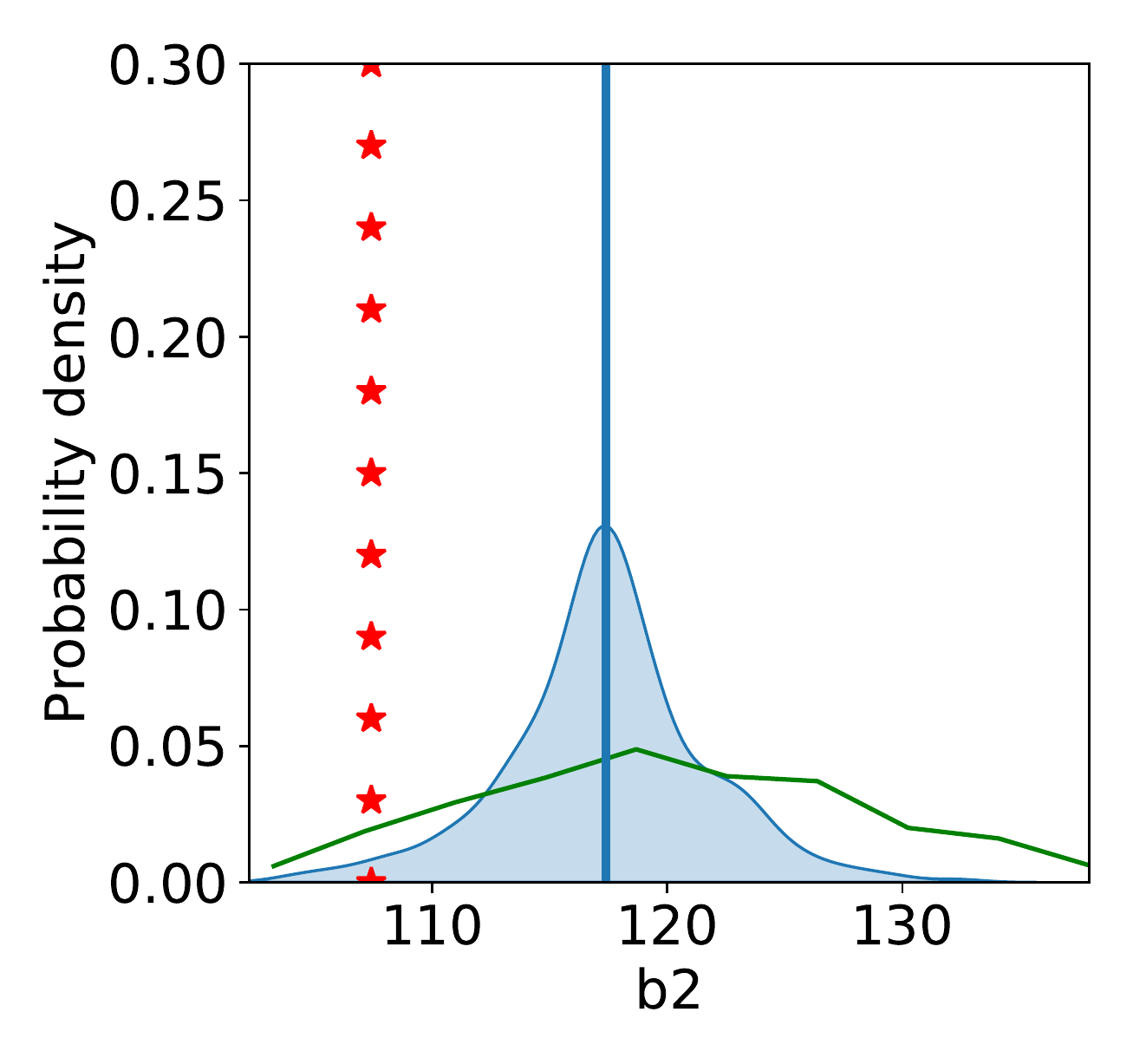} \\
		\end{tabular}
	\end{subfigure}

 \hspace{-0.5in}
 	\begin{subfigure}[normla]{0.5\textwidth}
		\begin{tabular}{ccc}
		\hspace{-0.2in}
		\includegraphics[scale=0.35]{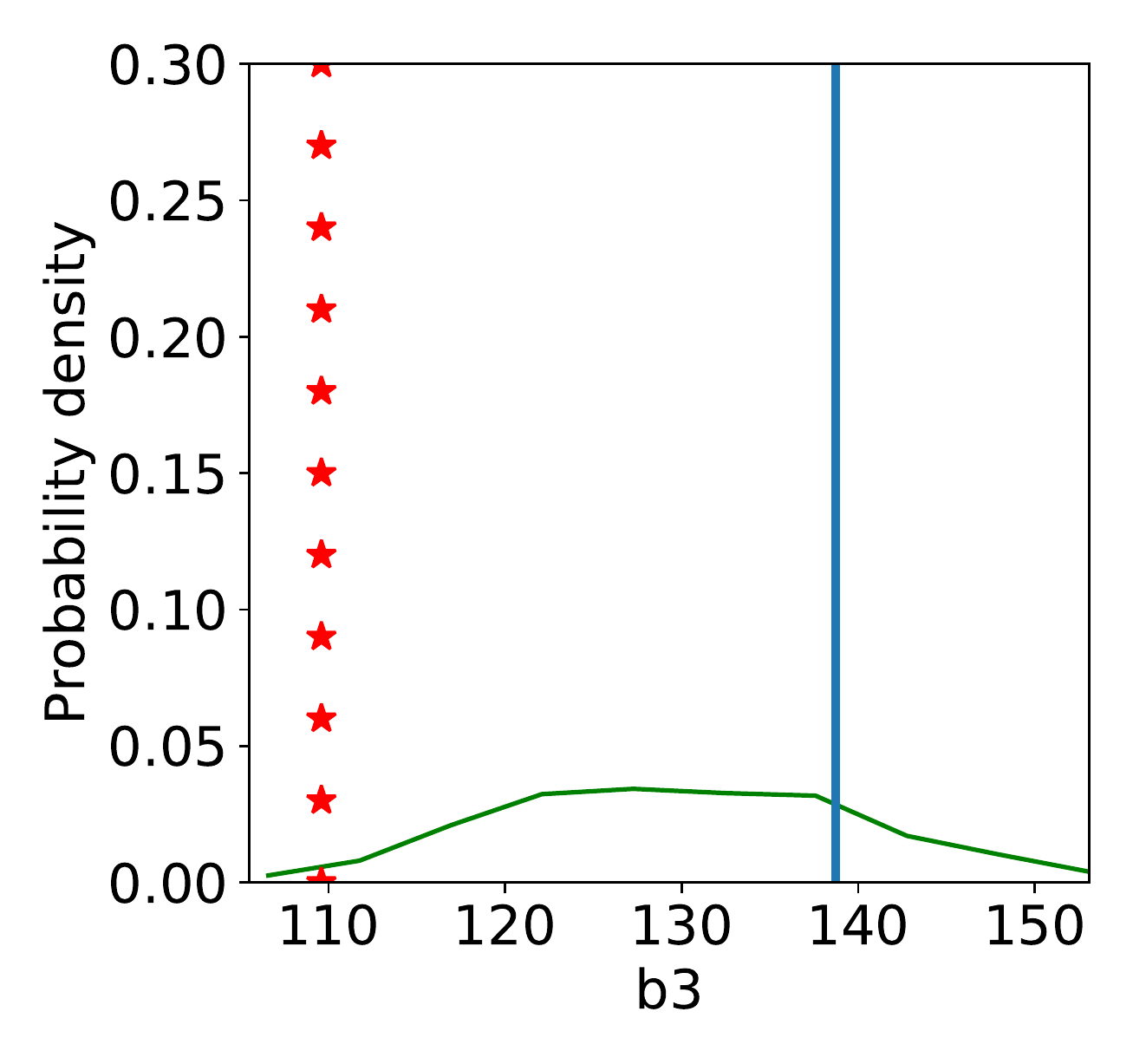} &
		\hspace{-0.2in}
		\includegraphics[scale=0.35]{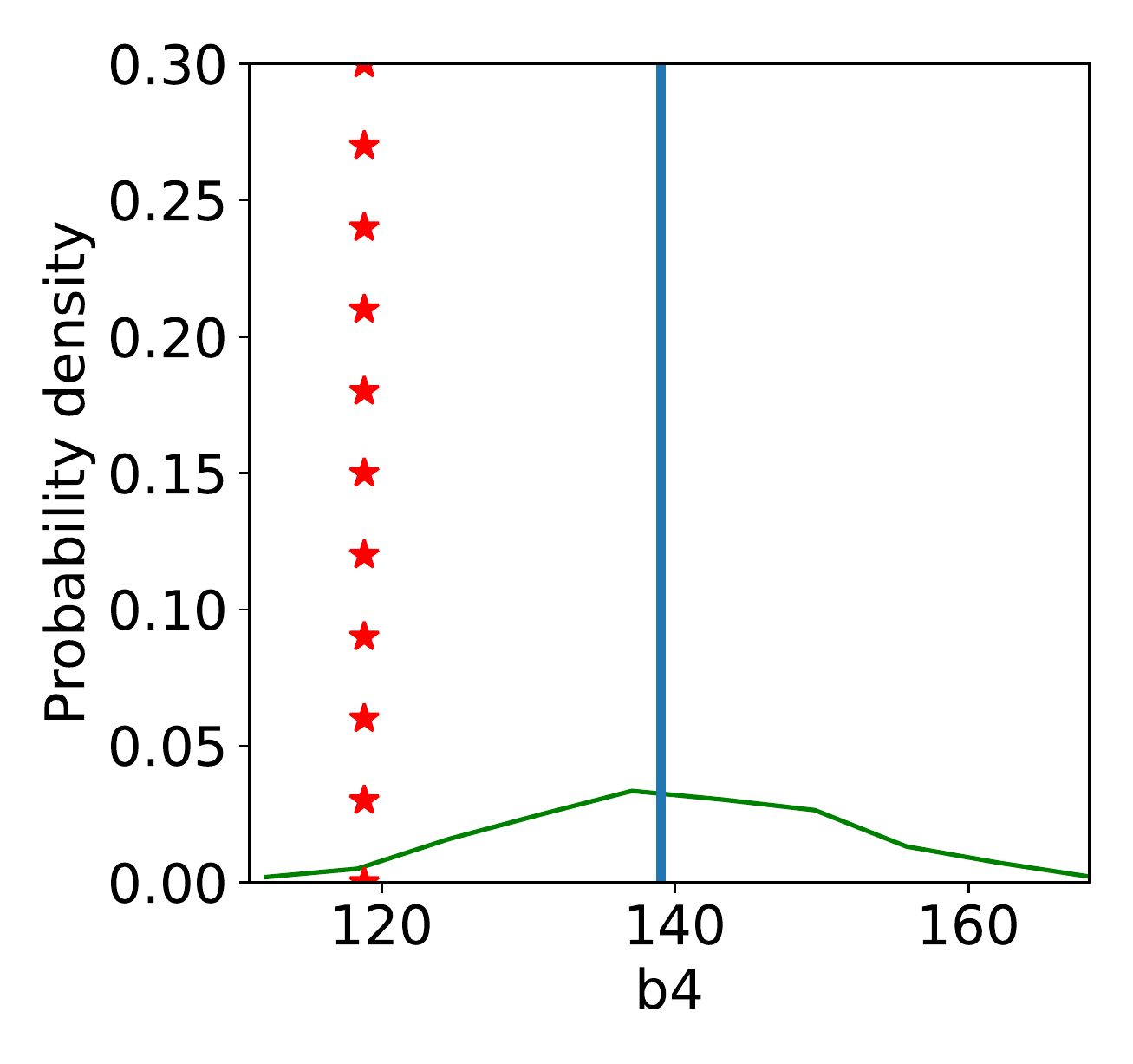} \\ 
		\end{tabular}
	\end{subfigure}
 \hspace{0.2in}	
	\begin{subfigure}[normla]{0.5\textwidth}
	   \begin{tabular}{cc}
		\includegraphics[scale=0.35]{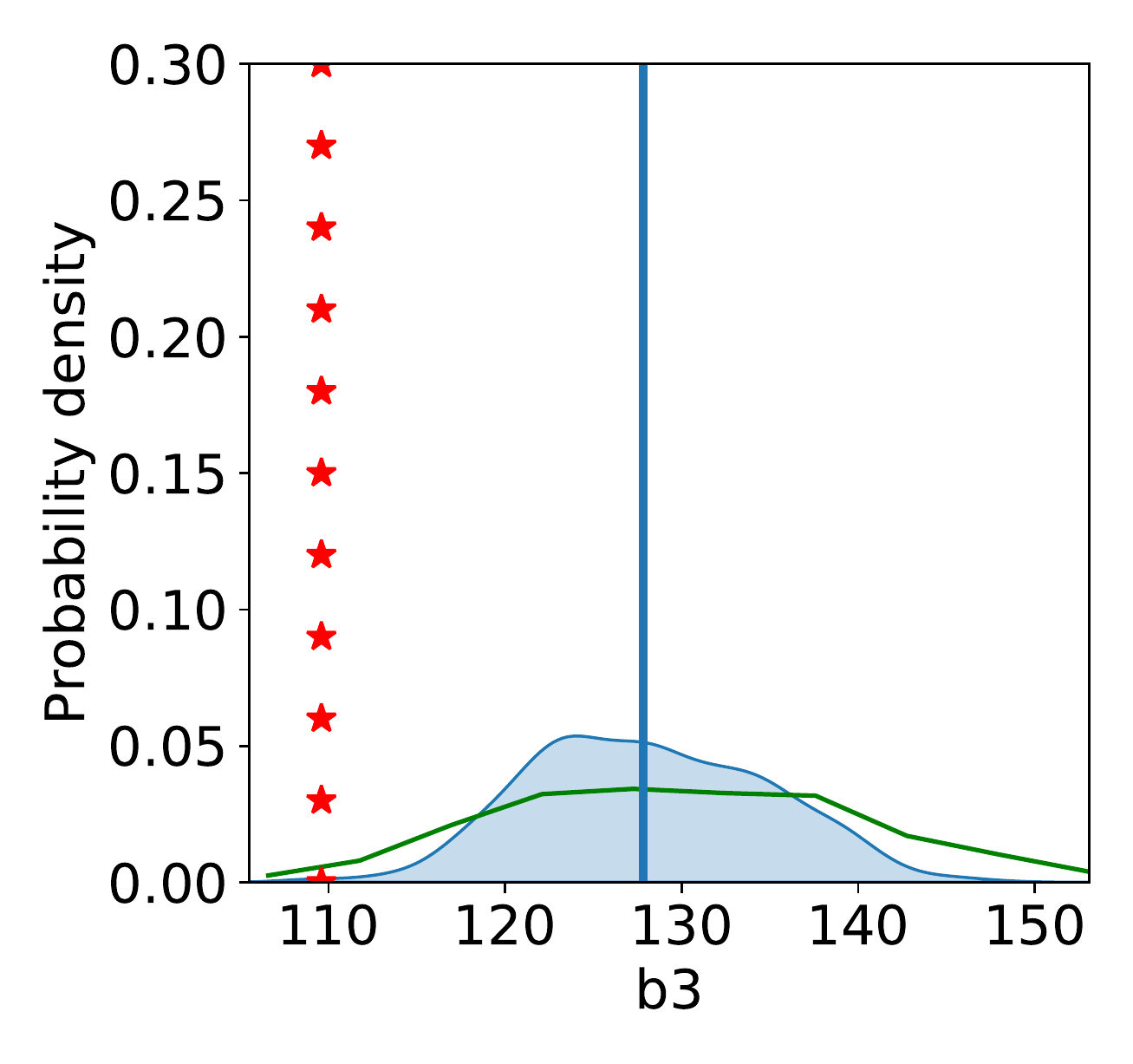} &
		\hspace{-0.2in}
		\includegraphics[scale=0.35]{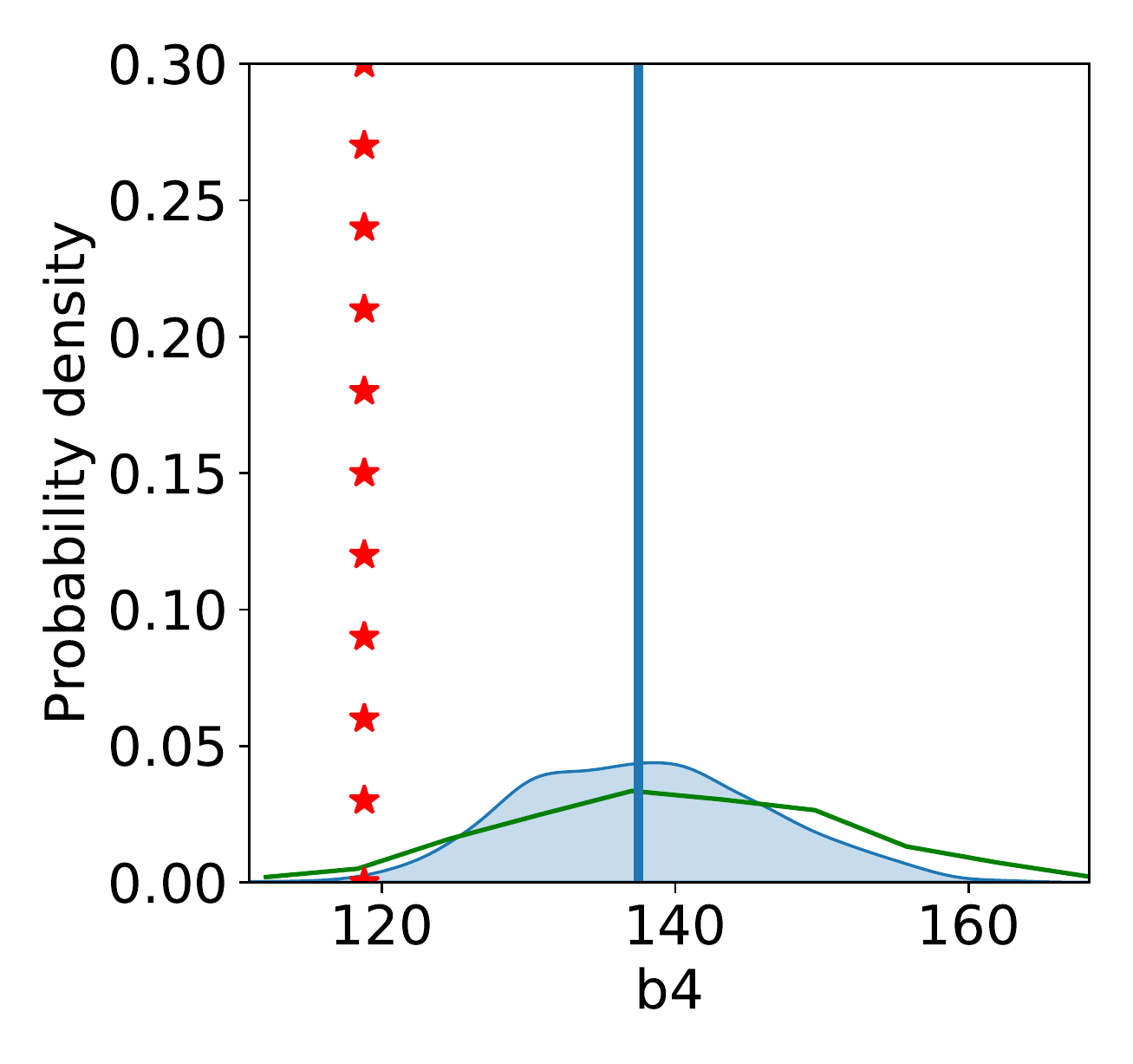} \\
		\end{tabular}
	\end{subfigure}

 \hspace{-0.5in}
 	\begin{subfigure}[normla]{0.5\textwidth}
		\begin{tabular}{ccc}
		\hspace{-0.2in}
		\includegraphics[scale=0.35]{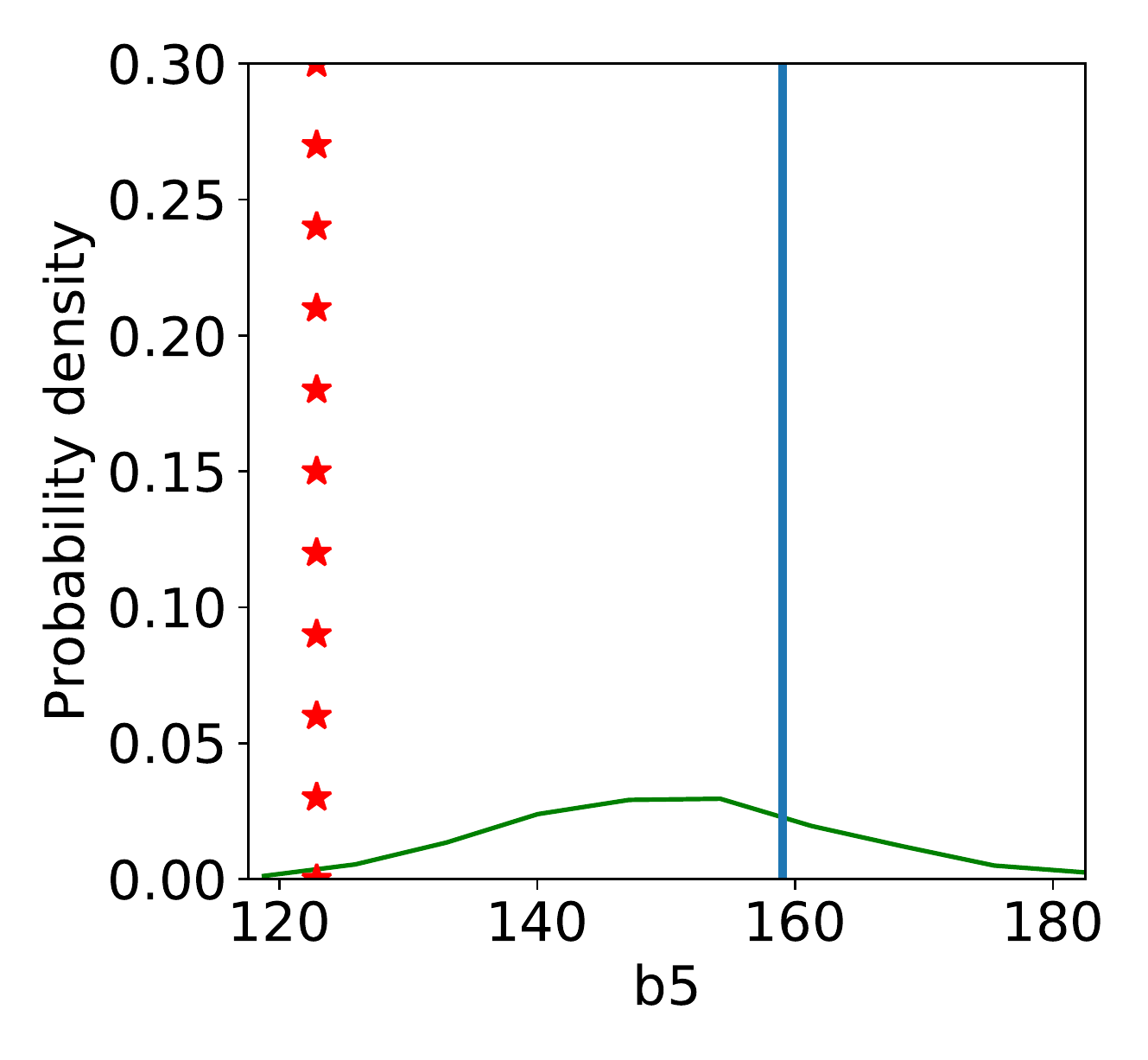} &
		\hspace{-0.2in}
		\includegraphics[scale=0.35]{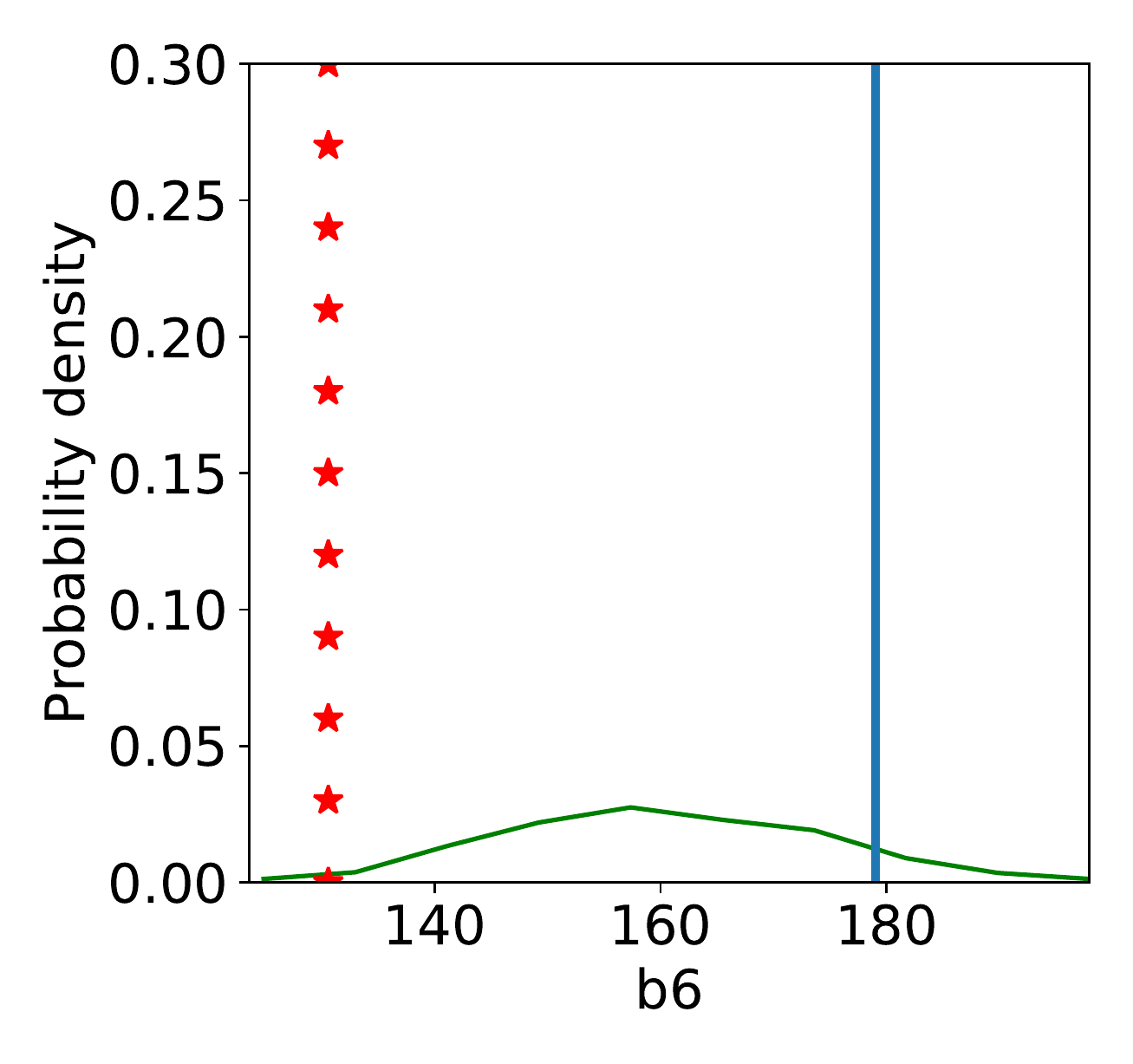} \\ 
		\end{tabular}
	\end{subfigure}
 \hspace{0.2in}	
	\begin{subfigure}[normla]{0.5\textwidth}
	   \begin{tabular}{cc}
		\includegraphics[scale=0.35]{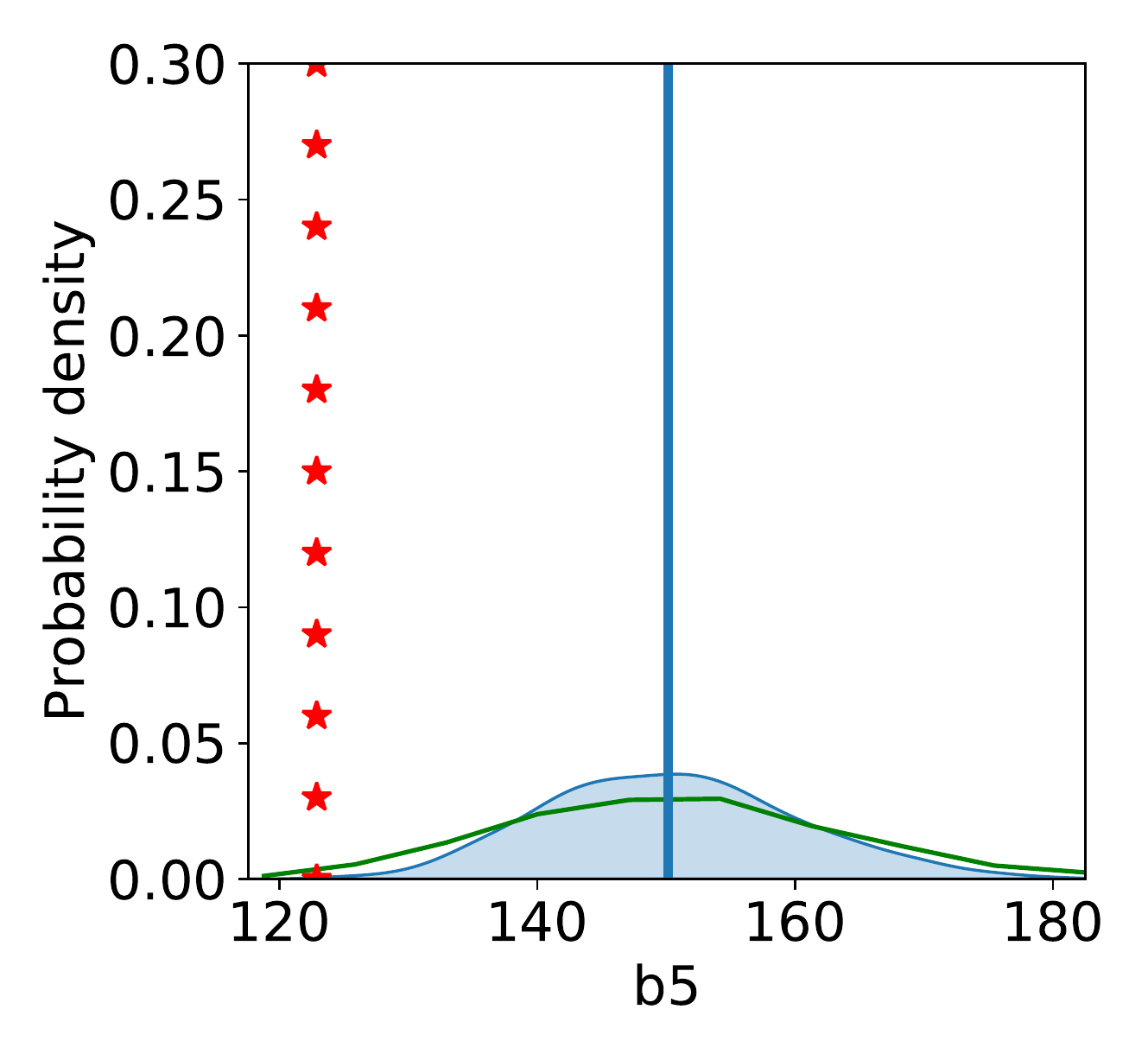} &
		\hspace{-0.2in}
		\includegraphics[scale=0.35]{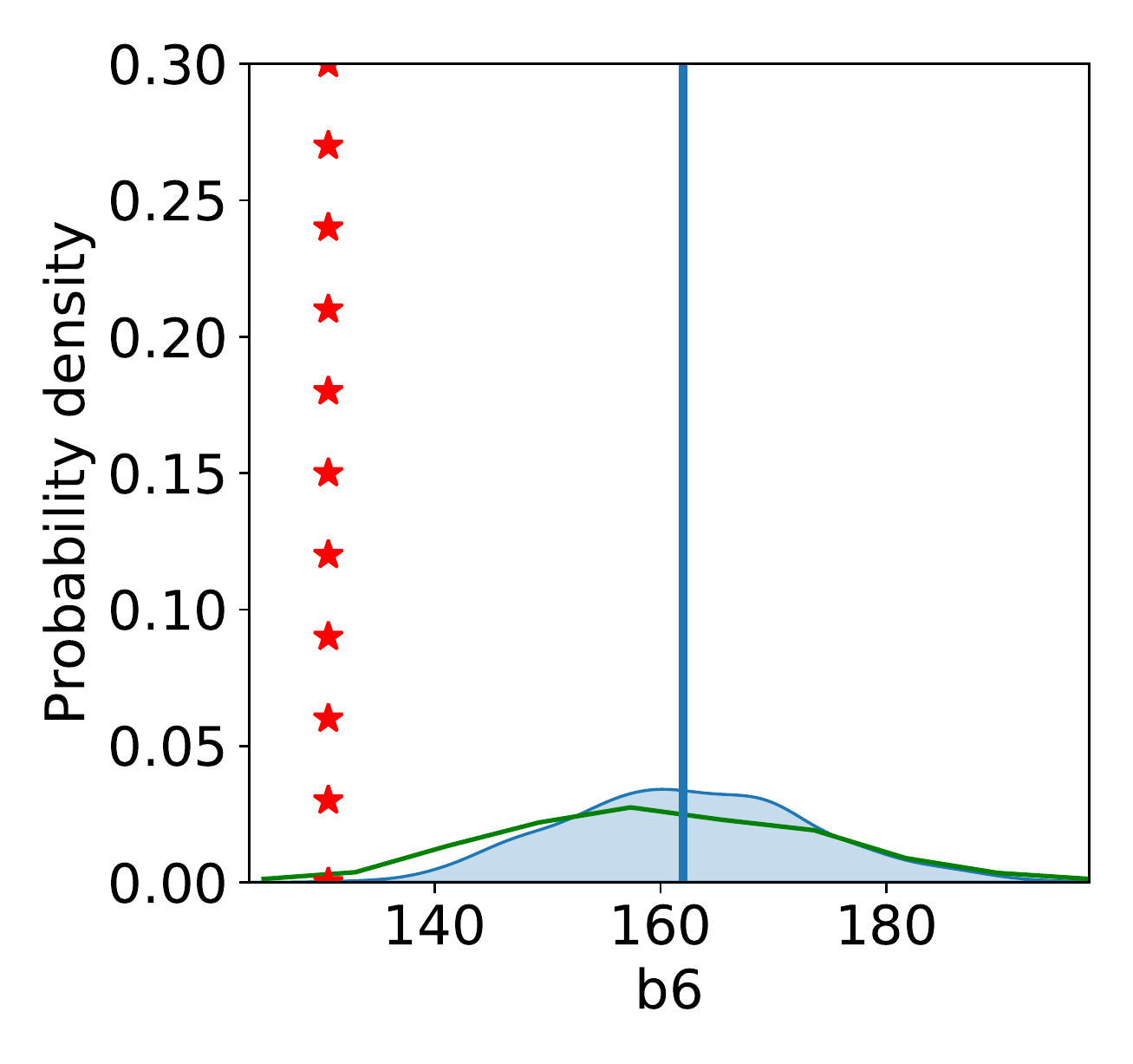} \\
		\end{tabular}
	\end{subfigure}
		
  \hspace{-0.1in}
		 ESMDA
 \hspace{2.5in}
 		 FlexIES
 \end{center}

\caption{Prior and posterior distribution of layer boundary positions  (true depth converted from thicknesses) in case of multi-modality caused by an uninformed (wide) prior. Green and blue lines show ensemble approximation of the prior and posterior distribution respectively and red stars show reference truth. Solid blue lines show $p50$ of the posterior distribution. The sub-figures in the first and second columns show results obtained from the ESMDA algorithm and the sub-figures in the third and fourth columns show results from the FlexIES algorithm. The posterior distributions appear as the point estimate in the first and second columns of the sub-figures.}
\label{post_outputs_3_3}
\end{figure}

In this setup without any model-error or measurement errors, we should get a perfect match. 
However, due to the multi-modality of the joint estimation problem of layers resistivities and thicknesses, we couldn’t obtain a match using ESMDA at the lower depths and the ensemble collapses as shown in Figure~(\ref{post_outputs_3_1}). 
The data mismatch from ESMDA inversion (Figure~(\ref{post_outputs_3_1})) shows that the estimated layers resistivities and thicknesses are trapped into a single local mode (local minimum) of the posterior. 
This results in a completely wrong estimation of the layers resistivities and thicknesses as shown in Figures~(\ref{post_outputs_3_2}) and (\ref{post_outputs_3_3}). 

\begin{figure}[H]
\begin{center}

 \hspace{-0.5in}
    \begin{subfigure}[normal]{0.4\textwidth}
	\includegraphics[scale=0.5]{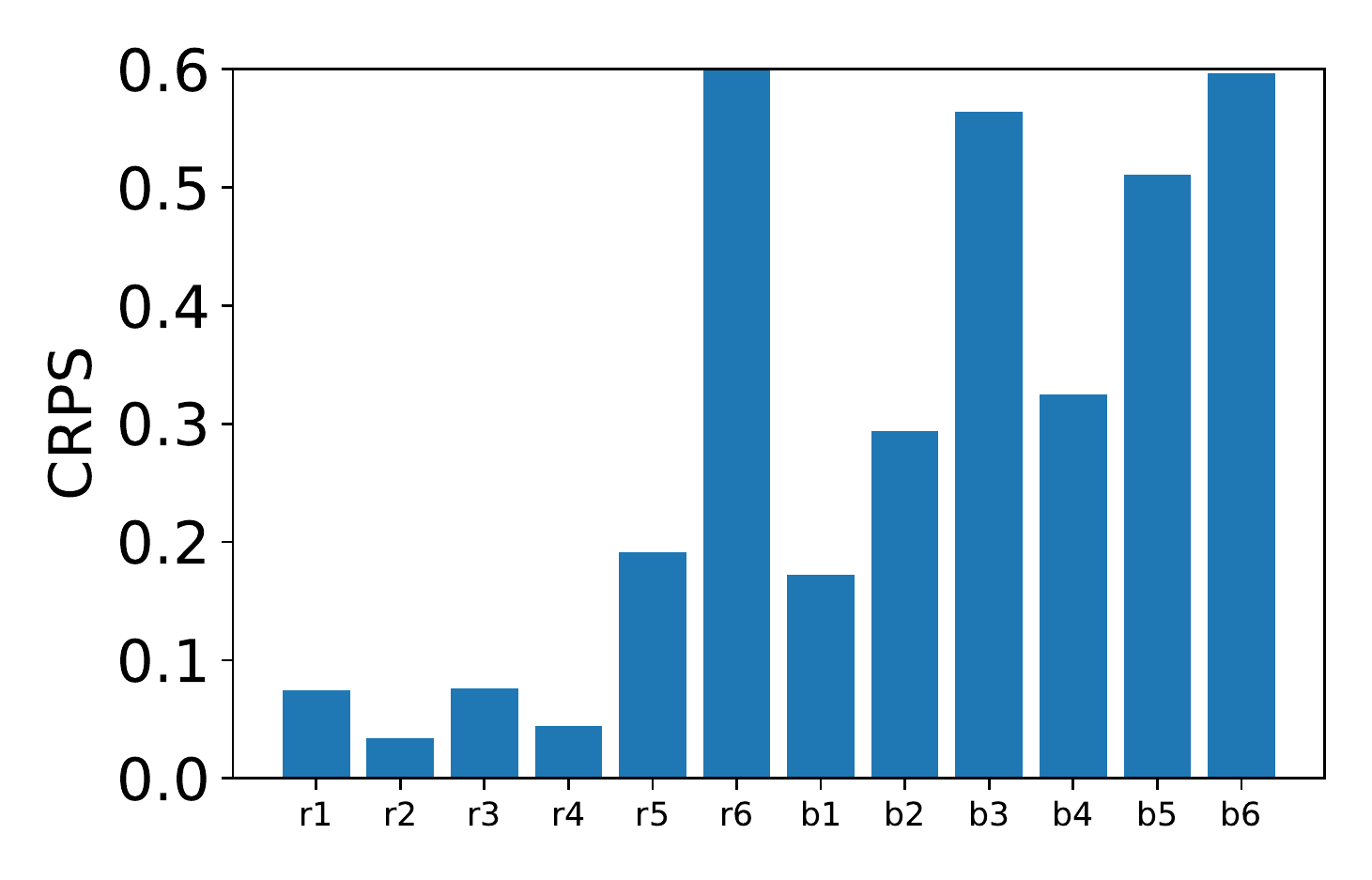}	
	\end{subfigure}
 \hspace{0.5in}
	\begin{subfigure}[normal]{0.4\textwidth}
	\includegraphics[scale=0.5]{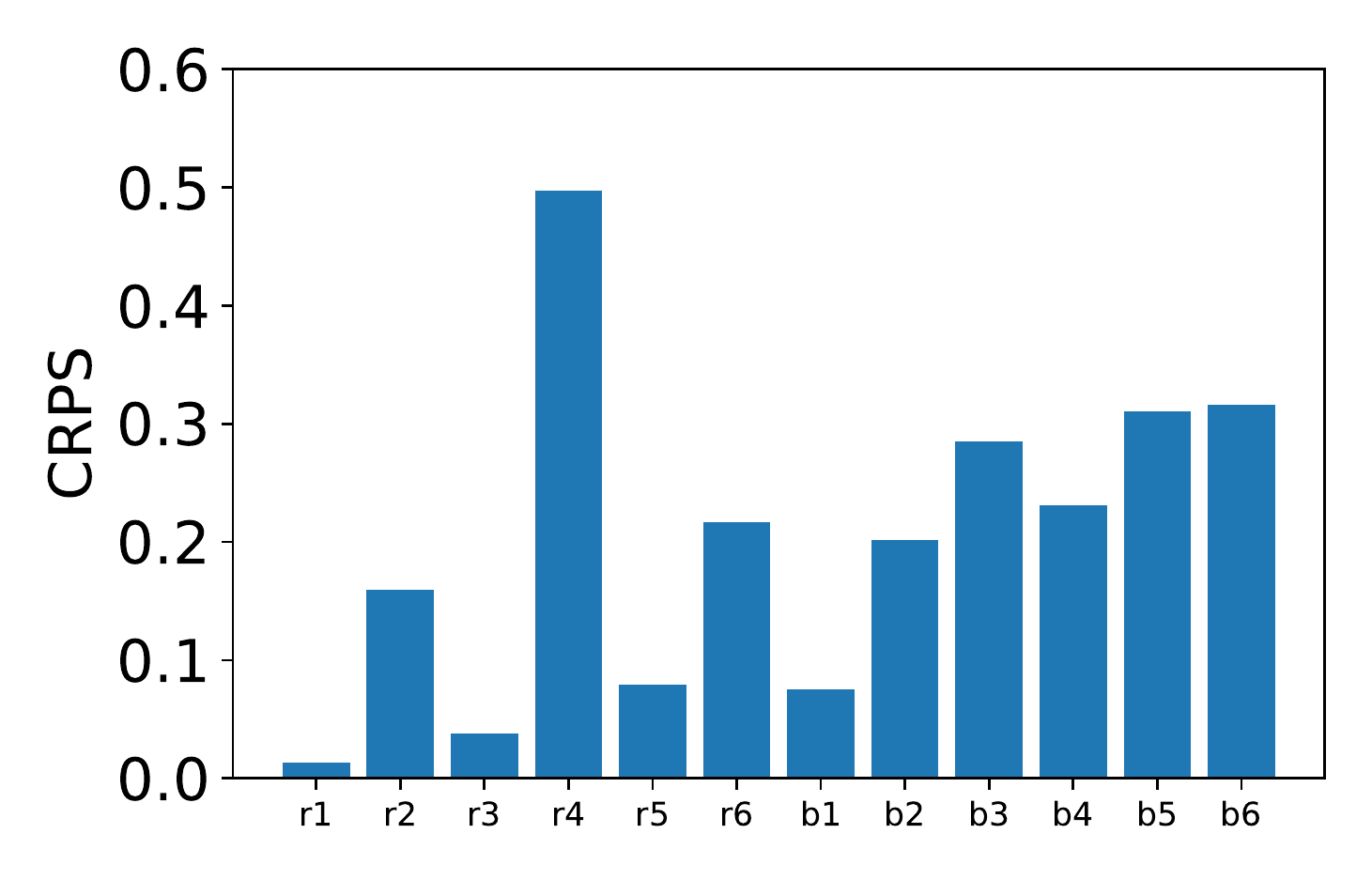}	
	\end{subfigure}

  \hspace{-0.1in}
		 ESMDA
 \hspace{2.5in}
 		 FlexIES
 \end{center}

\caption{Continuous ranked probability score (CRPS) obtained from inversion of EM measurements using ESMDA and FlexIES in case of multi-modality and uninformed (wide) prior. Lower values are better.}
\label{CRPS_3}
\end{figure}

FlexIES shows completely different behaviour of the data mismatch as compared to the ESMDA as shown in Figure~(\ref{post_outputs_3_1}). The ensemble of the EM output obtained from FlexIES are not trapped into the single mode of the posterior, which is evident from a very wide interval of the output of EM measurements. 
{\color{black} The wide interval may indicate that} different ensemble members are trapped in the different local modes of the posterior, which could be an advantage compared to the completely biased estimation with the erroneously low uncertainty. 
{\color{black} The wide interval could avoid completely wrong estimation} of the layers resistivities and thicknesses as shown in Figures~(\ref{post_outputs_3_2}) and (\ref{post_outputs_3_3}).
Furthermore, the CRPS of the estimated resistivities and thicknesses also improves significantly using FlexIES as compared to ESMDA as shown in Figure~(\ref{CRPS_3}) {\color{black} due to the coverage of the reference solution by the posterior distribution.} 
These improvements show that the FlexIES can be useful to alleviate the problems associated with the multi-modality or local modes of the real-time inversion.

\section{Conclusions}   
\label{sec:conclusions}

In this paper we presented a framework for real-time probabilistic inversion and model-error assessment of DNN models for borehole EM measurements with FlexIES.
When the model-error is not present, the results of the proposed method coincide with the classical ESMDA, which is a popular method for probabilistic inversion. 
Furthermore, in realistic scenarios,
FlexIES has the capability to capture the unknown model-error, which is useful to avoid over-confidence in the estimated parameters that might be inaccurate (convergence to wrong solution).
This increases the reliability of the real-time inversion, and results in a less biased estimation of the model parameters without sacrificing the real-time performance. 
The reliability comes at a cost of wider posterior distribution of the parameters.
We argue that the wider posterior is preferable when compared to a biased estimate with narrow posterior, because the full probabilistic information can be utilized in decision analysis during future developments.

In the presented numerical results, we invert borehole EM measurements
to estimate only the subsurface properties (layers resistivities),
and then to jointly estimate the properties and the boundary locations (layers thicknesses).
When hindered by model-errors, for both cases the FlexIES has more reliable estimation (low CRPS).
In a general case with an uninformed prior 
the joint inversion is an ill-posed problem admitting multiple local modes and has several local minima. 
Ensemble smothers, including ESMDA, can collapse to a single local mode and cannot handle such problems directly. {\color{black} However, FlexIES effectively detects multi-modality and results in wider uncertainty ranges for the estimated parameters, which can converge to multiple local modes indicated by the split parameter (i.e.~$s_p = 1$). Comparing the inversion results obtained by FlexIES with classical ESMDA allows us to detect that multi-modality is the source of the uncertainty, which can be minimized by utilizing an informed (reduced/narrow or intelligent) prior in practical applications. If we don't have an informed prior during real-time inversion, then FlexIES can be used to alleviate the problem of multi-modality by preserving the estimated parameters uncertainties.}

{\color{black} In the future we would try to estimate the anisotropic resistivities (i.e.,~different horizontal or traverse resistivities) during the inversion of EM measurements by utilizing DNN as a forward model. The possibility of real-time probabilistic inversion which accounts for model-errors greatly increases the utility for data-driven forward models in practical applications. 
We note the generic formulation of FlexIES is not specific to the presented DNN forward model. 
A direct extension of this work can be a generalization to an arbitrary number of possibly tilted layers with anisotropic properties, that will require retraining of the DNN model for these environments.}
Ultimately, we foresee that the method can be applied in other applications where machine learning, particularly DL models, are available.

\section{Acknowledgments}
Muzammil Hussain Rammay acknowledges the support from the research project 'Geosteering for IOR' (NFR-Petromaks2 project no. 268122), which is funded by the Research Council of Norway, Aker BP, Equinor, V{\aa}r Energi and Baker Hughes Norway. Sergey Alyaev acknowledges the support of the Center for Research-based Innovation DigiWells: Digital Well Center for Value Creation, Competitiveness and Minimum Environmental Footprint (NFR SFI project no. 309589, DigiWells.no). The center is a cooperation of NORCE Norwegian Research Centre, the University of Stavanger, the Norwegian University of Science and Technology (NTNU), and the University of Bergen, and funded by the Research Council of Norway, Aker BP, ConocoPhillips, Equinor, Lundin, TotalEnergies, and Wintershall Dea.

This is a pre-copyedited, author-produced PDF of an article accepted for publication in Geophysical Journal International following peer review. 
The version of record 
"\textbf{Muzammil Hussain Rammay, Sergey Alyaev, Ahmed H Elsheikh,
Probabilistic model-error assessment of deep learning proxies: an application to real-time inversion of borehole electromagnetic measurements,}
Geophysical Journal International, Volume 230, Issue 3, September 2022, Pages 1800–1817"
is available online at:  \url{https://doi.org/10.1093/gji/ggac147}.

\section{Statement regarding the availability of any code/data}

The study includes a proprietary code from the tool vendor, that currently cannot be disclosed. However, the codes related to the classical and flexible iterative ensemble smoothers can be shared on request.

\begin{appendices}

\section{Pseudocode of the ESMDA and FlexIES} \label{sec:AppendixA}
\setcounter{equation}{0}
\renewcommand{\theequation}{\thesection.\arabic{equation}}
\renewcommand{\thetable}{\thesection.\arabic{table}}
\def\NoNumber#1{{\def\alglinenumber##1{}\State #1}\addtocounter{ALG@line}{-1}}

\begin{algorithm}[H]
\renewcommand{\thealgorithm}{}
\caption{ESMDA function for EM inversion using DNN model $\rhd$ {\citep{emerick2013}}}
\begin{algorithmic}

\State

\Function {ESMDA}{$\mathbf{M}$, $\mathbf{EM}_{obs}$, $\mathbf{C}_D$, $DNN$, $\mathbf{t}$}
\State \textbf{Inputs:} $\mathbf{M} \in \mathbb{R}^{N_m \times N_e}$ is the ensemble of input parameters,
$\mathbf{C}_D$ is the covariance matrix of the measurement error, $\mathbf{t}$ is the given well trajectory.
\State Choose $N_a$ \quad $\rhd$ Number of data assimilations/iterations
\State $ i \gets 1$
\State $\alpha = N_a$
\While {$ i <= N_a$ }

\State $\mathbf{D}_{uc} = \mathbf{EM}_{obs}\mathbf{\overrightarrow{1_{N_e}}} + \sqrt{\alpha}\mathbf{C}_D^{1/2}\mathbf{Z}_d$ $\ \ \rhd$ Observation perturbations, $\mathbf{Z}_{d} = [\mathbf{z}_{d1}\; \mathbf{z}_{d2} \; \mathbf{z}_{d3} \; ...... \; \mathbf{z}_{dN_e}]$ $\in\mathbb{R}^{N_d \times N_e}$, $\mathbf{z}_d \sim \mathcal{N}(0, \mathbf{I}_{N_d}) \in \mathbb{R}^{N_d \times 1}$, ${\mathbf{\overrightarrow{1_{N_e}}} \in \mathbb{R}^{1 \times N_e}}$ is a row vector of ones.

\State $ \mathbf{D} = DNN(\mathbf{M, t}) $ \quad $\rhd$ Generate ensemble of model outputs $\mathbf{D} \in \mathbb{R}^{N_d \times N_e}$ from $\mathbf{M}$
\State	$\mathbf{C}_{MD} = \frac{1}{N_e - 1} (\mathbf{M}-\mathbf{\bar{M}}\mathbf{\overrightarrow{1_{N_e}}})(\mathbf{D}-\mathbf{\bar{D}} \mathbf{\overrightarrow{1_{N_e}}})^{^\intercal}$ \quad $\rhd$ $\mathbf{\bar{M}} \in \mathbb{R}^{N_m \times 1}$ is ensemble mean of $\mathbf{M}$
\State	$\mathbf{C}_{DD} = \frac{1}{N_e - 1} (\mathbf{D}-\mathbf{\bar{D}} \mathbf{\overrightarrow{1_{N_e}}})(\mathbf{D}-\mathbf{\bar{D}} \mathbf{\overrightarrow{1_{N_e}}})^{^\intercal}$ \quad $\rhd$ $\mathbf{\bar{D}} \in \mathbb{R}^{N_d \times 1}$ is ensemble mean of $\mathbf{D}$

\State $\mathbf{M} \gets \mathbf{M} + \mathbf{C}_{MD}~(\mathbf{C}_{DD} + \alpha~\mathbf{C}_D)^{-1}(\mathbf{D}_{uc}-\mathbf{D})$ \quad $\rhd$ Update ensemble

\State $ i \gets i + 1 $
\EndWhile
\State $\mathbf{M}_{post} = \mathbf{M}$ \quad $\rhd$ $\mathbf{M}_{post} \in \mathbb{R}^{N_m \times N_e}$ is the posterior ensemble of input parameters
\State $\mathbf{D}_{post} = DNN(\mathbf{M}_{post},\mathbf{t}) $ \quad $\rhd$ Generate posterior ensemble of model outputs $\mathbf{D}_{post}$
\State \textbf{return} $\mathbf{M}_{post}$, $\mathbf{D}_{post}$
\EndFunction

\end{algorithmic}
\end{algorithm}

\begin{algorithm}[H]
\renewcommand{\thealgorithm}{}
\caption{FlexIES function for EM inversion using DNN model $\rhd$ \citep{rammay2020flexible}
}
\begin{algorithmic}

\State

\Function {FlexIES}{$\mathbf{M}$, $\mathbf{EM}_{obs}$, $\mathbf{C}_D$, $DNN$, $\mathbf{t}$} 
\State \textbf{Inputs:} $\mathbf{M} \in \mathbb{R}^{N_m \times N_e}$ is the ensemble of input parameters, $\mathbf{C}_D$ is the covariance matrix of the measurement error, $\mathbf{t}$ is the given well trajectory.
\State Choose $N_a$ \quad $\rhd$ Number of data assimilations/iterations
\State $ i \gets 1$
\State $\alpha = N_a$

\While {$ i <= N_a$ }
\State $\mathbf{D}_{uc} = \mathbf{EM}_{obs}\mathbf{\overrightarrow{1_{N_e}}} + \sqrt{\alpha}\mathbf{C}_D^{1/2}\mathbf{Z}_d$ $\ \ \rhd$ Observation perturbations, $\mathbf{Z}_{d} = [\mathbf{z}_{d1}\; \mathbf{z}_{d2} \; \mathbf{z}_{d3} \; ...... \; \mathbf{z}_{dN_e}]$ $\in\mathbb{R}^{N_d \times N_e}$, $\mathbf{z}_d \sim \mathcal{N}(0, \mathbf{I}_{N_d}) \in \mathbb{R}^{N_d \times 1}$, ${\mathbf{\overrightarrow{1_{N_e}}} \in \mathbb{R}^{1 \times N_e}}$ is a row vector of ones.

\State $ \mathbf{D} = DNN(\mathbf{M,t}) $  \quad  $\rhd$ Generate ensemble of model outputs $\mathbf{D} \in \mathbb{R}^{N_d \times N_e}$ from $\mathbf{M}$

\State $ \mathbf{R}  = \mathbf{EM}_{obs} \mathbf{\overrightarrow{1_{N_e}}} - \mathbf{D}$

\If {$i = 1$}
\State ${s_p}^{(i)} = \frac{\|{\boldsymbol{\sigma}_{m}}^{(i)}\|}{\|\boldsymbol{\sigma}_{max}\|}$ \quad $\rhd$ Compute split parameter, $\|.\|$ is the Euclidean (L2) norm,  
\State $\rhd$ $\boldsymbol{\sigma}_{m} = \operatorname{mean}(\mathbf{R}) \in \mathbb{R}^{N_d \times 1}$, $\boldsymbol{\sigma}_{max} = \operatorname{max}(\operatorname{abs}(\mathbf{R})) \in \mathbb{R}^{N_d \times 1}$
\Else
\State ${s_p}^{(i)} = \frac{\|{\boldsymbol{\sigma}_{m}}^{(i)}\|}{\|{\boldsymbol{\sigma}_{m}}^{(i-1)}\|}$ 
\EndIf

\State $\mathbf{E} =  {s_p}^{(i)} \mathbf{R}$ \quad  $\rhd$ Compute ensemble of approximate model-error, $\mathbf{E}  \in \mathbb{R}^{N_d \times N_e}$

\State	$\mathbf{C}_{EE} = \frac{1}{N_e - 1} (\mathbf{E}-\mathbf{\bar{E}} \mathbf{\overrightarrow{1_{N_e}}})(\mathbf{E}-\mathbf{\bar{E}} \mathbf{\overrightarrow{1_{N_e}}})^{^\intercal}$ \quad $\rhd$ $\mathbf{\bar{E}} \in \mathbb{R}^{N_d \times 1}$ is ensemble mean of $\mathbf{E}$

\State	$\mathbf{C}_{MD} = \frac{1}{N_e - 1} (\mathbf{M}-\mathbf{\bar{M}}\mathbf{\overrightarrow{1_{N_e}}})(\mathbf{D}-\mathbf{\bar{D}} \mathbf{\overrightarrow{1_{N_e}}})^{^\intercal}$ \quad $\rhd$ $\mathbf{\bar{M}} \in \mathbb{R}^{N_m \times 1}$ is ensemble mean of $\mathbf{M}$
\State	$\mathbf{C}_{DD} = \frac{1}{N_e - 1} (\mathbf{D}-\mathbf{\bar{D}} \mathbf{\overrightarrow{1_{N_e}}})(\mathbf{D}-\mathbf{\bar{D}} \mathbf{\overrightarrow{1_{N_e}}})^{^\intercal}$ \quad $\rhd$ $\mathbf{\bar{D}} \in \mathbb{R}^{N_d \times 1}$ is ensemble mean of $\mathbf{D}$

\State $\mathbf{M}  \gets \mathbf{M} + \mathbf{C}_{MD} ~(\mathbf{C}_{DD} + \mathbf{C}_{EE} + \alpha~\mathbf{C}_D)^{-1}(\mathbf{D}_{uc}-\mathbf{D} - \mathbf{E})$  \quad $\rhd$ Update ensemble $\mathbf{M}$

\State $ i \gets i + 1 $
\EndWhile

\State $\mathbf{M}_{post} = \mathbf{M}$ \quad $\rhd$ $\mathbf{M}_{post}  \in \mathbb{R}^{N_m \times N_e}$ is the posterior ensemble of input parameters

\State $\mathbf{D}_{post} = DNN(\mathbf{M}_{post},\mathbf{t}) $ \quad $\rhd$ Generate posterior ensemble of model outputs, $\mathbf{D}_{post}$
\State \textbf{return} $\mathbf{M}_{post}$, $\mathbf{D}_{post}$
\EndFunction

\end{algorithmic}
\end{algorithm}

\section{Inversion assessment metrics}	\label{sec:AppendixB}
\setcounter{equation}{0}
\renewcommand{\theequation}{\thesection.\arabic{equation}}
\renewcommand{\thetable}{\thesection.\arabic{table}}

\subsection{Continuous Ranked Probability Score (CRPS)}

Mathematically, CRPS can be described as follows, \citep{hersbach2000decomposition}

\begin{equation}
CRPS = \int_{-\infty}^{\infty}[p(x) - H(x-x_{obs})]^2 dx,
\end{equation}
where $p(x) = \int_{-\infty}^{x} \rho(y) dy$  Cumulative distribution of quantity of interest, $H(x-x_{obs})$ = Heaviside function (Step function), i.e.,
\[
  H(x) =
  \begin{cases}
   0            & \text{if $x<0$} \\
   1            & \text{if $x\geq0$}. \\
  \end{cases}
\]

For an ensemble system with $N_e$ realizations, the CRPS can be written as follows,
\begin{equation}
CRPS = \sum_{r=0}^{N_e}{c_r},
\end{equation}

\begin{equation}
c_r = \alpha_r p_r^2 + \beta_r (1-p_r)^2,
\end{equation}
where $p_r = P(x) = r/N_e$,   for $x_r < x < x_{r+1}$  (Cumulative distribution is a piecewise constant function), and
\[
  \alpha_r =
  \begin{cases}
  0                                  & \text{if $x_{obs}<x_r$} \\
  x_{obs} - x_{r}                    & \text{if $x_r<x_{obs}<x_{r+1}$} \\
  x_{r+1} - x_{r}                    & \text{if $x_{obs}>x_{r+1}$}   \\
  x_{obs} - x_{N_e}                  & \text{if $x_{obs}>x_{N_e}$},   \\
  0                                  & \text{if $x_{obs}<x_1$}   \\
  \end{cases}
\]

\[
  \beta_r =
  \begin{cases}
  x_{r+1} - x_{r}                       & \text{if $x_{obs}<x_r$} \\
  x_{r+1} - x_{obs}                     & \text{if $x_r<x_{obs}<x_{r+1}$} \\
  0                                     & \text{if $x_{obs}>x_{r+1}$}.   \\
  0                                     & \text{if $x_{obs}>x_{N_e}$}   \\
  x_1 - x_{obs}                         & \text{if $x_{obs}<x_1$}   \\
  \end{cases}
\]

\end{appendices}

\bibliographystyle{apalike}
\bibliography{bibfile2}

\end{document}